\documentclass[11pt,a4paper]{article}
\usepackage{jstyle}
\usepackage[utf8]{inputenc}
\usepackage{enumitem}
\usepackage{youngtab,simplewick}
\usepackage{marginnote}
\usetikzlibrary{decorations.pathreplacing,angles,quotes,cd}
\usepackage{tikz-cd}
\usepackage{dsfont}
\usepackage{bm,cancel}
\usepackage{genyoungtabtikz}
\Yboxdim{8pt} 
\Ylinethick{1pt}
\usepackage{mdframed}
\usepackage{lscape}
\usepackage{pdflscape}
\usepackage{float}
\usepackage{arydshln} 
\DeclareFontFamily{U}{BOONDOX-calo}{\skewchar\font=45 }
\DeclareFontShape{U}{BOONDOX-calo}{m}{n}{
  <-> s*[1.05] BOONDOX-r-calo}{}
\DeclareFontShape{U}{BOONDOX-calo}{b}{n}{
  <-> s*[1.05] BOONDOX-b-calo}{}
\DeclareMathAlphabet{\lowercal}{U}{BOONDOX-calo}{m}{n}
\SetMathAlphabet{\lowercal}{bold}{U}{BOONDOX-calo}{b}{n}
\newcommand{\lowerbcal}[1]{\boldsymbol{\lowercal{#1}}}

\newcommand{\so}{\mathfrak{so}}
\newcommand{\su}{\mathfrak{su}}

\newcommand{\uu}{\mathfrak{u}}
\newcommand{\syp}{\mathfrak{sp}}
\newcommand{\id}{\mathds{1}}

\usepackage[colorinlistoftodos,draft]{todonotes}

\author[a]{Thomas Basile}
\author[a]{\quad Euihun Joung}
\author[b]{\quad Karapet Mkrtchyan}
\author[c]{\quad Matin Mojaza}

\affiliation[a]{Department of Physics, Kyung Hee University, Seoul 02447, Korea}
\affiliation[b]{Scuola Normale Superiore and INFN, Piazza dei Cavalieri 7, 56126 Pisa, Italy}
\affiliation[c]{Albert-Einstein-Institut, Max-Planck-Institut f\"ur Gravitationsphysik, \mbox{14476 Potsdam,} Germany}

\emailAdd{thomas.basile@khu.ac.kr}
\emailAdd{euihun.joung@khu.ac.kr}
\emailAdd{karapet.mkrtchyan@sns.it}
\emailAdd{matin.mojaza@aei.mpg.de}

\title{\centering 
Dual Pair Correspondence in Physics:\\
{\Large Oscillator Realizations and 
 Representations}
}

\abstract{We study
general aspects of
the reductive dual pair correspondence, also known as Howe duality.
We make an explicit and systematic treatment, where
we first derive the oscillator realizations
of all irreducible dual pairs:
 {\small $(GL(M,\mathbb R), GL(N,\mathbb R))$,
$(GL(M,\mathbb C), GL(N,\mathbb C))$,
$(U^*(2M), U^*(2N))$, $(U(M_+,M_-), U(N_+,N_-))$,
$(O(N_+,N_-),Sp(2M,\mathbb R))$,
$(O(N,\mathbb C), Sp(2M,\mathbb C))$ and 
$(O^*(2N), Sp(M_+,M_-))$.}
Then, we decompose
the Fock space into irreducible representations
of each group in the dual pairs
for the cases where one member  
of the pair
is compact as well as the first non-trivial cases of
where it is
non-compact.
We discuss the relevance of these representations
in several physical applications throughout this analysis.
In particular, we discuss peculiarities of their branching
properties.
Finally, 
closed-form expressions relating all Casimir operators
of two groups in a pair are established.
}

\begin{document}

\maketitle

\section{Introduction}

The reductive dual pair correspondence,
also known as Howe duality \cite{Howe1989i,Howe1989ii},
provides a useful mathematical framework
to study a physical system, allowing for
a straightforward analysis of its symmetries
and spectrum. It finds applications in a wide
range of subjects from low energy physics ---
such as in condensed matter physics, quantum
optics and quantum information theory --- to
high energy physics --- such as in twistor theory,
supergravity, conformal field theories,
scattering amplitudes and higher spin field theories.
Despite its importance, the dual pair correspondence,
as we will simply refer to it, is not among the most
familiar mathematical concepts in the theoretical
physics community, and particularly not in those of
field and string theory, where it nevertheless
appears quite often, either implicitly, or as an
outcome of the analyses. For this reason, many times
when the dual pair correspondence makes its occurrence
in the physics literature, its role often goes unnoticed
or is not being emphasized, albeit it is actually acting
as an underlying governing principle.

In a nutshell, the dual pair correspondence is
about the oscillator realization of Lie algebras.
In quantum physics, oscillators realizations 
---
that is, the 
description of physical systems
in terms of creation and annihilation operators, $a$ and $a^\dagger$ 
--- 
are ubiquitous, and their use in algebraically
solving the quantum harmonic oscillator
is a central and standard part of any undergraduate
physics curriculum. Oscillator realizations 
are more generally encountered in many 
contemporary physical problems.
It can therefore be valuable to study their 
properties in a more general and broader
mathematical framework. The dual pair correspondence
provides exactly such a framework, 
applicable to many unrelated physical problems.

The first oscillator realization 
of a Lie algebra 
was given by Jordan in 1935~\cite{Jordan1935}
to describe the relation between the symmetric and the linear groups.
This
so-called Jordan--Schwinger map
was used much later by Schwinger to 
realize the $\mathfrak{su}(2)$ obeying angular momentum operators\footnote{In this section, we mention several important works which went before the dual pair correspondence.
Our intention, however, is not to select the most important contributions, but rather to illustrate the historical development of oscillator realizations and the dual pair correspondence in physics.
}
in terms of two pairs of 
ladder operators, $a, a^\dagger$ and $b, b^\dagger$, 
as follows
\be
    J_3=\frac12(a^\dagger a-b^\dagger b)\,,
    \qquad J_+=a^\dagger\,b\,,
    \qquad J_-=b^\dagger\,a\,,
\ee 
thereby finding a gateway to construct all irreducible representation of $\su(2)$
in terms of two 
uncoupled quantum harmonic oscillators.
The spin-$j$ representations then simply become the states with 
excitation number $2j$\,:
\be
    N=a^\dagger\,a+b^\dagger\,b\,,
    \qquad 
    N\,|\Psi_j\ra=2j\,|\Psi_j\ra\,.
\ee  
When $\mathfrak{su}(2)$ is 
realized as above, 
there emerges another Lie algebra, $\uu(1)$ --- generated by the number operator $N$ 
--- which is not a subalgebra of the $\su(2)$.
Uplifting this to Lie groups, we find that
the spin-$j$ representations of $SU(2)$
are in one-to-one correspondence
with the $U(1)$ representations;
i.e. those labelled by the oscillator number $2j$. 
This interplay between the two groups $SU(2)$ and $U(1)$ is 
the first example of the dual pair correspondence.
A generalization of this mechanism to the duality 
between $U(4)$ and $U(N)$ 
was made 
by Wigner \cite{Wigner:1936dx}
to model a system in nuclear physics.
Replacing the bosonic oscillators by fermionic ones,
Racah extended the duality to the one between $Sp(M)$
and $Sp(N)$ in \cite{Racah1943}.
Subsequently, the method was widely used in various models of nuclear physics,
such as the nuclear shell model and the interacting boson model (see e.g. \cite{French1960,Arima1987}).

Concerning non-compact Lie groups,
an oscillator-like
representation,
named \emph{expansor}, 
was introduced for the Lorentz group by Dirac in 1944 \cite{Dirac:1945cm},
and its fermionic counterpart, the \emph{expinor}, was
 constructed by Harish-Chandra \cite{Harish-Chandra1947}
(see also \cite{Dirac:1949cp}
and an historical note \cite{Mukunda1993}).\footnote{These representations are closely related to 
Majorana's equation for 
infinite component spinor carrying a unitary representation of the Lorentz group, published in 1932 \cite{Majorana:1968zz}.
The complete classification of unitary and irreducible representations of the four-dimensional Lorentz group was later obtained in \cite{Harish-Chandra1947, Bargmann:1946me, Gelfand1947}.} 
In 1963, Dirac
also introduced an oscillator representation
for the 3d conformal group,
which he referred to as a ``remarkable representation'' \cite{Dirac:1963ta}.
The analogous representation for the 4d conformal group 
was constructed in \cite{kursunoglu1962modern}  (see also \cite{Mack:1969dg,Ruehl:1973nj}).
The oscillator realization
and the dual pair correspondence
are also closely related to twistor theory
\cite{Penrose:1967wn, Penrose:1968me, Penrose:1972ia},
where the dual group of the 4d conformal group 
was interpreted as an internal symmetry group in 
 the `naive twistor particle theory' 
(see e.g. the first chapter of \cite{Bailey1990}).

Around the same time, 
oscillator representations 
were studied in mathematics by Segal \cite{Segal1959},
Shale \cite{Shale1962}
and by Weil \cite{Weil1964} (hence, often referred to as Weil representations).
Based on these earlier works 
(and others that we omit to mention),
Howe eventually came up with
the reductive dual pair correspondence 
in 1976 (published much later in \cite{Howe1989i,Howe1989ii}).
In short, he showed that there exists a one-to-one correspondence
between oscillator representations of two mutually commuting subgroups of the symplectic group.
Since then, many mathematicians contributed to
the development of the subject 
(see e.g. reviews \cite{Prasad1993, Kudla1996, Adams2007} 
and references therein).
The dual pair correspondence is 
also referred to as the \emph{(local)} theta correspondence
and it has an intimate connection
to the theory of automorphic forms (see, e.g., \cite{Pioline:2003bk,Fleig:2015vky}).

In physics,
oscillator realizations were studied
also in
the context of coherent states \cite{Barut1971, Perelomov1972, Gilmore1972}.
In the 80-90's, 
G\"unaydin and collaborators extensively used  oscillators  to realize various super Lie algebras 
arising in 
supergravity theories; i.e. those of $SU(2,2|N)$ \cite{Gunaydin:1984fk, Gunaydin:1998sw}, 
$OSp(N|4,\mathbb R)$ \cite{Gunaydin:1985tc, Gunaydin:1990ag} and 
$OSp(8^*|N)$ \cite{Gunaydin:1984wc, Gunaydin:1999ci} 
(see also \cite{Gunaydin:1981yq, Gunaydin:1981dc, Bars:1982ep}). 
Part of the analysis of representations contained in this paper can be found already in these early references.
However, in those papers, the (role of the) dual group was not 
(explicitly) considered,\footnote{Note that 
the isometry and R-symmetry groups follow a similar pairing pattern,
but they are not reductive dual pairs.}
even though it implicitly appears in the tensor product decompositions.
(Its relevance was, however, alluded to in~\cite{Bars:1982ep}.)

Higher spin field theory is another 
area where the oscillator realizations
and the dual pair correspondence
were fruitfully employed.
Spinor oscillators were used to construct
4d higher spin algebra by Fradkin and Vasiliev \cite{Fradkin:1987ah}
and to identify the underlying representations by Konstein and Vasiliev \cite{Konstein:1989ij}.
Fradkin and Linetsky found that 
the extension of the work to 4d conformal higher spin theory, or equivalently 5d higher spin theory, requires a dual $\uu(1)$ algebra \cite{Fradkin:1989yd, Fradkin:1989ywa}.
The same mechanism was used by Sezgin and Sundell
in extending Vasiliev's 4d theory \cite{Vasiliev:1990en, Vasiliev:1990vu, Vasiliev:1992av}
to 5d \cite{Sezgin:2001zs} and 7d \cite{Sezgin:2001ij},
with the dual algebras $\mathfrak{u}(1)$
and $\mathfrak{su}(2)$ (as we shall later see, the latter algebra is more appropriately interpreted as $\mathfrak{sp}(1)$).
In 2003, Vasiliev generalized his theory to any dimensions using vector oscillators \cite{Vasiliev:2003ev},
and revisited its representation theory in \cite{Vasiliev:2004cm}. In both of these works, the dual pair correspondence played a crucial role, and since then, it has been used 
several times within the context of higher spin theories.\footnote{In fact, almost all of the higher spin literature is related to
the dual pair correspondence in one way or another. 
For this reason, we mention just a few papers
in which the duality is directly and explicitly used 
\cite{Vasiliev:2004cm,Alkalaev:2008gi, Boulanger:2008kw, Alkalaev:2009vm, Bekaert:2009fg, Alkalaev:2011zv, Bekaert:2013zya, Joung:2014qya, Alkalaev:2014nsa, Alkalaev:2014qpa,
Chekmenev:2015kzf,
Joung:2015jza, Alkalaev:2017hvj, Grigoriev:2018wrx,Vasiliev:2012tv,Vasiliev:2018zer,Alkalaev:2019xuv}.}

Despite the abundance of relevant works in physics, 
and reviews in mathematics, 
it is hard to find accessible references on the dual pair correspondence 
for physicists,
with a notable exception of
\cite{Rowe:2012ym}.\footnote{
The historical accounts 
given here
are indebted
to this review.}
In physics heuristic approaches rather than systematic ones are common, whereas in the mathematics literature, the treatment is formal and explicit examples are rare.
The current work 
is an attempt
to close the gap between the 
physics and math literature. 
Specifically,
we provide a systematic derivation of the
oscillator realizations for essentially all dual pairs,
as well as, in the relatively simple cases, 
an explicit decomposition of
the corresponding Fock spaces 
into irreducible representations of each group of the pairs.
To be self-contained, we included many
basic elements of representation theory
related to oscillators.
Thus, there will be
a considerable overlap with 
many earlier works mentioned previously.
Nevertheless, we believe that our treatment  
could help in filling
several gaps in the current understanding of 
the use and role of the dual pair correspondence in physics.
As the subject of the dual pair correspondence is rich, 
we will not 
cover all 
that we intend to in a single paper.
We instead complete our program 
in (at least) two follow-up papers;
one of the same kind as this,
but on more advanced issues,
and another dedicated to physical applications.
Although physical applications will be visited in depth
in a sequel,
we include brief comments on them 
throughout the current paper, as summarized below.

\subsection*{Brief summary of the paper}

In Section \ref{sec:generalities}, we review
generalities of
 the dual pair correspondence,
starting with the precise statement of the duality. 
In Section
\ref{sec:metaplectic}, we recall the definition
of the metaplectic representation.
In Section \ref{sec:irred_pairs}, we outline the
classification of irreducible reductive dual pairs.
In Section
\ref{sec:seesaw_pairs}, we
introduce the concept of
seesaw pairs, which will be crucial in the
rest of the paper.

In Section \ref{sec:osc_model_GL},
we derive the oscillator realizations
of all real forms of the 
complex Lie group duality
$(GL_M, GL_N)$,
that is,
$(GL(M,\mathbb R), GL(N,\mathbb R))$,
$(GL(M,\mathbb C), GL(N,\mathbb C))$,
$(U^*(2M), U^*(2N))$
and $(U(M_+,M_-), U(N_+,N_-))$ dualities,
starting from the definitions
of the real forms.

In Section \ref{sec:osc_model_O-Sp},
we derive the oscillator realizations
of all real forms of the 
complex Lie group duality
$(O_N, Sp_{2M})$,
that is,
$(O(N_+,N_-),Sp(2M,\mathbb R))$,
$(O(N,\mathbb C), Sp(2M,\mathbb C))$ and 
$(O^*(2N), Sp(M_+,M_-))$ dualities,
starting from the definitions
of the real forms.

In Section \ref{sec:rep_compact}, 
we derive
the correspondence for 
``compact dual pairs'', i.e.
the dual pairs in which
at least one member is compact. 
In doing so, we will recover the
familiar oscillator realizations of
finite-dimensional representations of compact Lie groups, as well as the lowest weight
representations of non-compact Lie groups.
We also briefly comment on the
application of this correspondence
to AdS$_{d+1}$/CFT$_d$ for $d=3, 4$
and $6$ (which will be discussed in
more details in a follow-up paper).

In Section \ref{sec:rep_excep_compact},
we derive the correspondence between
representations of what will be referred to as
``exceptionally compact pairs'', i.e.
dual pairs
in which one member becomes either compact or discrete
due to an exceptional isomorphism.
We briefly comment on its role
in dS$_d$ representations for $d=3$ and 4.

In Section \ref{sec:rep_non-compact},
we derive the correspondence between
representations of dual pairs where
both groups are non-compact, but one is
Abelian or simple enough. 
These cases are 
different from the compact ones
in that the representations are 
of non-polynomial excitations type.
We also present the Schr\"odinger
realizations for these
representations (i.e. their realizations on
$\cL^{2+\e}$ spaces).

In Section \ref{sec:singletons}, we discuss  
interesting aspects of the branching rules
of the representations appearing
in the dual pair correspondence.
We comment on the special
cases where one of the groups is the three-
or four-dimensional conformal group.
In such cases, these representations are
known as ``singletons'' and correspond
to free conformal fields.

In Section \ref{sec:casimir}, we
derive the relation between
the Casimir operators (of arbitrary order)
of the two groups of a dual  pair.

We leave concluding remarks for the sequel papers.

In order to be as self-contained as possible,
we included three
appendices detailing textbook
material, used in the bulk of the paper,
as well as one detailing the derivations of the
Casimir relations:
In Appendix \ref{app:irrep_compact}, we review
the realization of finite-dimensional representations
of the compact form of the classical groups on
spaces of tensors.
In Appendix \ref{sec: real form}, we summarize
the definition of the various real forms of the
classical groups.
In Appendix \ref{app:seesaw_diagrams},
we provide the seesaw pair diagrams
involving maximal compact subgroups and their dual
for all irreducible dual pairs. 
Finally, Appendix \ref{app:CasimiRR}
contains the technical details
on the derivations of the Casimir relations.

\subsection*{Conventions}
\label{sec:convention}

\paragraph{Classical Lie groups and their real forms.}
We denote the complex classical Lie groups by $GL_N$, $O_N$ and $Sp_{2N}$,
i.e. the general linear, the orthogonal and the symplectic group,
respectively.
When these Lie groups
are viewed as real Lie groups, they are denoted by
$GL(N,\mathbb C)$, $O(N,\mathbb C)$
and $Sp(2N,\mathbb C)$.
The real forms of $GL_N$ and $O_N$ will be denoted by
$GL(N,\mathbb R)$, $U^*(2N)$, $U(N_+,N_-)$,
$O(N_+,N_-)$, $O(N,\mathbb C)$ and $O^*(2N)$
with standard definitions and notations. 
There is no standard notation for the symplectic groups;
we use the notation $Sp(2N,\mathbb R)$, $Sp(2N,\mathbb C)$ and $Sp(N_+,N_-)$ for the groups with rank $N, 2N$ and $N_++N_-$, respectively.
See Appendix \ref{sec: real form} for the details.
The multiplicative group of non-zero 
real, complex and quaternionic numbers
are denoted by
$\mathbb R^\times$, $\mathbb C^\times$ and
$\mathbb H^\times$,
whereas the additive group of real numbers,
isomorphic 
to the multiplicative group of 
strictly positive real numbers,
is denoted by $\mathbb R^+$.
For brevity,
the phrase
``generators of the Lie group'' (also in symbolic form) will
be used when actually referring to
generators of the associated Lie algebra.

\paragraph{Young diagrams.}
A Young diagram of the form,
\begin{equation}
    \begin{tikzpicture}[scale=0.7]
    \draw[thick] (-3,3) -- (6,3) -- (6,2.5) -- (-3,2.5) -- (-3,3);
    \node at (0.5,2.75) {\footnotesize$\ell_1$};
    \draw[thick] (4,2.5) -- (4,2) -- (-3,2) -- (-3,2.5);
    \node at (0.5,2.25) {\footnotesize$\ell_2$};
    \draw[thick] (3,2) -- (3,1.5) -- (-3,1.5) -- (-3,2);
    \node at (0.5,1.75) {\footnotesize$\ell_3$};
    \draw[thick] (2,1.5) -- (2,1.25);
    \draw[thick, dashed] (2,1.25) -- (0,0.25);
    \draw[thick] (0,0.25) -- (0,0);
    \draw[thick] (0,0) -- (-1,0) -- (-3,0) -- (-3,2);
    \draw[thick] (-3,0) -- (-3,-0.5) -- (-1,-0.5) -- (-1,0);
    \node at (-2,-0.25) {\footnotesize$\ell_p$};
    \end{tikzpicture}
    \label{eq:example_Young}
\end{equation}
will be denoted by
\begin{equation}
    \bm\ell = (\ell_1, \dots, \ell_p)\,,
\end{equation}
i.e. $\ell_k \in \mathbb N$
corresponds to the length
of the $k$-th row of the diagram and $p$ is the
number of rows. It will sometimes be useful
to present a Young diagram in terms of its
columns instead of its rows. In this case,
we use the same bold symbol $\bm\ell$, but write
its components within square brackets and
with upper indices, i.e.
\begin{equation}
    \bm\ell = [\ell^1, \dots, \ell^q]\,,
\end{equation}
where $\ell^k\in\mathbb N$
corresponds to the height of the $k$-th
column, and $q$ is the number of columns.
Clearly, the two descriptions are equivalent.
In particular, $q=\ell_1$ and
$p=\ell^1$. 
For instance,
the diagram,
\begin{equation}
    \bm\ell = \gyoung(;;;;;,;;;,;;,;;,;,;)\,,
\end{equation}
can be presented as
\begin{equation}
    \bm\ell = (5,3,2,2,1,1) 
    \quad \text{and} \quad
    \bm\ell =[6,4,2,1,1]\,.
\end{equation}

\paragraph{Tensors and indices.}
We will use round and square brackets
for symmetrization and antisymmetrization of indices, respectively. For instance,
\begin{equation}
    T^{(a_1a_2)} = \tfrac12\, (T^{a_1a_2}
    + T^{a_2a_1})\,,
    \qquad 
     T^{[a_1a_2]} = \tfrac12\, (T^{a_1a_2}
    - T^{a_2a_1})\,,
\end{equation}
and more generally
\begin{equation}
    T^{(a_1 \dots a_n)} = \frac1{n!}\,
    \sum_{\sigma\in\cS_n}
    T^{a_{\sigma(1)} \dots a_{\sigma(n)}}\,,
    \qquad
     T^{[a_1 \dots a_n]} = \frac1{n!}\,
    \sum_{\sigma\in\cS_n} {\rm sgn}(\s)\, 
    T^{a_{\sigma(1)} \dots a_{\sigma(n)}}\,,
\end{equation}
where ${\rm sgn}(\s)$ is the signature of the permutation $\s$,
which is equal to $+1$ if $\s$ is an even permutation and $-1$ if it is odd.

Given a non-degenerate antisymmetric matrix $\Omega$
with components satisfying
\begin{equation}
    \Omega_{AB} = - \Omega_{BA}\,, \qquad
    \Omega_{AC}\, \Omega^{CB} = \delta^B_A\,,
\end{equation}
the indices of 
$Sp_{2N}$ tensors 
will be  raised and lowered as follows
\begin{equation}
    T^A = \Omega^{AB}\, T_B\,,
    \qquad
    T_A = \Omega_{AB}\, T^B\,.
    \label{eq:convention_sp_indices}
\end{equation}
In particular, this implies that
\be 
    \O_{AB}\,V^A\,W^B=V^A\, W_A = -V_A\, W^A\,.
    \label{symp contraction}
\ee

\paragraph{Oscillators.}
In this paper, 
annihilation operators  will be 
denoted by $a_A$ or $a_A^I$ with one or two indices.
Their Hermitian conjugate, i.e. the creation operators,
will be denoted with a tilde and opposite index position, i.e.
\be 
    (a_A^I)^\dagger=\tilde a^A_I\,.
\ee 
 We will refer to the creation and annihilation operators as \emph{oscillators}.
The letters $b, c$ and $d$ will also be used to denote (additional) oscillators.

\paragraph{Representations.}

A generic irreducible representation of a Lie group $G$ will be denoted by
\begin{equation}
    \pi_G(\zeta_1, \dots, \zeta_p)\,,
\end{equation}
where $\zeta_1, \dots, \zeta_p$ are the
(continuous or discrete)
parameters which label the representation.\footnote{In general,
a representation of $G$ can be induced from 
a representation of one of its parabolic subgroups $P$.
For this reason, a generic irreducible representation is usually denoted by 
\be 
    {\rm Ind}^G_P(\zeta_1,\ldots, \zeta_p)\,.
\ee 
The parameters $\zeta_i$
indicate a representation of $P$ which
induces the representation of $G$. 
The couple $P$ and $(\zeta_1, \dots, \zeta_p)$ are called the Langlands parameters.}
For two particular types of representations, the following specific notation will be used:
\begin{itemize}
\item Finite-dimensional irreducible representations of
the (double-cover of the) compact Lie groups
$G = U(N)$, $O(N)$ or $Sp(N)$,  will be
denoted by $[\bm\ell, \delta]_G$, where $\bm\ell$
is a Young diagram, and
$\delta\in\tfrac12 \mathbb Z$
is a half-integer such that
\begin{equation}
    (\ell_1+\delta, \dots, \ell_N+\delta)\,,
\end{equation}
is the highest weight of the representation.
Here, $\ell_{p+1}=\cdots =\ell_N=0$ for
the Young diagram $\bm\ell$ with height $p$.
The correspondence between these highest weight representations
and tensors with symmetry of the Young diagram
$\bm\ell$ is reviewed in
Appendix \ref{app:irrep_compact}.

\item Infinite-dimensional irreducible representations of a non-compact
Lie group $G$ of lowest weight type will be denoted by 
\begin{equation}
    \cD_G\big(\ell_1, \dots, \ell_{{\rm rank}(G)})\,,
\end{equation}
where  $\ell_k$
are the components of the lowest weight defining the
representation. If the group $G$ has the same
rank as its maximal compact subgroup $K$, we will combine
this notation with the previous one as
\begin{equation}
    \cD_G\big([\bm\ell,\delta]_K\big)\,.
\end{equation}

\end{itemize}

\section{Generalities of the
reductive dual pair correspondence}
\label{sec:generalities}
A reductive dual pair $(G, G')\subset Sp(2N,\mathbb R)$
consists of two
subgroups, $G$ and $G'$, of $Sp(2N,\mathbb R)$ 
which are centralizers of each other
and act reductively on $\mathbb{R}^{2N}$,
meaning that they consist of those
elements in $Sp(2N,\mathbb R)$, which
reduce completely $\mathbb R^{2N}$ into 
irreducible parts invariant under the subgroup elements.
In other words, when realized as $2N\times 2N$ 
real matrices, 
reductive subgroups are those having block-diagonal elements,
with each block being an irreducible representation of the subgroup.
(Recall that $Sp(2N,\mathbb R)$ is the
group of linear transformations of $\mathbb{R}^{2N}$
preserving the symplectic form $\Omega_{AB}\,x^A\,y^B$,
and hence the space $\mathbb{R}^{2N}$
carries the defining representation of $Sp(2N,\mathbb R)$.)

A central result due to Howe 
\cite{Howe1989i, Howe1989ii} is that the restriction of the metaplectic 
representation $\cW$ of $Sp(2N,\mathbb R)$,
to be discussed below,
to $G \times G'$ establishes 
a bijection between representations of $G$ and those of $G'$ appearing 
in the decomposition of $\cW$. More precisely, 
$\cW$ 
can be 
decomposed as
\begin{equation}
    \cW\,|_{G\times G'}
    = \bigoplus_{\zeta\in\Sigma^G_\cW}
    \pi_G(\zeta ) \otimes
    \pi_{G'}\big(\theta(\zeta)\big)\,,
\end{equation}
where 
$\pi_G(\zeta)$ and $\pi_{G'}\big(\theta(\zeta)\big)$
are irreducible representations (irreps) of $G$ and $G'$
labeled by $\zeta$ and $\theta(\zeta)$\,,
respectively.
The  map,
\ba 
    \theta\ :\quad &\Sigma^G_\cW \quad &\longrightarrow
    \quad \Sigma^{G'}_\cW\,,\nn 
    &\zeta\quad  &\longmapsto
    \quad \theta(\zeta)\,,
\ea  
defines a bijection  between a set $\Sigma^G_\cW$ of 
 $G$ irreps and 
the corresponding set $\Sigma^{G'}_\cW$ of $G'$ irreps.
In other words, each $G$ representation appears only once in $\Sigma^G_\cW$ and is paired with a unique $G'$ 
representation in $\Sigma^{G'}_\cW$,
and vice-versa.
As a consequence, the representation $\pi_G(\zeta)$ occurs with 
multiplicity ${\rm dim}\,\pi_{G'}\big(\theta(\zeta)\big)$,
and similarly for $\pi_{G'}\big(\theta(\zeta)\big)$:
\begin{equation}
    {\rm mult}_\cW(\pi_G(\zeta)\big) = \dim\, \pi_{G'}\big(\theta(\zeta)\big)\,, 
    \qquad 
    {\rm mult}_\cW\Big(\pi_{G'}\big(\theta(\zeta)\big)\Big) = \dim\,\pi_G(\zeta)\,.
\end{equation}
More details can be found, e.g. in the review articles
\cite{Adams2007, Rowe:2012ym}, or in the textbooks 
\cite{Goodman2009, Cheng2012} (where Howe duality is 
presented within the broader context of duality in representation theory).

\subsection{Metaplectic representation}
\label{sec:metaplectic}

The metaplectic representation $\cW$ of $Sp(2N,\mathbb R)$ is known under different names, such as the harmonic 
representation, the Segal--Shale--Weil representation or simply 
the oscillator representation \cite{Segal1959,Shale1962,Weil1964}.
This representation can be realized
as a Fock space, which is
generated by the free action of 
creation operators 
$a_i^\dagger$ with $i=1,\dots,N$ on a vacuum state $\lvert 0 \rangle$,
which, by definition, is annihilated by the annihilation operators $a_i$\,:
\begin{equation}
    \cW = {\rm span}_{\mathbb C}\big\{ a_{i_1}^\dagger 
    \dots a_{i_k}^\dagger\, \lvert 0 \rangle, 
    \ k \in \mathbb N \big\}\,, \qquad [a_i, a_j^\dagger] 
    = \delta_{ij}\,, \qquad a_i \lvert 0 \rangle = 0\,.\label{w}
\end{equation}
This infinite-dimensional space carries an irreducible and unitary 
representation of the order $N$ Heisenberg algebra, whose generators 
are represented by the $N$ pairs of creation and annihilation operators, 
as well as the identity (representing its center). 
Moreover, the operators 
bilinear in $a_i$ and $a_i^\dagger$ provide a representation of the Lie 
algebra $\mathfrak{sp}(2N, \mathbb R)$ of the symplectic group $Sp(2N, \mathbb R)$. More precisely, the operators,
\begin{equation}
    K^{ij} = a_i^\dagger\, a_j^\dagger\,, \qquad 
    K^i{}_j = \tfrac12\, \{a_i^\dagger, a_j\} 
    = a_i^\dagger\, a_j + \tfrac12\, \delta_{ij}\,, \qquad 
    K_{ij} = - a_i\, a_j\,,
    \label{eq:sp_in_w}
\end{equation}
satisfy the commutation relations
of $\mathfrak{sp}(2N, \mathbb R)$, i.e.
\begin{equation}
    [K^i{}_j, K^k{}_l] = \delta^k_j\, K^i{}_l
    - \delta^i_l\, K^k{}_j\,, \qquad
    [K^{ij}, K_{kl}] =  4\, \delta^{(i}_{(k}\, K^{j)}{}_{l)}\,,
\end{equation}
\begin{equation}
    [K^i{}_j, K_{kl}] = -2\, \delta^i_{(k}\, K_{l)j}\,, 
    \qquad [K^i{}_j, K^{kl}] = 2\, \delta^{(k}_j\, K^{l)i}\,.
\end{equation}

In this basis, the 
operators $K^i{}_i$ (where no summation is implied) 
generate a Cartan subalgebra while the operators $K_{ij}$ 
and $K^k{}_l$ with $k<l$ correspond to lowering operators, 
and the remaining ones to raising operators. 
The metaplectic 
representation is a direct sum of two $\mathfrak{sp}(2N, \mathbb R)$ 
lowest weight representations,
\begin{equation}
    \cW=
    \cD_{Sp(2N,\mathbb R)}(\tfrac12, \tfrac12, \dots, \tfrac12)
    \oplus 
    \cD_{Sp(2N,\mathbb R)}(\tfrac32, \tfrac12, \dots, \tfrac12)\,.
    \label{eq:highest_weight_metaplectic}
\end{equation}
The vacuum $\lvert 0 \rangle$
and $a_1^\dagger\, \lvert 0 \rangle$ 
are the lowest weight vectors
of the above two representations
with weight $(\tfrac12,\tfrac12, \dots, \tfrac12)$
and $(\tfrac32, \tfrac12, 
\dots, \tfrac12)$, respectively.\footnote{The seemingly privileged role of $a_1^\dagger$
comes from the choice of lowering operators,
$K^k{}_{l}$ with $k<l$\,. }
Repeated action of raising operators on these vectors
makes up the representations
$\cD_{Sp(2N,\mathbb R)}(\tfrac12, \tfrac12, \dots, \tfrac12)$ and $\cD_{Sp(2N,\mathbb R)}(\tfrac32, \tfrac12, \dots, \tfrac12)$,
which are therefore composed of the states
with even and odd excitation numbers, respectively.

The metaplectic representation can also be decomposed
in terms of the maximal compact subalgebra $\uu(N)
\cong \uu(1)\oplus \su(N)$,
where 
the $\uu(1)$ and $\su(N)$ subalgebras are generated respectively by
$K^i{}_i=\sum_{k=1}^N\, a_k^\dagger\, a_k+\tfrac N2$
and $K^i{}_j-\tfrac1N\,\delta^i_j\,K^k{}_k
    = a^\dagger_j\,a_i-\tfrac1N\,\delta^i_j\,
    \sum_{k=1}^N\, a_k^\dagger\, a_k\,.$
The vacuum $|0\ra$ 
carries the irrep $[(0),\tfrac12]_{U(N)}$ 
(i.e. it has the $\uu(1)$ eigenvalue $\frac N2$ 
and carries the trivial representations of  $\su(N)$),
whereas the $N$ dimensional space generated by 
$a^\dagger_i|0\ra$ 
carries the irrep $[(1),\tfrac12]_{U(N)}$ 
(i.e. it has the  $\uu(1)$ eigenvalue $\frac{N}2+1$
and carries  the fundamental representation
of $\su(N)$).
These $\uu(N)$ representations are annihilated by
the lowering operators $K_{ij}$
and they induce the $\syp(2N,\mathbb R)$ 
irreps
$\cD_{Sp(2N,\mathbb R)}([(0),\tfrac12]_{U(N)})
=\cD_{Sp(2N,\mathbb R)}(\tfrac12, \tfrac12, \dots, \tfrac12)$,
and 
$\cD_{Sp(2N,\mathbb R)}([(1),\tfrac12]_{U(N)})=\cD_{Sp(2N,\mathbb R)}(\tfrac32, \tfrac12, \dots, \tfrac12)$
by the action of the raising operators $K^{ij}$.

At this point, we have seen that $\cW$ carries both a representation of 
the order $N$ Heisenberg algebra and of the symplectic algebra 
$\mathfrak{sp}(2N, \mathbb R)$. A natural question is then to ask 
whether this can be uplifted to a group representation in both cases. 
For the Heisenberg group, the answer is positive: a direct computation 
shows that the operators 
$U(\alpha,\beta,\gamma) := e^{\alpha^i\, a_i 
+ \beta^i\, a_i^\dagger + \gamma}$ satisfy
\begin{equation}
    U(\alpha_1,\beta_1,\gamma_1)\, U(\alpha_2,\beta_2,\gamma_2)
     = U(\alpha_1+\alpha_2,\beta_1+\beta_2,\gamma_1+\gamma_2
     +\tfrac12\,(\alpha_1 \cdot \beta_2 - \beta_1 \cdot \alpha_2))
\end{equation}
where $\alpha \cdot \beta = \sum_{i=1}^N \alpha^i\, \beta^i$. 
The above reproduces the group multiplication of the Heisenberg 
group $H_N$ and hence provides its representation on $\cW$.

The situation is a bit more subtle for the symplectic group. 
It turns out that
the exponentiation of the $\mathfrak{sp}(2N,\mathbb R)$ representation \eqref{eq:sp_in_w} is double-valued.
To see whether a representation is single-valued or not, we need to examine the representation 
along a loop in the group manifold.
Since the loops
contractible to a point will always give single-valued
representations, we must
consider only
the other kind of loops, which are classified by the  fundamental group of the group manifold.
The fundamental group of $Sp(2N,\mathbb R)$
is that of its maximal compact subgroup
$U(N)\cong U(1)\ltimes SU(N)$,
where the $U(1)$ can be chosen
to have an element 
${\rm diag}(e^{i\,\phi}, \underbrace{1,\ldots,1}_{N-1})$.\footnote{As a topological space, $U(N)$ is 
homeomorphic to the product space $U(1)\times SU(N)$,
where the $U(1)$ subgroup cannot be taken
as the diagonal one, $U(1)_{\rm diag}=\{e^{i\,\phi}\,I_N\,|\,0\le \phi<2\pi\}$\,.
In fact, as a group, 
$U(N)\cong (U(1)_{\rm diag}\times SU(N))/{\mathbb Z_N}$ where $\mathbb Z_N=\{e^{i\,2\pi\,\frac{n}{N}}\,I_N\,|\, 
n=1,\ldots, N-1\}$\,.}
This 
element of $U(1)$ is embedded in $Sp(2N,\mathbb R)$ as 
\be 
    g(\phi)=
    \begin{pmatrix}
        \cos\phi & 0 &\sin\phi &0 \\
        0& I_{N-1} & 0& I_{N-1} \\
        -\sin\phi & 0 & \cos\phi & 0 \\
        0 & I_{N-1} & 0& I_{N-1}
    \end{pmatrix}\,.
\ee 
Since $SU(N)$ has a trivial first fundamental group,
we can just consider this $U(1)$ subgroup, generated by
\be 
    K^1{}_1=a^\dagger_1\,a_1+\frac12\,,
\ee 
and exponentiating $K^1{}_1$, one finds 
the representation,
\begin{equation}
    U_\cW(\phi)=e^{i\, \phi\, K^1{}_1}=e^{i\,\phi\,
    (a^\dagger_1\,a_1+\frac 12)}\,.
    \label{U(1) rep}
\end{equation}
Since $a^\dagger_1\,a_1$ has  integer eigenvalues, the $U(1)$ representation \eqref{U(1) rep} 
has the property,
\be 
    U_\cW(\phi+2\pi)=-U_\cW(\phi)\,,
\ee 
that is, the representation map
$U_\cW$ is double-valued.
One can replace the $U(1)$ subgroup by its double cover to render the representation map $U_\cW$ single-valued.  
This would require to replace the symplectic group $Sp(2N,\mathbb R)$ by its double cover, namely  the metaplectic
group, often denoted by $\widetilde{Sp}(2N, \mathbb R)$.
In principle, of course, one can also consider an arbitrary even cover of $Sp(2N,\mathbb R)$.

As we discussed just above, the essential point resides in the $U(1)$ part of $Sp(2N,\mathbb R)$,
but it will still be instructive
to consider the action of the entire $Sp(2N,\mathbb R)$ group.
For simplicity, let us focus on 
the $N=1$ case, $Sp(2,\mathbb R)$,
whose element can be parameterized as 
\be 
g(\phi,\psi,\tau)=
\begin{pmatrix}
    \cos\phi\,\cosh\t-\sin\psi\,\sinh\t & 
    \ \sin\phi\,\cosh\t+\cos\psi\,\sinh\t \\
    -\sin\phi\,\cosh\t+\cos\psi\,\sinh\t & 
    \ \cos\phi\,\cosh\tau+\sin\psi\,\sinh\t 
    \end{pmatrix},
    \label{g elem}
\ee 
with $g(\phi+2\pi,\psi,\tau)=g(\phi,\psi,\tau)$
and $g(\phi,\psi+\pi,\tau)=g(\phi,\psi,-\tau)$
and $\tau \in \mathbb R$\,.
The oscillator representation for the
group element $g(\phi,\psi,\tau)$ can be
obtained again by exponentiation and 
expressed as
\be 
    U_\cW(\phi,\psi,\tau)=e^{i\,\phi\,(a^\dagger\,a
    +\frac12)}\,e^{\beta\, a^2 
    - \beta^*\, (a^\dagger)^2}\,, 
    \label{U_w}
\ee 
where $\beta= \frac12\,e^{i\,(\psi-\phi)}\, \tau$.
Again, one can check the double-valued-ness of the map $U_\cW$\,:
\be 
    U_\cW(\phi+2\pi,\psi,\t)=-U_\cW(\phi,\psi,\t)\,.
    \label{double value}
\ee 
We may avoid this problem by viewing $U_\cW$ 
as a representation of the double cover 
$\widetilde{Sp}(2, \mathbb R)$\,;
it is parametrized by the same variables as 
$Sp(2, \mathbb R)$, 
but the period of $\phi$ is extended to $4\pi$.  
Since $U_\cW$ is faithful for the double cover (but
not for other even covers) of $Sp(2,\mathbb R)$, 
it can serve as a definition of $\widetilde{Sp}(2,\mathbb R)$,
which does not admit any matrix realization.
Instead of considering the double cover, one may also try to cure the problem \eqref{double value} by 
modifying the representation to
\be 
    V_\cW(\phi,\psi,\tau)=e^{i\,\phi\,a^\dagger\,a}\,e^{\beta\, a^2 
    - \beta^*\, (a^\dagger)^2}
    =U_\cW(\phi,\psi,\tau)\,e^{-\frac i2\,\phi}\,, 
\ee 
which is single-valued.
But, then the above 
has a modified composition rule,
\be 
    V_\cW(\phi_1,\psi_1,\tau_1)\,
    V_\cW(\phi_2,\psi_2,\tau_2)
    =e^{\frac i2\,(\phi_3-\phi_1-\phi_2)}\,
    V_\cW(\phi_3,\psi_3,\tau_3)\,,
\ee
where $\phi_3, \psi_3,\tau_3$
satisfy $g(\phi_1,\psi_1,\tau_1)\,g(\phi_2\,\psi_2\,\tau_2)
=g(\phi_3,\psi_3,\tau_3)$\,.
Therefore, 
$V_\cW$ is a projective representation
of $Sp(2,\mathbb R)$\,.

A more standard way
to introduce the metaplectic representation is as follows: 
given a pair of creation and annihilation operators 
$(a^\dagger, a)$, one can define a new pair 
$(b^\dagger, b)$ through a Bogoliubov transformation,
which is simply
a linear transformation,
\begin{equation}
    \begin{pmatrix} b \\ b^\dagger \end{pmatrix} = 
    A
    \begin{pmatrix} a \\ a^\dagger \end{pmatrix}\,,
    \label{bogol}
\end{equation}
preserving
the canonical commutation relation, $[b,b^\dagger] = 1$. 
The latter condition 
implies that $A \in SU(1,1)$, which can be
parametrized by
\be  
    A=\begin{pmatrix}
    e^{-i\,\phi}\, \cosh\tau & e^{i\,\psi}\, \sinh\tau\\
    e^{-i\,\psi}\, \sinh\tau & e^{i\,\phi}\, \cosh\tau
    \end{pmatrix}.
\ee 
Since $SU(1,1)\cong Sp(2,\mathbb R)$, the two-dimensional symplectic group 
$Sp(2,\mathbb R)$ is
isomorphic to the center-preserving automorphism group $SU(1,1)$ of the Heisenberg algebra.
Therefore, the Bogoliubov transformation \eqref{bogol} 
defines an action of $Sp(2, \mathbb R)$ on $\cW$\,,
which can be expressed as
\begin{equation}
    \begin{pmatrix} b\\ b^\dagger \end{pmatrix} 
    = U_\cW(\phi,\psi,\tau)\, \begin{pmatrix} a \\ a^\dagger 
    \end{pmatrix}\, U_\cW(\phi,\psi,\tau)^{-1}\,.
    \label{R R action}
\end{equation}
where $U_\cW$ is given in \eqref{U_w}\,,
and $A$ and $g(\phi,\psi,\tau)$ in \eqref{g elem} 
are related by
\be 
    g(\phi,\psi,\tau)=U^{-1}\,A\,U\,,
    \qquad U=\frac1{\sqrt{2}}\begin{pmatrix} 1& i \\ 
    i & 1 \end{pmatrix}.
\ee 
Since the formula \eqref{R R action} is invariant
under redefinition of $U_\cW$ with a phase factor,
it provides a projective representation of $Sp(2,\mathbb R)$. 

To conclude, let us 
consider the reflection subgroup of $Sp(2N,\mathbb R)$,
which will be used
in identifying representations of
dual pairs.
They are contained again in
the maximal compact subgroup $U(N)$ as the diagonal elements,
\be 
    \cR_i=e^{i\,\pi\,K^i{}_i}
    =e^{i\,\pi\,(a^\dagger_i\,a_i+\frac12)}\,,
\ee  
where no summation for the $i$ index is implied.
The reflection $\cR_i$ flips the sign of the $i$-th pair of oscillators:
\be 
    \cR_i\,\binom{a_i}{a^\dagger_i}\,\cR_i{}^{-1}
    =-\binom{a_i}{a^\dagger_i}\,,
\ee 
so it has a $\mathbb Z_2$ action on the oscillators.
Notice, however, that the reflection does not form a $\mathbb Z_2$ group, but a $\mathbb Z_4$\,:
\be 
    \cR_i^2=-1\,,\qquad \cR_i^4=1\,.
\ee 
This is because we are considering 
the metaplectic group where the maximal compact subgroup $U(N)$
should be understood as the double cover of the latter.
Since these double cover groups are not matrix groups,
``reflections'' do not work in the way 
we know for the matrix group. 
Viewed as an automorphism group of Heisenberg algebra $H_N$,
the reflections form $\mathbb Z_2$ groups. 
Besides the aforementioned reflections, there are many other 
$\mathbb Z_2$ or $\mathbb Z_4$ automorphism groups of $H_N$.
Obviously, the permutation group $S_N$ 
lying in the subgroup of $O(N)\subset Sp(2N,\mathbb R)$
contains many $S_2\cong \mathbb Z_2$ groups, i.e. the permutations of two oscillators:
\be 
    \cS_{ij}\,\begin{pmatrix}
        a_i\\ a_i^\dagger \\ a_j \\ a_j^\dagger \end{pmatrix}
        \cS_{ij}{}^{-1}
        =\begin{pmatrix}
        a_j\\ a_j^\dagger \\ a_i \\ a_i^\dagger \end{pmatrix}\,.
\ee 
Less obvious $\mathbb Z_2$ automorphisms are 
\be 
    \cP_{ij}\,\begin{pmatrix}
        a_i\\ a_i^\dagger \\ a_j \\ a_j^\dagger \end{pmatrix}
        \cP_{ij}{}^{-1}
        =\begin{pmatrix}
        a_j^\dagger\\ -a_j \\ -a_i^\dagger \\ a_i \end{pmatrix}\,,
\ee 
which can be decomposed as
\be 
    \cP_{ij}=\cS_{ij}\,\cP_i\,\cP_j{}^{-1}\,.
\ee 
Here, 
$\cP_i$ acts on $H_N$ as 
\be 
    \cP_i\,\binom{a_i}{a^\dagger_i}\,\cP_i{}^{-1}
    =\binom{-a_i^\dagger}{a_i}\,,
\ee 
forming a $\mathbb Z_4$ automorphism of $H_N$, though it is not compatible with
the reality structure of $H_N$.
However,
one can still view $\cP_i$
as $\mathbb Z_2$ automorphisms
of $Sp(2N,\mathbb R)$ 
preserving the reality condition
of the latter.
Basically, it interchanges the raising and lowering operators 
and flips the signs of
all generators of the maximal compact subalgebra 
$\uu(N)$,
and hence corresponds
to the composition
of the Cartan and Chevalley involutions.
From the identity,
\be 
    e^{i\,\frac{\theta}2\left((a_i^\dagger)^2-
    (a_i)^2\right)}\,\binom{a_i}{a^\dagger_i}\,
    e^{-i\,\frac{\theta}2\left((a_i^\dagger)^2-
    (a_i)^2\right)}
    =\begin{pmatrix}
        \cos\theta & -\sin\theta \\
        \sin\theta & \cos\theta 
    \end{pmatrix}
    \binom{a_i}{a^\dagger_i}\,,
\ee 
we can realize $\cP_i$ as
an exponentiation of quadratic oscillators,
\be 
    \cP_i=e^{i\,\frac{\pi}4\left(K_{ii}
    -K^{ii}\right)}
    =e^{i\,\frac{\pi}4\left((a_i^\dagger)^2-
    (a_i)^2\right)}\,.
\ee 
Note however that such a realization of $\cP_i$ 
is not unitary, and hence they are outer-automorphism
of $Sp(2N,\mathbb R)$.
In analyzing dual subgroups of $Sp(2N,\mathbb R)$, it will be often useful to consider
the automorphisms $\cS_{ij}$, $\cP_{ij}$ and $\cP_i$.

As we have discussed, 
the metaplectic representation 
can be easily understood 
using the Fock realization (also referred to as the Fock model).
Another simple description
of it is the wave function in the configuration space ---
the Schr\"odinger realization.
The relation between oscillator realization and wave function realization is
nothing but the typical relation
between $(\hat x_i, \hat p_i)$
and $(a_i,a_i^\dagger)$ of the quantum
harmonic oscillator,
\be 
    \hat x_i=\tfrac1{\sqrt{2}}\left(a_i+a_i^\dagger\right),
    \qquad 
    \hat p_i=\tfrac{i}{\sqrt{2}}\left(a_i^\dagger-a_i\right).
\ee 
Even though the above relation is a very familiar one, it helps to
simplify the description of representations
in several cases: a
certain complicated state $|\Psi\ra$ in the Fock space
becomes simple as wave function $\langle x|\Psi\ra$, and
vice-versa.
Another useful description is provided by
the Bargmann--Segal realization, wherein the states
are holomorphic functions in $\mathbb C^N$,
which are square integrable with respect to
a Gaussian measure. In this realization, the
creation operators simply act as multiplication
$z$ while the annihilation operators act as
derivatives with respect to these variables,
i.e.
\begin{equation}
    a_i^\dagger = z_i\,, \qquad
    a_i = \frac{\partial}{\partial z_i}\,.
\end{equation}
The connection to the Schr\"odinger realization is
made by an integral transform, called
the Bargmann--Segal transform
\cite{Bargmann1961, Segal1963}.

\subsection{Irreducible reductive dual pairs}
\label{sec:irred_pairs}
For a given embedding group $Sp(2N,\mathbb R)$,
we can find many inequivalent dual pairs.  
Any reductive dual pair $(G,G')$
has the form, 
\begin{equation}
    G = G_1 \times G_2 \times \dots \times G_p\,, 
    \qquad
    G' = G_1' \times G_2' \times \dots \times G_p'\,,
\end{equation}
where each couple $(G_k, G_k')$ is one
of the irreducible  dual pairs
listed in Table \ref{table:list_pairs} and embedded in a
symplectic group $Sp(2N_k, \mathbb R)$ such
that $N_1+\dots+N_p=N$. 
Moreover, one can
distinguish between two types of dual pairs:
the first one is a real form of
$(GL_M, GL_N) \subset Sp(2MN,\mathbb R)$
while the second one is a real form of
$(O_N, Sp_{2M})\subset Sp(2MN,\mathbb R)$.
For instance, in $Sp(2N,\mathbb R)$ one can
find the irreducible pair
$\big(U(N), U(1)\big)$, which is an example
of the first type, or the irreducible pair
$\big(O(N), Sp(2,\mathbb R)\big)$,
which is an example of the second type.
Accordingly, the pair $\big(O(N-m) \times U(m),
Sp(2,\mathbb R) \times U(1)\big)$
is an example of a reducible dual pair in $Sp(2N, \mathbb R)$ for $m=1, \dots, N$.

\renewcommand{\arraystretch}{1.25}
\setlength{\arrayrulewidth}{1pt}
\begin{table}[H]
    \centering\footnotesize
    \begin{tabular}{|c|c|c|}
        \hline
        Embedding group & $(G,G')$ & $(K,K')$ \\ \hline\hline
        $Sp(2MN, \mathbb R)$ & $\big(GL(M, \mathbb R),
        GL(N, \mathbb R)\big)$ & $\big(O(M), O(N)\big)$ \\ \hline
        $Sp(4MN, \mathbb R)$ & $\big(GL(M, \mathbb C),
        GL(N, \mathbb C)\big)$ & $\big(U(M), U(N)\big)$ \\ \hline
        $Sp(8MN, \mathbb R)$ & $\big(U^*(2M), U^*(2N)\big)$ &
        $\big(Sp(M), Sp(N)\big)$ \\ \hline
        $Sp(2(M_++M_-)(N_++N_-), \mathbb R)$ & $\big(U(M_+,M_-), U(N_+,N_-)\big)$
        & $\big(U(M_+) \times U(M_-), U(N_+) \times U(N_-)\big)$ \\ \hline
        $Sp(2M(N_++N_-), \mathbb R)$ & $\big(O(N_+,N_-),
        Sp(2M, \mathbb R)\big)$ &
        $\big(O(N_+) \times O(N_-), U(M)\big)$ \\ \hline
        $Sp(4MN, \mathbb R)$ & $\big(O(N, \mathbb C),
        Sp(2M, \mathbb C)\big)$ & $\big(O(N), Sp(M)\big)$ \\ \hline
        $Sp(4N(M_++M_-), \mathbb R)$ & $\big(O^*(2N), Sp(M_+,M_-)\big)$
        & $\big(U(N), Sp(M_+) \times Sp(M_-)\big)$ \\ \hline
    \end{tabular}
    \caption{List of all possible irreducible
    reductive dual pairs $(G,G')$ embedded in
    metaplectic groups and their respective
    maximal compact subgroups $(K,K')$.}
    \label{table:list_pairs}
\end{table}

Often irreducible dual pairs are classified, 
using another criterion, again
into two types: the type II if 
there exists a Lagrangian subspace in $\mathbb R^{2N}$
left invariant under both of $G$ and $G'$,
and the type I otherwise.
In the above table, the first three cases are type II,
while the rest are type I.

We can extend the idea of the dual pair
correspondence to other groups /
representations than the symplectic group
/ metaplectic representation. For instance,
one can also consider supersymmetric
dual pairs embedded in $OSp(N|2M, \mathbb R)$
(see e.g. \cite{Cheng2009, Cheng2012}
and references therein).
Another possible extension is to replace the
symplectic group with any simple group $G$
and the metaplectic representation with the
minimal representation.\footnote{
The minimal representation of a simple group
$G$ is the representation whose annihilator in
the universal enveloping algebra
$\mathcal{U}(\mathfrak{g})$ is the Joseph
ideal, which is the maximal primitive and completely prime ideal of
$\mathcal{U}(\mathfrak{g})$ 
(see e.g. \cite{Todorov:2010md, Joung:2014qya}).}
The classification of such pairs was obtained
in \cite{Rubenthaler1994} (see also
\cite{Magaard:1997,Schmidt1999} for complementary work).

\subsection{Seesaw pairs}
\label{sec:seesaw_pairs}

A useful concept is that of a seesaw pair
introduced in \cite{Kudla1986} (reviewed e.g. in \cite{Prasad1993, Adams2007}).
Consider the situation wherein we have two reductive
dual pairs $(G,G')$ and $(\tilde G,\tilde G')$ in the same
symplectic group, say $Sp(2N,\mathbb R)$. If the groups forming
the second pair $(\tilde G, \tilde G')$ satisfy
$\tilde G \subset G$ and $G'\subset \tilde G'$\,,
then the pair of dual pairs $(G,G')$ and 
$(\tilde G,\tilde G')$ is called a ``seesaw pair'',
and the situation is depicted as
\be 
\parbox{65pt}{
\begin{tikzpicture}
\draw [<->] (0,0) -- (1,0.8);
\draw [<->] (0,0.8) -- (1,0);
\node at (-0.3,1) {$G$};
\node at (-0.3,0.4) {$\cup$};
\node at (1.35,-0.2) {$G'$};
\node at (-0.3,-0.2) {$\tilde G$};
\node at (1.3,0.4) {$\cup$};
\node at (1.35,1) {$\tilde G'$};
\end{tikzpicture}}\,.
\ee 
Seesaw pairs satisfy the property,
\begin{equation}
    {\rm Hom}_{\tilde G} 
    \big(\pi_{\tilde G},
    \pi_G\rvert_{\tilde G}\big) \cong 
    {\rm Hom}_{G'}\big(\pi_{G'}, 
    \pi_{\tilde G'}\rvert_{G'}\big)\,,
\end{equation}
which means 
the space of $\tilde G$-equivariant linear maps (or intertwiners)
from $\pi_{\tilde G}$ to $ \pi_G\rvert_{\tilde G}$,
namely $ {\rm Hom}_{\tilde G} 
    \big(\pi_{\tilde G},
    \pi_G\rvert_{\tilde G}\big)$,
is isomorphic
to the space of $G'$-equivariant linear maps
from $\pi_{G'}$ to $ \pi_{\tilde G'}\rvert_{G'}$,
namely ${\rm Hom}_{G'}\big(\pi_{G'}, 
    \pi_{\tilde G'}\rvert_{G'}\big)$.
    Here, $\rvert_{\tilde G}$ (resp. $\rvert_{G'}$)
denotes the restriction to $\tilde G$ (resp. $G'$).
In particular,
this implies
\begin{equation}
	{\rm mult}_{\pi_G}(\pi_{\tilde G}) = 
	{\rm mult}_{\pi_{\tilde G'}}(\pi_{G'})\,,
\end{equation}
i.e. the multiplicity of the $\tilde G$-representation
$\pi_{\tilde G}$ in the branching rule of $\pi_G$ is the same
as the multiplicity of its  dual $\pi_{G'}$ in the
branching rule of the $\tilde G'$-representation
$\pi_{\tilde G'}$. This situation is usually depicted as
\be
\parbox{65pt}{
\begin{tikzpicture}
\draw [<->] (0,0) -- (1,0.8);
\draw [<->] (0,0.8) -- (1,0);
\node at (-0.3,1) {$\pi_G$};
\draw [->] (-0.3,0.7) -- (-0.3,0.1);
\node at (1.35,-0.2) {$\pi_{G'}$};
\node at (-0.3,-0.2) {$\pi_{\tilde G}$};
\draw [->] (1.3,0.7) -- (1.3,0.1);
\node at (1.35,1) {$\pi_{\tilde G'}$};
\end{tikzpicture}}\,,
\ee
where the
downward arrows denote the restriction of a $G$-irrep
(resp. $\tilde G'$-irrep) to a $\tilde G$-irrep (resp.
$G'$-irrep).
Seesaw pairs are particularly useful when trying
to derive the explicit correspondence between
representations of $G$ and $G'$, assuming that
the correspondence is known for representations
of the pair $(\tilde G, \tilde G')$, or to obtain
the decomposition of a representation into its
maximal compact subgroup.

Let us illustrate this with a simple example.
Both $\big(Sp(2,\mathbb R), O(2)\big)$ and
$\big(U(1), U(2)\big)$ are dual pairs in
$Sp(4,\mathbb R)$, which form a seesaw pair:
\begin{equation}
\parbox{95pt}{
\begin{tikzpicture}
\draw [<->] (0,0) -- (1,0.8);
\draw [<->] (0,0.8) -- (1,0);
\node at (-0.7,1) {$Sp(2,\mathbb R)$};
\node at (-0.7,0.4) {$\cup$};
\node at (1.5,-0.2) {$O(2)$};
\node at (-0.7,-0.2) {$U(1)$};
\node at (1.5,0.4) {$\cup$};
\node at (1.5,1) {$U(2)$};
\end{tikzpicture}}\,.
\end{equation}
Note that the generator of the $O(2)$ subgroup
of $U(2)$ correspond to the $U(1)$ generator
of $SU(2)$, and not that of the diagonal $U(1)$.
We will see later in Section \ref{sec:rep_compact}
the following correspondences,
\begin{equation}
\parbox{220pt}{
\begin{tikzpicture}
\draw [<->] (0,0) -- (1,0.8);
\draw [<->] (0,0.8) -- (1,0);
\node at (-1.3,1) {$\cD_{Sp(2,\mathbb R)}(\ell+1)$};
\node at (1.7,-0.2) {$[\ell]_{O(2)}$};
\node at (-1.3,-0.2) {$[\lowercal{m}+1]_{U(1)}$};
\node at (3,1) {$[\lowercal{m}+1]_{U(1)}\otimes [\lowercal{m}]_{SU(2)}$};
\end{tikzpicture}}\,,
\end{equation}
where $\ell$ and $\lowercal{m}$ are non-negative integers.
We can use
the seesaw pair property to find the
restriction of $\cD_{Sp(2,\mathbb R)}(\ell+1)$
under its maximal compact subgroup $U(1)$:
these representations will be dual to the $U(2)$
representations whose restriction under $O(2)$
contains $[\ell]_{O(2)}$. 
The branching rule $U(2) \downarrow O(2)$
amounts to finding those $SU(2)$ irreps which
contain a pair of states with weight $\pm \ell$.
As the irrep
$[\lowercal{m}]_{SU(2)}$ contains
states with weight $k=-\lowercal{m},
-\lowercal{m}+2,\dots,
\lowercal{m}$, we can conclude that
the relevant $U(2)$ irreps are those labelled by
$\lowercal{m}=\ell+2n$ 
with $n\in\mathbb N$. 
As a
consequence, we find
\begin{equation}
    \cD_{Sp(2,\mathbb R)}(\ell+1)\rvert_{U(1)}
    = \bigoplus_{n=0}^\infty\, [\ell+1+2n]_{U(1)}\,,
\end{equation}
which is indeed the correct decomposition.

We will see in Section \ref{sec:rep_compact}
that an important subgroup of $G$
to consider in the context of seesaw pairs is the maximal
compact subgroup $K \subset G$. Indeed, one can show that
the centralizer of $K$ in $Sp(2N, \mathbb R)$, that we will
denote $M'$, contains $G'$  
so that $(K, M')$ is a dual pair and it forms
a seesaw pair with $(G, G')$. The same property is true
if we reverse $G$ and $G'$: the centralizer $M$ of a maximal
compact subgroup $K' \subset G'$ 
also contains $G$. 
 As a consequence, $(G, G')$
and $(K', M)$ also form a seesaw pair. In general, the groups
$M$ and $M'$ are not compact, 
in which case one can also consider their 
maximal compact subgroup, denoted by $K_M$ and
$K_{M'}$, respectively. It turns out that also they form a dual pair.
We therefore end up with 
four dual pairs.
The situation is depicted as
\begin{equation}
\parbox{270pt}{
\begin{tikzpicture}[scale=0.9]
\node at (-1.5,1.5) {$M$};
\draw [<->] (-0.5,1.5) -- (0.5,1.5);
\node at (1.5,1.5) {$K'$};
\node [rotate=-45] at (-2.5,0.75) {$\cup$};
\node at (-1.5,0.75) {$\cup$};
\node at (1.5,0.75) {$\cap$};
\node [rotate=-125] at (2.5,0.75) {$\cup$};
\draw[<-] (-4.25,0) -- (-4.75,0);
\draw (-4.75, 0.5) arc(90:270:0.25);
\draw[dashed] (-4.75, 0.5) -- (-4,0.5);
\node at (-3.5,0) {$K_M$};
\node at (-1.5,0) {$G$};
\draw [<->] (-0.5,0) -- (0.5,0);
\node at (1.5,0) {$G'$};
\node at (3.5,0) {$K_{M'}$};
\draw[<-] (4.25,0) -- (4.75,0) arc(-90:90:0.25);
\draw[dashed] (4.75, 0.5) -- (4,0.5);
\node [rotate=45] at (-2.5,-0.75) {$\cup$};
\node at (-1.5,-0.75) {$\cup$};
\node at (1.5,-0.75) {$\cap$};
\node [rotate=125] at (2.5,-0.75) {$\cup$};
\node  at (-1.5,-1.5) {$K$};
\draw [<->] (-0.5,-1.5) -- (0.5,-1.5);
\node at (1.5,-1.5) {$M'$};
\end{tikzpicture}}.
\end{equation}
The application of the above diagram
to each irreducible dual pair 
in Table~\ref{table:list_pairs}
is collected in
Appendix \ref{app:seesaw_diagrams}.

Seesaw pairs are also useful 
for tensor products of the relevant representations and their plethysms.
Say, we have a dual pair $(G,G')\subset Sp(2N,\mathbb R)$
and their representations $\pi_G$
and $\pi_{G'}$.
The tensor product of the representation
$\pi_G$ and its contragredient one $\bar\pi_G$
can be obtained by considering the seesaw pair,
\be
\parbox{180pt}{
\begin{tikzpicture}
\draw [<->] (0,0) -- (1,0.8);
\draw [<->] (0,0.8) -- (1,0);
\node at (-1.5,1.2) {$\underbrace{G\times\cdots \times G}_{p+q}$};
\node at (-1.5,0.4) {$\cup$};
\node  at (-1.5,-0.4) {$G$};
\node at (2.5,-0.4) {$\underbrace{G'\times \cdots \times G'}_{p+q}$};
\node at (2.5,0.4) {$\cup$};
\node at (2.5,1.2) {$G'_{p,q}$};
\end{tikzpicture}}.
\ee
Here, the two pairs $(G,G'_{p,q})$
and $(G^{p+q},G'{}^{p+q})$ are embedded in $Sp(2(p+q)N,\mathbb R)$.
By examining the seesaw pairs for each irreducible pair $(G,G')$, one can find that the groups $G'_{p,q}$ 
can be taken as in Table \ref{table:list_tensor}.
\begin{table}[H]
    \centering\footnotesize
    \begin{tabular}{|c|c|}
        \hline
         $G'$ & $G'_{p,q}$ \\ \hline\hline
        $GL(n,\mathbb R)$ & $GL((p+q)n,\mathbb R)$ \\\hline
        $GL(n,\mathbb C)$ & $GL((p+q)n,\mathbb C)$ \\\hline
        $U^*(2n)$ & $U^*(2(p+q)n)$ \\\hline
        $U(m,n)$ & $U(pm+qn,pn+qm)$ \\\hline
        $O(m,n)$ & $O(pm+qn,pn+qm)$ \\\hline
        $O(n,\mathbb C)$ & $O((p+q)n,\mathbb C)$\\\hline
        $O^*(2n)$ & $O^*(2(p+q)n)$ \\\hline
        $Sp(2n,\mathbb R)$ & $Sp(2(p+q)n,\mathbb R)$\\\hline
        $Sp(2n,\mathbb C)$ & $Sp(2(p+q)n,\mathbb C)$\\\hline
        $Sp(m,n)$ & $Sp(pm+qn,pn+qm)$\\
        \hline
    \end{tabular}
    \caption{The list of $G'_{p,q}$ for each irreducible $G'$.}
    \label{table:list_tensor}
\end{table}
From the seesaw duality, we find 
that tensor product decompositions of $G$ representations
can be obtained by
the decomposition of $G'_{p,q}$ representations:
\be
\parbox{250pt}{
\begin{tikzpicture}
\draw [<->] (0,0) -- (1,0.8);
\draw [<->] (0,0.8) -- (1,0);
\node at (-2,1.2) {$(\pi_G{}^{\otimes p}\otimes \bar \pi_G{}^{\otimes q})\rtimes \Gamma_{p,q}$};
\draw [<-] (-2,0) -- (-2,0.8);
\node  at (-2,-0.4) {$\pi_G$};
\node at (3,-0.4) {$(\pi_{G'}{}^{\otimes p}\otimes \bar \pi_{G'}{}^{\otimes q})\rtimes \Gamma_{p,q}$};
\draw [<-] (2.7,0) -- (2.7,0.8);
\node at (2.7,1.2) {$\pi_{G'_{p,q}}$};
\end{tikzpicture}},
\ee
where the plethysm
can be also controlled by
a discrete group $\Gamma_{p,q}$
which is a subgroup of 
the symmetric group $\cS_p\times \cS_q$ 
permuting the same representations.
In particular, every irreducible dual pair $(G,G')$, wherein $G'$
is not the smallest group dual to $G$,
can be used for
the tensor product of the irreps of $G$
dual to the smallest $G'$.

\section{Dual pairs of type $(GL_M,GL_N)$}
\label{sec:osc_model_GL}

Let us describe the dual pair $(GL_M,GL_N)\subset Sp(2MN,\mathbb R)$ in detail.
Since it is embedded in $Sp(2MN,\mathbb R)$,
we introduce $MN$ pairs of conjugate operators
\be 
    ( \omega_A^I\,, \ \tilde \omega^A_I )\,,
    \qquad A = 1, \ldots, M \, , 
    \quad I = 1, \ldots, N \, , 
    \label{osc split}
\ee
which obey the canonical commutation relations
\be 
    [\omega_A^I, \tilde \omega^B_J]=\delta_A^B\,\delta^I_J\,,\qquad 
    [ \omega_A^I\,, \omega_B^J]=0=[\tilde \omega^A_I,\tilde \omega^B_J]\,.
\ee 
The $GL_M$ and $GL_N$ groups are generated
respectively by 
\ba
    && X^A{}_B=
    U_\cW(\mathsf{X}^A{}_B)=\frac12\,\{\tilde \o^A_I,\o^I_B\}
    =\tilde \o^A_I\,\o^I_B+\frac{N}2\, \delta^A_B\,,\nn
    &&
    R_I{}^J=
    U_\cW(\mathsf{R}_I{}^J)=\frac12\,\{\tilde \o^A_I,\o^J_A\} = \tilde \omega_I^A\, \omega_A^J+\frac{M}2\, \delta^J_I\,,
    \label{eq:gen-GL}
\ea
where $U_\cW$ is the metaplectic representation,
and $\mathsf X^A{}_B$ and $\mathsf R_I{}^J$
are the matrices with components 
$(\mathsf X^A{}_B)_C{}^D=\delta^A_C\,\delta_B^D$
and $(\mathsf R_I{}^J)^K{}_L
=\delta_I^K\,\delta^J_L$
generating $GL_M$ and $GL_N$ respectively.

Notice that when $GL_M$ and $GL_N$ are realized as above,
a specific ordering of $\o$ and $\tilde \o$ 
is chosen, namely the Weyl ordering.
If we do not 
embed the pair $(GL_M,GL_N)$ in $Sp(2MN,\mathbb R)$,
one could take different orderings
in each of the groups,
leading to different factors of $\delta^A_B$ and $\delta^J_I$
(shift constants)
in \eqref{eq:gen-GL}.
Here, the embedding of the pair in $Sp(2MN,\mathbb R)$
singles out a unique ordering,
which fixes the shift constants, such that
the diagonal $GL_1$ subgroups 
of the $GL_M$
and $GL_N$ coincide; i.e.
\be 
    Z:=X^A{}_A=R_I{}^I=\frac12\,\{\o^I_A,\tilde \o^A_I\}
    =\tilde \o^A_I\,\o^I_A+\frac{M\,N}2\,,
\ee
which is a number operator with a constant shift.

Various real forms of $GL_M$ and $GL_N$ can be chosen
by assigning different reality conditions to the operators $\o_A^I$
and $\tilde \o^A_I$.
Such conditions can be straightforwardly deduced
from the anti-involution $\sigma$
associated with each real form by asking that
\be 
    U_\cW(\mathsf{X}^{A}{}_B)^\dagger=
    U_\cW(\sigma(\mathsf X^{A}{}_B))\,,
    \qquad 
     U_\cW(\mathsf{R}_I{}^J)^\dagger=U_\cW(\sigma(\mathsf{R}_{I}{}^J))\,.
\ee 
See Appendix \ref{sec: real form} for 
the expressions of $\sigma$ 
for each real forms.
In the following, we describe
all such real forms
by introducing their oscillator realizations.

\subsection{$\big(GL(M,\mathbb R), GL(N, \mathbb R)\big) \subset Sp(2MN, \mathbb R)$}

To single out the real forms $GL(M, \mathbb R)$ and $GL(N, \mathbb R)$, we impose
the following reality conditions  
on $X^{A}{}_B$ and $R_I{}^J$ 
\begin{equation}
    (X^A{}_B)^\dagger = -X^A{}_B\,, \qquad (R_I{}^J)^\dagger = -R_I{}^J\,,
\end{equation}
which can be realized by requiring the $\omega$-operators to obey the reality conditions,
\begin{equation}
    (\omega^I_A)^\dagger = \omega^I_A\,, \qquad (\tilde \omega^A_I)^\dagger = - \tilde \omega^A_I\,.
    \label{omega glR}
\end{equation}
Let us define
\begin{equation}
    a_A^I := \tfrac1{\sqrt 2}\, \big(\omega_A^I -\tilde \omega_I^A\big)\,, \qquad \tilde a_I^A := \tfrac1{\sqrt 2}\, \big(\tilde \omega_I^A+\omega_A^I \big)\,,
\end{equation}
so that 
\begin{equation}
    (a^I_A)^\dagger = \tilde a^A_I\,, \qquad \text{and} \qquad [a^I_A, \tilde a_J^B] = \delta_A^B\, \delta^I_J\,,
\end{equation}
 In terms of these creation and annihilation operators, the generators of
$GL(M, \mathbb R)$ and $GL(N, \mathbb R)$ now read
\begin{equation}
    X^A{}_B = \tfrac12\, \big(\tilde a^A_I\, a_B^I - 
    \tilde a_I^B\, a_A^I +  \tilde a^A_I\, \tilde a^B_I - 
    a_A^I\, a_B^I\big)\,,
\end{equation}
and
\begin{equation}
    R_I{}^J = \tfrac12\, \big(\tilde a^A_I\, a_A^J - 
    \tilde a_J^A\, a_A^I +  \tilde a^A_I\, \tilde a^A_J - 
    a_A^I\, a_A^J\big)\,,
\end{equation}
and the common Abelian subgroup $GL(1,\mathbb R)\cong \mathbb R^\times$ is generated by
\be 
    Z= \tfrac12\,\big(\tilde a^A_I\, \tilde a^A_I - 
    a_A^I\, a_A^I\big).
\ee 
Note that in the above expressions, the repeated indices are summed over
even when they are positioned both up or both down.
Since $GL(M,\mathbb R)$ and $GL(N,\mathbb R)$ are disconnected,\footnote{
The group $GL(N,\mathbb R)$ is isomorphic to $\mathbb R^{\times}\ltimes SL(N,\mathbb R)$, where $\mathbb R^\times$ is 
the multiplicative group of real numbers.
Since $\mathbb R^\times$ has two connected components,
positive and negative real numbers,
$GL(N,\mathbb R)$
has two connected components,
whose elements have positive and negative determinant, respectively.}
we need also to take into account 
the reflection group elements $\cR_{A}$ and $\cR^I$,
\be 
    \cR_A=\prod_{I=1}^N\cR_A^I\,,
    \qquad 
    \cR^I=\prod_{A=1}^M\,\cR_A^I\,,
\ee 
where $\cR_A^I$ are the reflections in $Sp(2MN,\mathbb R)$ introduced in Section \ref{sec:metaplectic},
\be 
    \cR_A^I\,\binom{a_B^J}{\tilde a^B_J}\,(\cR_A^I)^{-1}
    =(-1)^{\delta_{AB}\,\delta^{IJ}}\,\binom{a_B^J}{\tilde a^B_J}\,.
\ee 
Let us remark also that the elements,
\begin{equation}
    M_{AB} := 2\, \delta_{C[A}\, X^C{}_{B]} = \tilde a^A_I\, a_B^I - 
    \tilde a_I^B\, a_A^I\,, \qquad N_{IJ} := 2\, R_{[I}{}^K\, \delta_{J]K} = \tilde a^A_I\, a_A^J - \tilde a_J^A\, a_A^I\,,
\end{equation}
generate the maximal compact subgroups $O(M) \subset GL(M,\mathbb R)$ and $O(N) \subset GL(N, \mathbb R)$
respectively. Note that 
the generators of these compact groups
take the form (creation) $\times$ (annihilation),
so that they preserves the total oscillator number.

\subsection{$\big(GL(M,\mathbb C), GL(N, \mathbb C)\big) \subset Sp(4MN, \mathbb R)$}
The real Lie groups $GL(M, \mathbb C)$ and $GL(N, \mathbb C)$ are obtained by starting with complex operators
$\omega$ and $\tilde \omega$ 
with complex conjugates $\omega^*$ and $\tilde\omega^*$,
obeying the commutation relations,
\be 
    [\o^I_A{}^*,\tilde \o^B_J{}^*]=\delta_A^B\,\delta^I_J\,,
\ee
and 
\be 
    [\o^I_A,\o^J_B{}^*]=[\o^I_A,\tilde\o_J^B{}^*]=
    [\tilde \o_I^A,\o^J_B{}^*]=[\tilde \o_I^A,\tilde \o_J^B{}^*]=0\,.
\ee 
We require the reality conditions,
\begin{equation}
    (\omega^I_A)^\dagger = \omega^I_A{}^*\,, \qquad (\tilde\omega^A_I)^\dagger = - \tilde\omega^A_I{}^*\,,
    \qquad 
     (\omega^I_A{}^*)^\dagger = \omega^I_A{}\,, \qquad (\tilde\omega^A_I{}^*)^\dagger = - \tilde\omega^A_I{}\,,
     \label{omega GLC}
\end{equation}
so that the generators,
\begin{equation}
    X_\pm^A{}_B = X^A{}_B\pm (X^A{}_B)^*\,, \qquad R_\pm{}_I{}^J =R_I{}^J\pm (R_I{}^J)^*\,,
     \label{XR C}
\end{equation}
satisfy
\begin{equation}
    (X_\pm^A{}_B)^\dagger = \mp\, X_\pm^A{}_B\,, \qquad 
    (R_\pm{}_I{}^J)^\dagger = \mp\, R_\pm{}_I{}^J\,.
\end{equation}
Notice that $X_+^A{}_B$ and $R_+{}_I{}^J$ generate
the $GL(M,\mathbb R)$ and $GL(N,\mathbb R)$ subgroup
of $GL(M,\mathbb C)$ and $GL(N,\mathbb C)$\,, 
respectively. For the oscillator realization,
we introduce 
\begin{eqnarray}
    & a_A^I := \tfrac1{\sqrt 2}\, \big(\omega_A^I -\tilde \omega_I^A{}^*\big)\,, \qquad 
    \tilde a_I^A := \tfrac1{\sqrt 2}\, \big(\tilde \omega_I^A+\omega_A^I {}^*\big)\,,\nn 
    & b^A_I := \tfrac1{\sqrt 2}\, \big(\omega_A^I{}^* -\tilde \omega_I^A\big)\,, \qquad \tilde b^I_A := \tfrac1{\sqrt 2}\, \big(\tilde \omega_I^A{}^*+\omega_A^I \big)\,,
\end{eqnarray}
so that 
\be 
    (a_A^I)^\dagger=\tilde a_I^A\,,\qquad
    (b^A_I)^\dagger=\tilde b^I_A\,,
\ee 
and the only non-zero commutators are
\be 
    [a^I_A, \tilde a_J^B] = \delta_A^B\, \delta^I_J\,,
    \qquad 
    [b_I^A, \tilde b^J_B] = \delta^A_B\, \delta_I^J\,.
\ee 
Note that under complex conjugation, the $a$ and $b$
oscillators are mapped to one another:
\be  
    (a_A^I)^*=b^A_I\,,
    \qquad 
    (\tilde a^A_I)^*=\tilde b_A^I\,.
    \label{ab map}
\ee 
In terms of these oscillators, the generators
$X^A{}_B$ and $R_I{}^J$ read
\begin{equation}
    X^A{}_B = \tfrac12\, \big(\tilde a^A_I\, a_B^I - 
    \tilde b^I_B\, b^A_I +  \tilde a^A_I\, \tilde b_B^I - 
    b^A_I\, a_B^I\big)\,,
\end{equation}
and 
\begin{equation}
    R_I{}^J = \tfrac12\, \big(\tilde a^A_I\, a_A^J - 
    \tilde b^J_A\, b^A_I +  \tilde a^A_I\, \tilde b_A^J - 
    b^A_I\, a_A^J\big)\,.
\end{equation} 
From the above, one can easily find the generators $X_\pm^A{}_B$
and $R_\pm{}_I{}^J$ using \eqref{XR C} and \eqref{ab map}.
The common Abelian subgroup $GL(1,\mathbb C)\cong \mathbb C^\times \cong 
\mathbb R^+\times U(1)$
is generated by
\be 
    Z_+=\tilde a^A_I\, \tilde b_A^I - 
    a_A^I\, b^A_I\,,
    \qquad {\rm and} \qquad 
    Z_-=\tilde a^A_I\, a_A^I - 
    \tilde b^I_A\, b^A_I\,,
\ee 
which can be interpreted as the radial and angular 
part of the complex plane without origin $\mathbb C^\times$, respectively.
The maximal compact subgroups $U(M)$ and $U(N)$
of $GL(M,\mathbb C)$ and $GL(N,\mathbb C)$ are 
generated respectively by
$X^A{}_B-(X^B{}_A)^*$ and $R_I{}^J-(R_J{}^I{})^*$,
and they have the form (creation)$\times$(annihilation).
Their common $U(1)$ generator coincides with $Z_-$\,.

\subsection{$\big(U^*(2M), U^*(2N)\big) \subset Sp(8MN, \mathbb R)$}
\label{sec:U*-U*}

To single out the real forms $U^*(2M)$ and $U^*(2N)$,\footnote{Notice 
that  the determinant of an element 
of $U^*(2N)$ is always positive,
and hence the group $U^*(2N)$  is isomorphic to $\mathbb R^+\times SU^*(2N)$.} we require 
$X^A{}_B$ and $R_I{}^J$ to satisfy
\begin{equation}
    (X^A{}_B)^\dagger = 
    -\Omega_{AC}\,X^C{}_D\,\Omega^{DB}\,, \qquad (R_I{}^J)^\dagger = -\Omega^{IK}\, R_K{}^L\,\Omega_{LJ}\,.
\end{equation}
This can be realized by requiring the $\omega$ 
operators to obey the reality conditions
\begin{equation}
    (\omega^I_A)^\dagger = \Omega_{IJ}\, \Omega^{AB}\, \omega^J_B\,, \qquad (\tilde \omega_I^A)^\dagger = -\Omega^{IJ}\, \Omega_{AB}\, \tilde \omega_J^B\,,
    \label{U* real}
\end{equation}
where the $A, B$, and $I, J$ indices take 
respectively $2M$ and $2N$ values.
The two matrices $\Omega_{AB}$ and $\Omega^{IJ}$
are both antisymmetric and square to minus the
identity matrix, and $\Omega^{AB}$ and $\Omega_{IJ}$
are their respective inverses:
$\Omega^{AB}\,\Omega_{BC}=\delta^{A}_C$\,,
$\Omega_{IJ}\,\Omega^{JK}=\delta_{I}^K$\,.
As a consequence, $\Omega_{AB}=-\Omega^{AB}$ and $\Omega_{IJ}=-\Omega^{IJ}$. Upon defining the
oscillators,
\begin{equation}
    a_A^I := \tfrac1{\sqrt2}\, \big(\omega_A^I 
    -\Omega_{AB}\,\Omega^{IJ}\,\tilde \omega_J^B\big)\,,
    \qquad 
    \tilde a_I^A := (a_A^I)^\dagger\,,
\end{equation}
with
\begin{equation}
    [a_A^I, \tilde a_J^B] 
    = \delta_A^B\, \delta_J^I\,.
\end{equation}
the generators $X^A{}_B$ and $R_I{}^J$
can be expressed as
\begin{equation}
    X^A{}_B = \tfrac12\,\Big(\tilde a^A_I\, a^I_B
    -\tilde a^I_{B}\, a^{A}_I- a^{A}_{I}\, a_B^I
    +\tilde a^A_I\,
    \tilde a^I_B \Big)\,,
\end{equation}
\begin{equation}
    R_I{}^J = \tfrac12\,\Big(\tilde a^A_I\, a^J_A
    - \tilde a_A^J\, a^A_I
    -a_I^A\, a_A^J
    +\tilde a^A_I\,
    \tilde a_A^J \Big)\,,
\end{equation}
where we used $\Omega_{AB}$ and $\Omega_{IJ}$ (and their inverse) to raise and lower indices as in \eqref{eq:convention_sp_indices}.
The common center $\mathbb R^+$ of $U^*(2M)$ 
and $U^*(2N)$ is generated by
\begin{equation}
    Z   =\tilde a^A_I\,\tilde a_A^I
    -a^I_A\,a_I^A\,.
\end{equation}
The Lie algebras of the maximal compact subgroups 
$Sp(M) \subset U^*(2M)$
and $Sp(N) \subset U^*(2N)$ are generated respectively by
\begin{equation}
    X^{(AB)}= \tilde a^{(A}_I\,a^{B)I}_{\phantom{A}}\,,
    \qquad 
    R_{(IJ)}
    =\tilde a^A_{(I}\,a_{J)A}^{\phantom{A}}\,.
\end{equation}
Again, one can see
that the generators of these compact 
subgroups have the form of (creation) $\times$ (annihilation).

\subsection{$\big(U(M_+,M_-),U(N_+,N_-)\big) \subset Sp(2(M_++M_-)(N_++N_-), \mathbb R)$}
To single out the real forms $U(M_+,M_-)$ and $U(N_+,N_-)$, we impose 
\begin{equation}
    (X^A{}_B)^\dagger = \eta^{AD}\, \eta_{BC}\, X^C{}_D\,, \qquad (R_I{}^J)^\dagger = \eta_{IL}\, \eta^{JK}\, R_K{}^L\,,
\end{equation}
where  $\eta_{AB}$  and $\eta^{IJ}$ 
are diagonal matrices
of signature
$(M_+,M_-)$ and $(N_+,N_-)$ respectively, and
where $X^A{}_B$ and $R_I{}^J$ are defined in \eqref{eq:gen-GL}.
This can be achieved by requiring the $\omega$ operators to obey the reality condition,
\begin{equation}
    (\omega^I_A)^\dagger = \eta_{AB}\,\eta^{IJ}\, \tilde \omega_J^B\,.
\end{equation}
To introduce the oscillator realization,
we split the $A$ and $I$ indices
into $(a,\mathtt{a})$ and $(i,\mathtt{i})$
and set $\eta_{AB}=(\delta_{ab}, -\delta_{\mathtt{ab}})$
and $\eta^{IJ} = (\delta^{ij}, -\delta^{\mathtt{ij}})$.
Here, $a$ and $\mathtt{a}$ take
respectively $M_+$ and $M_-$ values,
whereas $i$ and $\mathtt{i}$ take
respectively $N_+$ and $N_-$ values.
Upon defining 
\begin{equation}
    a^i_a := \omega^i_a\,, \qquad 
    b_i^{\mathtt a} := \tilde \omega_i^{\mathtt a}\,, \qquad c_{\mathtt i}^a := \tilde \omega_{\mathtt i}^a\,, \qquad d^{\mathtt i}_{\mathtt a} := \omega^{\mathtt i}_{\mathtt a}\,,
\end{equation}
and
\begin{equation}
    \tilde a^a_i = (a_a^i)^\dagger\,,
    \qquad 
    \tilde b^i_{\mathtt a} = (b^{\mathtt a}_i)^\dagger\,,
    \qquad
    \tilde c_a^{\mathtt i} = (c^a_{\mathtt i})^\dagger\,,
    \qquad
    \tilde d_{\mathtt i}^{\mathtt a} = (d^{\mathtt i}_{\mathtt a})^\dagger\,,
\end{equation}
we end up with $(M_++M_-)(N_++N_-)$ canonical pairs of creation and annihilation operators, as these oscillators obey
\begin{equation}
	[a^i_a, \tilde a^b_j] = \delta^i_j\, \delta^b_a\,, \qquad [b_j^{\mathtt a}, \tilde b^i_{\mathtt b}] = \delta^i_j\, \delta^{\mathtt a}_{\mathtt b}\,,
	\qquad
	[c^a_{\mathtt i}, \tilde c_b^{\mathtt j}] = \delta^a_b\, \delta_{\mathtt i}^{\mathtt j}\,, \qquad [d_{\mathtt a}^{\mathtt i}, \tilde d^{\mathtt b}_{\mathtt j}] = \delta^{\mathtt b}_{\mathtt a}\, \delta^{\mathtt i}_{\mathtt j}\,,
\end{equation}
with all other commutators vanishing.
In terms of the $a,b,c$ and $d$ oscillators, the generators $X^A{}_B$ and $R_I{}^J$ read
\begin{equation}
    \begin{aligned}
	X^a{}_b & = \tilde a^a_i\, a_b^i
	-\tilde c_b^{\mathtt i}\, c^a_{\mathtt i} + \tfrac{N_+-N_-}2\, \delta^a_b\,,\\
	X^a{}_{\mathtt b} & = 
	-\tilde a^a_i\, \tilde b_{\mathtt b}^i
	+ c_{\mathtt i}^a\, d^{\mathtt i}_{\mathtt b}\,,
	\end{aligned}
	\qquad\,
	\begin{aligned}
	X^{\mathtt a}{}_b & = a^i_b\, b_i^{\mathtt a}
	-\tilde c^{\mathtt i}_b\, \tilde d^{\mathtt a}_{\mathtt i}\,,\\
	X^{\mathtt a}{}_{\mathtt b} & = -\tilde b_{\mathtt b}^i\, b^{\mathtt a}_i + \tilde d^{\mathtt a}_{\mathtt i}\, d_{\mathtt b}^{\mathtt i} - \tfrac{N_+-N_-}2\, \delta^{\mathtt a}_{\mathtt b}\,,
	\end{aligned}
	\label{eq:genUmn}
\end{equation}
\begin{equation}
    \begin{aligned}
    R_j{}^i & = \tilde a^a_j\, a^i_a - 
    \tilde b_{\mathtt a}^i\, b_j^{\mathtt a} + \tfrac{M_+-M_-}2\, \delta^i_j\,,\\
    R_{\mathtt j}{}^i & = a^i_a\, c^a_{\mathtt j} - \tilde b^i_{\mathtt a}\, \tilde d^{\mathtt a}_{\mathtt j}\,,
    \end{aligned}
    \qquad\,
    \begin{aligned}
    R_j{}^{\mathtt i} & = -\tilde a^a_j\, \tilde c^{\mathtt i}_a + 
    b^{\mathtt a}_j\, d^{\mathtt i}_{\mathtt a}\,,\\
    R_{\mathtt j}{}^{\mathtt i} & = -\tilde c^{\mathtt i}_a\, c^a_{\mathtt j} + \tilde d^{\mathtt a}_{\mathtt j}\, d^{\mathtt i}_{\mathtt a} - \tfrac{M_+-M_-}2\, \delta^{\mathtt i}_{\mathtt j}\,,
    \end{aligned}
    \label{eq:genUpq}
\end{equation}
and the common $U(1)$ center of $U(M_+,M_-)$ and 
$U(N_+,N_-)$ is generated by\footnote{Notice 
that the operators in \eqref{eq:genUmn} and \eqref{eq:genUpq}
do 
not have integer eigenvalues
when $N_+-N_-$ and $M_+-M_-$ are both odd.
In this case, 
the realizations of $U(N_+,N_-)$ and $U(M_+,M_-)$
are double-valued.}
\be 
    Z=\tilde a^a_i\,a^i_a-
    \tilde b_{\mathtt a}^i\,b_i^{\mathtt a}-
    \tilde c_a^{\mathtt i}\,c^a_{\mathtt i}+\tilde d^{\mathtt a}_{\mathtt i}\,d_{\mathtt a}^{\mathtt i} +\tfrac12\,(N_+-N_-)(M_+-M_-)\,.
\ee
The  maximal compact subgroups $U(M_+) \times U(M_-) \subset U(M_+,M_-)$ and $U(N_+) \times U(N_-) \subset U(N_+,N_-)$ are respectively generated by $X^a{}_b$ and $X^{\mathtt a}{}_{\mathtt b}$, and $R_j{}^i$ and $R_{\mathtt j}{}^{\mathtt i}$.
Note also that when $N_-=0$ or $M_-=0$, that is, when one
of the group, say $G$, in the pair $(G,G')$ is compact,
the Lie algebra of the dual group $G'$ 
can be decomposed into eigenspaces of the total number operator
with eigenvalues $+2, 0, -2$.
In other words, they do not have the mixed form of
(creation)$^2$+(annihilation)$^2$.
This property will prove useful when 
decomposing the Fock space into 
irreducible representations.

\section{Dual pairs of type $(O_N,Sp_{2M})$}
\label{sec:osc_model_O-Sp}

To describe the
dual pairs $(O_N,Sp_{2M})\subset Sp(2NM,\mathbb R)$ in detail,
we consider again
$MN$ pairs of operators.
In this case, it is more convenient to
use $2M N$ operators
without 
explicitly pairing them,
which we denote by
\be
y^I_A \, , \qquad 
A= 1, \ldots , N \, ,
\quad
I = 1, \ldots, 2M \, ,
\ee
and which satisfy
\begin{equation}
    [y^I_A, y^J_B] = E_{AB}\, \Omega^{IJ}\,,
\end{equation}
where $E_{AB}$ and $\O^{IJ}$ 
are symmetric and antisymmetric 
invertible matrices of 
dimension $N$ and $2M$, respectively.
Then, the reductive subgroups $O_N$ and $Sp_{2M}$ are generated respectively by 
\be 
    M_{AB}=U_\cW(\mathsf{M}_{AB})=\O_{IJ}\,y_{[A}^I\,y_{B]}^{J}\,,
    \qquad 
    K^{IJ}=U_\cW(\mathsf{K}^{IJ})=E^{AB}\,y^{(I}_A\,y^{J)}_B\,,
    \label{eq:gen O-Sp}
\ee 
where $U_\cW$ is the metaplectic representation,
and $\mathsf{M}_{AB}$ and $\mathsf{K}^{IJ}$
are the matrices with components
$(\mathsf{M}_{AB})^{CD}
=2\,\delta_{[A}^C\,\delta_{B]}^D$ and $(\mathsf{K}^{IJ})_{KL}=2\,\delta^{(I}_K\,\delta^{J)}_L$ generating $O_N$ and $Sp_{2M}$.
The commutation relations
of $M_{AB}$ are
\be 
    [M_{AB},M_{CD}]=
   2\left( E_{D[A}\,M_{B]C}
    -E_{C[A}\,M_{B]D}\right),
    \label{O CR}
\ee
whereas those of $K^{IJ}$ are
\be 
    [K^{IJ},K^{KL}]=
   2\, \big(\Omega^{I(K}\,K^{L)J}
    +\O^{J(K}\,K^{L)I}\big)\,.
\ee
Note here that we have not specified 
the signature of the flat metric $E_{AB}$.
Clearly, at the level of the complex Lie algebra, different $E_{AB}$ can
be all brought to the form $\delta_{AB}$ 
by suitably redefining the $y$-operators.
We leave the ambiguity of $E_{AB}$
at this stage since 
there is a preferable choice of $E_{AB}$
in each real form of $O_N$
to obtain a compact expression of their generators.

Various real forms of $O_N$ and $Sp_{2M}$ can be chosen
by assigning different reality conditions to the operators
$y_A^I$.
Such conditions can be straightforwardly deduced
from the anti-involution $\tilde\sigma$ 
associated with each real form by asking
\be 
    U_\cW(\mathsf{M}_{AB})^\dagger=
    U_\cW(\tilde\sigma(\mathsf{M}_{AB}))\,,
    \qquad 
    U_\cW(\mathsf{K}^{IJ})^\dagger=
    U_\cW(\tilde\sigma(\mathsf{K}^{IJ}))\,.
\ee 
See Appendix \ref{sec: real form} for 
the expression of $\tilde \sigma$
in each real forms.
In the following, we describe
all such real forms
by introducing their oscillator realizations.

\subsection{$\big(O(N_+,N_-), Sp(2M, \mathbb R)\big) \subset Sp(2M(N_++N_-), \mathbb R)$}

To identify the real forms $O(N_+,N_-)$ and $Sp(2M, \mathbb R)$, 
we set $E_{AB}=\eta_{AB}$, the flat metric of signature $(N_+,N_-)$,
and 
require
that the generators $M_{AB}$ and $K^{IJ}$
satisfy the reality conditions,
\begin{equation}
    (M_{AB})^\dagger = -M_{AB}\,,
    \qquad 
   (K^{IJ})^\dagger = J_{IK}\,K^{KL}\,J_{LJ},
\end{equation}
where the matrix $J_{IJ}$ satisfy $J_{IJ}=J_{JI}$, $J_{IJ}\,J_{JK}=\delta_{IK}$
and $J_{IJ}\,\O_{JK}=-\O_{IJ}\,J_{JK}$\,.\footnote{Note here that we could use the more familiar reality condition, 
\be 
    (K^{IJ})^\dagger = -K^{IJ}\,,
\ee 
which can be realized by the operators satisfying
\be 
    (y_A^I)^\dagger=i\,y_A^I\,.
\ee 
Even though the above seems simpler, it actually leads
to slightly more involved and unfamiliar expression for $K^{IJ}$
in terms of oscillators.
That is why we chose another, but equivalent, reality condition for $K^{IJ}$.}
The above can be realized by requiring the $y^I_A$ operators to obey 
the reality condition,
\begin{equation}
    (y^I_A)^\dagger = J_{IJ}\,y^J_A\,.
    \label{Opq real}
\end{equation}
In order to introduce the oscillator realization, we split the $A$ and $I$ indices
into $A=(a,\mathtt a)$ and $I=({\st+}i,{\st-}i)$ where
$a$ and $\mathtt a$ take respectively $N_+$ and $N_-$ values.
In this convention,
$\eta_{AB}=(\delta_{ab},-\delta_{\mathtt{ab}})$\,, $\Omega^{{\sst+}i\,{\sst-}j}=\delta^{ij}$
and $J_{{\sst+}i\,{\sst-}j}=\delta_{ij}$, and 
the condition \eqref{Opq real} becomes
\begin{equation}
    (y^{\pm i}_A)^\dagger = y^{\mp i}_A\,.
    \label{real y split}
\end{equation}
We therefore introduce 
\begin{equation}
    a^i_a = y^{{\sst +}i}_a\,,
    \qquad 
    \tilde a_i^a = y^{{\sst -}i}_a\,,
    \qquad 
    b_i^{\mathtt a} 
    = y^{{\sst -}i}_{\mathtt a}\,,
    \qquad 
    \tilde b_i^{\mathtt a} = y^{{\sst +}i}_{\mathtt a}\,,
\end{equation}
which obey 
\begin{equation}
    \tilde a_i^a = (a^i_a)^\dagger\,, \qquad 
    \tilde b^i_{\mathtt a} = (b_i^{\mathtt a})^\dagger\,,
\end{equation}
and the canonical commutation relation
\begin{equation}
    [a^i_a, \tilde a_j^b] = \delta^i_j\, \delta_a^b\,, \qquad 
    [b_i^{\mathtt a}, \tilde b^j_{\mathtt b}] 
    = \delta_i^j\, \delta^{\mathtt a}_{\mathtt b}\,.
\end{equation}
In terms of the canonical pairs $a$ and $b$,  
the generators $M_{AB}$ and $K^{IJ}$ read
\begin{equation}
    M_{ab} = \tilde a^a_i\,a^i_b
    - \tilde a^b_i\,a^i_a\,, \qquad 
    M_{a\mathtt b} = 
    \tilde a_i^a\, \tilde b^i_{\mathtt b} 
    -  a^i_a\, b^{\mathtt b}_i\,, \qquad 
    M_{\mathtt a\mathtt b} = \tilde b^i_{\mathtt b}\,b_i^{\mathtt a}
    -\tilde b^i_{\mathtt a}\,b_i^{\mathtt b}\,,
\end{equation}
and
\begin{equation}
    K^{{\sst+}i\,{\sst+}j} = a^i_a\, a^j_a - \tilde b^i_{\mathtt a}\, \tilde b^j_{\mathtt a}\,, 
    \qquad 
    K^{{\sst-}i\,{\sst-}j} = \tilde a^a_i\, \tilde a^a_j - b_i^{\mathtt a}\, b_j^{\mathtt a}\,,
\end{equation}
\begin{equation}
    K^{{\sst+}i\,{\sst-}j} = 
    \tilde a^a_j\, a^i_a - \tilde b_{\mathtt a}^i\, b_j^{\mathtt a} + \tfrac{N_+-N_-}2\, \delta^i_j\,.
\end{equation}
Since $O(N_+,N_-)$ has 4 connected components (2 components when $N_+$ or $N_-$ vanishes), we need to take into account 
the reflections,
\be 
    \cR_a=\prod_{i=1}^M \cR_{a}^i\,,
    \qquad \cR_{\mathtt a}
    =\prod_{i=1}^M \cR_{\mathtt a}^i\,,
\ee 
where $\cR_a^i$ and $\cR_{\mathtt a}^i$
flip the sign of $a_a^i$ and $b_{\mathtt a}^i$ respectively.

Notice that  $M_{ab}$ and $M_{\mathtt{ab}}$ generate the maximal compact subgroup $O(N_+) \times O(N_-)$ of $O(N_+,N_-)$, while 
$K^{{\sst+}i{\sst-}j}$ generate
the maximal compact
subgroup $U(M)$ of $Sp(2M, \mathbb R)$.
When $N_-=0$, no generator of $Sp(2M,\mathbb R)$ 
has the mixed form of (creation)$^2$+(annihilation)$^2$.
Again, this will prove useful
in decomposing the Fock space into
irreducible representations.

\subsection{$\big(O(N,\mathbb C), Sp(2M, \mathbb C)\big) \subset Sp(4NM, \mathbb R)$}
The real Lie groups $O(N, \mathbb C)$ and $Sp(2M, \mathbb C)$,
are obtained by starting with complex operators $y$
with complex conjugates $y^*$, obeying 
the commutation relations,
\begin{equation}
    [y^I_A{}^*, y_B^J{}^*] = \delta_{AB}\, \Omega^{IJ}\,,
    \qquad [y^I_A, y^J_B{}^*] = 0\,.
\end{equation} 
We require the reality condition,
\begin{equation}
    (y^I_A)^\dagger = J_{IJ}\, y^J_A{}^*\,, 
    \label{ONC real}
\end{equation}
so that
\begin{equation}
    (M_{AB})^\dagger = -(M_{AB})^*\,, \qquad 
    (K^{IJ})^\dagger = J_{IK}\, (K^{KL})^*\, J_{LJ}\,,
\end{equation}
and hence the generators of $O(N,\mathbb C)$, which are
given by
\begin{equation}
    M^\pm_{AB} := M_{AB} \pm (M_{AB})^*\,, \qquad
    K_\pm^{IJ} := K^{IJ} \pm (K^{IJ})^*\,,
    \label{eq:def_M_K_complex}
\end{equation}
satisfy
\begin{equation}
    (M^\pm_{AB})^\dagger = \mp\, M^\pm_{AB}\,, \qquad 
    (K_\pm^{IJ})^\dagger = \pm\, J_{IK}\, K_\pm^{IJ}\, J_{LJ}\,,
\end{equation}
and $M^+_{AB}$ and $K_+^{IJ}$ generate the subgroups $O(N)$ and
$Sp(2M,\mathbb R)$ of $O(N,\mathbb C)$ and
$Sp(2M,\mathbb C)$ respectively\,. To define the oscillator realization, we introduce 
\begin{equation}
    a_A^I := \tfrac{1}{\sqrt 2}\, \big(y_A^I +
    \eta_{IJ}^{\,}\, y_A^J{}^*\big)\,, \qquad 
    \tilde a^A_I = \tfrac{1}{\sqrt 2}\, \Omega_{IJ}\,
    \big(-y^J_A + \eta_{JK}^{\,}\, y^K_A{}^*\big)\,,
    \label{eq:conjugation_Sp2NC}
\end{equation}
where $\eta_{IJ} = (\delta_{{\sst+}i{\sst+}j},
-\delta_{{\sst-}i{\sst-}j})$ and satisfies
$J_{IK}\, \eta_{KJ} = \Omega_{IJ}$ so that 
\begin{equation}
    (a_A^I)^\dagger = \tilde a_I^A\,,\qquad
    (a^I_A)^* = \eta_{IJ}^{\,}\, a^J_A\,,\qquad
    [a^I_A, \tilde a_J^B] = \delta_A^B\, \delta^I_J\,.
\end{equation}
In terms of these oscillators, 
the generators $M_{AB}$ and $K^{IJ}$ read
\begin{equation}
    M_{AB} = \tfrac12\, \big(\tilde a^A_I\, a_B^I - 
    \tilde a^B_I\, a^I_A -  \tilde a^{A}_I\,
    \tilde a^{BI} - a_{AI}\, a^I_{B}\big)\,,
\end{equation}
and 
\begin{equation}
    K^{IJ} = \tfrac12\,\big(a^I_A\, a_A^J +
   \tilde a^{AI}\, \tilde a^{AJ} -
    2\, \tilde a^{A(I}_{\phantom K}\, a_A^{J)}\big)\,.
\end{equation}
Since $O(N,\mathbb C)$ has two connected components,
corresponding to elements of determinant $\pm1$ respectively,
we need to take into account the reflections,
\be 
    \cR_A=\prod_{I=1}^{2M}\,\cR_A^I\,,
\ee
where $\cR^I_A$ flips the sign of $a_A^I$.
The generators of the maximal compact subgroup
$O(N) \subset O(N, \mathbb C)$ read
\begin{equation}
    M_{AB}^+ = \tilde a^A_I\, a_B^I - \tilde a^B_I\, a^I_A\,,
\end{equation}
while the generators of $Sp(M) \subset Sp(2M, \mathbb C)$ are given by
\begin{equation}
    K^{IJ} - \eta_{IK}\,\eta_{JL}\, (K^{KL})^* = 
    \Omega^{K(I}\, \tilde a_K^A\, a_A^{J)}\,.
\end{equation}
Once again, we can notice that the maximal compact subgroups $O(N)$ and $Sp(M)$
are generated by operators of the form (creation) $\times$
(annihilation).

\subsection{$\big(O^*(2N), Sp(M_+,M_-)\big) \subset Sp(4N(M_++M_-), \mathbb R)$}

To identify the real forms $O^*(2N)$ and $Sp(M_+,M_-)$, we begin
with the  metric $E_{AB}=J_{AB}$
where $J_{AB}$ is the symmetric off-diagonal matrix defined, in terms of $A=({\sst+}a,{\sst-}a)$,  as 
\be 
    J_{{\sst\pm}a\,{\sst\pm}b}=0\,,
    \qquad 
    J_{{\sst \pm}a\,{\sst\mp}b}=\delta_{ab}\,.
\ee 
The reality conditions to be imposed
on $M_{AB}$ and $K^{IJ}$  are
\begin{equation}
    (M_{AB})^\dagger = -\O^{AC}\,M_{CD}\,\O^{DB}\,,
    \qquad 
   (K^{IJ})^\dagger = \Psi_{IK}\,K^{KL}\,\Psi_{LJ}\,,
   \label{eq:reality_condition_O*Sp}
\end{equation}
where $\Psi_{IJ}$ is an antisymmetric matrix given by
\be 
    \qquad \Psi_{{\sst\pm}i\,{\sst\pm}j}=0\,,
    \qquad 
     \Psi_{{\sst\mp}i\,{\sst\pm}j}=\pm\eta_{ij}\,.
     \label{MK III}
\ee 
with the flat metric $\eta_{ij}$ of 
signature $(M_+,M_-)$.
Here, we used  $I=({\st+}i,{\st-}i)$ 
where $\Omega^{{\sst+}i\,{\sst-}j}=\delta^{ij}$.
The conditions \eqref{eq:reality_condition_O*Sp}
can be realized by requiring the $y$ operators
to obey the reality condition,
\begin{equation}
    (y^I_A)^\dagger =\O^{AB}\,\Psi_{IJ}\,y^J_B\,.
\end{equation}
Splitting again
the $A$ and $i$ indices into
$A=({\st +}a,{\st-}a)$ and $i=(r,\mathtt r)$
so that 
$\O^{{\st+}a\,{\st-}b}=\delta^{ab}$
and $\eta_{ij}=(\delta_{rs},-\delta_{\mathtt{rs}})$,
the reality condition reads
\begin{equation}
    (y^{{\sst +}r}_{{\sst+}a})^\dagger =
    -y^{{\sst-}r}_{{\sst-}a}\,,
    \qquad 
    (y^{{\sst +}r}_{{\sst-}a})^\dagger =
    y^{{\sst-}r}_{{\sst+}a}\,, 
    \qquad 
    (y^{{\sst +}\mathtt r}_{{\sst+}a})^\dagger =
    y^{{\sst-}\mathtt r}_{{\sst-}a}\,,
    \qquad 
    (y^{{\sst +}\mathtt r}_{{\sst-}a})^\dagger = -y^{{\sst-}\mathtt r}_{{\sst+}a}\,.
\end{equation}
By repairing the symplectic indices as
$R=({\st+}r,{\st-}r)$ and $\mathtt R=({\st+}\mathtt r,{\st-}\mathtt r)$, 
we introduce
\be 
    a^a_R = \O_{RS}\, y^{S}_{{\sst-}a}\,,
    \qquad 
    \tilde a^R_a = y^R_{{\sst +}a}\,,
    \qquad 
    b_{\mathtt R}^a = \O_{\mathtt{RS}}\,y^{\mathtt S}_{{\sst+a}}\,,
    \qquad 
    \tilde b^{\mathtt R}_a = y^{\mathtt R}_{{\sst -}a}\,,
\ee 
where $\O_{{\sst-}r\,{\sst+}s}=\delta_{rs}$
and $\O_{{\sst-}\mathtt r\,{\sst+}\mathtt s}=\delta_{\mathtt{rs}}$.
These $2N(M_++M_-)$
pairs of oscillators 
satisfy
\be 
    [a_R^a,\tilde a^S_b]=\delta_{R}^S\,\delta^a_b\,,
    \qquad 
    [b_{\mathtt R}^a,\tilde b^{\mathtt S}_b]=\delta_{\mathtt R}^{\mathtt S}\,\delta^a_b
    \qquad 
    \tilde a^R_a=(a_R^a)^\dagger\,,
    \qquad 
    \tilde b^{\mathtt R}_a=(b_{\mathtt R}^a)^\dagger\,.
\ee
In terms of $a, b$ oscillators, 
the generators $M_{AB}$ read
\ba 
    & M_{{\sst+}a\,{\sst -}b} = \tilde a^R_a\, a_R^b
    -\tilde b^{\mathtt R}_b\, b_{\mathtt R}^a
    + (M_+-M_-)\, \delta_{ab}\,, \nn 
    & M_{{\sst+}a\,{\sst+}b} = 2\, \tilde a^R_{[a}\, \tilde a_{b]R}^{\,}
    + 2\, b^{\mathtt R[a}_{\,}\, b_{\mathtt R}^{b]}\,,
    \qquad 
    M_{{\sst-}a\,{\sst-}b} = 2\, a^{R[a}_{\,}\, a_R^{b]}
    + 2\, \tilde b^{\mathtt R}_{[a}\, \tilde b_{b]\mathtt{R}}^{\,}\,,
\ea
and $K^{IJ}$ has altogether 3 types of components given by
\be 
    K^{RS}=2\,
    \tilde a_a^{(R}\,a^{S)a}\,,
    \qquad 
    K^{\mathtt{RS}}=2\,
    \tilde b_a^{(\mathtt R}\,b^{\mathtt S)a}\,,
    \qquad 
    K^{R\mathtt R}=
    a^{Ra}\,b^{\mathtt Sa}+
    \tilde a^R_a\,\tilde b^{\mathtt R}_a\,,
\ee
where 
the $R, S$ and $\mathtt R, \mathtt S$ indices
are lowered and raised by
$\O_{RS}$, 
 $\O^{RS}$, and $\O_{\mathtt{RS}}$, 
$\O^{\mathtt{RS}}$ as in \eqref{eq:convention_sp_indices}.
Notice that $M_{{\sst+}a{\sst-}b}$ 
generate the maximal compact subgroup $U(N)$ of $O^*(2N)$, while $K^{RS}$ and 
$K^{\mathtt{RS}}$ 
generate the 
maximal compact subgroup $Sp(M_+) \times Sp(M_-)$ of $Sp(M_+, M_-)$.
When $M_-=0$, no generator of $O^*(2N)$ 
is of the form
(creation)$^2$ + (annihilation)$^2$.

\section{Representations of
compact dual pairs}
\label{sec:rep_compact}

In the previous sections, we have introduced
oscillator realizations for various dual pairs,
and made two observations:
\begin{enumerate} 

\item \label{1st pt} Generators of compact
(sub)groups $K \subseteq G$ of dual pairs $(G, G')$
are of the form
(creation)$\times$(annihilation). In other
words, the Lie algebra of $K$ is the
eigenspace of the total number operator
with eigenvalue 0.

\item \label{2nd pt} When $K=G$, the dual
group $G'$ does not have any
generators of the mixed form,
(creation)$^2$+(annihilation)$^2$.\footnote{The only exception
is the duality $(O^*(2)\cong U(1), Sp(M_+,M_-))$,
detailed in Section \ref{sec:O*2-Sp}.}
In other words, the Lie algebra of $G'$ 
can be decomposed into eigenspaces
of the total number operator with eigenvalues 
$+2,0,-2$:
\be 
    \mathfrak{g}'=\mathfrak{g}'_{+2}\oplus
    \mathfrak{g}'_{0}\oplus\mathfrak{g}'_{-2}\,.
    \label{g' decomp}
\ee 
The subalgebra $\mathfrak{g}'_0$ 
corresponds to the Lie algebra of the maximal compact subgroup 
$K'$ of $G'$.

\end{enumerate}

To analyze the representations of the dual pairs,
we first construct the Fock space $\cW$ using the oscillators in the usual manner:
\be 
    \cW=\bigoplus_{n=0}^\infty W_n\,,
    \qquad 
    W_n={\rm span}_{\mathbb C}\left\{\tilde a_{\mathcal I_1}\cdots \tilde a_{\mathcal I_n}\,|0\ra\right\}\,,
    \qquad 
    a_\mathcal I\,|0\rangle =0\,.
\ee 
Here, $\mathcal I$ is a collective index which
can be specified in each dual pairs,
and the number $n$ of $W_n$ stands for the total oscillator number. 
Since all the states of $\cW$ are positive definite,
all irreducible representations appearing in the decomposition of $\cW$ are unitary.
Obtaining this decomposition becomes
particularly simple when $K=G$ due to
the properties \ref{1st pt} and \ref{2nd pt}. From now on,
we assume that $G$ is compact.

From the property \ref{1st pt}, we know that
$G$ commutes with  the total number operator,
and hence the eigenspace $W_n$ can be
decomposed into several irreducible
representations $\pi_G(\zeta)$:
\begin{equation}
    W_n = \bigoplus_\zeta
    \bigoplus_{i=1}^{m_n^\zeta}\,
    V_{n,i}^\zeta\,,
\end{equation}
where the first sum runs over a finite
number of labels $\zeta$, $m_n^\zeta$ is
the multiplicity of $\pi_G(\zeta)$
in $W_n$, and $V^\zeta_{n,i}$ is the
representation space of $\pi_G(\zeta)$
in $W_n$ with the multiplicity label $i$.
The restriction to $\pi_G(\zeta)$
in $W_n$ therefore reads
\begin{equation}
    W_n\lvert_{\pi_G(\zeta)}
    = \bigoplus_{i=1}^{m_n^\zeta}
    V_{n,i}^\zeta\,.
\end{equation}
Consequently, a finite-dimensional
irreducible representation 
$\pi_G(\zeta)$ is realized in
infinitely many subspaces $W_n$:
\be 
    \cW|^G_\zeta = \bigoplus_{n\in N_\zeta}
    W_n\lvert_{\pi_G(\zeta)}
    =\bigoplus_{n\in N_\zeta}\bigoplus_{i=1}^{m_n^\zeta}
    V_{n,i}^\zeta\,,
\ee 
where $N_\zeta$ is the set of integers $n$
such that $\pi_G(\zeta)$ appears in
$W_{n}$, or equivalently the possible
numbers of oscillators with which a state
in $\pi_G(\zeta)$ can be realized.
All $V_{n,i}^\zeta$ with $n\in N_\zeta$
and $i=1,\ldots,m_n^\zeta$ have to be related
to one another by the dual group $G'$, since
the decomposition of $\cW$ under $G\times G'$
is multiplicity-free (i.e. $G'$ acts
irreducibly on $\cW\lvert_\zeta^G$). In particular,
the space $\bigoplus_{i=1}^{m_n^\zeta}
V_{n,i}^\zeta$ carries a $m^\zeta_n$-dimensional
representation of the maximal compact subgroup
$K'$ of $G'$. 

When $G'=K'$, its generators also commute
with the total number operator, and hence 
they preserve the subspaces $W_n$. Consequently,
the set $N_\zeta$ has only one element and 
the corresponding $m^\zeta_n$-dimensional representation $\zeta'$ of $G'$ is irreducible.
In this way, we find the duality between the representations $\zeta$ and $\zeta'$
for the dual pairs of the type $(K,K')$. 
In fact, the only such case is the $(U(M),U(N))$ duality.

When $G'\neq K'$, the set $N_\zeta$ will have
the form,
\be 
    N_\zeta = \{n_\zeta+2\,m\,|\,m\in \mathbb N\},
\ee 
since the generators $G'$ are quadratic in
the oscillators. Here, $n_\zeta$ is the
minimum oscillator number with which $\zeta$
is realized. Now, from the property
\ref{2nd pt}, we find that the Lie algebra of
$G'$ admits the decomposition \eqref{g' decomp}
and the lowering operators in $\mathfrak{g}'_{-2}$
annihilate the states $V_{n_\zeta,i}^\zeta$.
In this way, we demonstrate that
the dual representation $\zeta'$ of $G'$ 
is a lowest weight representation.\footnote{
Notice that in general, $\mathfrak{g}_{-2}$
does not contain all the lowering operators of
$\mathfrak{g}$, as some of them are also included
in $\mathfrak{g}_0$. However,
$\bigoplus_{i=1}^{m_{n_\zeta}^\zeta}
V_{n_\zeta,i}^\zeta$ forms a finite-dimensional
irreducible representation of the compact group
$K$, and is therefore itself a lowest weight
representation. This implies that it contains
a lowest weight vector, which is also a lowest
weight vector for $\mathfrak{g}$ due to the
fact that the Cartan subalgebra of the latter
coincides with the Cartan subalgebra of $K$.}

In the following, we will 
review the correspondences
between the representations of dual
pairs, in which at least one member
admits a compact real form.
This correspondence was originally derived in  \cite{Kashiwara1978} (see e.g. \cite{Adams2007}
for a review). Here, we rederive the same results
using the oscillator realizations presented above.
Notice that a similar treatment can be found
in Appendix B of \cite{Boulanger:2008kw},
though the discussion was at the level of
complex Lie algebras. Let us also mention that
in \cite{Gunaydin:1981yq, Gunaydin:1981dc, Bars:1982ep, Gunaydin:1984fk, Gunaydin:1998sw, Gunaydin:1985tc, Gunaydin:1990ag, Gunaydin:1984wc, Gunaydin:1999ci}, representations of
$U(N_+,N_-), Sp(2N,\mathbb R), O^*(2N)$
are studied using the same oscillator
realizations as the ones used in this paper.
However, in these references, the role of
the dual pairs and the correspondences of
their representations were not explored.

\subsection{$\big(U(M),U(N)\big)$}

Let us begin the analysis with the simplest
case, the $(U(M),U(N))$ duality. In this case,
the groups $U(M)$ and $U(N)$ are respectively 
generated by
\begin{equation}
	X^a{}_b  = \tilde a^a_i\, a_b^i
	+ \tfrac N2\, \delta^a_b\,,
	\qquad 
    R_j{}^i  = \tilde a^a_j\, a^i_a
    + \tfrac M2\, \delta^i_j\,,
    \label{XR compact}
\end{equation}
where $a,b=1,\ldots,M$ and $i,j=1,\ldots, N$.
The Fock space $\cW$ is generated by polynomials
in $\tilde a_i^a$, namely
\begin{equation}
    \cW = \bigoplus_{L \in \mathbb N} W_{L}\,,
    \qquad 
    W_{L} := {\rm span}_{\mathbb C}
    \big\{\tilde a^{a_1}_{i_1} 
    \dots \tilde a^{a_L}_{i_L}\,
    \lvert 0 \rangle\big\}\,.
\end{equation}
Any state  $\lvert \Psi \rangle \in W_{L}$ 
can be expressed as
\begin{equation}
    \lvert \Psi \rangle :=     
    \Psi^{i_1 \cdots  i_L}_{a_1 \cdots a_L}\, 
    \tilde a^{a_1}_{i_1} \cdots\,
    \tilde a^{a_L}_{i_L}\, \lvert 0 \rangle\,,
\end{equation}
in terms of a tensor $\Psi$ 
with the symmetry,
$\Psi^{\cdots i_k\cdots i_l\cdots}_{\cdots a_k\cdots a_l\cdots}
       =\Psi^{\cdots i_l\cdots i_k\cdots}_{\cdots a_l\cdots a_k\cdots}$\,.
The upper and lower indices of $\Psi$ carry tensor representations 
$\Box^{\otimes L}$ of $U(M)$ and $U(N)$ respectively,
where $\Box$ denotes their fundamental representation.

Let us first consider the $U(M)$ representation:
$\Box^{\otimes L}$ can be decomposed into
irreducible representations corresponding to the
Young diagrams with height not greater than $M$,
and we pick up the irreducible representation 
corresponding to the Young diagram,
\be 
    \bm\ell = (\ell_1, \dots, \ell_p)\,,
\ee 
where $\ell_k$ is the length of the $k$-th row
and $p$ is the height $h(\bm\ell)$ of $\bm\ell$,
and $L=|\bm\ell|:=\ell_1+\cdots+\ell_p$\,.
Since the $U(M)$ generators
can be realized in a Fock space 
with any constant shift, such as $\frac{N}2\,\delta^a_b$ in \eqref{XR compact},
we need to specify it together with  a Young diagram
when labeling the $U(M)$ representations.
Note that this shift only affects the representation 
of the diagonal $U(1)$ in $U(M)$,
but not the $SU(M)$ part.
Taking this into account, let us
label the $U(M)$ representation as
\be 
    [\bm\ell,\tfrac N2]_{U(M)},
\ee
which can be decomposed into $U(1)\times SU(M)$ representation as
\be 
    [\bm\ell,\tfrac N2]_{U(M)}
    =
    [|\bm\ell|+\tfrac{NM}2]_{U(1)}\otimes [\bm\ell]_{SU(M)}\,,
\ee 
where $|\bm\ell|+NM/2$ is the eigenvalue of
the $U(1)$ generator. 
 In this way, we can produce 
 all $U(M)$ representations
 using an oscillator realization.
The highest
weight of  $[\bm\ell,\tfrac{N}{2}]_{U(M)}$
simply reads
\be 
    (\ell_1+\tfrac N2,\ell_2+\tfrac N2,\ldots,\ell_M+\tfrac N2)\,.
\ee 
Now, let us see to which $U(N)$ representation 
the $[\bm\ell,\tfrac N2]_{U(M)}$ representation 
corresponds. We can do that by picking up a
particular state in the $[\bm\ell,\tfrac N2]_{U(M)}$ representation --- which consists of multiple states $|\Psi\ra$ in $W_L$ ---
and reading off the $U(N)$ representation from such states $|\Psi\ra$.
Let us pick up the lowest weight state 
in the $[\bm\ell,\tfrac N2]_{U(M)}$ representation.
By acting with the lowering operators ($X^a{}_b$ with $a<b$) of $U(M)$\,, 
we bring $|\Psi\ra$ to the lowest weight state of $[\bm\ell,\tfrac N2]_{U(M)}$, corresponding to the Young tableau,
\be
    \scriptsize
    \gyoung({1}_6{\cdots};1;1,{2}_6{\cdots};2,_6{\cdots},_6{\cdots},{p}_3{\cdots};p)\,.
    \label{Yngtab}
\ee 
Such states $|\Psi\ra$ satisfy 
\be 
    X^a{}_b\,|\Psi\ra=0\qquad 
    [1\le a<b\le p]\,.
    \label{LW cond}
\ee
To identify the solution space of the above condition,
it will be convenient to first re-express 
 the state $|\Psi\ra$ or the tensor $\Psi$ as
\begin{equation}
    \lvert \Psi \rangle = 
    \Psi^{i_1{\sst (\ell_1)}, 
    \ldots,i_l{\sst(\ell_p)}}\, 
    \tilde a^1_{i_1{\sst(\ell_1)}} 
    \cdots\, \tilde a^p_{i_p{\sst(\ell_p)}}\, 
     \lvert 0 \rangle\,,
\end{equation}
where  $\Psi^{\cdots\, i_k({\sst \ell_k})\,\cdots}$
and $\tilde a^k_{i_k({\sst \ell_k})}$ are defined by
\ba 
    && \Psi^{\cdots\, i_k({\sst \ell_k})\,\cdots}
    := \Psi^{\cdots\, \overbrace{\st i_k\cdots i_k}^{\ell_k}\,\cdots}:=\Psi^{\cdots\,(i_k^{1}\, \cdots\, i_k^{\ell_k})\,\cdots}\,,
    \nn  
    && \tilde a^k_{i_k({\sst \ell_k})}:=
    \overbrace{\tilde a^k_{i_k}\cdots \,\tilde a^k_{i_k}}^{\ell_k}:=
    \tilde a^k_{(i_k^{1}}\cdots\, \tilde a^k_{i_k^{\ell_k})}\,.
\ea
Then, the condition \eqref{LW cond} is translated to
\begin{equation}
  \Psi^{i_1{\sst(\ell_1)},\dots,
    i_{a}{\sst(\ell_{a})},
    \dots,i_{a}\, i_{b}{\sst(\ell_{b}-1)},
    \dots, i_p{\sst(\ell_p)}} = 0
    \qquad [1\le a<b\le p]\,.
\end{equation}
To rephrase the above condition in words, the symmetrization of the $\ell_a$ indices
in the $a$-th group with one index $i_b$ in the $b$-th 
group  vanishes identically.
This 
is nothing but the definition of 
the same Young diagram $\bm\ell$,
but this time it designates $U(N)$ representation,
\be 
   [\bm\ell,\tfrac M2]_{U(N)}\,.
\ee 
Therefore, we find the correspondence of the representations:
\be 
    [\bm\ell,\tfrac M2]_{U(N)} \qquad \longleftrightarrow
    \qquad 
     [\bm\ell,\tfrac N2]_{U(M)}\,,
     \label{eq:correspondence_U-U}
\ee 
where the height $h(\bm\ell)$ of the Young diagram should be bounded by both $M$ and $N$: $h(\bm\ell)\le {\rm min}\{M,N\}$.

\subsection{$\big(U(M_+,M_-),U(N)\big)$}
The dual pair $\big(U(M_+,M_-), U(N)\big)$ is realized as
\be
	X^a{}_b  = \tilde a^a_i\, a_b^i
	 + \tfrac N2\, \delta^a_b\,,\quad 
	X^{\mathtt a}{}_{\mathtt b}  = -\tilde b_{\mathtt b}^i\, b^{\mathtt a}_i -
	\tfrac{N}2\,\delta^{\mathtt a}_{\mathtt b}
	\,,
	 \quad 
	X^a{}_{\mathtt b} = 
	-\tilde a^a_i\, \tilde b_{\mathtt b}^i\,,
    \quad 
	X^{\mathtt a}{}_b  = a^i_b\, b_i^{\mathtt a}\,,
\ee 
\be 
 R_j{}^i  = \tilde a^a_j\, a^i_a - 
    \tilde b_{\mathtt a}^i\, b_j^{\mathtt a} + 
    \tfrac{M_+-M_-}2\, \delta^i_j\,,
\ee 
where $a,b=1,\ldots,M_+$, $\mathtt a, \mathtt b=\mathtt 1,\ldots,\mathtt{M}_-$ 
and $i,j=1,\ldots, N$.
The Fock space $\cW$ is 
generated by polynomials in two families of oscillators, namely
\begin{equation}
    \cW = \bigoplus_{L,M \in \mathbb N} W_{L,M}\,, \qquad 
    W_{L,M} := {\rm span}_{\mathbb C}\big\{\tilde a^{a_1}_{i_1} 
    \dots \tilde a^{a_L}_{i_L}\, \tilde b^{j_1}_{\mathtt a_1} \dots 
    \tilde b^{j_M}_{\mathtt a_M} \lvert 0 \rangle\big\}\,.
\end{equation}
Recall for a dual pair $(G, K')$ with $K'$
a compact Lie group, the realization of a
finite-dimensional irrep of $K'$ with the minimal
number of oscillators in $\cW$ also forms an
irreducible representation of the maximal compact
subgroup $K \subset G$. For $G=U(M_+,M_-)$, the
maximal compact subgroup is $U(M_+) \times U(M_-)$,
represented by $X^a{}_b$ and
$X^{\mathtt a}{}_{\mathtt b}$.
Let us pick up a $K=U(M_+) \times U(M_-)$ representation,
\be 
    [\bm\ell, \tfrac N2]_{U(M_+)} \otimes
    [\lowerbcal{m}, \tfrac N2]_{U(M_-)}\,.
    \label{Um Un}
\ee
According to the seesaw duality,
\be
\parbox{180pt}{
\begin{tikzpicture}
\draw [<->] (0,0) -- (1,0.8);
\draw [<->] (0,0.8) -- (1,0);
\node at (-1.5,1.2) {$U(M_+,M_-)$};
\node at (-1.5,0.4) {$\cup$};
\node  at (-1.5,-0.4) {$U(M_+) \times U(M_-)$};
\node at (2.2,-0.4) {$U(N)$};
\node at (2.2,0.4) {$\cup$};
\node at (2.2,1.2) {$U(N)\times U(N)$};
\end{tikzpicture}},
\ee
where $U(N) \times U(N)$ are 
generated by 
$\tilde a^a_j\, a^i_a + 
    \tfrac{M_+}2\, \delta^i_j$
    and $\tilde b_{\mathtt a}^i\, b_j^{\mathtt a} + 
    \tfrac{M_-}2\, \delta^i_j$\,,
the representation of $U(N) \times U(N)$
dual to \eqref{Um Un}
will be
\be 
    [\bm\ell,\tfrac{M_+}2]_{U(N)} \otimes
    [\lowerbcal{m},\tfrac{M_-}2]_{U(N)}\,.
\ee 
A lowest weight state $|\Psi\ra$ of 
\eqref{Um Un} can be expressed as
\be 
    |\Psi\ra 
    =\Psi^{i_1{\sst(\ell_1)},\dots, 
    i_p{\sst(\ell_p)}}_{j_1{\sst (\lowercal{m}_1)}, 
    \dots, 
  j_q{\sst(\lowercal{m}_q)}}\,
 \tilde a^{1}_{i_1{\sst (\ell_1)}}\cdots\, \tilde a^p_{i_p{\sst (\ell_p)}}\,
  \tilde b_{\mathtt 1}^{j_1{\sst (\lowercal{m}_1)}}\cdots \,\tilde b_{\mathtt q}^{j_q{\sst (\lowercal{m}_q)}} |0\ra\,,
\ee 
where the upper and lower indices of $\Psi$ 
have the symmetries of the Young diagram
$\bm\ell$ and $\lowerbcal{m}$ respectively,
with $h(\bm\ell)=p$ and $h(\lowerbcal{m})=q$.
Such a state is annihilated by
$X^{\mathtt b}{}_a$, as these generators
decrease the total oscillator number,
and $\lvert \Psi \rangle$ is a state with
the lowest number of oscillator in the irreducible
representations of $U(M_+,M_-)$ and $U(N)$.
This condition is translated to 
\begin{equation} 
    X^{\mathtt b}{}_a \lvert \Psi \rangle = 0 
    \qquad \Leftrightarrow \qquad
    \delta^{j_b}_{i_a}\, \Psi^{i_1{\sst(\ell_1)}, \dots, 
    i_p{\sst(\ell_p)}}_{j_1{\sst (\lowercal{m}_1)}, \dots,
  j_q{\sst(\lowercal{m}_q)}} = 0\,.
\end{equation}
In words, the $\Psi$ tensor is traceless 
in the sense that any contraction between
upper and lower indices vanishes identically.
This defines the irreducible representation $[\bm\ell \oslash
\lowerbcal{m}, \tfrac{M_+-M_-}2]_{U(N)}$ of $U(N)$ 
whose highest weight is
\begin{equation}
    (\ell_1, \dots, \ell_p, 0, \dots, 0, -\lowercal{m}_q,
    \dots, -\lowercal{m}_1) + \tfrac{M_+-M_-}2\, (1, \dots, 1)\,.
\end{equation}
See Appendix \ref{app:irrep_compact} for more details on general $U(N)$ representations.
In this way, we see that
the $U(M_+) \times U(M_-)$ representation $[\bm\ell,\tfrac N2]_{U(M_+)}
\otimes [\lowerbcal{m},\tfrac N2]_{U(M_-)}$
induces a lowest weight $U(M_+,M_-)$ representation,
while
the $U(N) \times U(N)$ representation $[\bm\ell,\tfrac{M_+}2]_{U(N)}
\otimes [\lowerbcal{m}, \tfrac{M_-}2]_{U(N)}$
is restricted to the $U(N)$ representation $[\bm\ell \oslash
\lowerbcal{m}, \tfrac{M_+-M_-}2]_{U(N)}$.
In the end, we find 
the following correspondence,
\be 
   \cD_{U(M_+,M_-)}\big(
    [\bm\ell,\tfrac N2]_{U(M_+)} \otimes
    [\lowerbcal{m},\tfrac N2]_{U(M_-)}\big)
   \qquad 
   \longleftrightarrow \qquad
   [\bm\ell \oslash \lowerbcal{m},\tfrac{M_+-M_-}2]_{U(N)}\,.
   \label{eq:corresp_U-U}
\ee 
For $M_+=M_-=1$, due to the isomorphism between
$U(1,1)$ and $U(1)\times SL(2,\mathbb R)$ representations,
we can write
\be 
    \cD_{U(1,1)}([\ell+\tfrac N2]_{U(1)} \otimes
    [\lowercal{m}+\tfrac N2]_{U(1)})
    = [\ell-\lowercal{m}]_{U(1)} \otimes
    \cD_{SL(2,\mathbb R)}(\ell+\lowercal{m}+N)\,,
\ee 
where $\cD_{SL(2,\mathbb R)}(h)$ denote the positive discrete
series representation of $SL(2,\mathbb R)$ with lowest weight
$h \in \mathbb N$.

Let us briefly comment on
an example, which is important in physics.
The double-cover of the four-dimensional
conformal group $\widetilde{SO}{}^+(2,4)$ is isomorphic to
$SU(2,2)$ so that the representation
$\cD_{U(2,2)}\big([(\ell_1,\ell_2),
\tfrac N2]_{U(2)}\otimes
[(\lowercal{m}_1, \lowercal{m}_2),
\tfrac N2]_{U(2)}\big)$
corresponds to the representation $[2h]_{U(1)} \otimes
\cD_{\widetilde{SO}{}^+(2,4)}\big(\Delta;s_1,s_2\big)$ with
\begin{equation}
    \begin{aligned}
    h & = & \tfrac12\,(\ell_1+\ell_2
    -\lowercal{m}_1-\lowercal{m}_2)\,, \\
    s_1 & = & \tfrac12\,(\ell_1-\ell_2
    +\lowercal{m}_1-\lowercal{m}_2)\,,
    \end{aligned}
    \qquad
    \begin{aligned}
    \Delta & = & \tfrac12\,(\ell_1+\ell_2
    +\lowercal{m}_1+\lowercal{m}_2)+N\,, \\
    s_2 & = & \tfrac12\,(\ell_1-\ell_2
    -\lowercal{m}_1+\lowercal{m_2})\,. \qquad
    \end{aligned}
\end{equation}
The $\widetilde{SO}{}^+(2,4)$ representations appearing in this duality
describe CFT$_4$ operators or
equivalently fields in AdS$_5$.
In light of the previous dictionary, it will be
convenient to parametrize the $U(N)$ irreps
$[\bm\ell\oslash\lowercal{m}]_{U(N)}$ appearing
in this correspondence as
\begin{equation}
    \bm\ell = (s_1+s_2+n,n)\,, \qquad
    \lowerbcal{m} = (s_1-s_2+k,k)\,,
    \label{eq:param}
\end{equation}
with $k,n \in \mathbb N$ and $s_i$ satisfying
\begin{equation}
    s_1, s_2 \in \tfrac12\, \mathbb Z\,,
    \qquad
    s_1 \ge \lvert s_2 \rvert\,.
\end{equation}
Moreover, $s_1$ and $s_2$ are either both integers
or bother half-integers.
Note that $s_1$ is positive
while $s_2$ can be positive or negative. The
correspondence \eqref{eq:corresp_U-U} then
reads
\ba
    &[(s_1+s_2+n,n) \oslash (s_1-s_2+k,k)]_{U(N)}&
    \nn 
    &\updownarrow &\nn
    &[2(s_2+n-k)]_{U(1)} \otimes
    \cD_{\widetilde{SO}{}^+(2,4)}\big(s_1+n+k+N;s_1,s_2\big)\,.&
\ea
For low values of $N$, the $U(N)$
labels \eqref{eq:param} 
are
restricted and the dual $\widetilde{SO}{}^+(2,4)$ 
representations
correspond to particular fields:
\begin{itemize}
\item $N=1$: We should require $k=n=0$
and $s_1=\pm s_2$, so that the
correspondence reads
\begin{equation}
    [\pm2s]_{U(1)} \qquad \longleftrightarrow
    \qquad [\pm 2s]_{U(1)} \otimes 
    \cD_{\widetilde{SO}{}^+(2,4)}\big(s+1;s,\pm s\big)\,.
\end{equation}
These $\widetilde{SO}{}^+(2,4)$ representations
describe four-dimensional massless
fields of helicity $\pm s$, also known
as singletons.

\item $N=2$: There are two types of $U(2)$ irreps
that can occur. 

The first one characterized by $n=k=0$, and for which
the correspondence reads
\begin{equation}
     [(s_1+s_2,s_2-s_1)]_{U(2)} \qquad \longleftrightarrow
    \qquad [2s_2]_{U(1)} \otimes \cD_{\widetilde{SO}{}^+(2,4)}(s_1+2;s_1,s_2)\,.
\end{equation}
These $\widetilde{SO}{}^+(2,4)$ representation correspond to CFT$_4$ conserved
current of spin $(s_1,s_2)$, or equivalently to
massless fields in AdS$_5$ with the same spin.

The second type is characterized by either
$k=0$ and $s_1=s_2=s$, or $n=0$ and $s_1=-s_2=s$,
and the correspondence reads
\ba
    [(2s+n,n)]_{U(2)}
    \quad &\leftrightarrow&
    \quad
    [2(s+n)]_{U(1)} \otimes
    \cD_{\widetilde{SO}^+(2,4)}
    \big(s+n+2;s,s\big)\,,\nn
    \,[(-n,-2s-n)]_{U(2)}
    \quad &\leftrightarrow&
    \quad
    [-2(s+n)]_{U(1)} \otimes
    \cD_{\widetilde{SO}^+(2,4)}
    \big(s+n+2;s,-s\big)\,.\qquad
\ea
These $\widetilde{SO}{}^+(2,4)$ representations correspond to all possible CFT$_4$ operators
with spin $(s,\pm s)$ and twist $\tau=\Delta-s_1\ge2$,
or equivalently massive AdS$_5$ fields with the
same spin.

As expected, these are all the possible
representations appearing in the decomposition
of the tensor product of two singletons,
which was given in \cite{Dolan:2005wy}.

\item $N=3$: In this case, the $U(3)$ 
irrep labelled by \eqref{eq:param}
appearing in the decomposition are
such that either $n=0$ or $k=0$,
and
the correspondence reads
\ba
 [(s_1+s_2+n,n,s_2-s_1)]_{U(3)}
\quad &\leftrightarrow&
    \quad
    [2(s_2+n)]_{U(1)} \otimes
    \cD_{\widetilde{SO}{}^+(2,4)}\big(s_1+n+3; s_1, s_2\big)\,,\nn 
 \,[(s_1+s_2,-n,s_2-s_1-n)]_{U(3)}
\quad &\leftrightarrow&
    \quad
    [2(s_2-n)]_{U(1)} \otimes
    \cD_{\widetilde{SO}{}^+(2,4)}\big(s_1+n+3; s_1, s_2\big)\,.\nn
\ea
These $\widetilde{SO}{}^+(2,4)$ representations
describe massive fields in AdS$_5$
with spin $(s_1,s_2)$ or CFT$_4$
operators with twist $\tau \ge 3$.

\item $N \ge 4$: No restriction is
to be imposed on the $U(N)$ labels
\eqref{eq:param} and the dual representations
correspond to massive AdS$_5$ fields of any
spin or CFT$_4$ operators with twist
$\tau \ge N$.
\end{itemize}
Notice that in \cite{Bars:1982ep,Gunaydin:1998sw}
the oscillator realization of $SU(2,2|4)$
was obtained and  
some tensor product decompositions of
singletons were analyzed.
This dual pair correspondence was also
used in the context of AdS$_5$/CFT$_4$
for higher spin gauge theory in
\cite{Sezgin:2001ij}. 
More recently
its relevance as a possible tool
for computing scattering amplitudes
was pointed out in
\cite{Henning:2019mcv, Henning:2019enq}.

\subsection{$\big(O(N), Sp(2M, \mathbb R)\big)$}
\label{sec:O-Sp_correspondence}
The dual pair $\big(O(N), Sp(2M, \mathbb R)\big)$ 
is realized as 
\begin{equation}
    M_{ab} = \tilde a^a_i\,a^i_b
    - \tilde a^b_i\,a^i_a\,,
\end{equation}
\begin{equation}
    K^{{\sst+}i\,{\sst+}j} = a^i_a\, a^j_a \,, 
    \qquad 
    K^{{\sst-}i\,{\sst-}j} = \tilde a^a_i\, \tilde a^a_j \,,
    \qquad 
    K^{{\sst+}i\,{\sst-}j} = 
    \tilde a^a_j\, a^i_a+\tfrac{N}2\,\delta^i_j \,,
\end{equation}
where $a,b=1,\ldots N$ and $i,j=1,\ldots M$,
and
the Fock space $\cW$ is spanned 
by polynomials in one family of oscillators, namely
\begin{equation}
    \cW = \bigoplus_{L\in\mathbb N} W_L\,, \qquad 
    W_L := {\rm span}_{\mathbb C}\big\{\tilde a^{a_1}_{i_1} \dots \tilde a^{a_L}_{i_L}\, 
    \lvert 0 \rangle\,\}\,.
\end{equation}
Similarly to the previous section,
we can use the seesaw duality,
\be
\parbox{135pt}{
\begin{tikzpicture}
\draw [<->] (0,0) -- (1,0.8);
\draw [<->] (0,0.8) -- (1,0);
\node at (-1,1.2) {$U(N)$};
\node at (-1,0.4) {$\cup$};
\node  at (-1,-0.4) {$O(N)$};
\node at (2,-0.4) {$U(M)$};
\node at (2,0.4) {$\cup$};
\node at (2,1.2) {$Sp(2M,\mathbb R)$};
\end{tikzpicture}},
\ee
where $U(N)$ is generated by $\tilde a^a_i\,a_b^i$
and $U(M)$ is the maximal compact subgroup of $Sp(2M,\mathbb R)$ generated by $K^{{\sst+}i{\sst-}j}$.
The dual pair $(U(N), U(M))$ gives 
the correspondence of the representations,
\be 
    [\bm\ell,0]_{U(N)}
    \qquad \longleftrightarrow \qquad 
    [\bm\ell,\tfrac{N}2]_{U(M)}\,.
\ee 
This means that 
by considering the lowest weight state
of $[\bm\ell,0]_{U(M)}$ like \eqref{Yngtab},
the state $|\Psi\ra$ can be expressed as 
\begin{equation}
    \lvert \Psi \rangle = \Psi_{a_1{\sst(\ell_1)},
    \dots, a_p{\sst(\ell_p)}}\, 
    \tilde a_1^{a_1{\sst(\ell_1)}} \cdots \,\tilde a_p^{a_p{\sst(\ell_p)}}\, 
    \lvert 0 \rangle\,,
\end{equation}
and the lower indices of $\Psi$ 
have the symmetries of the Young diagram
$\bm\ell$  with $h(\bm\ell)=p$,
indicating the $[\bm\ell,0]_{U(N)}$ representation.
The lowest $K=U(M)$ condition on these states
reads
\begin{equation} 
    K^{{\sst+}i{\sst+}j}\,|\Psi\ra=0
    \qquad \Leftrightarrow \qquad 
    \delta^{a_ia_j}\, 
    \Psi_{a_1{\sst(\ell_1)},\dots, 
    a_p{\sst(\ell_p)}} = 0\,,
\end{equation} 
meaning that $\Psi$ is a traceless tensor.
The traceless condition imposes 
$\ell^1+\ell^2\le N$ besides $\ell^1\le N$.
The traceless Young diagram $\bm\ell$ specifies a unique
irreducible representation of $O(N)$:
\be 
    \cD_{Sp(2M,\mathbb R)}\big([\bm\ell,\tfrac N2]_{U(M)}\big)\qquad \longleftrightarrow \qquad 
    [\bm\ell]_{O(N)}\,,
    \label{eq:correspondence_O-Sp}
\ee 
where $\ell^1\le {\rm min}(M,N)$
and $\ell^1+\ell^2\le N$.

For $M=1$, we can use
the isomorphism $Sp(2,\mathbb R)\cong SL(2,\mathbb R)$ and find
the discrete series,
\be 
    \cD_{SL(2,\mathbb R)}(\ell+\tfrac N2)\qquad \longleftrightarrow \qquad 
    [(\ell)]_{O(N)}\,.
    \label{eq:correspondence_O-Sp}
\ee 
The $N=1,2$ and $3$ cases cover all available
discrete series representations.
In contrast with $(U(1,1),U(M))$ dual pairs,
the $(Sp(2,\mathbb R),O(N))$ pairs also give the
irreps with
half-integral lowest weights
for odd $N$,
which represent the double cover  $\widetilde{SL}(2,\mathbb R)$.

Notice that 
the above correspondence remains valid even for $N=1$,
where the $O(1) \cong \mathbb Z_2$ representation
just reduces to a sign $[\pm1]_{\mathbb Z_2}$,
and we find the duality,
\be 
    \cD_{Sp(2M,\mathbb R)}\big([(\bar n), \tfrac12]_{U(M)}\big)
    \qquad \longleftrightarrow 
    \qquad [(\bar n)]_{O(1)}=[(-1)^{\bar n}]_{\mathbb Z_2}\,,
\ee 
where $\bar n=0$ or $1$.
In terms of highest weight,  
the former gives 
\be 
    \cD_{Sp(2M,\mathbb R)}\big(\tfrac12,\tfrac12,\ldots,\tfrac12\big)
    \quad{\rm or}\quad 
    \cD_{Sp(2M,\mathbb R)}\big(\tfrac32,\tfrac12,\ldots,\tfrac12\big)\,,
    \label{eq:decompo_metaplectic}
\ee 
which is in accordance with the discussion of Section \ref{sec:metaplectic}, 
and in particular \eqref{eq:highest_weight_metaplectic}.

Let us briefly comment on
an example, which is important in physics.
The double cover of the three dimensional conformal group 
$\widetilde{SO}{}^+(2,3)$ is isomorphic to $Sp(4,\mathbb R)$. 
Consequently, the lowest weight representation of 
$\cD_{Sp(4,\mathbb R)}\big(\big[(\ell_1, \ell_2),
\tfrac N2]\big)$ corresponds to the $\widetilde{SO}{}^+(2,3)$
representation $\cD_{\widetilde{SO}{}^+(2,3)}(\Delta;s)$ with conformal
dimension/minimal energy $\Delta$ and
spin-$s$ given by 
\begin{equation}
    \Delta = \tfrac12\,(\ell_1+\ell_2+N)\,,
    \qquad
    s = \tfrac12\,(\ell_1-\ell_2)\,.
    \label{eq:translation_Sp_weights}
\end{equation}
Using this relation, we can interpret the
$Sp(4, \mathbb R)$ irreps appearing in
the dual pair correspondence
$\big(Sp(4,\mathbb R), O(N)\big)$ as a
spectrum of conformal fields in three
dimensions, or fields in AdS$_4$.
A convenient way to parametrize the
irreps $[\bm\ell]_{O(N)}$
appearing in
\eqref{eq:correspondence_O-Sp} is
\begin{equation}
    \bm\ell = (2s+n,n)\,,
\end{equation}
with $s\in\tfrac12\mathbb N$ and
$n\in\mathbb N$, so that the
correspondence reads
\begin{equation}
    [(2s+n,n)]_{O(N)}
    \qquad \longleftrightarrow \qquad
    \cD_{\widetilde{SO}{}^+(2,3)}\big(s+n+\tfrac N2;
    s\big)\,.
\end{equation}
For low values of $N$, not all two-row
Young diagrams define a representation
of $O(N)$:
\begin{itemize}
\item $N=1$: We find the correspondence,
\begin{equation}
    [(0)]_{O(1)}\quad \leftrightarrow \quad \cD_{\widetilde{SO}{}^+(2,3)}\big(\tfrac12;0\big)
    \qquad \text{and} \qquad
     [(1)]_{O(1)}\quad \leftrightarrow \quad
    \cD_{\widetilde{SO}{}^+(2,3)}\big(1;\tfrac12\big)\,,
\end{equation}
where the $\widetilde{SO}{}^+(2,3)$ representations
are
respectively isomorphic to the spin-$0$
singleton (or Rac) and spin-$\tfrac12$
singleton (or Di). Originally discovered
by Dirac \cite{Dirac:1963ta}, they
describe a free conformal scalar and
spinor respectively, in three dimensions.

\item $N=2$: The $O(2)$ irreps are
labelled by a Young diagram which is
either a single row of length $2s$  (i.e. $n=0$),
or a single column of height two (i.e.
$s=n=1$). For a single row diagram,
we have
\begin{equation}
    [(2s)]_{O(2)}
    \qquad \longleftrightarrow \qquad
    \cD_{\widetilde{SO}{}^+(2,3)}\big(s+1;s\big)\,,
\end{equation}
and the $\widetilde{SO}{}^+(2,3)$ representation describes a spin-$s$ conserved
current in three dimensions or a spin-$s$
massless field in AdS$_4$, for
$s > \tfrac12$. For $s=0$ (resp.
$\tfrac12$), this representation describes a
scalar (resp. spinor) CFT$_3$
operator / AdS$_4$ field with conformal
dimension / minimal energy $1$ (resp.
$\tfrac32$). For the single column diagram, we have 
\be 
 [(1,1)]_{O(2)}
    \qquad \longleftrightarrow \qquad
  \cD_{\widetilde{SO}{}^+(2,3)}\big(2;0\big)\,,
\ee
and the $\widetilde{SO}{}^+(2,3)$ representation describes
a scalar CFT$_3$ operator / AdS$_4$ field
with conformal dimension / minimal energy
$2$. As expected, this spectrum is that of
the decomposition of the tensor product of
two singletons, first derived in
\cite{Flato:1978qz} (later revisited and
extended to arbitrary dimensions in
\cite{Laoues:1998ik, Vasiliev:2004cm,
Dolan:2005wy}).

\item $N = 3$: In this case, we can only
consider $n=0$ or $n=1$ so that the
correspondence reads
\begin{equation}
    [(2s)]_{O(3)} 
    \qquad
    \longleftrightarrow \qquad
    \cD_{SO(2,3)}\big(s+\tfrac32;s\big)
\end{equation}
or
\begin{equation}
    [(2s+1,1)]_{O(3)}
   \qquad
    \longleftrightarrow \qquad
    \cD_{\widetilde{SO}{}^+(2,3)}\big(s+\tfrac52;s\big)\,.
\end{equation}
In other words, all operators of twist
$\tau=\tfrac32$ and $\tau=\tfrac52$ / 
massive fields with spin-$s$ and minimal
energy $s+\tfrac32$ or $s+\tfrac52$
appear.

\item $N \ge 4$: No restriction is to be imposed 
on the $O(N)$ labels and 
the corresponding $\widetilde{SO}{}^+(2,3)$ representations describe all
operators with twist $\tau=k+\tfrac N2$
/ massive fields with spin-$s$ and minimal
energy $s+k+\tfrac N2$.
\end{itemize}
Notice that the spectrum of the $N$-th
tensor product of three-dimensional
singletons was obtained 
already in
\cite{Bae:2016rgm} using the same oscillator
realization (see also
\cite{Gunaydin:1985tc, Govil:2013uta}
and references therein for earlier works).

\subsection{$\big(O^*(2N), Sp(M)\big)$}
The dual pair $\big(O^*(2N),Sp(M)\big)$
is realized as
\be
    M_{{\sst+}a\,{\sst -}b} = 
    \tilde a^R_{a}\,a_R^{b}
    +M\,\delta_{ab}\,, \qquad 
     M_{{\sst +}a\,{\sst +}b} = 2\,\tilde a^R_{[a}\,
     \tilde a_{b]R}\,,
    \qquad 
    M_{{\sst -}a\,{\sst -}b} = 2\, a^{R[a}\, a_R^{b]}\,,
    \nonumber
\ee
\be 
    K^{RS}=2\,\tilde a_a^{(R}\,a^{S)a}\,.
\ee
where $a,b=1,\ldots,n$ and $R,S=1,\ldots,2M$.
The Fock space $\cW$ is spanned by 
polynomials in one family of oscillators, namely
\begin{equation}
    \cW  = \bigoplus_{L\in\mathbb N} W_{L}\,, \qquad 
    W_{L} := {\rm span}_{\mathbb C}\big\{\tilde a^{R_1}_{a_1} \dots \tilde a^{R_L}_{a_L}\, 
    \lvert 0 \rangle\,\}\,.
\end{equation}
We use again the seesaw duality,
\be
\parbox{130pt}{
\begin{tikzpicture}
\draw [<->] (0,0) -- (1,0.8);
\draw [<->] (0,0.8) -- (1,0);
\node at (-1,1.2) {$O^*(2N)$};
\node at (-1,0.4) {$\cup$};
\node  at (-1,-0.4) {$U(N)$};
\node at (2,-0.4) {$Sp(M)$};
\node at (2,0.4) {$\cup$};
\node at (2,1.2) {$U(2M)$};
\end{tikzpicture}},
\ee
where $U(N)$ and $U(2M)$ are generated by $M_{{\sst+}a{\sst -}b}$
and $\tilde a^R_a\, a^a_S$ respectively.
The dual pair $(U(N),U(2M))$ gives
the correspondence,
\be 
    [\bm\ell,M]_{U(N)} \qquad 
    \longleftrightarrow 
    \quad [\bm\ell,0]_{U(2M)}\,.
\ee 
This means that by considering the lowest weight state
of $[\bm\ell,M]_{U(N)}$ like \eqref{Yngtab},
the state $|\Psi\ra$ can be expressed as 
\begin{equation}
    \lvert \Psi \rangle = \Psi_{R_1{\sst(\ell_1)},
    \dots, R_p{\sst(\ell_p)}}\,
    \tilde a_1^{R_1{\sst(\ell_1)}} \cdots \,\tilde a_p^{R_p{\sst(\ell_p)}}\, 
    \lvert 0 \rangle\,,
\end{equation}
and the indices of $\Psi$ 
have the symmetries of the Young diagram
$\bm\ell$  with $h(\bm\ell)=p$,
indicating the $[\bm\ell,0]_{U(2M)}$ representation.
The lowest $K=U(N)$ condition on these states
reads
\begin{equation}
    M_{{\sst-}a\,{\sst-}b}\, \lvert \Psi \rangle=0
   \qquad \Leftrightarrow \qquad 
    \O^{R_aR_b}\, \Psi_{R_1{\sst(\ell_1)},\dots, R_p{\sst(\ell_p)}} = 0\,,
\end{equation}
meaning that the tensor $\Psi$ is traceless
in the symplectic sense. The symplectic traceless
condition imposes the stronger condition $\ell^1\le M$
than the usual one $\ell^1\le 2M$. The symplectic
traceless Young diagram $\bm\ell$ specifies a
unique irreducible representation of $Sp(M)$,
hence we find the correspondence,
\be 
    \cD_{O^*(2N)}\big([\bm\ell,M]_{U(N)}\big)
    \qquad \longleftrightarrow \qquad 
    [\bm\ell]_{Sp(M)}\,,
    \label{eq:correspondence_O*Sp}
\ee 
where $p\le {\rm min}\{N,M\}$.

Let us consider the $N=1,2$ cases.
When $N=1$, we have the correspondence,
\be 
    [\ell+M]_{O^*(2)\cong U(1)}
    \qquad \longleftrightarrow \qquad 
    [(\ell)]_{Sp(M)}\,,
\ee
between the finite-dimensional representations,
due to the isomorphism $O^*(2)\cong U(1)$\,.
When $N=2$, we have  the isomorphism
$O^*(4)\cong (SU(2)\times SL(2,\mathbb R))/\mathbb Z_2$
where $SU(2)$ is the special subgroup of the
maximal compact subgroup $U(2)$,
and $SL(2,\mathbb R)$ is generated by
\ba  
    &&H=\left(M_{{\sst +}1{\sst-}1}
    +M_{{\sst +}2{\sst-}2}\right)
    =\tilde a^R_1\,a^1_R+\tilde a^R_2\,a^2_R+2M\,,
    \nn  
    &&E=M_{{\sst+}1{\sst+}2}
    =\tilde a^R_1\,\tilde a_{R2}\,,
    \nn  
    &&F=-M_{{\sst-}1{\sst-}2}
    =-a^{R1}\,a_R^2\,,
    \label{SL2 O*4}
\ea 
with the usual Lie brackets,
\be 
    [H,E]=2\,E\,,\qquad 
    [H,F]=-2\,F\,,
    \qquad 
    [E,F]=H\,.
    \label{sl2 LB}
\ee 
In terms of $SU(2)\times SL(2,\mathbb R)$ irreps,
the $O^*(4)$ representation reduces to
\be 
    \cD_{O^{*}(4)}\big([(\ell_1,\ell_2),M]_{U(2)}\big)
    =[\ell_1-\ell_2]_{SU(2)} \otimes 
    \cD_{SL(2,\mathbb R)}(\ell_1+\ell_2+2M)\,,
\ee 
where $[\ell_1-\ell_2]_{SU(2)}$
is the $(\ell_1-\ell_2+1)$-dimensional representation,
and with the understanding that $\ell_2=0$ when $M=1$.

Let us briefly comment on
an example, which is important in physics.
The double-cover of the
six-dimensional conformal group $\widetilde{SO}{}^+(2,6)$
and $O^*(8)$ are isomorphic. The lowest
weight representation $O^*(8)$ corresponds
to that of $\widetilde{SO}{}^+(2,6)$ as
\ba
 &[(s_1+s_2+s_3+n,
    s_1+n,
    s_2+n,
    s_3+n)]_{Sp(M)}&\nn 
    &\updownarrow& \nn
    &
    \cD_{\widetilde{SO}{}^+(2,6)}
    \big(s_1+s_2+s_3+2n+2M;s_1,s_2,\pm s_3\big)\,,&
    \label{O* iso}
\ea
where $n, s_i$ satisfy
\be
     n, s_1, s_2, s_3\in \tfrac12\,\mathbb{Z}\,,
    \qquad s_1 \ge s_2 \ge |s_3|\,,
    \qquad n\ge -s_3\,.
\ee 
and either $n,s_1,s_2,s_3$ are all integers or all half-integers.
The $\pm$ sign represents 
two possible isomorphisms 
between $O^*(8)$ and $\widetilde{SO}{}^+(2,6)$\,,
which reflects the possibility of parity transformation.
Note however
that an oscillator realization of $\widetilde{SO}{}^+(2,6)$
provide representations with $+$ or $-$ sign
but not both.

For low values of $M$, 
a few representations are allowed for $Sp(M)$:
\begin{itemize}
\item $M=1$: Only one row Young diagram is allowed for $\bm\ell$, and we find 
the correspondence,
\be 
    [(2s)]_{Sp(1)}\qquad 
    \longleftrightarrow\qquad 
    \cD_{\widetilde{SO}{}^+(2,6)}
    \big(s+2;s,s,\pm s\big)\,.
\ee
The $\widetilde{SO}{}^+(2,6)$
representations
describe the six-dimensional spin-$s$
singletons. Notice that only positive helicity
singletons appear in the decomposition of the Fock
space.

\item $M=2$: The height of $\bm\ell$ is at most two, and we find the correspondence,
\be 
[(s_1+s_2,s_1-s_2)]_{Sp(2)}\qquad 
    \longleftrightarrow\qquad 
\cD_{\widetilde{SO}{}^+(2,6)}\big(s_1+4;s_1,s_2,\pm s_2\big)\,.
\ee
The $\widetilde{SO}{}^+(2,6)$ representation describes a conserved current with spin
$(s_1,s_2)$ in six dimensions or a massless field
with the same spin in AdS$_7$. 
Here again, this
coincides with the spectrum of the tensor product
of two higher-spin singletons (of the same
chirality), which was worked out in
\cite{Dolan:2005wy}.

\item $M=3$: The height of $\bm\ell$ is at most three, and we find the correspondence,
\be 
    [(s_1+s_2,s_1-s_3,s_2-s_3)]_{Sp(3)}
    \qquad 
    \longleftrightarrow\qquad 
    \cD_{\widetilde{SO}{}^+(2,6)}\big(s_1+s_2-s_3+6;s_1,s_2,\pm s_3\big).
\ee 
The $\widetilde{SO}{}^+(2,6)$ representation describe CFT$_6$ operators with spin
$(s_1,s_2,\pm s_3)$ and conformal weight $s_1+s_2-s_3+6$
or massive fields in AdS$_7$ with the same spin.

\item $M \ge 4$: No restriction is to be imposed
on the $Sp(M)$ labels,
and the corresponding $\widetilde{SO}{}^+(2,6)$
describes all 6d operators with twist $\t\ge 2M$
or the corresponding massive AdS$_7$ fields.

\end{itemize}
The oscillator realization of
$O^*(8)$ in relation with AdS$_7$/CFT$_6$
was previously considered in e.g.
\cite{Gunaydin:1999ci, Fernando:2001ak,
Sezgin:2001ij}.

\section{Representation of
exceptionally compact dual pairs}
\label{sec:rep_excep_compact}

In this section, 
we consider two dual pairs 
wherein one member becomes compact or discrete
for an exceptional reason:
the Lie groups $O^*(2N)$ and $O(N,\mathbb C)$
become compact and discrete, respectively, only for $N=1$.
As we will discuss next, 
these cases require to consider
an outer automorphism of the embedding symplectic group
in order to make sense of the reductive dual pair correspondence.

\subsection{$\big(O^*(2), Sp(M_+,M_-)\big)$}
\label{sec:O*2-Sp}

The dual pair $\big(O^*(2), Sp(M_+,M_-)\big)\subset Sp(4(M_++M_-),\mathbb R)$
 is realized as 
 \begin{equation}
    M_{\sst+-} = \tilde a^R\, a_R
    -\tilde b^{\mathtt R}\, b_{\mathtt R} 
    + (M_+-M_-)\,.
\end{equation}
\be 
    K^{RS}=2\,
    \tilde a^{(R}\,a^{S)}\,,
    \qquad 
    K^{\mathtt{RS}}=2\,
    \tilde b^{(\mathtt R}\,b^{\mathtt S)}\,,
    \qquad 
    K^{R\mathtt R}=
    a^R\,b^{\mathtt R}+
    \tilde a^R\,\tilde b^{\mathtt R}\,,
\ee
where $R,S=1,\cdots,2M_+$ and $\mathtt R,\mathtt S=\mathtt1,\ldots, \mathtt{2M_-}$.
As noted above,
this is a rather exceptional case, 
where we still have a compact group in
the dual pair, due 
to the isomorphism $O^*(2) \cong U(1)$.
At first glance, this case
does not seem to suffice the conditions
of a reductive dual pair,
since the centralizer
of $U(1)\cong O^*(2)$ in $Sp(4(M_++M_-),\mathbb R)$
is not $Sp(M_+,M_-)$, but $U(2M_+,2M_-)$,
even though $U(1)$ is the centralizer
of $Sp(M_+,M_-)$.
We can distinguish
the dual pair $\big(O^*(2),Sp(M_+,M_-)\big)$ 
from $\big(U(1),U(2M_+,2M_-)\big)$ by
enlarging the embedding group $Sp(4(M_++M_-),\mathbb R)$
to $Sp(4(M_++M_-),\mathbb R)\rtimes \mathbb Z_2$,
where $\mathbb Z_2$ acts as
\be 
    \cP\,\begin{pmatrix}a_R\\ \tilde a^R
    \\ b_{\mathtt R}\\ \tilde b^{\mathtt R}\end{pmatrix}\,\cP^{-1}
    =\begin{pmatrix}-\tilde a_R\\ a^R
    \\ -\tilde b_{\mathtt R}\\ b^{\mathtt R}\end{pmatrix}\,,
    \label{Z2}
\ee 
(see the relevant discussion in Section \ref{sec:metaplectic}).
Then, we have the dual pair
$\big(O^*(2)\rtimes \mathbb Z_2, Sp(M_+,M_-)\big)
\subset Sp(4(M_++M_-),\mathbb R)\rtimes \mathbb Z_2$\,,
where $Sp(M_+,M_-)$ is the centralizer
of $O^*(2)\rtimes \mathbb Z_2$
as $U(2M_+,2M_-)$ is not invariant
under the $\mathbb Z_2$ action \eqref{Z2}.

To analyze the relevant representations, we 
begin with the seesaw duality,
\be
\parbox{180pt}{
\begin{tikzpicture}
\draw [<->] (0,0) -- (1,0.8);
\draw [<->] (0,0.8) -- (1,0);
\node at (-1,1.2) {$U(1)\times U(1)$};
\node at (-1,0.4) {$\cup$};
\node  at (-1,-0.4) {$U(1)$};
\node at (2.5,-0.4) {$Sp(M_+)\times Sp(M_-)$};
\node at (2.5,0.4) {$\cup$};
\node at (2.5,1.2) {$Sp(M_+,M_-)$};
\end{tikzpicture}},
\label{eq:seesaw_SpM+M-}
\ee
where $U(1)\times U(1)$ is generated
by $\tilde a^R\, a_R +M_+$
and $-\, (\tilde b^{\mathtt R}\, 
b_{\mathtt R}+M_-)$\,,
whereas $Sp(M_+)$ and $Sp(M_-)$ are
generated by
$K^{RS}$ and $K^{\mathtt{RS}}$\,.
The $(U(1)\times U(1), Sp(M_+)\times Sp(M_-))$ duality gives the correspondence,
\be 
    [\ell+M_+]_{U(1)}\otimes [\lowercal{m}+M_-]_{U(1)}
    \qquad \longleftrightarrow \qquad 
    [(\ell)]_{Sp(M_+)}\otimes [(\lowercal{m})]_{Sp(M_-)}\,,
    \label{eq:corresp_SpM+M-}
\ee 
where $\ell,\lowercal{m}\in \mathbb N$.
Any $U(1)\cong O^*(2)$ irrep simply consists
of an eigenvector of $M_{+-}$, 
\begin{equation}
    M_{\sst+-}\,|\Psi_{\pm k}\ra =(\pm k+M_+-M_-)\,
    \lvert \Psi_{\pm k} \rangle\,,
    \qquad [k \in \mathbb N]\,,
\end{equation}
and can therefore result from the tensor product
$[\ell+M_+]_{U(1)}\otimes [\lowercal{m}+M_-]_{U(1)}$ with
\be 
    \ell-\lowercal{m}=\pm\, k\,.
\ee
From the seesaw pair \eqref{eq:seesaw_SpM+M-}
and correspondence \eqref{eq:corresp_SpM+M-},
we therefore deduce that the dual 
$Sp(M_+,M_-)$ representation decomposes into
\be 
    \bigoplus_{m=0}^{\infty}
    [(k+m)]_{Sp(M_\pm)} \otimes [(m)]_{Sp(M_\mp)}\,,
\ee 
with Fock states,
\begin{equation}
    \lvert \Psi_{+k}
    \rangle = \sum_{m=0}^\infty\,
    \Psi_{R_1 \dots R_{k+m}|
    \mathtt{R}_1 \dots \mathtt{R}_m}\,
    \tilde a^{R_1} \dots \tilde a^{R_{k+m}}\,
    \tilde b^{\mathtt{R}_1} \dots
    \tilde b^{\mathtt{R}_m}\, \lvert 0 \rangle\,,
\end{equation}
and a similar expression for $\lvert \Psi_{-k}
\rangle$ upon exchanging the role of $\tilde a$
and $\tilde b$.
This $Sp(M_+,M_-)$ representation can be induced from 
$[(k)]_{Sp(M_\pm)}\otimes [0]_{Sp(M_\mp)}$ by the actions
of $K^{R\mathtt R}$,
\begin{eqnarray}
    K^{R\mathtt R}\ :\ 
    [(m_1)]_{Sp(M_+)}\otimes [(m_2)]_{Sp(M_-)}
    & \to & 
   \left([(m_1+1)]_{Sp(M_+)}\otimes [(m_2+1)]_{Sp(M_-)}\right)
    \\ && \qquad \oplus
   \left([(m_1-1)]_{Sp(M_+)}\otimes [(m_2-1)]_{Sp(M_-)}\right).
   \nonumber
\end{eqnarray}
As a consequence, we find the correspondence,
\be 
    \pi_{Sp(M_+,M_-)}([(k)]_{Sp(M_\pm)} \otimes  [(0)]_{Sp(M_\mp)})
    \qquad \longleftrightarrow \qquad 
    [\pm k+M_+-M_-]_{O^*(2)\cong U(1)} \qquad [k \in \mathbb N]\,,
\ee 
where $[(0)]_{Sp(M_\mp)}$
is the trivial
representation of $Sp(M_\mp)$.

For $M_+=M_-=1$, we have $\widetilde{SO}{}^+(1,4) \cong
Sp(1,1)$, so that we can
interpret the various representations
appearing in the previous correspondence
as fields around dS$_4$. The irreducible
representation $[\pm 2s]_{O^*(2)}$ with
$s \in \tfrac12\,\mathbb{N}$ corresponds
to the $Sp(1,1)$ representation induced from
$[(s\pm s)]_{Sp(1)} \otimes [(s\mp s)]_{Sp(1)}$.
Due to the isomorphism
$Sp(1) \times Sp(1) \cong \widetilde{SO}(4)$,
the latter irrep corresponds to
$[(s,\pm s)]_{\widetilde{SO}(4)}$,
and we find the correspondence,
\be 
    [\pm 2s]_{O^*(2)}
    \qquad\longleftrightarrow\qquad
    \pi_{\widetilde{SO}{}^+(1,4)}\big([(s,\pm s)]_{\widetilde{SO}(4)}\big)\,.
\ee 
The $\widetilde{SO}{}^+(1,4)$ representation
describes a massless field of helicity $\pm s$ 
in dS$_4$ \cite{Garidi:2003bg, Behroozi:2005md, Gazeau:2010mn}.
In fact, this representation 
lies in the discrete series:
it corresponds to
the irrep $\pi^\pm_{p,q}$ with $p=q=s$ of \cite{Dixmier1961}
and the irrep $D_{\ell\nu}^\pm$ with $\ell=s-1$ and $\nu=1$ in \cite{Dobrev:1977qv}.

\subsection{$\big(Sp(2N,\mathbb C),O(1,\mathbb C)\big)$ }

The groups of the dual pair $\big(Sp(2N,\mathbb C), O(1,\mathbb C)\big)\subset Sp(4N,\mathbb R)$
 are generated by 
\be 
    K^{IJ}=\tfrac12
    \left(\tilde a^I\,\tilde a^J
    +a^I\,a^J-2\,\tilde a^{(I}\,a^{J)}\right),
\ee
and its complex conjugate\footnote{Recall that $(a^I)^* = \eta_{IJ}\, a^J$, see \eqref{eq:conjugation_Sp2NC}.} $(K^{IJ})^*$ with $I,J=1,\cdots,2N$, and 
the reflection,
\be 
    \cR=\prod_{I=1}^{2N} \cR_I\,,
    \qquad 
    \cR \begin{pmatrix}a^I\\ \tilde a^I
   \end{pmatrix}\,\cR^{-1}
    =-\begin{pmatrix}a^I\\ \tilde a^I
   \end{pmatrix}\,.
\ee
Again the centralizer of $O(1,\mathbb C)$ is not 
$Sp(2N,\mathbb C)$ but $Sp(4N,\mathbb R)$ itself.
We can resolve this problem like the previous case of $(Sp(p,q),O^*(2))$ by 
enlarging the embedding group $Sp(4M,\mathbb R)$ 
with the $\mathbb Z_2$ action, 
\be 
    \cP\,\begin{pmatrix}a^I\\ \tilde a^I
   \end{pmatrix}\,\cP^{-1}
    =\begin{pmatrix}-\tilde a^I\\ a^I
   \end{pmatrix}\,.
    \label{Z2'}
\ee
In this way, we get
the dual pair $(Sp(2N,\mathbb C),O(1,\mathbb C)\times \mathbb Z_2)\subset 
Sp(4M,\mathbb R)\rtimes\mathbb Z_2$\,,
and $Sp(2N,\mathbb C)$
is the centralizer of $O(1,\mathbb C)\times \mathbb Z_2$\,.

Let us consider
the seesaw duality,
\be
\parbox{155pt}{
\begin{tikzpicture}
\draw [<->] (0,0) -- (1,0.8);
\draw [<->] (0,0.8) -- (1,0);
\node at (-1,1.2) {$Sp(2N,\mathbb C)$};
\node at (-1,0.4) {$\cup$};
\node  at (-1,-0.4) {$Sp(N)$};
\node at (2.2,-0.4) {$O(1,\mathbb C)\cong \mathbb Z_2$};
\node at (2.2,0.4) {$\cup$};
\node at (2.2,1.2) {$O^*(2)\cong U(1)$};
\end{tikzpicture}},
\ee
where $Sp(N)$ and $O^*(2)\cong U(1)$ is realized 
respectively by $\tilde a^{(I}\,a^{J)}$ and
$\tilde a^I\,a_I+N$. The duality
$(Sp(N),O^*(2)\cong U(1))$
gives
\be 
    [(\ell)]_{Sp(N)} \qquad \longleftrightarrow \qquad [\ell+N]_{U(1)}\,,
\ee 
where the $U(1)$ irrep can be branched into $O(1,\mathbb C)
\cong \mathbb Z_2$ irrep $\pm$ as
\be 
    [\ell+N]_{U(1)} |_{\mathbb Z_2}= [(-1)^\ell]_{\mathbb Z_2}\,.
\ee 
In the end, for a given 
$O(1,\mathbb C)$ irrep $[(\bar n)]_{O(1,\mathbb C)}= [(-1)^{\bar n}]_{\mathbb Z_2}$, with $\bar n=0$ or $1$,
the dual $Sp(2N,\mathbb C)$ irrep $\pi_{Sp(2N,\mathbb C)}(\bar n)$
can be decomposed into
\be 
    \pi_{Sp(2N,\mathbb C)}(\bar n)
    = \bigoplus_{n=0}^\infty
    [(2n+\bar n)]_{Sp(N)}\qquad [\bar n=0,1]\,,
\ee 
and we find the correspondence,
\be
    [(\bar n)]_{O(1,\mathbb C)}
    =[(-1)^{\bar n}]_{\mathbb Z_2}
    \qquad \longleftrightarrow \qquad 
    \pi_{Sp(2N,\mathbb C)}(\bar n)\,.
\ee

Notice that when $N=1$, we have the
isomorphism $Sp(2,\mathbb C)
\cong \widetilde{SO}{}^+(1,3)$\,,
and the irreps $\pi_{Sp(2,\mathbb C)}(0)$
and $\pi_{Sp(2,\mathbb C)}(1)$
correspond to the following irreps
in the classification \cite{Harish-Chandra1947} of Harish-Chandra:  
\be 
    \pi_{Sp(2,\mathbb C)}(0)\ \, :\ \,  
    (\mu,j)=(i\,\tfrac12\,,0)\,,\qquad\quad 
    \pi_{Sp(2,\mathbb C)}(1)\ \,  :\ \,     
    (\mu,j)=(0\,,\tfrac12)\,.
\ee
They sit in the complementary 
and the principal series, 
describing respectively a
conformally coupled scalar and spinor in dS$_3$.
Their tensor product will be considered
in the next section, where we analyze
the duality $(O(2,\mathbb C),Sp(2N, \mathbb C))$.
These representations
are also the relevant ones in
Majorana's infinite component
spinor equation \cite{Majorana:1968zz}.

\section{Representations of
the simplest non-compact dual pairs}
\label{sec:rep_non-compact}

When both of the groups in dual pairs are
non-compact, their representations have very
different features compared to the case where
at least one member is compact. In the latter
case, we have seen that
the representation space of $G$ or $G'$ is
spanned by 
states in the Fock space
with a certain excitation number, or in other
words, a state is produced by the action of 
a homogeneous polynomial of the creation
operators on the vacuum state. However,
in the case of non-compact dual pairs, none
of the vectors in the representation space are
realized as such polynomials. 
They instead consists of infinite linear
combinations of excitation states,
i.e. they are a certain kind of coherent states,
and  the norms of these states are divergent 
implying the underlying representations are tempered ones.

\subsection{$\big(GL(N,\mathbb R),GL(1,\mathbb R)\big)$
and $\big(GL(N,\mathbb R),GL(M,\mathbb R)\big)$}
The groups of the dual pair $\big(GL(N,\mathbb R),
GL(1,\mathbb R)\big)$ are respectively generated by
\begin{equation}
    X^A{}_B = \tfrac12\, \big(\tilde a^A\, a_B - 
    \tilde a^B\, a_A +  \tilde a^A\, \tilde a^B - 
    a_A\, a_B\big)\,,
\end{equation}
and
\be 
    Z= \tfrac12\,\big(\tilde a^A\, \tilde a^A - 
    a_A\, a_A\big),
\ee 
where $A,B=1,\ldots, N$.
For the analysis of the representations associated with this duality, let us consider 
the seesaw pair,
\be
\parbox{220pt}{
\begin{tikzpicture}
\draw [<->] (0,0) -- (1,0.8);
\draw [<->] (0,0.8) -- (1,0);
\node at (-1.2,1.2) {$GL(N,\mathbb R)$};
\node at (-1.2,0.4) {$\cup$};
\node  at (-1.2,-0.4) {$O(N)$};
\node at (3.5,-0.4) {$GL(1,\mathbb R)\cong \mathbb R^\times \cong 
\mathbb R^+\times \mathbb Z_2$};
\node at (3.5,0.4) {$\cup$};
\node at (3.5,1.2) {$Sp(2,\mathbb R)\cong SL(2,\mathbb R)$};
\end{tikzpicture}}\,,
\ee
where the maximal compact subgroup $O(N)$
is generated by $X^{A}{}_{B}-X^{B}{}_{A}=\tilde a^{A}\,a_{B}-\tilde a^{B}\,a_{A}$
and its dual $Sp(2,\mathbb R)\cong SL(2,\mathbb R)$ is generated by
\be 
    H=\tilde a^A\,a_A+\tfrac{N}2\,,
    \qquad E=\tfrac{1}{\sqrt{2}}\,\tilde a^A\,\tilde a^A\,,
    \qquad F=-\tfrac{1}{\sqrt{2}}\,a_A\,a_A\,,
\ee 
with the standard Lie bracket \eqref{sl2 LB}.
We begin with the correspondence,
\be 
    [(\ell)]_{O(N)}
    \qquad \longleftrightarrow \qquad 
    \cD_{SL(2,\mathbb R)}(\ell+\tfrac N2)
    \,,
    \label{ON SL2R}
\ee 
and consider the restriction
$SL(2,\mathbb R) \downarrow \mathbb R^+\times \mathbb Z_2$
where the $\mathbb R^+$ factor is generated by
\be 
    Z=\tfrac1{\sqrt{2}}(E+F)\,,
\ee 
and the $\mathbb Z_2$ is the reflection group
with the reflection element $\cR$ acting as
\be 
    \cR\,\binom{a_A}{\tilde a^A}\,\cR^{-1}
    =-\binom{a_A}{\tilde a^A}\,.
\ee 
The restriction of $SL(2,\mathbb R)$
to the $\mathbb Z_2$ determines
the parity of $\ell$.
The restriction of $SL(2,\mathbb R)$
to the $\mathbb R^+$ part amounts to finding 
the spectrum of $Z$ in the positive discrete
series representation $\cD_{SL(2,\mathbb R)}$.
To tell the conclusion first,
the spectrum is the entire set of pure imaginary numbers (as $Z$ is anti-Hermitian),
and any $Z$-eigenstate is an infinite linear combinations of $H$-eigenstates, 
namely a coherent state.
As is usual for a coherent state,
$Z$-eigenstates in $\cW$ do not have finite norms
and can be considered only as tempered distributions.
More detailed discussions
on this point
will be provided
below.
To summarize, for a fixed 
eigenvalue $i\,\zeta$ of the $Z$ generator of $\mathbb R^+$ with $\zeta \in \mathbb R$ and 
the sign $\pm$ for $\mathbb Z_2$,
the dual $GL(N,\mathbb R)$ representation
consists of all even/odd $O(N)$ tensors:
\be 
   \pi_{GL(N,\mathbb R)}(\zeta,\bar n):= \bigoplus^\infty_{m=0}
    [(2m+\bar n)]_{O(N)}
    \qquad \longleftrightarrow \qquad 
    [\zeta,\bar n]_{\mathbb R^\times}=[\zeta]_{\mathbb R^+}\otimes [(-1)^{\bar n}]_{\mathbb Z_2}\,.
\ee 
Note that the representation space of $GL(N,\mathbb R)$ does not depend on $\zeta$,
but only on $\bar n$ (which is 0 or 1). The $\zeta$ dependence can be seen only 
from the actions of $GL(N,\mathbb R)$.
This is a generic feature of a principal (or complementary) series representation
in contrast to the discrete series ones that appeared
for the dual pairs involving at least one compact group.

Let us find the explicit $GL(N,\mathbb R)$ action by
solving the $GL(1,\mathbb R)$ conditions,
\be 
    Z\,|\Psi_{\zeta,\bar n}\ra=\,i\,\zeta\,|\Psi_{\zeta,\bar n}\ra\,,
    \qquad 
    \cR\,|\Psi_{\zeta,\bar n}\ra=(-1)^{\bar n}\,|\Psi_{\zeta,\bar n}\ra\,.
    \label{GL1 cond}
\ee 
According to \eqref{ON SL2R}, a
vector in the representation $[(\ell)]_{O(N)}\otimes \cD_{SL(2,\mathbb R)}(\ell+\tfrac{N}2)$,
\be 
    \Yboxdim{10pt}
    {\tiny\gyoung(;{A_1}_4{\cdots};{A_\ell})}_{\sst O(N)}
    \otimes 
   |2n+\ell+\tfrac N2\ra,
\ee 
is realized by the following state in $\cW$,
\be 
    (\tilde a^B\,\tilde a^B)^n\,\tilde a^{\{A_1}\cdots \tilde a^{A_\ell\}}\,|0\ra\,,
\ee 
up to a normalization constant.
Here, $\{A_1\cdots A_\ell\}$ denotes the traceless part of the symmetrization $(A_1\cdots A_\ell)$.
In order to solve the conditions \eqref{GL1 cond},
we consider an infinite linear combination of Fock states,
namely a coherent state,
\be 
    |T^{A_1\cdots A_\ell}_\zeta \ra=f_{\zeta,\ell}(\tfrac12\,\tilde a^B\,\tilde a^B)\,\tilde a^{\{A_1}\cdots \tilde a^{A_\ell\}}\,|0\ra\,,
    \qquad f_{\zeta,\ell}(z)=\sum_{n=0}^\infty c_n^{\zeta,\ell}\,z^n\,,
\ee 
corresponding to an infinite linear combination of $H$-eigenstates,
\be 
    \Yboxdim{10pt}
    {\tiny\gyoung(;{A_1}_4{\cdots};{A_\ell})}_{\sst O(N)}\otimes 
   \left(\sum_{n=0}^\infty c_n^{\zeta,\ell}\,|2n+\ell+\tfrac N2\ra\right),
\ee
and ask them to satisfy
\be 
    Z\,|T^{A_1\cdots A_\ell}_\zeta\ra=i\,\zeta\,|T^{A_1\cdots A_\ell}_\zeta\ra\,,
    \qquad 
    \cR\,|T^{A_1\cdots A_\ell}_\zeta\ra=(-1)^\ell\,|T^{A_1\cdots A_\ell}_\zeta\ra\,.
    \label{Z egs}
\ee 
As a consequence, a general state $|\Psi_{\zeta,\bar n}\ra$ satisfying \eqref{GL1 cond}
will be given as a linear combination,
\be 
    |\Psi_{\zeta,\bar n}\ra=\sum_{n=0}^\infty
    \Psi_{A_1\cdots A_{2n+\bar n}}\,|T^{A_1\cdots A_{2n+\bar n}}_\zeta\ra\,.
\ee 
The $Z$-eigenstate condition \eqref{Z egs} reads
\be 
    \tfrac12(\tilde a^B\,\tilde a^B-a_B\,a_B)\,f_{\zeta,\ell}(\tfrac12\,\tilde a^C\,\tilde a^C)\, \tilde a^{\{A_1}\cdots \tilde a^{A_\ell\}}\,|0\ra
    =i\,\zeta\,f_{\zeta,\ell}(\tfrac12\,\tilde a^C\,\tilde a^C)\,\tilde a^{\{A_1}\cdots \tilde a^{A_\ell\}}\,|0\ra\,.
\ee 
The above can be translated into a differential equation\footnote{One may
also try to 
solve the $Z$-eigenstate condition starting from a generic state,
\be 
    |\Psi_{\zeta,\bar n} \ra=
    \Psi_{\zeta,\bar n}(\tilde a)\,|0\ra\,,
\ee 
where $\Psi_{\zeta,0/1}(\vec x)$ is a even/odd scalar function of an $N$ dimensional vector $\vec x$.
Then the condition of \eqref{Z egs}
reduces to the differential equation,
\be 
    (-\nabla^2+\vec x^2-2\,i\,\zeta)\,\Psi_{\zeta,\bar n}(\vec x)=0\,,
\ee 
having the form of the time independent Schr\"odinger equation
for $N$-dimensional harmonic oscillator with imaginary energy.
Since $\Psi_{\zeta,\bar n}(\vec x)$ is not a wave function
with $\cL^{2}$ scalar norm but a (coherent) Fock state,
the ``imaginary energy'' does not violate
unitarity.} for $h_{\zeta,\ell+\frac N2}(z)=f_{\zeta,\ell}(z)$,
\begin{equation}
    \left(z\, \partial_z^2+\alpha\, \partial_z
    -z + i\, \zeta \right)h_{\zeta,\alpha}(z)=0\,,
    \label{h eq}
\end{equation}
whose solution can be uniquely determined
upon imposing $h_{\zeta,\alpha}(0) = 1$ as
\ba
    h_{\zeta,\alpha}(z) \eq  e^{-z}\, _1F_1\big(\tfrac{\alpha-i\,\zeta}2;
    \alpha;2z\big)\nn 
    \eq \frac{\Gamma(\a)}{2^{\a-1}\,\Gamma(\frac{\alpha-i\,\zeta}2)\,\Gamma(\frac{\alpha+i\,\zeta}2)}
    \int_{-1}^1dt\, e^{tz}\,(1+t)^{\frac{\alpha-i\,\zeta}2-1}\,(1-t)^{\frac{\alpha+i\,\zeta}2-1}
    \,.
\ea
Note that the above function has the symmetry,
\be 
    h_{\zeta,\alpha}(-z)=h_{-\zeta,\alpha}(z)\,.
    \label{h fun sym}
\ee 
Let us examine the norm of the states 
$|T_\zeta^{A_1\cdots A_\ell}\ra$
with the simplest example of the scalar 
states with $\ell=0$\,:
\ba  
    \langle T_{\zeta'}|T_\zeta\ra
    \eq  \langle 0|\,h^*_{\zeta',\frac N2}(\tfrac12\,
    a_A\,a_A)\,h_{\zeta,\frac N2}(\tfrac12\,
    \tilde a_B\,\tilde a_B)\,|0\ra\nn 
    \eq \, 
    \frac{\Gamma(\frac N2)^2}{2^{N-2}\,
    |\Gamma(\frac N4+i\,\frac\zeta2)|^2\,|\Gamma(\frac N4+i\,\frac{\zeta'}2)|^2}
    \int_{-1}^1dt \int_{-1}^1ds\,
    (1+ts)^{-\frac N2}\,\times  \nn 
    &&\qquad \times (1+t)^{\frac N4+i\,\frac{\zeta'}2-1}\,(1-t)^{\frac N4-i\,\frac{\zeta'}2-1}
    (1+s)^{\frac N4-i\,\frac{\zeta}2-1}\,(1-s)^{\frac N4+i\,\frac{\zeta}2-1}\,,
\ea 
where we used 
\be 
    \langle 0|\,e^{ \frac t2\,
    a_A\,a_A}\,e^{ \frac s2\,
    \tilde a^B\,\tilde a^B}\,|0\ra
    =(1+ts)^{-\frac N2}\,.
\ee 
The above integral can be simplified by
performing the change of variables
$t=\tanh\tau$ and $s=\tanh\sigma$,
then $u=\s+\t$ and $v=(\s-\t)/2$. 
This leads to
\begin{eqnarray}
    \langle T_{\zeta'}|T_\zeta \rangle
    & = & \frac{\Gamma(\frac N2)^2}{2^{N-2}\,
    \lvert\Gamma(\frac N4+i\,\frac\zeta2)\rvert^2\,
    \lvert\Gamma(\frac N4+i\,\frac{\zeta'}2)\rvert^2}\,
    \int_{-\infty}^\infty  du\
    \frac{e^{i\,(\zeta+\zeta')\,
    \frac u2}}{\cosh^{\frac N2}u}\,
    \int_{-\infty}^\infty dv\
    e^{i\,v(\zeta-\zeta')} \nn  & = &
    \frac{2\,\pi\,\Gamma(\frac N2)}{2^{\frac{N-2}2}\,
    \lvert\Gamma(\frac N4+i\,\frac\zeta2)\rvert^2}\, \delta(\zeta-\zeta')\,.
\end{eqnarray}
The interested readers may consult
\cite{Nagel:1997rx} for additional details where the $N=1$ case
was treated explicitly
within the context of the $SL(2,\mathbb R)$ coherent states. 
The state 
$|T^{A_1\cdots A_\ell}_\zeta\ra$
has a divergent norm in the Fock space,
due to the factor 
$\delta(\zeta-\zeta')$ appearing in the
product of two states
$\lvert T^{A_1\cdots A_\ell}_\zeta \rangle$
and $\lvert T^{A_1\cdots A_{\ell'}}_{\zeta'} \rangle$.
This reflects the fact that states in different representations
are orthogonal to one another, but the Dirac distribution 
blows up when the two $GL(1,\mathbb R)$ representations coincides.
This is a usual property
of a coherent state,
and from the representation point of view, this means that the relevant representation is 
a tempered one.

Let us now spell out the $GL(N,\mathbb R)$ 
action on the basis $|T^{A_1\cdots A_\ell}_\zeta\ra$.
First of all, the action of
$GL(1, \mathbb R) \times O(N) \subset GL(N,\mathbb R)$
(where $GL(1,\mathbb R)$ is the center generated by $Z$)
is clear. The remaining generators are given by the
symmetric and traceless part of the $GL(N,\mathbb R)$ ones,
\begin{equation}
    X^{\{A}{}_{B\}}
    =\tilde a^{\{A}\,\tilde a^{B\}} -a_{\{A}\,a_{B\}}\,.
\end{equation}
Using the differential equations of  $f_{\zeta,\ell}(z)$
and the decomposition,
\be 
    \tilde a^{B}\,\tilde a^{\{A_1}\cdots \tilde a^{A_n\}}
    =\tilde a^{\{B}\,\tilde a^{A_1}\cdots \tilde a^{A_n\}}
    +\tfrac{n}{N+2(n-1)}\,
    \tilde a^2\,\delta^{B\{A_1}\,\tilde a^{A_2} \cdots
    \tilde a^{A_n\}}\,,
\ee 
one can derive the action of $X^{\{A}{}_{B\}}$
on $|T^{A_1\cdots A_\ell};\zeta\ra$ as
\ba 
   && X^{\{B_1}{}_{B_2\}}\,|T^{A_1\cdots A_\ell}_\zeta\ra
   =  p_\ell\,
   |T^{B_1\,B_2\,A_1\cdots A_\ell}_\zeta\ra \nn 
   &&\qquad \qquad +\, q_\ell\,
    \delta^{\{A_1|\{B_1}\,|T_{\zeta}^{B_2\}|A_2\cdots A_\ell\}}\ra
   +r_\ell\,\delta^{\{A_1|\{B_1}\,\delta^{B_2\}|A_2}\,|T^{A_3\cdots A_\ell\}}_\zeta\ra\,,
\ea 
where the coefficients $p_\ell, q_\ell$ and $r_\ell$ are functions of $\ell, N$ and $\zeta$\,: 
\be 
 p_\ell=  \tfrac{(N+\ell)(N+2\ell-4)+4\,\zeta}{(N+2\ell)(N+2\ell-2)}\,,
 \qquad 
 q_\ell=
 \tfrac{2\ell\,\zeta}{N+2\ell}\,,
 \qquad 
 r_\ell=-\tfrac{\ell^2(\ell-1)^2}{2(N+2\ell-2)(N+2\ell-4)}\,.
\ee 
In this way, we find explicit expression of the $GL(N,\mathbb R)$-representation
realized on the space of even or odd symmetric $O(N)$ tensors.

For $N=2$, the restriction
of $\pi_{GL(2,\mathbb R)}(\zeta,\bar n)$ to $SL(2,\mathbb R)$ gives 
the principal series representations, 
describing scalar/spinor tachyons in AdS$_2$
or massive scalar/spinor in dS$_2$.
Together with the discrete series
representations $\cD_{SL(2,\mathbb R)}(h+1)$ with $h\in \mathbb N$
appearing in the $(U(1,1),U(1))$ pair,
these are all
oscillator realizations
of $SL(2,\mathbb R)$
in the $(GL_2,GL_1)$ dual pairs.
Note  that the complementary 
series representations of $SL(2,\mathbb R)$
do not appear
in this oscillator realization.

In fact, the representation we described above becomes
much simpler by moving back 
to the $\omega_A$ and $\tilde \omega^A$ operators with the reality condition \eqref{omega glR},
namely,  the Schr\"odinger realization with real variables $x_A$\,:
\be 
    \omega_A=\frac{\partial}{\partial x^A}\,,\qquad \tilde\omega^A=x^A\,.
\ee 
In this realization, the $GL(N,\mathbb R)$ generators 
have the form,
\be 
    X^A{}_B=x^A\,\frac{\partial}{\partial x^B}+\frac12\,\delta_B^A\,,
    \qquad 
    Z=x^A\,\frac{\partial}{\partial x^A}+\frac{N}2\,.
\ee 
The exponentiation of the above gives the action of a group element $g\in GL(N,\mathbb R)$ as
\be 
    \langle x\,|\,
    U_\cW(g)\,|\Psi_{\zeta,\bar n}\ra 
    =|\det g|^{\frac 12}\,\langle x\,g\,|\Psi_{\zeta,\bar n}\ra
    \qquad 
    [\,g\in GL(N,\mathbb R)\,]\,,
    \label{right action G}
\ee 
whereas the dual $GL(1,\mathbb R)=\mathbb R^\times$ acts for an element $a$ as
\be 
 \langle x\,|\,U_\cW(h)\,|\Psi_{\zeta,\bar n}\ra 
    =h^{\frac N2}\,\langle h\,x\,|\Psi_{\zeta,\bar n}\ra
    \qquad [\,h\in GL(1,\mathbb R)\,]\,.
    \label{left action G}
\ee 
In the above, $x\,g$ should be understood as the right multiplication of the $N\times N$ matrix $g$ on the $N$-vector $x$, i.e.
$(x\,g)^A=x^B\,g_B{}^A$,
where
$g_A{}^B$ are the components of $g$.
The $Z$-eigenstate condition becomes the homogeneity condition,
\be 
    \langle x\,|\,U_\cW(h)\,|\Psi_{\zeta,\bar n}\ra ={\rm sgn}(h)^{\bar n}\,|h|^{i\,\zeta}\, \langle  x\,|\Psi_{\zeta,\bar n}\ra\,.
\ee 
Using the above condition, we can reduce the representation space to the space of  functions on $S^{N-1}$\,,
\be 
    \langle x\,|\Psi_{\zeta,\bar n}\ra 
    =|x|^{-\frac{N}2+i\,\zeta}\,\psi_{\bar n}(\hat x)\,, \qquad [\,\hat x^A=x^A/|x|\,]\,,
\ee 
where $\psi_{\bar n}(-\hat x)=(-1)^{\bar n} \psi_{\bar n}(\hat x)$\,.
In other words,
the representation space is the space of functions on
$\mathbb{RP}^{N-1}$\,.
The scalar product is inherited from that of $\cL^2(\mathbb R^N)$ as
\be  
    \langle \Psi_{\zeta,\bar n}|\Phi_{\rho,\bar m}\rangle
    = \int d^{N}x\, \langle x\,|\,\Psi_{\zeta,\bar n}\ra^*\,\langle x\,|\,\Phi_{\rho,\bar m}\rangle 
    = 2\pi\,\delta(\zeta-\rho)\,\delta_{\bar n\bar m}
    \int_{S^{N-1}}\!\!\! d^{N-1}\Omega\ 
    \psi_{\e}(\hat x)^*\,\phi_{\e}(\hat x)\,,
\ee
where the delta function arises from the radial integral,
\be 
    \int_0^\infty \frac{dr}{r}\,r^{-i(\zeta-\rho)}=
    \int_{-\infty}^\infty dt\,e^{-i(\zeta-\rho)\,t}
    =2\,\pi\,\delta(\zeta-\rho)\,.
\ee 
Recall that the same delta function appears also
in the Fock realization where we use the basis of
symmetric $O(N)$ tensors.
Let us also recall that the relation between 
a $\cL^2(\mathbb R^N)$ wave function 
and a Fock state $|\Psi\ra=\Psi(a^\dagger)\,|0\ra$
is simply
\be 
    \Psi(a^\dagger)\,|0\ra 
    \qquad \longleftrightarrow 
    \qquad 
    \langle q|\Psi\ra=\Psi\left(\frac{q-\partial_q}{\sqrt{2}}\right)\,e^{-\frac12\,q^2}\,,
\ee
and hence one can move from one realization to the other
using the above expression.
The Schr\"odinger realization of the $GL(N,\mathbb R)$
representation dual to $GL(1,\mathbb R)$ is
in fact  what is referred to as the most
degenerate principal series representation
(see e.g. \cite{Vilenkin2013,Kobayashi2011}). 

It is worth noting
that the most
degenerate principal series representation
belongs to a more general class of representations
 realized as functions on the space of
$M\times N$ real matrices $x$
with $M\le N$, endowed with the right action, 
\be 
    \langle x|\,U_\cW(g)\,|\Psi_{\zeta,\bar n}\ra 
    =|\det g|^{\frac M2}\,\langle x\,g\,|\Psi_{\zeta,\bar n}\ra
    \qquad [\,g\in GL(N,\mathbb R)\,]\,,
    \label{GL N act}
\ee 
and the determinant-homogeneity condition 
\cite{Vilenkin2013,Howe1999},
\be 
    \langle h^t\,x |\Psi_{\zeta,\bar n}\ra ={\rm sgn}(\det h)^{\bar n}\,|\det h|^{-\frac N2+i\,\zeta}\, \langle x|\Psi_{\zeta,\bar n}\ra\qquad  [\,h\in GL(M,\mathbb R)\,]\,.
    \label{GL M act}
\ee 
Here, $x\,g$ and $h^t\,x$ are the products
of the $M\times N$ matrix  $x$ with respectively 
the $N\times N$ matrix $g$, and 
the $M\times M$ matrix $h^t$
(where $h^t$ denotes the transpose of $h$),
i.e. $(x\,g)^I_A=x_I^B\,g_B{}^A$
and $(h^t\,x)_I^A=(h^t)_I{}^J\,x_J^A=x_J^A\,h^J{}_I$.
The above representation admits a natural interpretation in the dual pair correspondence $(GL(N,\mathbb R), GL(M,\mathbb R))$\,,
where $GL(N,\mathbb R)$ and $GL(M,\mathbb R)$ are 
generated by
\be 
    X^A{}_B=x^A_I\,\frac{\partial}{\partial x^B_I}+\frac M2\,
    \delta_B^A\,\qquad 
     R_I{}^J=x^A_I\,\frac{\partial}{\partial x^A_J}+\frac N2\,
    \delta_I^J\,,
    \qquad 
    Z=x^A_I\,\frac{\partial}{\partial x^A_I}+\frac{MN}2\,.
\ee 
Each of $GL(N,\mathbb R)$ and $GL(M,\mathbb R)$
are realized, respectively, by right and left actions on the space of $M\times N$ matrices,
though the action of $GL(M,\mathbb R)$ is trivial
except for its $GL(1,\mathbb R)$ subgroup:
\ba
	\langle x|\,U_\cW(h)\,|\Psi_{\zeta,\bar n}\ra\eq
   |\det h|^{\frac N2}\, \langle h^t\,x |\Psi_{\zeta,\bar n}\ra \nn
   \eq |\det h|^{i\,\zeta}\,{\rm sgn}(\det h)^{\bar n}\,\langle x|\Psi_{\zeta,\bar n}\ra
   \qquad [\,h\in GL(M,\mathbb R)\,]\,.
    \label{GL M left act}
\ea 
In other words, the degenerate principal series representation of $GL(N,\mathbb R)$
given by the determinant-homogeneity condition \eqref{GL M act} 
is dual to the $GL(M,\mathbb R)$ representation
where $SL(M,\mathbb R)$ part is trivial.
Similarly to the $(GL(N,\mathbb R), GL(1,\mathbb R))$ case,
we can reduce the representation space using the determinant-homogeneity condition \eqref{GL M act} 
to the space of functions on the real Grassmann manifold $Gr_{M,N}(\mathbb R)$, isomorphic to
\be 
    Gr_{M,N}(\mathbb R)\cong O(N)/(O(M)\times O(N-M))\,.
\ee
These $GL(N,\mathbb R)$ representations can be
induced by
the parabolic subgroup whose Levi factor is $GL(M,\mathbb R)\times GL(N-M,\mathbb R)\subset GL(N,\mathbb R)$\,,
with the one-dimensional  representation $(\zeta,\bar n)$ of $GL(M,\mathbb R)$
and the trivial representation of $GL(N-M,\mathbb R)$
\cite{Vilenkin2013}.
The restriction
of the degenerate principal series 
$\pi_{GL(N,\mathbb R)}(\zeta,\bar n;M)$
to $O(N)$ is \cite{Howe1999}
\be
    \pi_{GL(N,\mathbb R)}(\zeta,\bar n;M)=\bigoplus_{h(\bm\ell)\le M}
     [ (2\,\bm\ell+\bar n\,\bm 1_M) ]_{O(N)}\,,
\ee 
where $\bm 1_M=(\underbrace{1,\ldots,1}_M)$.
In this notation, 
the most degenerate principal series
 is $\pi_{GL(N,\mathbb R)}(\zeta,\bar n)=\pi_{GL(N,\mathbb R)}(\zeta,\bar n;1)$\,.

Finally, let us briefly mention that
the relevance of the above representations
for 3d scattering amplitude:
the scattering amplitude of $N_+$ incoming and 
$N_-$ outgoing 3d conformal fields
can be viewed as
the Poincar\'e singlet in
the tensor product of $N_+$ singletons
and $N_-$ contragredient singletons,
(or, $N_+$ lowest (energy) weight and $N_-$ highest (energy) weight singletons).
The situation can be controlled by the seesaw pair,
\be
\parbox{170pt}{
\begin{tikzpicture}
\draw [<->] (0,0) -- (1,0.8);
\draw [<->] (0,0.8) -- (1,0);
\node at (-1,1.2) {$Sp(4,\mathbb R)$};
\node at (-1,0.4) {$\cup$};
\node  at (-1,-0.4) {$GL(2,\mathbb R)$};
\node at (2.5,-0.4) {$O(N_+,N_-)$};
\node at (2.5,0.4) {$\cup$};
\node at (2.5,1.2) {$GL(N_++N_-,\mathbb R)$};
\end{tikzpicture}}\,,
\label{3d amplitude seesaw}
\ee
where $GL(2,\mathbb R)$ decomposes into the dilatation
$GL(1,\mathbb R)$ and the 3d Lorentz group $SL(2,\mathbb R)$,
wherein we require the latter to take the trivial 
representation (so that the amplitude is Lorentz invariant).
This is precisely the representation dual to the 
determinant-homogeneous degenerate principal series representation 
$\pi_{GL(N_++N_-,\mathbb R)}(\zeta,\bar n)$\,,
which should be restricted to the 
irrep of $O(N_+,N_-)$,
whose dual representation 
is translation invariant.
This irrep should be responsible for the form factors of the scattering amplitudes.
We shall provide more detailed analysis of
this point in a follow-up paper.

\subsection{$(GL(N,\mathbb C),GL(1,\mathbb C))$
and
$(GL(N,\mathbb C),GL(M,\mathbb C))$}

The groups of the dual pair $\big(GL(N,\mathbb C),GL(1,\mathbb C)\big)$
are respectively generated by
\begin{equation}
    X^A{}_B = \tfrac12\, \big(\tilde a^A\, a_B - 
    \tilde b_B\, b^A +  \tilde a^A\, \tilde b_B- 
    b^A\, a_B\big)\,,
\end{equation}
and
\be 
    Z_+=\tilde a^A\, \tilde b_A - 
    a_A\, b^A\,,
    \qquad  
    Z_-=\tilde a^A\, a_A - 
    \tilde b_A\, b^A\,,
\ee 
where $A,B=1,\dots, N$.
As $GL(1,\mathbb C)\cong \mathbb R^+\times U(1)$ is Abelian, its representations are
simply given by eigenvectors of $Z_+$ and $Z_-$,
generating respectively $\mathbb R^+$ and $U(1)$\,.
The
decomposition of the $GL(N,\mathbb C)$ representation
under its maximal compact subgroup $U(N)$ is given by
the seesaw pair,
\be
\parbox{205pt}{
\begin{tikzpicture}
\draw [<->] (0,0) -- (1,0.8);
\draw [<->] (0,0.8) -- (1,0);
\node at (-1.1,1.2) {$GL(N,\mathbb C)$};
\node at (-1.1,0.4) {$\cup$};
\node  at (-1.1,-0.4) {$U(N)$};
\node at (3.5,-0.4) {$GL(1,\mathbb C)\cong \mathbb C^\times \cong \mathbb R^+ \times U(1)$};
\node at (3.5,0.4) {$\cup$};
\node at (3.5,1.2) {$U(1,1)\cong SL(2,\mathbb R)\rtimes U(1)$};
\end{tikzpicture}}\qquad\,,
\label{eq:seesaw_GLC}
\ee
where $U(N)$ is generated by
$X^A{}_B-(X^B{}_A)^*
    =\tilde a^A\, a_B - 
    \tilde b_B\, b^A$ 
and 
the $SL(2,\mathbb R)$ and $U(1)$ subgroups of 
$U(1,1)\cong SL(2,\mathbb R)\rtimes U(1)$  are
generated respectively by
\be 
    H=\tilde a^A\,a_A+\tilde b_A\,b^A+N\,,
    \qquad 
    E=\tilde a^A\,\tilde b_A\,,
    \qquad 
    F=-a_A\,b^A\,,
\ee 
and
\be 
    J=\tilde a^A\, a_A - 
    \tilde b_A\, b^A\,.
\ee 
They are related to the
$GL(1,\mathbb C)$ generators through
\be
    Z_+=E+F\,,\qquad Z_-=J\,.
\ee
Let us consider an irreducible representation of $GL(1,\mathbb C)$
in the Fock space,
\begin{equation}
    Z_+\,|\Psi_{\zeta,m}\ra =i\,\zeta\,|\Psi_{\zeta,m}\ra,
    \qquad 
    Z_-\, \lvert \Psi_{\zeta,m} \rangle = m\,|\Psi_{\zeta,m}\ra,
\end{equation}
where $\zeta \in \mathbb R$ and $m \in \mathbb Z$.
Henceforth we assume $m\ge 0$ as the $m<0$ case can be treated analogously. 
According to the correspondence
\begin{equation}
    [(m)]_{U(1)} \otimes \cD_{SL(2,\mathbb R)}(m+2n+N) \qquad
    \longleftrightarrow \qquad [(m+n)\oslash(n)]_{U(N)}\,,
\end{equation}
between $U(1,1) \cong SL(2,\mathbb R) \rtimes U(1)$ and
$U(N)$ irreps and the seesaw pair \eqref{eq:seesaw_GLC},
the $GL(N,\mathbb C)$ representation $\pi_{GL(N,\mathbb C)}(\zeta,m)$
dual to $[\zeta,m]_{GL(1,\mathbb C)}$ can be decomposed into 
\begin{equation}
    \pi_{GL(N,\mathbb C)}(\zeta,m):=\bigoplus_{n=0}^\infty [(m+n)\oslash(n)]_{U(N)}\,.
\end{equation}
As a consequence, any state $\lvert \Psi_{\zeta,m} \rangle$ in 
$\pi_{GL(N,\mathbb C)}(\zeta,m)$ can be written as
\begin{equation}
    \lvert \Psi_{\zeta,m} \rangle = \sum_{n=0}^\infty
     \Psi_{A_1\cdots A_{m+n}}^{B_1\cdots B_n}\,\left(
    f_{\zeta,m,n}(\tilde a^C\,\tilde b_C)\,
   \tilde a^{A_1} \dots \tilde a^{A_{m+n}}\,
    \tilde b_{B_1} \dots \tilde b_{B_n}\, \lvert 0\rangle\right),
\end{equation}
where $\delta^{A_1}_{B_1}\, \Psi_{A_1\cdots A_{m+n}}^{B_1\cdots B_n}=0$
 and the functions 
 $h_{\zeta,m+2n+N}(z)=f_{\zeta,m,n}(z)$ satisfy the differential equation \eqref{h eq}, and can be uniquely determined with the condition $f_{\zeta,m,n}(0)=1$\,.
In this way, we find the coherent states carrying a $GL(N,\mathbb C)$ representation
dual to $[\zeta,m]_{GL(1,\mathbb C)}$.
In the case $m<0$, the representation is constructed in the
exact same way, except that the $a$ and $b$ oscillators are
exchanged.

Similarly to the $(GL(N,\mathbb R),GL(1,\mathbb R))$ case, 
the scalar product of the states $|\Psi_{\zeta,m}\ra$
and $\lvert \Psi_{\zeta',m'} \rangle$ is proportional to $\delta(\zeta-\zeta')$
implying that the relevant representations are tempered ones.
They are again known as the most degenerate principal series representations,
and we can see this either
by directly computing the norm using the precise form of $f_{\zeta,m,n}$ or
by switching back to the $\o_A$ and $\tilde \o^A$ operators 
with the reality condition \eqref{omega GLC}.
The latter amounts to using the Schr\"odinger realization with complex variables $z_A$\,:
\be
    \o_A=\frac{\partial}{\partial z^A}\,,\quad 
    \o_A{}^*=\frac{\partial}{\partial \bar z^A}\,,\quad 
    \tilde \o^A=z^A\,,
    \quad 
    \tilde \o^A{}^*=\bar z^A\,.
\ee 
In this realization,
the $Z_\pm$ generators read
\begin{equation}
    Z_+ = z^A \frac{\partial}{\partial z^A} + \bar z^A \frac{\partial}{\partial \bar z^A} + N\,,
    \qquad Z_- = z^A \frac{\partial}{\partial z^A} - \bar z^A \frac{\partial}{\partial \bar z^A}\,,
\end{equation}
while the $GL(N,\mathbb C)$ generators
are given by
\begin{equation}
    X^A{}_B = z^A\, \frac{\partial}{\partial z_B} + \frac12\, \delta^A_B\,,
\end{equation}
and their complex conjugate.
The associated group action reads
\ba 
    &\langle z \,|\,U_\cW(g)\,|\Psi_{\zeta,m}\ra 
    =|\det g|\,\langle z\,g |\Psi_{\zeta,\pm}\ra
    \qquad &[\,g\in GL(N,\mathbb C)\,]\,,\nn
   & \langle z \,|\,U_\cW(h)\,|\Psi_{\zeta,m}\ra 
    =|h|\,\langle h\,z\,|\Psi_{\zeta,\pm}\ra
    \qquad &[\,h\in GL(1,\mathbb C)\,]\,,
\ea
where $z\in \mathbb C^N$.
 The $Z$-eigenstate condition gives the homogeneity condition,
\be 
    \langle z\,| U_\cW(h)\,|\Psi_{\zeta,m}\ra =h^{\frac{i\,\zeta+m}2}\,\bar h^{\frac{i\,\zeta-m}2}\, \langle z\,|\Psi_{\zeta,m}\ra
    =|h|^{i\,\zeta}\,\left(\frac{h}{|h|}\right)^m\, \langle z|\Psi_{\zeta,m}\ra\,.
\ee 
Using the above condition, we can reduce the representation space to the space of functions on $S^{2N-1}$\,,
\be 
    \langle z|\Psi_{\zeta,m}\ra 
    =|z|^{-N+i\,\zeta}\,\psi_{m}(\hat z)\,, \qquad [\, |z|=\sqrt{z^A\,\bar z^A}\,,\quad \hat z^A=z^A/|z|\,]\,,
\ee 
where $\psi_m$ satisfies the $U(1)$ irrep condition,
\be 
    \psi_m(e^{i\,\phi}\,\hat z)=e^{i\,m\,\phi}\,\psi_m(\hat z)\,.
\ee 
If we mod out this $U(1)$ symmetry,
then the representation space is reduced to the space of functions on $\mathbb{CP}^{N-1} \cong S^{2N-1}/U(1)$\,,
The scalar product is inherited from that of $\cL^2(\mathbb C^N)$,
and reads
\ba  
    \langle \Psi_{\zeta,m}|\Phi_{\rho,n}\rangle
    \eq  \int d^{N}zd^N\bar z\, \langle z|\Psi_{\zeta,\e}\ra^*\,\langle z|\Phi_{\rho,\s}\rangle \nn 
    \eq  2\pi\,\delta(\zeta-\rho)\,\delta_{mn}
    \int_{S^{2N-1}}\!\!\! d^{2N-1}\Omega\ 
    \psi_{m}(\hat z)^*\,\phi_{m}(\hat z)\,.
\ea
This representation can be generalized to the space  of $M\times N$ complex matrices $z$
with $M\le N$, endowed with the right action of $GL(N,\mathbb C)$\,, 
\be 
    \langle z|\,U_\cW(g)\,|\Psi_{\zeta,m}\ra 
    =|\det g|^{M}\,\langle z\,g\,|\Psi_{\zeta,m}\ra
    \qquad [\,g\in GL(N,\mathbb C)\,]\,,
    \label{GL C N act}
\ee 
and the left action of $GL(M,\mathbb C)$\,,
\ba
	\langle z|\,U_\cW(h)\,|\Psi_{\zeta,m}\ra 
	\eq   |\det h|^N\,\langle h^t\,z\,|\Psi_{\zeta,m}\ra  \nn
	\eq |\det h|^{i\,\zeta}\, \left(\frac{\det h}{|\det h|}\right)^{\!m}\,\langle z\,|\Psi_{\zeta,m}\ra
	\qquad [\,h\in GL(M,\mathbb C)\,]\,,
    \label{GL C M act}
\ea
which can be interpreted as a determinant-homogeneity condition (see e.g. \cite{Howe1999}).
The $GL(N,\mathbb C)$ representation given in \eqref{GL C N act}
is dual to the one-dimensional representation $[\zeta,m]_{GL(1,\mathbb C)}\otimes \bm1_{SL(M,\mathbb C)}$  of the dual group $GL(M,\mathbb C)$ given in \eqref{GL C M act}.
At the same time, it can be also
induced by
the parabolic subgroup with Levi factor $GL(M,\mathbb C)\times GL(N-M,\mathbb C)\subset GL(N,\mathbb C)$
acting in the aforementioned one-dimensional representation for $GL(M,\mathbb C)$
and the trivial representation for $GL(N-M,\mathbb C)$.
Using the determinant-homogeneity condition \eqref{GL C M act},
we can reduce the representation space 
to the space of functions on the complex Grassmannian  $Gr_{M,N}(\mathbb C)$\,, isomorphic to
\be 
    Gr_{M,N}(\mathbb C)\cong U(N)/(U(M)\times U(N-M))\,.
\ee 
The restriction of the degenerate principal series $\pi_{GL(N,\mathbb C)}(\zeta,m,M)$ to $U(N)$ is 
given in \cite{Howe1999} for $M\le \frac{N}2$ as
\be
    \pi_{GL(N,\mathbb C)}(\zeta,m;M)=\bigoplus_{h(\bm\ell)\le M}
     [ (\bm\ell+\tfrac{|m|+m}2\,\bm1_M)\oslash (\bm\ell
     +\tfrac{|m|-m}2\,\bm1_M)]_{U(N)}\,.
\ee 
In this notation, 
the most degenerate principal series
 is $\pi_{GL(N,\mathbb C)}(\zeta,m)=\pi_{GL(N,\mathbb C)}(\zeta,m;1)$\,.

Like the $(GL(N,\mathbb R), GL(2,\mathbb R))$ case, the degenerate principal representations appearing
in the dual pair $(GL(N,\mathbb C), GL(2,\mathbb C))$
is relevant for the scattering amplitudes of 4d conformal fields. Consider the seesaw pair,
\be
\parbox{170pt}{
\begin{tikzpicture}
\draw [<->] (0,0) -- (1,0.8);
\draw [<->] (0,0.8) -- (1,0);
\node at (-1,1.2) {$U(2,2)$};
\node at (-1,0.4) {$\cup$};
\node  at (-1,-0.4) {$GL(2,\mathbb C)$};
\node at (2.5,-0.4) {$U(N_+,N_-)$};
\node at (2.5,0.4) {$\cup$};
\node at (2.5,1.2) {$GL(N_++N_-,\mathbb C)$};
\end{tikzpicture}}\,,
\ee
where we can apply exactly the same logic as 
the seesaw pair
for 3d amplitudes \eqref{3d amplitude seesaw}.
Again, more detailed analysis of this point will be provided in one of our follow-up papers.

\subsection{$\big(U^*(2N), U^*(2)\big)$
and $\big(U^*(2N), U^*(2M)\big)$}

The Lie algebra of $U^*(2N)$ is generated by
$X^A{}_B$ which can be split to the maximal
compact subgroup part,
\begin{equation}
    X^{(AB)} = \tilde a^{(A}_+\, a^{B)+}
    + \tilde a^{(A}_-\, a^{B)-}\,,
\end{equation}
and the rest,
\be 
    X^{[AB]} = \tilde a^{[A}_+\, \tilde a^{B]}_-
    -a^{[A+}\, a^{B]-}\,,
    \label{X[ab]}
\ee 
with $A,B=1,\dots,2N$, while the Lie algebra
of the dual group $U^*(2)$ is generated by
 \begin{equation}
    Z = \tilde a_+^A\, \tilde a_{A-}
    - a^{A+}\, a^{-}_A\,,
\end{equation}
\begin{equation}
    R_{++} =-J_+= -\tilde a^A_+\, a^-_A\,,
    \qquad 
    R_{--} =J_-= \tilde a^A_-\, a^+_A\,,
    \qquad 
    R_{+-} =J_3=\tfrac12\, \big(\tilde a^A_+\, a^+_A
    - \tilde a^A_-\, a^-_A\big)\,,
\end{equation}
where $Z$, and $J_3, J_\pm$ generate
respectively $\mathbb R^+$ and $SU(2)$ of 
$U^*(2)\cong \mathbb R^+\times SU(2)$\,.

Let us consider the seesaw dual pairs,
\be
\parbox{220pt}{
\begin{tikzpicture}
\draw [<->] (0,0) -- (1,0.8);
\draw [<->] (0,0.8) -- (1,0);
\node at (-1,1.2) {$U^*(2N)$};
\node at (-1,0.4) {$\cup$};
\node  at (-1,-0.4) {$Sp(N)$};
\node at (3,-0.4) {$U^*(2)\cong SU(2)\times \mathbb R^+$};
\node at (3,0.4) {$\cup$};
\node at (3.5,1.2) {$O^*(4)\cong (SU(2)\times SL(2,\mathbb R))/\mathbb Z_2$};
\end{tikzpicture}},
\ee
where $O^*(4)$
is generated by the $SU(2)$ generators $R_{IJ}$, 
and the $SL(2,\mathbb R)$ ones given by
\be   
    H = \tilde a^A_+\, a^+_A + \tilde a^A_-\, a^-_A + 2N\,,
    \qquad  
    E = \tilde a^A_+\,\tilde a_{A-}\,,
    \qquad 
    F =- a^{A+}\, a_A^{-}\,,
\ee 
with the standard Lie bracket \eqref{sl2 LB}.
We begin with the correspondence,
\be 
    [(m+n,n)]_{Sp(N)}
    \qquad \longleftrightarrow \qquad 
     [m]_{SU(2)}
     \otimes 
     \cD_{SL(2,\mathbb R)}(m+2n+2N)\,,
\ee 
where $n=0$ for $N=1$.
Since the two $SU(2)$'s in $O^*(4)$ 
and $U^*(2)$ are the same, 
the non-trivial part of the restriction $O^*(4) \downarrow U^*(2)$
is the restriction $SL(2,\mathbb R) \downarrow \mathbb R$\,,
where $\mathbb R$ is embedded in $SL(2,\mathbb R)$ as 
\be 
    Z=2\,(E + F)\,.
\ee 
Since the spectrum of $Z$ in $\cD_{SL(2,\mathbb R)}(h)$
is the entire set of
pure imaginary numbers,
we find that the 
$U^*(2N)$ representation $\pi_{U^*(2N)}(\zeta,m)$,
dual to $[\zeta]_{\mathbb R^+}\otimes[m]_{SU(2)}$, 
is decomposed as
\be 
    \pi_{U^*(2N)}(\zeta,m):=\bigoplus^\infty_{n=0}
    [(m+n,n)]_{Sp(N)}\,,
    \label{eq:decompo_U*}
\ee 
and it can be induced from
$[(m,0)]_{Sp(N)}$ by the actions of $X^{[AB]}$,
\be 
    X^{[AB]}\ :\ 
    [(\ell_1,\ell_2)]_{Sp(N)}
    \quad \longrightarrow 
    \quad 
    [(\ell_1+1,\ell_2+1)]_{Sp(N)}
    \oplus
    [(\ell_1-1,\ell_2-1)]_{Sp(N)}\,.
\ee 
In this way, we find 
the correspondence,
\be 
    \pi_{U^*(2N)}(\zeta,m)
    \qquad 
    \longleftrightarrow \qquad
    [\zeta]_{\mathbb R^+}\otimes[m]_{SU(2)}\,.
    \label{U*2n pi}
\ee 
Let us consider an irreducible 
representation of $U^*(2)$ in the Fock space,
\ba 
    & Z\,|\Psi^k_{\zeta,m}\ra=i\,\zeta\,|\Psi^k_{\zeta,m}\ra\,,\nn 
    & 
    J_3\,|\Psi^k_{\zeta,m}\ra=\tfrac{k}2\,|\Psi^k_{\zeta,m}\ra\,,
    \qquad 
    J_\pm\,|\Psi^k_{\zeta,m}\ra=\frac{\sqrt{(m\mp k)(m\pm k+2)}}2\,|\Psi^{k\pm2}_{\zeta,m}\ra\,.
\ea 
The decomposition \eqref{eq:decompo_U*} tells us that
$|\Psi^m_{\zeta,m}\ra$ can be written as
\begin{equation}
    \lvert \Psi^m_{\zeta,m} \rangle = \sum_{n=0}^\infty
    \Psi^m_{A_1 \dots A_{m+n},B_1 \dots B_n}
    \left(
    f_{\zeta,m,n}(\tilde a_+^C\,\tilde a_{C-})\,
    \tilde a^{A_1}_+\cdots \tilde a^{A_{m+n}}_+\,
    \tilde a^{B_1}_-\cdots \tilde a^{B_n}_- \lvert 0 \rangle\right),
\end{equation}
where $\Omega^{A_1\,B_1}\,\Psi^m_{A_1 \dots A_{m+n},B_1 \dots B_n}=0$
and the function $h_{\zeta,m+2n+2N}(z)=f_{\zeta,m,n}(z)$ satisfies the differential equation as \eqref{h eq}, and 
can be uniquely determined with the condition
$f_{\zeta,m,n}(0)=1$.
Other states $|\Psi^k_{\zeta,m}\ra$ with $k\neq m$ 
can be obtained by successive actions of $J_-$ on 
$|\Psi^m_{\zeta,m}\ra$.
Again, the scalar product between the states  $|\Psi^k_{\zeta,m}\ra$ is proportional to $\delta(\zeta-\zeta')$
and the relevant representations are the tempered ones.
They are the most degenerate principal series representations
of $U^*(2N)$,
similarly to the $GL(N,\mathbb R)$
and $GL(N,\mathbb C)$ cases.
We can see this by moving back to the $\o_A$ and $\tilde \o^A$ operators 
with the reality condition \eqref{U* real}.
The corresponding Schr\"odinger realization makes use of $4N$ complex variables $q^{\ve a}_{\e}$
where $\varepsilon, \epsilon=\pm$
and $a$ takes $N$ values:
\be 
    \o_{\varepsilon a}^{\e}=\frac{\partial}{\partial q^{\ve a}_{\e}}\,,
    \qquad 
    \tilde \o^{\varepsilon a}_{\e} = q^{\ve a}_{\e}\,,
\ee
with the reality conditions,
\be
	q^{+a}_{+}{}^*=q^{-a}_{-}\,,
	\qquad q^{+a}_{-}{}^*=-q^{-a}_{+}\,.
\ee
In terms of the above, the generators of $U^*(2N)$ and $U^*(2)$ are 
\begin{equation}
    X^{\varepsilon a}{}_{\epsilon b} = 
    q^{\ve a}_{\sigma}\, \frac{\partial}{\partial q^{\e b}_{\sigma}} + 
    \delta^a_b\,\delta^\varepsilon_{\epsilon}\,,
\qquad
    R_\varepsilon{}^{\epsilon} = q^{\s a}_{\varepsilon}\,
    \frac{\partial}{\partial q^{\s a}_{\epsilon}} + 
    N\,\delta_\varepsilon^{\epsilon}\,.
\end{equation}
Defining $2\times 2$ matrix,
\be
	q^a=\begin{pmatrix} q^{+a}_{+} & q^{-a}_{+} \\
	q^{+a}_{-} & q^{-a}_{-} \end{pmatrix},
\ee
 the $R_{\ve}{}^\e$ action on it can be written as 
\be
	p^{\ve}_{\epsilon}\, R_{\ve}{}^\e\, q^a
    =\begin{pmatrix} p^{+}_{+} & p^{-}_{+} \\
	p^{+}_{-} & p^{-}_{-} \end{pmatrix} \begin{pmatrix} q^{+a}_{+} & q^{-a}_{+} \\
	q^{+a}_{-} & q^{-a}_{-} \end{pmatrix}+N\,(p^+_++p^-_-)\,q^a
	=p\,q^a+N\,\tr(p)\,q^a,
\ee
Exponentiation of the above reads
\be
    \langle q\, |\,U_\cW(p)\,|\Psi^k_{\zeta,m}\ra 
    =|p|^{2N}\,\langle p\,q\,|\Psi^k_{\zeta,m}\ra
    \qquad [\,p\in GL(1,\mathbb H)\,]\,,
\ee
where $p\in U^*(2)=GL(1,\mathbb H)$,
and hence can be considered as a non-zero quaternion,
and $|p|=\sqrt{\det p}$ as the quaternionic norm.
Also the $U^*(2N)$ group action associated to $X^{\ve a}{}_{\e b}$
can be treated similarly,
 and it is more natural to view it as $GL(N,\mathbb H)$:
\be
    \langle q |\,U_\cW(g)\,|\Psi^k_{\zeta,m}\ra 
    =(\det g)^{2}\,\langle q\,g\, |\Psi^k_{\zeta,m}\ra
    \qquad [\,g\in GL(N,\mathbb H)\,]\,,
\ee
where $q\,g$ is the right multiplication of
the $N\times N$ quaternionic 
matrix $g$ 
on the $N$-dimensional quaternionic vector $q$, i.e.
$(q\,g)^a=q^b\,g_b{}^a$ where 
$g_b{}^a$ are quaternionic 
components of $g\in GL(N,\mathbb H)$\,. 
Note also that $\det g$ is the determinant of $g$ seen as 
an element of $U^*(2N)\subset GL(2N,\mathbb C)$,
which is a positive real number,
and $\langle  q\,\lvert \Psi^k_{\zeta,m} \rangle$ is a square integrable function on $\mathbb H^N$ taking values
in $\mathbb C$. 
The $Z$-eigenstate condition reads
\be 
     \langle q \,|\,U_\cW(p)\,|\Psi_{\zeta,m}^{k}\ra =
    |p|^{i\,\zeta}\,D_m(\hat p)^k_l\,
    \langle q\,|\Psi^l_{\zeta,m}\ra
    \qquad [\,p\in GL(1,\mathbb H)\,]\,,
\ee 
where $\hat p=p/|p|$ is a unit quaternion corresponding to an $SU(2)$ element
when viewed as a $2\times 2$ matrix,
and 
$D_m$ is the $(m+1)$-dimensional representation of $SU(2)$\,.
Using the above condition, we can reduce the representation space 
from the space of functions on $\mathbb H^N\cong \mathbb R^{4N}$
to the space of functions on $S^{4N-1}$\,,
\be 
    \langle q\,|\Psi^k_{\zeta,m}\ra 
    =|q|^{-2N+i\,\zeta}\,\psi^k_{m}(\hat q^a)\,, \qquad [\,|q|=\sqrt{q^a\,\bar q^a}\,,
    \quad \hat q^a=q^a/|q|\,]\,,
\ee 
where $\psi^k_m$ transforms as
\be 
  \psi^k_m(\hat p\,\hat q^a)=
    D_m(\hat p)^k_l\,\psi^l_m(\hat q^a)\,.
\ee 
If we mod out this $SU(2)$ symmetry,
the representation space is reduced to the space of functions on $\mathbb{HP}^{N-1} \cong S^{4N-1}/SU(2)$\,, 
The scalar product is inherited from that of $\cL^2(\mathbb H^N)$ as
\ba  
    \langle \Psi^k_{\zeta,m}|\Phi^l_{\rho,n}\rangle
    \eq  \int d^{4N}q\,\langle q\,|\Psi^{k}_{\zeta,m}\ra^*\,\langle q\,|\Phi^{l}_{\rho,n}\rangle \nn 
    \eq  2\pi\,\delta(\zeta-\rho)\,\delta_{mn}\,\delta^{kl}
    \int_{S^{4N-1}}\!\!\! d^{4N-1}\Omega\ 
    \psi^k_{m}(\hat q)^*\,\phi^l_{m}(\hat q)\,.
\ea
This can be generalized to the $(U^*(2N),U^*(2M))$ duality with $M\le N$,
whose representations are realized in the space  of $M\times N$ quaternionic matrices $q$\,,
endowed with the right action of $GL(N,\mathbb H)$\,, 
\be 
    \langle q\,|\,U_\cW(g)\,|\Psi_{\zeta}\ra 
    =(\det g)^{2M}\,\langle q\,g\,|\Psi_{\zeta}\ra
    \qquad [\,g\in GL(N,\mathbb H)\,]\,,
    \label{GL H N act}
\ee 
and the left action of $GL(M,\mathbb H)$\,,
\ba  
	\langle q\,|\,U_\cW(h)\,|\Psi_{\zeta}\ra 
    \eq (\det h)^{2N}\,\langle h^t\,q\,|\Psi_{\zeta}\ra\nn
    \eq     (\det h)^{i\,\zeta}\, 
    \langle q\,|\Psi_{\zeta}\ra
    \qquad [\,h\in GL(M,\mathbb H)\,]\,.
    \label{GL H M act}
\ea
The last equality can be interpreted as a determinant-homogeneity condition (see e.g. \cite{Lee2007}).
Note here that we do not have the additional label
$m$ besides $\zeta$ in contrast with the $(GL(N,\mathbb H),GL(1,\mathbb H))$ case.
This is due to the isomorphism
$GL(N,\mathbb H)\cong \mathbb R^+\times SL(N,\mathbb H)$,
where the last factor does not reduce to 
$SU(2)$ except for $GL(1,\mathbb H)$. 
The $GL(N,\mathbb H)$ representation given in \eqref{GL H N act}
is dual to the $[\zeta]_{GL(1,\mathbb H)}\otimes \bm 1_{SL(M,\mathbb H)}$
representation of the dual group $GL(M,\mathbb H)$ given in \eqref{GL H M act}.
The $GL(N,\mathbb H)$ representation can be also induced by
the parabolic subgroup with Levi factor $GL(M,\mathbb H)\times GL(N-M,\mathbb H)\subset GL(N,\mathbb H)$.
Using the determinant-homogeneity condition \eqref{GL H M act},
we can reduce the representation space 
to the space of functions on the quaternionic Grassmannian manifold $Gr_{M,N}(\mathbb H)$, isomorphic to
\be 
    Gr_{M,N}(\mathbb H)\cong Sp(N)/(Sp(M)\times Sp(N-M))\,.
\ee 
The restriction
of the degenerate principal series 
$\pi_{GL(N,\mathbb H)}(\zeta,M)$
to $Sp(N)$ is \cite{Lee2007}
\be
    \pi_{GL(N,\mathbb H)}(\zeta;M)=\bigoplus_{\substack{h(\bm \ell)\le 2M\\ \ell_{2i-1}=\ell_{2i}}}
     [ \bm\ell]_{Sp(N)}\,.
\ee 
In this notation, 
the most degenerate principal series
$\pi_{U^*(2N)}(\zeta,0)$ defined in \eqref{U*2n pi} corresponds to
$\pi_{GL(N,\mathbb H)}(\zeta;1)$\,.

Analogously to the $(GL(N,\mathbb R), GL(2,\mathbb R))$
and $(GL(N,\mathbb C), GL(2,\mathbb C))$ cases, the degenerate principal representations appearing
in the dual pair $(GL(N,\mathbb H), GL(2,\mathbb H))$
can be used for 
the scattering amplitudes of 6d conformal fields. 
The relevant seesaw pair is
\be
\parbox{170pt}{
\begin{tikzpicture}
\draw [<->] (0,0) -- (1,0.8);
\draw [<->] (0,0.8) -- (1,0);
\node at (-1,1.2) {$O^*(8)$};
\node at (-1,0.4) {$\cup$};
\node  at (-1,-0.4) {$GL(2,\mathbb H)$};
\node at (2.5,-0.4) {$Sp(N_+,N_-)$};
\node at (2.5,0.4) {$\cup$};
\node at (2.5,1.2) {$GL(N_++N_-,\mathbb H)$};
\end{tikzpicture}}\,,
\ee
where the same logic as for the seesaw pair
of the 3d amplitudes \eqref{3d amplitude seesaw} can be applied.
Again, we reserve more detailed analysis 
of this point for a follow-up paper.

\subsection{$\big(Sp(2N,\mathbb R),O(1,1)\big)$
and $\big(Sp(2N,\mathbb R),O(M,M)\big)$ }

The dual pair $(Sp(2N,\mathbb R),O(1,1))$
is realized as
\begin{equation}
    K^{{\sst+}i\,{\sst+}j} = a^i\, a^j - \tilde b^i\, \tilde b^j\,, 
    \qquad 
    K^{{\sst-}i\,{\sst-}j} = \tilde a_i\, \tilde a_j - b_i\, b_j\,,
    \qquad 
    K^{{\sst+}i\,{\sst-}j} = 
    \tilde a_j\, a^i - \tilde b^i\, b_j\,,
\end{equation}
\begin{equation}
   M_{\circ\bullet}
   = 
    \tilde a_i\, \tilde b^i
    -  a^i\, b_i\,,
\end{equation}
where $i,j=1,\ldots, N$.
We need to supplement  $O(1,1)$ by two reflections $\cR_a$ and $\cR_b$ which act as
\be
	\cR_a\,\binom{a^i}{\tilde a_i}\,\cR_a^{-1}=- \binom{a^i}{\tilde a_i}\,,
	\qquad 
	\cR_b\,\binom{b_i}{\tilde b^i}\,\cR_b^{-1}=- \binom{b_i}{\tilde b^i}\,,
\ee
and which form a finite group $\mathbb Z_2^a\times \mathbb Z_2^b$\,.
Let us consider
the seesaw pair,
\be
\parbox{220pt}{
\begin{tikzpicture}
\draw [<->] (0,0) -- (1,0.8);
\draw [<->] (0,0.8) -- (1,0);
\node at (-1,1.2) {$Sp(2N,\mathbb R)$};
\node at (-1,0.4) {$\cup$};
\node  at (-1,-0.4) {$U(N)$};
\node at (3.5,-0.4) {$O(1,1)\cong \mathbb R^+ \rtimes (\mathbb Z^a_2\times \mathbb Z^b_2)$};
\node at (3.5,0.4) {$\cup$};
\node at (3.5,1.2) {$U(1,1)\cong SL(2,\mathbb R)\rtimes U(1)$};
\end{tikzpicture}}\,.
\ee
The maximal compact subgroup
$U(N)$ of $Sp(2N,\mathbb R)$ is generated by $K^{{\sst+}i{\sst-}j}$,
while its dual $U(1,1)\cong SL(2,\mathbb R)\rtimes U(1)$
is generated by the $SL(2,\mathbb R)$ generators,
\be 
    H = \tilde a_i\, a^i + \tilde b^i\, b_i + N\,,
    \qquad E = \tilde a_i\, \tilde b^i\,,
    \qquad F = -a^i\, b_i\,,
\ee 
with the usual Lie bracket \eqref{sl2 LB},
and the diagonal $U(1)$ generator,
\be 
    Z=\tilde a_i\,a^i-
    \tilde b^i\,b_i\,,
\ee 
which is also the center $U(1)$
of $U(N)\subset Sp(2N,\mathbb R)$.
Note that $U(1)\subset SL(2,\mathbb R)$ and 
the diagonal $U(1)$\,, 
generated respectively by $H$ and $Z$\,, contain the discrete subgroups
$\mathbb Z^+_4$
and $\mathbb Z^-_4$ generated 
by the elements,
\be 
	\cP_+\,\begin{pmatrix}a^i\\ \tilde a_i \\ b_i \\ \tilde b^i \end{pmatrix}\,\cP_+^{-1}=\begin{pmatrix}i\,a^i\\ -i\,\tilde a_i \\ i\, b_i \\ -i\,\tilde b^i \end{pmatrix}\,,
	\qquad 
	\cP_-\,\begin{pmatrix} a^i\\ \tilde a_i \\ b_i \\ \tilde b^i \end{pmatrix}\,\cP_-^{-1}=\begin{pmatrix}i\,a^i\\ -i\,\tilde a_i \\ -i\,b_i \\ i\,\tilde b^i \end{pmatrix}\,.
\ee 
They are related to the finite subgroup $\mathbb Z_2^a\times \mathbb Z_2^b$ of $O(1,1)$ as 
\be 
    \cR_a=\cP_+\,\cP_-\,,\qquad \cR_b=\cP_+\,\cP^{-1}_-\,.
\ee 
The $(U(N),U(1,1))$ duality
has the correspondence of the representations,
\ba  
    [(m) \oslash (n)]_{U(N)}
    \qquad \longleftrightarrow \qquad 
    && \cD_{U(1,1)}\big([m+\tfrac N2]_{U(1)}
    \otimes [n+\tfrac N2]_{U(1)}\big)
    \nn 
    &&\cong\,\cD_{SL(2,\mathbb R)}(m+n+N)
    \otimes [m-n]_{U(1)}\,.
\ea
Noting that the isomorphism $O(1,1)\cong 
\mathbb R^+\rtimes (\mathbb Z_2^a\times
\mathbb Z_2^b)
\cong 
\mathbb E \times \mathbb Z_2$
where $\mathbb E$ is the one dimensional Euclidean group isomorphic to
$\mathbb R^+\rtimes \mathbb Z_2^a$
and $\mathbb Z_2$ is generated 
by $\cR=\cR_a\,\cR_b$\,,
we find that an irrep of $O(1,1)\cong \mathbb E \times \mathbb Z_2$ is
\be 
    \pi_{O(1,1)}(\mu,\bar n):=[\mu]_{\mathbb E}\otimes 
    [(-1)^{\bar n}]_{\mathbb Z_2}\qquad 
    [\mu\ge 0]\,,
\ee
where $[\mu]_{\mathbb E}$
is one or two dimensional space,
\ba 
    [0]_{\mathbb E}
    =[0]_{\mathbb R^+}\,,
    \qquad [\mu]_{\mathbb E}
    =[\mu]_{\mathbb R^+}
    \oplus [-\mu]_{\mathbb R^+}
    \quad [\mu\ge 0]\,.
\ea 
The  restriction
$U(1,1) \cong SL(2,\mathbb R)\rtimes U(1) \downarrow O(1,1)
\simeq \mathbb R^+ \rtimes 
(\mathbb Z_2^a\times \mathbb Z_2^b)$
amounts to embedding
$\mathbb R^+$ in $SL(2,\mathbb R)$ as
\be 
    M_{\circ\bullet}=E+F\,,
\ee 
then supplemented with the restriction 
to the subgroup $\mathbb Z_2$ generated by $\cR$.
The restriction $SL(2,\mathbb R) \downarrow \mathbb R^+$
again amounts to finding an
$M_{\circ\bullet}$-eigenstate in 
the positive discrete series representation
$\cD_{SL(2,\mathbb R)}(h)$.
For any $h$ and $M_{\circ\bullet}$-eigenvalue $i\,\mu$
with $\mu\in \mathbb R$,
one can find such a (coherent) state.
Taking into account the restriction
to the $\mathbb Z_2$ with eigenvalue $\bar n$,
we find that the representation space of $Sp(2N,\mathbb R)$ 
is 
\be 
    \pi_{Sp(2N,\mathbb R)}(\mu,\bar n):=\bigoplus_{
    \substack{m,n\ge 0\\ 
    {\rm even}\ m+n+\bar n}}
    [(m)\oslash(n)]_{U(N)}\,,
    \label{Sp U decomp}
\ee 
where the $\mu$-dependence does not appear explicitly.
For further specification, we can solve 
the $O(1,1)$ irrep condition,
\be 
   M_{\circ\bullet}\,|\Psi_{\mu,\bar n}^{\pm}\ra
    =\pm\,i\,\mu\,|\Psi_{\mu,\bar n}^{\pm}\ra\,,\qquad
     \cR_a\,|\Psi_{\mu,\bar n}^{\pm}\ra
    =|\Psi_{\mu,\bar n}^{\mp}\ra\,,
    \qquad 
    \cR\,|\Psi_{\mu,\bar n}^{\pm}\ra
    =(-1)^{\bar n}\,|\Psi_{\mu,\bar n}^{\pm}\ra\,,
    \label{O(1,1) irrep cond}
\ee 
by taking an infinite linear combination
of $U(N)$ representation states
as
\begin{equation}
    \lvert \Psi^+_{\m,\bar n} \rangle = \sum_{
    \substack{m,n\ge 0\\ 
    {\rm even}\ m+n+\bar n}}
    \Psi_{i_1\cdots i_{m}}^{j_1\cdots j_{n}}
    \left( f_{\m,m,n}(\tilde a^k\,\tilde b_k)\,
    \tilde a^{i_1}\cdots\tilde a^{i_{m}}\,
    \tilde b_{j_1}\cdots \tilde b_{j_{n}}\,|0\ra \right),
    \label{Sp2N result}
\end{equation}
where the $U(N)$ irrep condition imposes $\delta^{i_1}_{j_1}\,\Psi_{i_1\cdots i_{m}}^{j_1\cdots j_{n}}=0$
and the $M$-eigenstate condition is translated to
the differential equation
\eqref{h eq}
for $h_{\pm\mu,m+n+N}(z)=f_{\pm\m,m,n}(z)$, which can be uniquely fixed 
by $f_{\pm\mu,m,n}(0)=1$\,.
The other state
$|\Psi^-_{\mu,\bar n}\ra$
is, by definition,
 $\cR_a\,|\Psi^+_{\mu,\bar n}\ra$\,.

Note that 
the $(Sp(2N,\mathbb R),O(1,1))$ dual pair can serve also
for the tensor product between
a metaplectic representation 
and a contragredient metaplectic representation
of $Sp(2N,\mathbb R)$\,:
\be
    \cD_{Sp(2N,\mathbb R)}\big([(\bar m),\tfrac12]_{U(N)}\big)
    \otimes 
    \overline{\cD_{Sp(2N,\mathbb R)}}\big([(\bar n ),\tfrac12]_{U(N)}\big)
    =\int_{0}^\infty d\mu\     \pi_{Sp(2N,\mathbb R)}\big(\mu,1-\delta_{\bar m\bar n}\big)\,.
\ee
The above can be understood from the seesaw pair,
\be
\parbox{210pt}{
\begin{tikzpicture}
\draw [<->] (0,0) -- (1,0.8);
\draw [<->] (0,0.8) -- (1,0);
\node at (-2,1.2) {$Sp(2N,\mathbb R)\times Sp(2N,\mathbb R)$};
\node at (-2,0.4) {$\cup$};
\node  at (-2,-0.4) {$Sp(2N,\mathbb R)$};
\node at (2.2,-0.4) {$O(1)\times O(1)$};
\node at (2.2,0.4) {$\cup$};
\node at (2.2,1.2) {$O(1,1)$};
\end{tikzpicture}},
\ee
where the irrep
$[\mu]_{\mathbb E}\otimes [\s]_{\mathbb Z_2}$
of $O(1,1)$ 
branches to 
the irrep $[(\bar m)]_{O(1)}\otimes [(\bar n)]_{O(1)}$
with $(-1)^{\bar m+\bar n}=\s$\,.
Conversely,
we can also use the above seesaw diagram to construct 
the 
$(Sp(2N,\mathbb R),O(1,1))$ representations
out of the $(Sp(2N,\mathbb R),O(1))$ ones,
namely the metaplectic representations.
We begin with the tensor product of two
metaplectic representations, that is nothing but
the Fock space with doubled oscillators
(the doubling is 
from $(a_i,\tilde a^i)$ to $(a_i,b^i,\tilde a^i,\tilde b_i)$).
The same result \eqref{Sp2N result}
can be obtained simply imposing 
the $O(1,1)$ irrep conditions
\eqref{O(1,1) irrep cond} on this space.
The difference of the construction 
lies on the viewpoint
whether we consider the Fock space 
as the representation space of $(U(N),U(1,1))$
or the tensor product of the
$(Sp(2N,\mathbb R),O(1))$ representation space.

For $N=1$, the isomorphism
$Sp(2,\mathbb R)\cong SL(2,\mathbb R)$ tells that
the irrep
$\pi_{Sp(2,\mathbb R)}(\mu,\bar n)$ 
corresponds to
the principal series representation
$\pi_{SL(2,\mathbb R)}((-1)^{\bar n}\,\mu)$
of $SL(2,\mathbb R)$.
Together with 
the discrete representations
of $(Sp(2,\mathbb R), O(2))$,
they constitute 
all oscillator realizations of
$SL(2,\mathbb R)$
within all 
$(Sp_{2},O_2)$ dualities.
Again, the complementary series
representations do not appear in this construction.
Let us summarize here all
$SU(2)$ and $SL(2,\mathbb R)$ 
oscillator realizations within $Sp(4,\mathbb R)$\,:
\ba 
    & (U(2),U(1))\ \&\ (Sp(1),O^*(2))
    \quad &\longrightarrow \quad 
    [n]_{SU(2)}\quad [n\in\mathbb N]\,,\nn
    &(U(1,1),U(1))\ \&
    \ (Sp(2,\mathbb R), O(2))
    \quad &\longrightarrow \quad 
    \cD_{SL(2,\mathbb R)}(h+1)
    \quad [h\in \mathbb N]\,,\nn
    &(GL(2,\mathbb R),GL(1,\mathbb R))\ \&
    \ (Sp(2,\mathbb R), O(1,1))
    \quad &\longrightarrow \quad 
    \pi_{SL(2,\mathbb R)}(\zeta)
    \quad [\zeta\in \mathbb R]\,.
\ea  

When $N=2$, the isomorphism $Sp(4,\mathbb R)\cong 
\widetilde{SO}{}^+(2,3)$
suggests 
that the irreps, 
\be 
    \pi_{Sp(4,\mathbb R)}(\mu,\bar n)
    =\bigoplus^{\infty}_{m=-\infty}
    [2m+\bar n]_{U(1)}\otimes 
    \left(\bigoplus^{\infty}_{n=|m|}
    [2n+\bar n]_{SU(2)}\right),
\ee
may admit
a CFT$_3$ or AdS$_4$ interpretation.
In fact, it
describes the scalar or spinor
tachyon in AdS$_4$.
This is to be contrasted with the 
$(Sp(4,\mathbb R),O(2))$ duality,
where we find massless spin $s$ fields. By flipping the
signature of the dual orthogonal group, or to put it differently, flipping 
one of singletons to its contragredient one,
the resulting representation
has ``zero spin'' rather than ``higher spin''.
Analogous phenomenon occurs
in the AdS$_5$ case,
where the relevant dual pairs are
$(U(2,2),U(1,1))$
and $(U(2,2),U(2,2))$.
Their correspondence will
be described in one of the follow-up papers.

The representations that we have constructed 
are again the most degenerate
principal series representation of $Sp(2N,\mathbb R)$\,.
It will become more transparent if we move 
to 
the Schr\"odinger realization,
\be 
    y^{\pm i}_{A}=\frac{1}{\sqrt{2}}\left(x^i_A\pm \eta_{AB}\,\frac{\partial}{\partial x^i_B}\right).
\ee 
Let us work first with the dual pair $(Sp(2N,\mathbb R),
O(M_+,M_-))$ with any $M_\pm$
and metric $\eta_{AB}$ of signature $(M_+,M_-)$.
In this realization, the generators take the analogous form,
\be 
    M_{AB}=2\,x^i_{[A}\,\frac{\partial}{\partial x^{B]i}}\,,
    \qquad 
    K^{{\sst \e}i\,{\sst \ve}j}
    =\tilde K^{ij}+\e\,\ve\,\tilde K_{ij}
    +\tfrac{\e+\ve}2\,\tilde K^i{}_j\,,
\ee 
with
\be 
    \tilde K^{ij}=\frac12\,x^i_{A}\,x^{Aj}\,,
    \quad 
    \tilde K_{ij}=\frac12\,\frac{\partial}{\partial x^i_A}
    \frac{\partial}{\partial x^{Aj}}\,,
    \qquad 
    \tilde K^{i}{}_j=
    x^i_A\,\frac{\partial}{\partial x_{A}^{j}} +\frac{M_++M_-}2\,\delta^{i}_{j}\,.
\ee 
Here $\epsilon,\varepsilon=\pm$.
The $A,B$ indices above are lowered or raised by $\eta_{AB}$.
Now consider the $M_+=M_-=M$ case
with off-diagonal metric $\eta_{AB}$\,:
setting $A=(+a,-a)$,
the metric components are
$\eta_{\pm a\pm b}=0$ and 
$\eta_{+a-b}=\delta_{ab}$\,.
Then, the generators become
\be 
     M_{\pm a\pm b}=x^i_{\pm a}\,\frac{\partial}{\partial x^{i}_{\mp b}}-x^i_{\pm b}\,\frac{\partial}{\partial x^{i}_{\mp a}}
     \,,
     \qquad 
      M_{+a-b}=x^i_{+a}\,\frac{\partial}{\partial x_{+b}^i}
      -x^i_{-b}\,\frac{\partial}{\partial x_{-a}^i}\,,
\ee 
and 
\be 
    \tilde K^{ij}=x^{(i}_{+a}\,x^{j)}_{-a}\,,
    \quad 
    \tilde K_{ij}=\frac{\partial}{\partial x^{(i}_{+a}}
    \frac{\partial}{\partial x^{j)}_{-a}}\,,
    \qquad 
    \tilde K^{i}{}_j=
    x^i_{+a}\,\frac{\partial}{\partial x_{+a}^{j}}+x^i_{-a}\,\frac{\partial}{\partial x_{-a}^{j}}
    +M\,\delta^{i}_{j}\,.
\ee 
If we perform a Fourier transformation
on the variable $x^i_{-a}$, the $Sp(2N,\mathbb R)$
generators become homogeneous,
\be 
    \tilde K^{ij}=i\,x^{(i}_{+a}\,
    \frac{\partial}{\partial p_{-aj)}}
    \quad 
    \tilde K_{ij}=-i\,p_{-a(i}\,\frac{\partial}{\partial x^{j)}_{+b}}\,,
    \qquad 
     \tilde K^{i}{}_j=
    x^i_{+a}\,\frac{\partial}{\partial x_{+a}^{j}}+p_{-aj}\,\frac{\partial}{\partial p_{-ai}}\,.
\ee 
In this way, we find the natural action
of $(Sp(2N,\mathbb R), O(M,M))$
on the space of $M\times 2N$ real matrices
$(x_+,p_-)$ or 
 $2M\times N$ real matrices $\big(\substack{x_+\\x_-}\big)$
($x_\pm$ and $p_-$ are $M\times N$ real matrices
with components $x^i_{\pm a}$ and $p_{-ai}$):
\ba
    &\langle 
    (x_+,p_-)\,|
    \,U_\cW(g)\,|
    \Psi\ra
    =\langle  (x_+,p_-)\,g\,|
    \Psi\ra
    \qquad &[\,g\in Sp(2N,\mathbb R)\,]\,,
    \\ 
    & 
    \langle \big(\substack{x_+\\x_-}\big)\,|
    \,U_\cW(h)\,|\,
    \Psi\ra
    =\langle h^t\,\big(\substack{x_+\\x_-}\big)\,|
    \Psi\ra
    \qquad &[\,h\in O(M,M)\,]\,.
\ea
This space is reducible and
we can impose the determinant-homogeneity
condition,
\be 
    \langle k\,(x_+,p_-)\,|
    \Psi^\pm_{\mu,\bar n}\ra
    = |\det k|^{\pm\,i\,\mu}\,
    {\rm sgn}(\det k)^{\bar n}\,\langle (x_+,p_-)\,|
    \Psi^\pm_{\mu,\bar n}\ra
    \qquad [\,k\in GL(M,\mathbb R)\,]\,,
    \label{sp o hom}
\ee
or equivalently,
\be 
    \left\langle {\scriptsize\begin{pmatrix}
    k& 0 \\ 0 & (k^{-1})^t \end{pmatrix}\begin{pmatrix}x_+\\x_-\end{pmatrix}}\,\bigg|
    \Psi^\pm_{\mu,\bar n}\right \ra
    = |\det k|^{\pm\,i\,\mu+1}\,
    {\rm sgn}(\det k)^{\bar n}\,\langle \big(\substack{x_+\\x_-}\big)\,|
    \Psi^\pm_{\mu,\bar n}\ra
    \qquad [\,k\in GL(M,\mathbb R)\,]\,,
    \label{sp o hom 2}
\ee
to make it irreducible for both
of $Sp(2N,\mathbb R)$ and $O(M,M)$.
Note that the additional factor of $|\det k|$ arises from
the Jacobian in the Fourier transformation.
For $M=1$, one can easily recognize
that the condition \eqref{sp o hom}
is equivalent to the assignment
of the  eigenvalue $\pm i\,\mu$ to the generator $M_{+-}$ .
This condition coincides in fact with the first one in \eqref{O(1,1) irrep cond}
simply because $M_{+-}=M_{\circ\bullet}$.
The other two conditions can be 
also easily identified.
Therefore, 
we generalized the 
representations appearing in $(Sp(2N,\mathbb R),O(1,1))$
to the more general case of  $(Sp(2N,\mathbb R),O(M,M))$\,.
The resulting representations are
in fact
the degenerate principal series representation for $Sp(2N,\mathbb R)$ 
(it becomes the most degenerate one
if the dual group is $O(1,1)$),
and the most degenerate principal series ones
for $O(M,M)$.
The corresponding parabolic subgroups
are the ones with Levi factors
$GL(N-M,\mathbb R)\times GL(M,\mathbb R)$ for $Sp(2N,\mathbb R)$ 
and $GL(M,\mathbb R)$ for $O(M,M)$.
To put it differently,
the representations of $(Sp(2N,\mathbb R),O(M,M))$
are obtained by restricting and inducing
the $(GL(2N,\mathbb R),GL(M,\mathbb R))$ representations,
as summarized in the following seesaw diagram:
\be
\parbox{150pt}{
\begin{tikzpicture}
\draw [<->] (0,0) -- (1,0.8);
\draw [<->] (0,0.8) -- (1,0);
\node at (-1,1.2) {$GL(2N,\mathbb R)$};
\node at (-1,0.4) {$\cup$};
\node  at (-1,-0.4) {$Sp(2N,\mathbb R)$};
\node at (2,-0.4) {$GL(M,\mathbb R)$};
\node at (2,0.4) {$\cup$};
\node at (2,1.2) {$O(M,M)$};
\end{tikzpicture}}.
\ee

\subsection{$\big(Sp(2N,\mathbb C),O(2,\mathbb C)\big)$
and $\big(Sp(2N,\mathbb C),O(2M,\mathbb C)\big)$}

In order to describe $O(2,\mathbb C)$, it is convenient
to define
\be 
    a^I_\pm=\tfrac1{\sqrt{2}}\left(a_1^I\mp i\,a_2^I\right),
    \qquad
    \tilde a^{\pm}_I=
    \tfrac1{\sqrt{2}}\left(\tilde a^1_I\pm i\,\tilde a^2_I\right),
\ee 
where $I=1,\ldots,2N$ is the symplectic index.
In terms of these oscillators,
$O(2,\mathbb C)$ generators are given by
\begin{equation}
    M^+_{12}=i\left(\tilde a^+_I\, a_+^I - \tilde a^-_I\, a^I_-\right),
    \qquad M^-_{12} 
    =i\left(\tilde a^{+I}\, \tilde a^-_I-a^I_+\, a_{-I}\right),
\end{equation}
with the reflection 
flipping the sign of $a_2^I$
and $\tilde a^2_I$\,,
\be
    \cR\,\begin{pmatrix}
    a_\pm^I \\ \tilde a^\pm_I 
    \end{pmatrix}
    \cR^{-1}
    =\begin{pmatrix}
    a_{\mp}^I \\ \tilde a^{\mp}_I 
    \end{pmatrix}.
\ee 
Recall that
the symplectic indices
are contracted as in \eqref{symp contraction}.
The dual group  $Sp(2N,\mathbb C)$ is
generated by
\ba
    && T^{IJ} =K^{IJ}-\eta_{IK}\,\eta_{JL}\,(K^{KL})^*=
    -\left(\tilde a^{+(I}\, a_+^{J)}
    + \tilde a^{-(I}\, a_-^{J)}\right),\nn 
    && S^{IJ} =K^{IJ}+\eta_{IK}\,\eta_{JL}\,(K^{KL})^*=
    i\left(
    \tilde a^{+(I}\, \tilde a^{J)-}-a^{(I}_+\, a_-^{J)} 
    \right),
\ea
where $T^{IJ}$ generates the maximal compact subgroup $Sp(N)$.
The relevant seesaw pair is
\be
\parbox{225pt}{
\begin{tikzpicture}
\draw [<->] (0,0) -- (1,0.8);
\draw [<->] (0,0.8) -- (1,0);
\node at (-1,1.2) {$Sp(2N,\mathbb C)$};
\node at (-1,0.4) {$\cup$};
\node  at (-1,-0.4) {$Sp(N)$};
\node at (3.5,-0.4) {$O(2,\mathbb C)\cong (\mathbb R^+\times U(1)  )\rtimes \mathbb Z_2$};
\node at (3.5,0.4) {$\cup$};
\node at (3.5,1.2) {$O^*(4)\cong (SL(2,\mathbb R)\times SU(2))/\mathbb Z_2$};
\end{tikzpicture}},
\ee
where the correspondence of the  $(Sp(N), O^*(4))$ representations
is
\be 
    [(\ell_1,\ell_2)]_{Sp(N)}
    \quad 
    \longleftrightarrow\quad 
    [\ell_1-\ell_2]_{SU(2)}
    \otimes 
    \cD_{SL(2,\mathbb R)}(\ell_1+\ell_2+2N)\,.
\ee
Here, the $SL(2,\mathbb R)$ 
and  $SU(2)$ of $O^*(4)$ are generated by 
\begin{equation}
    H = \tilde a_I^+\, a_+^I + \tilde a_I^-\, a_-^I + 2N\,,
    \qquad  
    E = \tilde a_I^+\,\tilde a^{-I}\,,
    \qquad 
    F = - a_{+I}\, a_-^I\,,
\end{equation}
and 
\begin{equation}
    J_3 = \tfrac 12\, \big(\tilde a_I^+\, a_+^I
    - \tilde a_I^-\, a_-^I\big)\,,
    \qquad
    J_+ = \tilde a_I^+\, a_-^I\,,
    \qquad
    J_-=\tilde a_I^-\, a_+^I\,.
\end{equation}
Let us fix an irrep of $O(2,\mathbb C)$ as
\be 
    \pi_{O(2,\mathbb C)}(\mu,m):=
    \left([\mu]_{\mathbb R^+}\otimes
    [m]_{U(1)}\right)
    \oplus 
     \left([-\mu]_{\mathbb R^+}\otimes
    [-m]_{U(1)}\right)
    \qquad [\mu\ge 0]\,,
\ee 
which is two dimensional
unless $\mu=m=0$.
In the latter case, the irrep becomes
one dimensional:
\be 
    \pi_{O(2,\mathbb C)}(0,0):=
    [0]_{\mathbb R^+}\otimes
    [0]_{U(1)}\,.
\ee
The restriction $O^*(4)\downarrow O(2,\mathbb C)$ amounts
to embedding
\be     
    M_{12}^+=2\,i\,J_3\,,
    \qquad 
    M_{12}^-=i\,(E+F)\,,
\ee 
which determines
possible $Sp(N)$ irreps as
\be 
    \ell_1-\ell_2=|m|+2k\,, \qquad k\in\mathbb N\,,
\ee 
and hence defines
the vector space
of the corresponding $Sp(2N,\mathbb C)$ irrep as
\begin{equation}
   \pi_{Sp(2N,\mathbb C)}(\m,m):=\bigoplus_{n,k=0}^\infty [(|m|+2k+n,n)]_{Sp(N)}\,.
\end{equation}
For further specifications,
we solve
the $O(2,\mathbb C)$ irrep condition,
\be 
    M_{12}^+\,|\Psi^\pm_{\mu,m}\ra= \pm i\,m\,|\Psi^\pm_{\mu,m}\ra\,,
    \quad 
    M_{12}^-\,|\Psi^\pm_{\mu,m}\ra=
    \pm\,\mu\,|\Psi^\pm_{\mu,m}\ra\,,
    \quad 
    \cR\,|\Psi^{\pm}_{\mu,m}\ra=
    |\Psi^\mp_{\mu,m}\ra,
    \label{O2C cond}
\ee 
as the coherent state,
\be
    \lvert \Psi^+_{\mu,m} \rangle =  \sum_{n,k=0}^\infty
    \Psi^{I(|m|+2k+n),J(n)}
    \left( f_{\m,m,n,k}(\tilde a^+_K\,\tilde a^{-K})
    (\tilde a^+_I)^{k+n}
    (\tilde a^{{\rm sgn}(m)}_I)^{|m|}
    (\tilde a^-_{I})^k
   (\tilde a^-_{J})^n
   |0\ra \right),
   \label{Sp2C result}
\ee
where $\O_{IJ}\,\Psi^{I(\ell_1),J(\ell_2)}=0$ and 
$h_{\mu,m+2n+2k}(z)=f_{\m,m,n,k}(z)$
satisfies the differential equation
\eqref{h eq} and can be fixed uniquely with 
$f_{\m,m,n,k}(0)=1$.
The other state 
$\lvert \Psi^-_{\mu,m} \rangle$
can be obtained by acting with $\cR$
on the above, by definition.

Note that 
the $(Sp(2N,\mathbb C),O(2,\mathbb C))$ dual pair can serve also
for the tensor product
of the irreps $\pi_{Sp(2N,\mathbb R)}(\bar n)$
appearing 
in the $(Sp(2N,\mathbb C),O(1,\mathbb C))$ dual pair:
\be
    \pi_{Sp(2N,\mathbb C)}(\bar m)
    \otimes 
    \pi_{Sp(2N,\mathbb C)}(\bar n)
   =\int_0^\infty d\mu\ \bigoplus_{n=0}^\infty \
    \pi_{Sp(2N,\mathbb C)}(\mu,2n+1-\delta_{\bar m\bar n})\,.
\ee
The above can be understood from the seesaw pair,
\be
\parbox{230pt}{
\begin{tikzpicture}
\draw [<->] (0,0) -- (1,0.8);
\draw [<->] (0,0.8) -- (1,0);
\node at (-2,1.2) {$Sp(2N,\mathbb C)\times Sp(2N,\mathbb C)$};
\node at (-2,0.4) {$\cup$};
\node  at (-2,-0.4) {$Sp(2N,\mathbb C)$};
\node at (2.5,-0.4) {$O(1,\mathbb C)\times O(1,\mathbb C)$};
\node at (2.5,0.4) {$\cup$};
\node at (2.5,1.2) {$O(2,\mathbb C)$};
\end{tikzpicture}},
\ee
where 
the irrep
$[\mu,m]_{O(2,\mathbb C)}$ 
branches 
to the irreps $[(\bar m)]_{O(1,\mathbb C)}\otimes [(\bar n)]_{O(1,\mathbb C)}$ with $(-1)^{\bar m+\bar n}=(-1)^m$.
Conversely,
the above seesaw diagram can be used 
to construct 
the 
$(Sp(2N,\mathbb C),O(2,\mathbb C))$ representations
out of the $(Sp(2N,\mathbb C),O(1,\mathbb C))$ ones:
Begin with the tensor product Fock space with doubled oscillators
(the doubling is 
from $(a^I,\tilde a_I)$ to $(a^\pm_I,\tilde a^I_\pm)$).
The same result \eqref{Sp2C result}
can be obtained simply imposing 
the $O(2,\mathbb C)$ irrep conditions
\eqref{O2C cond} on this space.

When $N=1$,
the $Sp(2,\mathbb C)\cong 
\widetilde{SO}{}^+(1,3)$ irreps become
simpler since two-row Young diagrams are not allowed
for $Sp(1)\cong SU(2)$:
\begin{equation}
   \pi_{Sp(2,\mathbb C)}(\m,2s):=\bigoplus_{k=0}^\infty [(|2s|+2k)]_{Sp(1)}\,.
   \label{Sp2C irrep}
\end{equation}
These irreps match the entire principal series representations
of $\widetilde{SO}{}^+(1,3)$
(see  Harish-Chandra
\cite{Harish-Chandra1947}),
which describe
massive spin $s$ fields in dS$_3$.
Note that 
all these representations, that is,
for
any mass and any spin,
can be obtained 
by the tensor product of 
two conformal representations
$\pi_{SL(2,\mathbb C)}(0)$ or $\pi_{SL(2,\mathbb C)}(1)$\,,
while the latter irreps
uplift
to the singleton irreps of the conformal group $\widetilde{SO}{}^+(2,3)$,
mentioned in Section \ref{sec:O-Sp_correspondence}.
Recall that the tensor product of two singletons
give the tower of all massless spin $s$ fields in AdS$_4$.
Therefore, 
the direct sum
of \eqref{Sp2C irrep}
over all masses and spins
is the restriction of the aforementioned tower to dS$_3$.

The representations we have constructed 
is again the most degenerate
principal series representation of $Sp(2N,\mathbb C)$\,.
Let us see this point 
with the more general case of $(Sp(2N,\mathbb C),
O(2M,\mathbb C))$ duality  with any $M$.
We first move to the Schr\"odinger realization,
\be 
    y^{\pm i}_{A}=\frac{1}{\sqrt{2}}\left(z^i_A\pm \eta_{AB}\,\frac{\partial}{\partial z^i_B}\right),
    \qquad 
   y^{\pm i}_{A}{}^*=\frac{1}{\sqrt{2}}\left(\bar z^i_A\pm \eta_{AB}\,\frac{\partial}{\partial \bar z^i_B}\right),
    \label{yy zz def}
\ee 
where we introduced a metric $\eta_{AB}$ for $O(2M,\mathbb C)$,
which will also appear in the commutation relation of the $y_A^I$ operators. Since the group is complex, different signatures of $\eta_{AB}$ are all equivalent. In other words, we can move 
from one to the other by a suitable redefinition of the $y$ operators.
For our purpose, it will be convenient to fix 
the metric $\eta_{AB}$
as the off-diagonal one
with non-trivial components $\eta_{+a-b}=\delta_{ab}$
where $A=(+a,-a)$. 
Then, the $O(2M,\mathbb C)$ and $Sp(2N,\mathbb C)$ generators become
\be 
     M_{\pm a\pm b}=z^i_{\pm a}\,\frac{\partial}{\partial z^{i}_{\mp b}}-z^i_{\pm b}\,\frac{\partial}{\partial z^{i}_{\mp a}}
     \,,
     \qquad 
      M_{+a-b}=z^i_{+a}\,\frac{\partial}{\partial z_{+b}^i}
      -z^i_{-b}\,\frac{\partial}{\partial z_{-a}^i}\,,
\ee 
and
\be 
 K^{{\sst \e}i\,{\sst \ve}j}
    =\tilde K^{ij}+\e\,\ve\,\tilde K_{ij}
    +\tfrac{\e+\ve}2\,\tilde K^i{}_j\,,
\ee 
with
\be 
    \tilde K^{ij}=z^{(i}_{+a}\,z^{j)}_{-a}\,,
    \quad 
    \tilde K_{ij}=\frac{\partial}{\partial z^{(i}_{+a}}
    \frac{\partial}{\partial z^{j)}_{-a}}\,,
    \qquad 
    \tilde K^{i}{}_j=
    z^i_{+a}\,\frac{\partial}{\partial z_{+a}^{j}}+z^i_{-a}\,\frac{\partial}{\partial z_{-a}^{j}}
    +M\,\delta^{i}_{j}\,.
\ee 
Here, $\epsilon,\varepsilon=\pm$.
The rest of generators are simply the complex conjugate of the above.
We can also perform a Fourier transformation
on the half of the variable $z^i_{-a}$
to render the $Sp(2N,\mathbb C)$ generators homogeneous:
\be 
    \tilde K^{ij}=i\,z^{(i}_{+a}\,
    \frac{\partial}{\partial w_{-aj)}}
    \quad 
    \tilde K_{ij}=-i\,w_{-a(i}\,\frac{\partial}{\partial z^{j)}_{+b}}\,,
    \qquad 
     \tilde K^{i}{}_j=
    z^i_{+a}\,\frac{\partial}{\partial z_{+a}^{j}}+w_{-aj}\,\frac{\partial}{\partial w_{-ai}}\,.
\ee 
Here, $w_{-ai}$ are the variables in the Fourier space.
In the end, we can realize
the dual pair $(Sp(2N,\mathbb C), O(2M,\mathbb C))$
in the space of functions of $M\times 2N$ complex matrices
$(z_+,w_-)$ or $(z_+,z_-)$:
\ba
    &\langle 
    (z_+,w_-)\,|
    \,U_\cW(g)\,|
    \Psi\ra
    =\langle  (z_+,w_-)\,g\,|
    \Psi\ra
    \qquad &[\,g\in Sp(2N,\mathbb C)\,]\,,
    \\ 
    & 
   \langle \big(\substack{z_+\\z_-}\big)\,|
    \,U_\cW(h)\,|\,
    \Psi\ra
    =\langle h^t\,\big(\substack{z_+\\z_-}\big)\,|
    \Psi\ra
    \qquad &[\,h\in O(2M,\mathbb C)\,]\,.
\ea
The determinant-homogeneous
condition,
\be 
    \langle k\,(z_+,w_-)\,|
    \Psi^\pm_{\mu,m}\ra
    = |\det k|^{\pm\,i\,\mu}\,
    \big(\tfrac{\det k}{|\det k|}\big)^{m}\,\langle (z_+,w_-)\,|
    \Psi^\pm_{\mu,m}\ra
    \qquad [\,k\in GL(M,\mathbb C)\,]\,,
    \label{sp C o hom1}
\ee
or equivalently,
\be 
    \left\langle {\scriptsize\begin{pmatrix}
    k& 0 \\ 0 & (k^{-1})^t \end{pmatrix}\begin{pmatrix}z_+\\z_-\end{pmatrix}}\,\bigg|
    \Psi^\pm_{\mu,m}\right \ra
    = |\det k|^{\pm\,i\,\mu+2}\,
    \big(\tfrac{\det k}{|\det k|}\big)^{m}\,\langle \big(\substack{z_+\\z_-}\big)\,|
    \Psi^\pm_{\mu,m}\ra
    \qquad [\,k\in GL(M,\mathbb C)\,]\,,
    \label{sp C o hom2}
\ee
makes the representation irreducible for both
of $Sp(2N,\mathbb C)$ and $O(2M,\mathbb C)$.
Note that the additional factor  of $|\det k|^2$ arises from
the Jacobian in the Fourier transformation.
The resulting representations 
correspond to
the degenerate principal series representation for 
$Sp(2N,\mathbb C)$ 
and the most degenerate principal series one
for $O(2M,\mathbb C)$.
Their parabolic subgroups
are the ones with Levi factors
$GL(N-M,\mathbb C)\times GL(M,\mathbb C)$ for $Sp(2N,\mathbb C)$ 
and $GL(M,\mathbb C)$ for $O(2M,\mathbb C)$.
This parabolic induction can be understood from 
what is depicted by the seesaw diagram,
\be
\parbox{150pt}{
\begin{tikzpicture}
\draw [<->] (0,0) -- (1,0.8);
\draw [<->] (0,0.8) -- (1,0);
\node at (-1,1.2) {$GL(2N,\mathbb C)$};
\node at (-1,0.4) {$\cup$};
\node  at (-1,-0.4) {$Sp(2N,\mathbb C)$};
\node at (2,-0.4) {$GL(M,\mathbb C)$};
\node at (2,0.4) {$\cup$};
\node at (2,1.2) {$O(2M,\mathbb C)$};
\end{tikzpicture}}.
\ee

\section{Branching properties}
\label{sec:singletons}
The representation of $G$ appearing 
in the dual pair $(G,G')$, wherein $G'$ is 
the smallest in its family, has several interesting 
branching properties.
In fact, when the group $G$ is the 
conformal group $Sp(4,\mathbb R)\cong \widetilde{SO}{}^+(2,3)$,
$SU(2,2)\cong \widetilde{SO}{}^+(2,4)$ or $O^*(8)\cong \widetilde{SO}{}^+(2,6)$,
this representation becomes a conformal field, namely 
a singleton representation,
which possesses a number
of special properties (see e.g. \cite{Bekaert:2011js} 
for a short review).
Therefore, in a sense,
we can consider these representations as 
`generalized singletons'.
In the following,
we discuss their branching properties
along with a comparison with those of conformal fields.

\subsection{Restriction to maximal compact subgroup}

The $d$-dimensional singleton is the special representation 
of $\widetilde{SO}{}^+(2,d)$,
whose name is a reference
to a peculiarity, first derived
in \cite{Ehrman1957}, namely
that its restriction
 to the maximal compact subgroup
$SO(2) \times \widetilde{SO}(d)$ consists in a
single direct sum where each irrep appears with
multiplicity one.

In fact, all the $G$-representations
with the smallest dual group $G'$
treated in this paper
have such properties.
Let us gather the relevant formulas here:
\begin{itemize} 
\item 
$(GL(N,\mathbb R),GL(1,\mathbb R))$\,:\,
the $GL(N,\mathbb R)$ irrep
dual to the $GL(1,\mathbb R)\cong \mathbb R^+\otimes \mathbb Z_2$ irrep $[\zeta]_{\mathbb R^+}\otimes 
[(-1)^{\bar n}]_{\mathbb Z_2}$ decomposes as
\be 
     \pi_{GL(N,\mathbb R)}(\zeta,\bar n)\,\big|_{O(N)}=\bigoplus_{k=0}^\infty
    [(2k+\bar n)]_{O(N)}\,.
\ee
\item 
$(GL(N,\mathbb C),GL(1,\mathbb C))$\,:\,
the $GL(N,\mathbb C)$ irrep
dual to the $GL(1,\mathbb C)\cong \mathbb R^+\otimes U(1)$ irrep $[\zeta]_{\mathbb R^+}\otimes 
[n]_{U(1)}$ decomposes as
\be 
     \pi_{GL(N,\mathbb C)}(\zeta,n)\,\big|_{U(N)}=\bigoplus_{k=0}^\infty
    [(k+\tfrac{|n|+n}2)\oslash (k+\tfrac{|n|-n}2)]_{U(N)}\,.
\ee
\item 
$(U^*(2N),U^*(2))$\,:\,
the $U^*(2N)$ irrep
dual to the $U^*(2)\cong \mathbb R^+\otimes SU(2)$ irrep $[\zeta]_{\mathbb R^+}\otimes 
[n]_{SU(2)}$ decomposes as
\be 
     \pi_{U^*(2N)}(\zeta,n)\,\big|_{Sp(N)}=\bigoplus_{k=0}^\infty
    [(n+k,k)]_{Sp(N)}\,.
\ee
\item 
$(U(N_+,N_-),U(1))$\,:\,
the $U(N_+,N_-)$ irrep
dual to the $U(1)$ irrep $[\pm n]_{U(1)}$ with $n\ge 0$ decomposes as
\ba 
    && \cD_{U(N_+,N_-)}
     \big([(n),\tfrac12]_{U(N_\pm )} \otimes
    [(0), \tfrac12]_{U(N_\mp )}\big)\,\big|_{U(N_+)\times U(N_-)}\nn 
     && \quad =\bigoplus_{k=0}^\infty
    [(n+k),\tfrac12]_{U(N_\pm )}
    \otimes[(n+k),\tfrac12]_{U(N_\mp)} \,.
\ea
\item 
$(Sp(2N,\mathbb R),O(1))$\,:\,
the $Sp(2N,\mathbb R)$ irrep
dual to the $O(1)\cong \mathbb Z_2$ irrep $[(\bar n)]_{O(1)}=
[(-1)^{\bar n}]_{\mathbb Z_2}$ decomposes as
\be 
     \cD_{Sp(2N,\mathbb R)}
     \big([(\bar n),\tfrac12]_{U(N)}\big)\,\big|_{U(N)}
     =\bigoplus_{k=0}^\infty
    [(2k+\bar n ),\tfrac12]_{U(N)}\,.
\ee
\item 
$(Sp(2N,\mathbb C),O(1,\mathbb C))$\,:\,
the $Sp(2N,\mathbb C)$ irrep
dual to the $O(1,\mathbb C)\cong \mathbb Z_2$ irrep $[(\bar n)]_{O(1,\mathbb C)}=
[(-1)^{\bar n}]_{\mathbb Z_2}$ decomposes as
\be 
     \pi_{Sp(2N,\mathbb C)}(\bar n)\,\big|_{Sp(N)}
     =\bigoplus_{k=0}^\infty
    [2k+\bar n]_{Sp(N)}\,.
\ee
\item 
$(O^*(2N),Sp(1))$\,:\,
the $O^*(2N)$ irrep
dual to the $Sp(1)\cong SU(2)$ irrep $[n]_{SU(2)}$ decomposes as
\be 
     \cD_{O^*(2N)}\big([(n),1]_{U(N)}\big)\,\big|_{U(N)}
     =\bigoplus_{k=0}^\infty
    [(n+k,k)]_{U(N)}\,.
\ee
\item 
$(Sp(N_+,N_-),O^*(2))$\,:\,
the $Sp(N_+,N_-)$ irrep
dual to the $O^*(2)\cong U(1)$ irrep $[\pm n+N_+-N_-]_{U(1)}$ with $n\ge 0$ decomposes as
\be 
     \pi_{Sp(N_+,N_-)}(\pm n)\,\big|_{Sp(N_+)\times Sp(N_-)}
     =\bigoplus_{k=0}^\infty
    [(n+k)]_{Sp(N_\pm)}\otimes [(k)]_{Sp(N_\mp)}\,.
\ee
\end{itemize} 
The only cases of $G$-representations
with the smallest $G'$ that we have not explored are
the dual pairs $(O(N_+,N_-),Sp(2,\mathbb R))$
and $(O(N,\mathbb C), Sp(2,\mathbb C))$,
whose representations are
not treated in this  paper.
Let us mention that the scalar singleton of $O(2,d)$
has been explored within
the $(O(2,d), Sp(2,\mathbb R))$ duality
in \cite{Vasiliev:2004cm} (see also \cite{Bekaert:2009fg}).
It is also possible to generalize the above
properties to the other branching rules where
the maximal compact subgroup $K$ 
is replaced by a group whose
complexification coincides with that of $K$.
We shall provide the analysis of these cases
together with $(O(N_+,N_-),Sp(2,\mathbb R))$
and $(O(N,\mathbb C), Sp(2,\mathbb C))$
in our forthcoming paper.

\subsection{Irreducibility under restriction}
Other properties of the $d$-dimensional singletons are that
\begin{itemize} 
\item they decompose into
at most two irreps, when restricted to 
a $d$-dimensional isometry group, namely
the (anti-)de Sitter or
Poincar\'e subgroup (see e.g. \cite{Mack:1969dg,Angelopoulos:1999bz,
Bekaert:2011js}),
\item 
and, 
they are 
unique, or are one 
of 
two possible extensions of 
a $d$-dimensional isometry irrep to
the $d$-dimensional conformal group.
\end{itemize}
Once again, this property can be investigated using seesaw pairs,
\begin{equation}
\parbox{180pt}{
\begin{tikzpicture}
\draw [<->] (0,0) -- (1,0.8);
\draw [<->] (0,0.8) -- (1,0);
\node at (-0.5,1.2) {$G$};
\node at (-2,1.2) {$\pi_G(\zeta)$};
\draw [->] (-2,0.8) -- (-2,0);
\node at (-2,-0.4) {$\pi_{\tilde G}(\tilde \zeta)$};
\node at (-0.5,0.4) {$\cup$};
\node at (1.5,-0.4) {$G'$};
\node at (-0.5,-0.4) {$\tilde G$};
\node at (1.5,0.4) {$\cup$};
\node at (1.5,1.2) {$\tilde G'$};
\node at (3,1.2) {$\pi_{\tilde G'}(\theta(\tilde \zeta))$};
\draw [->] (3,0.8) -- (3,0);
\node at (3,-0.4) {$\pi_{G'}(\theta(\zeta))$};
\end{tikzpicture}}\,,
\end{equation}
where the conformal group and isometry group
can be placed at $G$ and $\tilde G$, respectively.
In order that a 
singleton-like irrep $\pi_G(\zeta)$
decomposes at most into two irreps
$\pi_{\tilde G}(\tilde \zeta)$,
on the dual side
there should exist
at most two irreps $\pi_{\tilde G'}(\theta(\tilde\zeta))$
which can branch to $\pi_{G'}(\theta(\zeta))$.
Similarly, in order for an irrep
$\pi_{\tilde G}(\tilde \zeta)$ to admit
an extension to at most two singleton-like irreps of $G$,
the restriction of the dual representation
$\pi_{\tilde G'}(\theta(\tilde\zeta))$
should contain at most two $G'$-irreps.
This property is guaranteed,
if the dual groups  $\tilde G'$ and $G'$ are
isomorphic, or isomorphic  up to $\mathbb Z_2$ finite group.
The simplest case of $\tilde G'$ and $G'$ is 
\be 
    O(1,\mathbb C)\cong O(1)\,.
\ee 
There are four more such groups among the irreducible ones,
\be
    O(2)\supset U(1)\,,
    \qquad 
    O^*(2)\cong U(1)\,,
    \qquad O(1,1)\supset GL(1,\mathbb R)\,,
    \qquad O(2,\mathbb C)\supset GL(1,\mathbb C),
\ee 
which are in fact different real forms of $O_2\supseteq GL_1$\,.
If one takes into account the reducible pairs,
there are two more options, 
\be
    \tilde G'=G'\times O(1)\supset G'\,,
    \qquad \tilde G'=G'\times O(1,\mathbb C)\supset  G'\,.
\ee 
Here, $G'$ can be any group
suitable for (reducible) dual pairing. 
In the following, we present 
the dual groups $G$ and $\tilde G$ and 
their representations for each of the above seven cases.

\begin{enumerate}

\item The dual of $O(1,\mathbb C)\downarrow O(1)$ is
the restriction $Sp(4N,\mathbb R)\downarrow Sp(2N,\mathbb C)$, and we find
\begin{equation}
\parbox{200pt}{
    \begin{tikzpicture}
    \node at (2,1.2) {$\cD_{Sp(4N,\mathbb R)}
    \big([(\bar n),\tfrac12]_{U(2N)}\big)$};
    \draw [<-] (2,0) -- (2,0.8);
    \node at (2,-0.4) 
    {$\pi_{Sp(2N,\mathbb C)}(\bar n)$};
    \draw [<->] (4,0) -- (5,0.8);
    \draw [<->] (4,0.8) -- (5,0);
    \node at (6,1.2) {$[(\bar n)]_{O(1,\mathbb C)}$};
    \draw [<-] (6,0) -- (6,0.8);
    \node at (6,-0.4) {$[(\bar n)]_{O(1)}$};
    \end{tikzpicture}},
\end{equation}
where $\bar n=0,1$.
The $N=1$ case corresponds
to the restriction of the three-dimensional singletons to
the dS$_3$ subgroup.

 \item The dual of $O(2)\downarrow U(1)$ 
 is the restriction of $U(N,N)\downarrow Sp(2N,\mathbb R)$,
 and we find 
\begin{equation}
    \parbox{280pt}{
    \begin{tikzpicture}
    \node at (-3,1.2) {$\cD_{U(N,N)}
    \big([(\tfrac{|n|+n}2),\tfrac12]_{U(N)} \otimes
    [(\tfrac{|n|-n}2),\tfrac12]_{U(N)}\big)$};
    \draw [<-] (-3,0) -- (-3,0.8);
    \node at (-3,-0.4) {$\cD_{Sp(2N, \mathbb R)}\big([(|n|),1]_{U(N)}\big)$};
    \draw [<->] (0.5,0) -- (1.5,0.8);
    \draw [<->] (0.5,0.8) -- (1.5,0);
    \node at (2.2,1.2) {$[(|n|)]_{O(2)}$};
    \draw [<-] (2.2,0) -- (2.2,0.8);
    \node at (2.2,-0.4) {$[(n)]_{U(1)}$};
    \end{tikzpicture}}\,,
\end{equation}
where $n \in \mathbb Z$. 
Notice that for $N\ge 2$, the trivial representation
of $U(1)$ can be found in two $O(2)$ representations,
 $[(0)]_{O(2)}$ and $[(1,1)]_{O(2)}$. As a
consequence, the $Sp(2N, \mathbb R)$ representation
dual to this latter $O(2)$ irrep also enters the
branching rule, and we find 
\begin{eqnarray}
    &&\cD_{U(N,N)}\big([(0),\tfrac12]_{U(N)} \otimes
    [(0),\tfrac12]_{U(N)}\big)\,\big|_{Sp(2N,\mathbb R)}\nn
    &&=\,
    \cD_{Sp(2N, \mathbb R)}\big([(0),1]_{U(N)}\big)\oplus
    \cD_{Sp(2N, \mathbb R)}\big([(1,1),1]_{U(N)}\big)\,.
\end{eqnarray}
The $N=2$ case corresponds to the restriction of four-dimensional
singletons of helicity $\tfrac n2$
to massless
fields of spin $\tfrac{|n|}2$ in AdS$_4$
and the restriction of  
four-dimensional scalar singleton
to conformal scalar
and pseudo scalar fields in AdS$_4$.

\item 
The dual of $O^*(2)\downarrow U(1)$ is 
the restriction of $U(2N_+,2N_-)\downarrow Sp(N_+,N_-)$,
and we find 
\begin{equation}
\parbox{380pt}{
    \begin{tikzpicture}
    \node at (-5,1.2) {$\cD_{U(2N_+,2N_-)}
    \big(\big[\big( \frac{|n|+n}2\big),\tfrac12\big]_{U(2N_+)} \otimes
    \big[\big(\frac{|n|-n}2\big), \tfrac12\big]_{U(2N_-)}\big)$};
    \draw [<-] (-5,0) -- (-5,0.8);
    \node at (-5,-0.4) {$\pi_{Sp(N_+,N_-)}
    \big(\big[\big( \frac{|n|+n}2\big)\big]_{Sp(N_+)}
    \otimes
    \big[\big(\frac{|n|-n}2\big)\big]_{Sp(N_-)}\big)$};
    \draw [<->] (-0.5,0) -- (0.5,0.8);
    \draw [<->] (-0.5,0.8) -- (0.5,0);
    \node at (2,1.2) {$[n+N_+-N_-]_{O^*(2)}$};
    \draw [<-] (2,0) -- (2,0.8);
    \node at (2,-0.4) {$[n+N_+-N_-]_{U(1)}$};
    \end{tikzpicture}},
\end{equation}
where $n\in \mathbb Z$\,.
The $N_+=N_-=1$ case corresponds to the restriction of
four-dimensional helicity $\tfrac n2$ singletons
to massless fields of helicity $\tfrac n2$ in dS$_4$.

\item 
The dual of $O(1,1)\downarrow GL(1,\mathbb R)$
is the restriction $GL(2N,\mathbb R)
\downarrow Sp(2N,\mathbb R)$, and we find
\begin{equation}
    \parbox{190pt}{
    \begin{tikzpicture}
    \node at (-1.5,1.2) {$\pi_{GL(2N,\mathbb R)}(\zeta,\bar n)$};
    \draw [<-] (-1.5,0) -- (-1.5,0.8); 
    \node at (-1.5,-0.4) {$\pi_{Sp(2N, \mathbb R)}(|\zeta|,\bar n)$};
    \draw [<->] (0,0) -- (1,0.8);
    \draw [<->] (0,0.8) -- (1,0);
    \node at (2.5,1.2) {$\pi_{O(1,1)}(|\zeta|,\bar n)$};
    \draw [<-] (2.5,0) -- (2.5,0.8); 
    \node at (2.5,-0.4) {$\pi_{GL(1,\mathbb R)}(\zeta,\bar n)$};
    \end{tikzpicture}},
\end{equation}
where $\zeta \in \mathbb R$ and $\bar n=0, 1$.
The $N=2$ case is to be compared with the 
$GL(4,\mathbb R)$-invariant equations for massless
fields in AdS$_4$ discussed in \cite{Vasiliev:2007yc}.
Note, however, that the representations here
do not describe massless spin $s$ fields, but scalar or spinor tachyons in AdS$_4$.

\item  The dual of $O(2,\mathbb C)\downarrow GL(1,\mathbb C)$ 
is the restriction $GL(2N,\mathbb C)\downarrow Sp(2N,\mathbb C)$, and we find
\begin{equation}
\parbox{190pt}{
    \begin{tikzpicture}
    \node at (-1.5,1.2) {$\pi_{GL(2N,\mathbb C)}(\zeta,n)$};
    \draw [<-] (-1.5,0) -- (-1.5,0.8); 
    \node at (-1.5,-0.4) {$\pi_{Sp(2N, \mathbb C)}(|\zeta|,n)$};
    \draw [<->] (0,0) -- (1,0.8);
    \draw [<->] (0,0.8) -- (1,0);
    \node at (2.5,1.2) {$\pi_{O(2,\mathbb C)}(|\zeta|,n)$};
    \draw [<-] (2.5,0) -- (2.5,0.8); 
    \node at (2.5,-0.4) {$\pi_{GL(1,\mathbb C)}(\zeta,n)$};
    \end{tikzpicture}},
\end{equation}
where $\zeta \in \mathbb R$ and $n\in\mathbb Z$.
The $N=1$ case may be interpreted as a lift 
of
the representations 
to
the principal series of
$\widetilde{SO}{}^+(1,3)$, which correspond
to lifting
massive fields of helicity $\tfrac n2$ in dS$_3$ to a representation of
$GL(2,\mathbb C)$.

\item 
Let $G_N$ be the family of groups dual to $G'$.
Then, the dual of $G'\times O(1)\downarrow G'$ 
is the restriction $G_{N+M}\downarrow G_N\times Sp(2M,\mathbb R)$, and we find 
\begin{equation}
\parbox{250pt}{
    \begin{tikzpicture}
    \node at (-2,1.2) {$\pi_{G_{M+N}}(\zeta)$};
    \draw [<-] (-2,0) -- (-2,0.8); 
    \node at (-2,-0.4) {$\pi_{G_M}(\zeta)\otimes 
    \cD_{Sp(2N, \mathbb R)}(\bar n)$};
    \draw [<->] (0,0) -- (1,0.8);
    \draw [<->] (0,0.8) -- (1,0);
    \node at (2.7,1.2) {$\pi_{G'}(\theta(\zeta))\otimes [(\bar n)]_{O(1)}$};
    \draw [<-] (2.7,0) -- (2.7,0.8); 
    \node at (2.7,-0.4) {$\pi_{G'}(\theta(\zeta))$};
    \end{tikzpicture}},
\end{equation}
where $\bar n=0,1$.
The case $M=N=1$ with $G_{2}=Sp(4,\mathbb R)$,
$G_{1}=Sp(2,\mathbb R)$ and $G'=O(1)$ 
corresponds to the restriction of the three-dimensional
singletons to the AdS$_3$ subgroup.

\item The dual of $G'\times O(1,\mathbb C)\downarrow G'$ 
is the restriction $G_{N+M}\downarrow G_N\times Sp(2M,\mathbb \mathbb C)$, and we find 
\begin{equation}
\parbox{250pt}{
    \begin{tikzpicture}
    \node at (-2,1.2) {$\pi_{G_{N+M}}(\zeta)$};
    \draw [<-] (-2,0) -- (-2,0.8); 
    \node at (-2,-0.4) {$\pi_{G_N}(\zeta)\otimes 
    \pi_{Sp(2M, \mathbb C)}(\bar n)$};
    \draw [<->] (0,0) -- (1,0.8);
    \draw [<->] (0,0.8) -- (1,0);
    \node at (2.7,1.2) {$\pi_{G'}(\theta(\zeta))\otimes [(\bar n)]_{O(1,\mathbb C)}$};
    \draw [<-] (2.7,0) -- (2.7,0.8); 
    \node at (2.7,-0.4) {$\pi_{G'}(\theta(\zeta))$};
    \end{tikzpicture}},
\end{equation}
where $\bar n=0,1$.

\end{enumerate}

\section{Casimir relations}
\label{sec:casimir}

One of the consequences of 
the dual pair correspondence
is that the Casimir operators of 
the two groups in a pair
are related \cite{Klink_1988,Leung_1993,Leung1994,itoh_2003}.
Since the Casimir relations
can be derived at the level of the metaplectic representation
of the embedding symplectic group,
it is possible to study 
them 
in terms of the previously introduced $\o$ or $y$ operators
without specifying the representations of
each group. 
However,
when a specific representation 
of one group
is chosen, 
the values of its Casimir operators will be fixed.
This, in turn, 
determines the values of the Casimir
operators of the dual group,
and hence can be used to identify
the dual representation.
In the following, we derive the relations between Casimir operators
for the complex dual pairs, i.e. $(GL_M,GL_N)$ and $(O_N,Sp_{2M})$.
The relations of the Casimir operators 
for real forms of these complex dual pairs
can be obtained 
by imposing the relevant reality conditions.

\subsection{Duality $(GL_M,GL_N)$}

We will here derive the relation between 
the Casimir operators of 
$GL_M$ and $GL_N$ in a dual pair.
Let us first introduce generating functions for the Casimir operators:
\be 
    x(t)=\sum_{n=0}^\infty\,t^{n}\,\cC_n[\bm X]\,,
    \qquad
    r(t)=\sum_{n=0}^\infty\,t^{n}\,\cC_n[\bm R]\,,\label{Casimir 1}
\ee 
where $\cC_n[\bm X]$ and $\cC_n[\bm R]$
are the Casimir operators of order $n$ defined by
\ba
&&\cC_n[\bm X]=\tr(\bm X^n)=X^{A_1}{}_{A_2}\,X^{A_2}{}_{A_3}\dots\,X^{A_{n}}{}_{A_1}\,,\nonumber\\
&&\cC_n[\bm R]=\tr(\bm R^n)=R_{I_1}{}^{I_2}\,R_{I_2}{}^{I_3}\dots\,R_{I_n}{}^{I_1}\,.\label{C}
\ea
Here, $\bm X$ and $\bm R$ stand for
matrix-valued operators with components $(\bm X)_{AB}=X^A{}_B$
and $(\bm R)_{IJ}=R_I{}^J$.
It is also convenient to introduce two other generating functions, 
\be
\tilde{x}(t)=\sum_{n=0}^{\infty}\, t^n\,
\cC_n[\bm X^t]
\,,\qquad \tilde{r}(t)=\sum_{n=0}^{\infty}\,t^n\,
{\cC}_{n}[\bm R^t]
\,,\label{Casimir 2}
\ee
where $\bm X^t$ and $\bm R^t$ are the 
``transpose''
of $\bm X$ and $\bm R$: $(\bm X^t)_{AB}=X^B{}_A$
and $(\bm R^t)_{IJ}=R_J{}^I$,
and hence the operators ${\cC_n}[\bm X^t]$
and ${\cC_n}[\bm R^t]$ have the form,
\ba
{\cC}_n[\bm X^t]\eq \tr((\bm X^t)^n)=
X^{A_1}{}_{A_n}\,X^{A_2}{}_{A_1}\,X^{A_3}{}_{A_2}\dots\,X^{A_n}{}_{A_{n-1}}\,,\nonumber\\
{\cC}_n[\bm R^t]\eq \tr((\bm R^t)^n)=R_{I_1}{}^{I_n}\,R_{I_2}{}^{I_1}\,R_{I_3}{}^{I_2}\dots\,R_{I_n}{}^{I_{n-1}}\,.\label{Ct}
\ea
Both $\cC_n[\bm X]$ and $\cC_n[\bm X^t]$
are $n$-th order invariants of $GL_M$,
and their difference 
shows up starting from $n=3$\,:
\begin{align}
    \cC_3[\bm X]=X^A{}_B\,X^B{}_C\,X^C{}_A\neq X^A{}_B\,X^C{}_A\,X^B{}_C=\cC_3[\bm X^t]\,.
\end{align}
The freedom in defining the $n$-th order Casimir operators
reflects the ambiguity in choosing a basis for the center of the universal enveloping algebra.
Below we will also derive a relation between
the above two choices,
which we will make essential
use of to derive 
a relation between the
generating functions in \eqref{Casimir 1}.

The operator $X^A{}_B$ and $R_I{}^J$
contain the $(\o,\tilde\o)$ operators in different orders: it is convenient to work with
the following order, 
\be 
    Y^A{}_B=\tilde\o^A_I\,\o^I_B\,,
    \qquad 
    S^I{}_J=\o^I_A\,\tilde \o^A_J\,,\label{YS}
\ee 
which are related to the $X^A{}_B$ and $R^I{}_J$ by
\be 
    X^A{}_B=Y^A{}_B+\tfrac{N}2\,\delta^A_B\,,
    \qquad 
    R_I{}^J= S^J{}_I-\tfrac{M}2\,\delta^J_I\,.
\ee 
By introducing generating functions for $\tr[\bm Y^n]$
and $\tr[\bm S^n]$ as
\be 
    y(t)=\sum_{n=0}^\infty t^{n}\,\tr(\bm Y^n)\,,
    \qquad 
    s(t)=\sum_{n=0}^\infty t^{n}\,\tr(\bm S^n)\,,
\ee
we can find the relation between $x(t)$ and $y(t)$,
\be 
    x(t)=\tr\left[(1-t\,\bm X)^{-1}\right]
    =\frac{1}{1-\frac{N}2\,t}\,\tr\left[\Big(1-\frac{t}{1-\frac{N}2\,t}\,\bm Y\Big)^{-1}\right]
    =\frac{y\big(\frac{t}{1-\frac{N}2\,t}\big)}{1-\frac{N}2\,t}\,.
    \label{tr geo series}
\ee 
Similarly, $s(t)$ is related to $\tilde{r}(t)$ as
\be 
     s(t)=\frac{\tilde r\big(\frac{t}{1-\frac{M}2\,t}\big)}{1-\frac{M}2\,t}\,.
\ee
Now, let us relate $y(t)$ and $s(t)$:
In the definition of $\tr(\bm Y^n)$, we 
move the $\tilde \o$ operator in the left end to the right end
and find the relation,
\be 
    \tr(\bm Y^n)=
    \tr(\bm S^n)-\sum_{k=0}^{n-1}\,
     \tr(\bm S^k)\, \tr(\bm Y^{n-1-k})
     \qquad [n\ge1]\,.
     \label{YS rel}
\ee 
This relation can be rephrased in terms of the generating functions as
\be
    y(t)-M=s(t)-N-t\,s(t)\,y(t)\,,
\ee 
which can be solved as
\be
    y(t)=\frac{s(t)+M-N}{1+t\,s(t)}\,.
\ee 
Finally combining the results so far obtained,
we find the relation,
\be 
    x(t)=\frac{\tilde r\big(\frac{t}{1-\frac{M+N}2\,t}\big)
    +(M-N)\frac{1-\frac{M+N}2\,t}{1-\frac{N}2\,t}}{1-\frac{M+N}2\,t+t\cdot  \tilde r\big(\frac{t}{1-\frac{M+N}2\,t}\big)}\,.
    \label{x rt gen rel}
\ee
The formula \eqref{x rt gen rel} generates the relations between $\cC_n[\bm X]$ and $\tilde\cC_n[\bm R]$.
Next we would like to relate the generating functions $r(t)$ and $\tilde r(t)$. This computation is done in Appendix \ref{app:Casimir} and the result is
\be
\tilde r(t)=\frac{r(\tfrac{t}{1+N\,t})}{1+N\,t-t\cdot r(\tfrac{t}{1+N\,t})}\,,
\ee
which can be inserted in \eqref{x rt gen rel} to arrive at
\be
x(t)=\frac{1}{1-\tfrac{N}2\, t}\left[\frac{1+\tfrac{N-2\,M}2\,t}{1+\tfrac{N-M}2\, t}\; r\left(\tfrac{ t}{1+\tfrac{N-M}2\, t}\right)+M-N\right].\label{x r gen rel}
\ee
The inverse relation is given by
\be
r(t)=\frac{1}{1-\tfrac{M}2\, t} \left[\frac{ 1+\tfrac{M-2\,N}2\, t}{1+\tfrac{M-N}2\, t}\;x\left(\tfrac{t}{1+\tfrac{M-N}2\,t}\right)+N-M\right].
\ee
This is the same formula as \eqref{x r gen rel} with  $x(t)\leftrightarrow r(t)$ and  $M\leftrightarrow N$, as expected.
The first few relations, generated by \eqref{x r gen rel}, read
\ba 
    && \cC_1[\bm X]=\cC_1[\bm R]=Z\,,\nn
    && \cC_2[\bm X]=\cC_2[\bm R]-\frac{M N (M-N)}4\,,
    \label{eq:C2_GL_relation} \\
    && \cC_3[\bm X]=\cC_3[\bm R]+\frac{(M-N)}8\,\left[4\,\cC_2[\bm R]
    -2\,(M+N)\,Z -M^2 N\right]\,,\label{eq:C3_GL_relation}\\
    && \cC_4[\bm X]=\cC_4[\bm R]+\frac{(M-N)}{16}\left[16\,\cC_3[\bm R]
    -8\,N\,\cC_2[\bm R]-4\,M^2\,Z-M\,N\,(M^2-M\,N+N^2)\right]\,.\nonumber
\ea
A straightforward check shows that our result for the Casimir relations at any order \eqref{x r gen rel} reproduce the relations of \cite{itoh_2003} (Theorem A).

The Casimir operators $\cC_n[\bm R]$ with $n>N$ are given by
the lower order Casimirs, and the relation can be extracted from
\be 
    \det(1-t\,\bm R)=\exp\left(-\sum_{n=1}^\infty \frac{t^n}{n}\,
    {\cal C}_n[\bm R]\right)=1-p_{N}(t)\,,
\ee
where $p_{N}(t)$ is an order $N$ polynomial with $p_N(0)=0$.
The coefficients of $p_N(t)$ parameterize the independent Casimir operators.
By differentiating the above equation, we obtain the relation between
$p_N(t)$ and the generating function $r(t)$ of $\cC_n[\bm R]$\,:
\be 
    r(t)=N+\frac{t\,p'_N(t)}{1-p_N(t)}\,.
\ee 
We can put this expression into \eqref{x r gen rel}
to obtain a rational function whose numerator and denominator
have degree $N+1$ and $N+2$, respectively.

One can also solve for $\tilde r(t)$ in \eqref{x rt gen rel} and get the following formula:
\be
    \tilde r(t)=\frac{x\big(\frac{t}{1+\frac{M+N}2\,t}\big)
    +(N-M)\frac{1+\frac{M+N}2\,t}{1+\frac{M}2\,t}}{1+\frac{M+N}2\,t-t\, x\big(\frac{t}{1+\frac{M+N}2\,t}\big)}\,.
    \label{r gen rel}
\ee
We could do exactly the same computation as when deriving \eqref{x rt gen rel}, but for the $r(t)$ in terms of $\tilde x(t)$, and get a similar relation with replaced $M$ and $N$:
\be
    r(t)=\frac{\tilde x\big(\frac{t}{1-\frac{M+N}2\,t}\big)
    +(N-M)\frac{1-\frac{M+N}2\,t}{1-\frac{M}2\,t}}{1-\frac{M+N}2\,t+t\, \tilde x\big(\frac{t}{1-\frac{M+N}2\,t}\big)}\,.
    \label{r gen rel 2}
\ee
One may notice that the relations \eqref{r gen rel} and \eqref{r gen rel 2} are interchanged 
upon 
$(t, \tilde r(t), x(t))\leftrightarrow (-t,r(-t),\tilde x(-t))$:
$r(-t)$ depends on $\tilde x(-t)$ exactly same way as $\tilde r(t)$ on $x(t)$.
This can be understood in the following way: the difference in the structure of the $r(t)$ and $\tilde r(t)$, and correspondingly $x(t)$ and $\tilde x(t)$ are given via commutators, which are changing their signs when we change the sign of the generators. On the other hand, changing the signs of the commutators is equivalent to taking the opposite product rule.

\subsection{Duality $(O_N,Sp_{2M})$}
We finally provide the 
duality relations between
the Casimir operators
 $\cC_n[\bm M]=\tr(\bm M^n)$ and $\cC_n[\bm K]=\tr(\bm K^n)$
 to all orders in $n$.
Compared to the $(GL_M,GL_N)$ case of the previous section,
the $(O_N,Sp_{2M})$ case requires
more steps of
computations,
and we leave most details to the Appendix~\ref{App:OnSp2M}.
We begin with defining the operators,
\be 
    N^A{}_B=y^A_I\,y_B^I\,,
    \qquad
    L^I{}_J=y^I_A\,y^A_J\,,
\ee 
where $y_I^A$ is defined as
\be 
    y_I^A=E^{AB}\,\O_{IJ}\,y_B^J\,.
\ee 
and satisfies 
\begin{equation}
    [y_I^A, y^J_B] = \delta^A_B\,\delta^J_I\,.
\end{equation}
The operators $N^A{}_B$ and $L^I{}_J$ are related to 
$M^A{}_B$ and $K^I{}_J$ by\footnote{The numbers $N$ and $M$ should not be confused with the generators $N^A{}_B$ and $M^A{}_B$ in this section.}
\be 
   N^A{}_B = -M^{A}{}_B+M\,\delta^A_B\,,
    \qquad 
    L^I{}_J = -\big(K^I{}_J+\frac{N}2\,\delta^I_J\big)\,,
    \label{NMLK relation}
\ee 
where $M^A{}_B := E^{AC} M_{CB}$ and
$K^I{}_J := K^{IK}\, \Omega_{KJ}$.

By defining generating functions,
\be 
    m(t)=\sum_{n=0}^\infty t^n\,\cC_n[\bm M]\,,
    \qquad
    k(t)=\sum_{n=0}^\infty t^n\,\cC_n[\bm K]\,,
\ee
\be 
    n(t)=\sum_{n=0}^\infty t^n\,\tr(\bm N^n)\,,
    \qquad
    l(t)=\sum_{n=0}^\infty t^n\,\tr(\bm L^n)\, \, ,
\ee 
it is possible to derive exact relations
between the Casimir operators to all orders through their generating functions.
To derive these relations we find it useful 
to expand the space of generating functions to the following 
set of two-parameter family of generating functions:
\ba
    & a(t,u)=\sum_{k,l=0}^\infty
    t^k\,u^l\,{}^I[2k+1]_A\,{}_I[2l+1]^A\,
    \nn 
     & b(t,u)=\sum_{k,l=0}^\infty
    t^k\,u^l\,{}^B[2k]_A\,{}_B[2l]^A\,,
  \qquad 
     d(t,u)=\sum_{k,l=0}^\infty
    t^k\,u^l\,{}^I[2k]_J\,{}_I[2l]^J\,,
\ea
where the brackets are short-hand for
\be 
    {}^{\alpha}[n]_\beta:=y^\alpha_{\gamma_2}\,y_{\gamma_3}^{\gamma_2} \cdots y^{\gamma_{n}}_\beta\,,
    \qquad
    {}_{\alpha}[n]^\beta:=y_\alpha^{\gamma_2}\,y^{\gamma_3}_{\gamma_2} \cdots y_{\gamma_{n}}^\beta\, ,
\ee 
with the greek indices being generic, and with $
{}^\alpha [0]_\beta := \delta^\alpha_\beta $ and ${}_\alpha[0]^\beta := \delta_\alpha^\beta
$. The $a,b,d$ system of generating functions is a direct uplift of the $n, l$ system,
as should be clear by setting $t=0$ or $u=0$. 
The usefulness of introducing these generalized generating functions is seen directly,
when trying to relate $n$ with $l$ in a similar way, as $y$ was related to $s$ from \eqref{YS rel} in the previous section: The equivalent expression of \eqref{YS rel} here reads
\be
\tr(\bm N^n)=
    \tr(\bm L^n)+\sum_{i=0}^{n-1}
    \tr(\bm L^i)\,\tr(\bm N^{n-1-i})
    -\sum_{i=0}^{n-2} 
    (-1)^{n-2-i}\,{}^I[2i+1]_A\,{}_I[2(n-2-i)+1]^A,
\ee 
valid for $n\geq 1$, and with the last term understood as zero for $n=1$.
It can be equivalently written as
\be
    n(t)-N=l(t)-2M+
    t\,l(t)\,n(t)
    -t^2\,a(t,-t)\, .
    \label{nla rel main}
\ee 
As described in Appendix~\ref{App:OnSp2M},
it is possible to relate the generators $a,b,d$ with each other,
through similar, but much more involved manner,
and as an outcome it is possible relate $n$ entirely in terms of $l$ (or vice-versa).
The result is that
\begin{align}
n(t) = 
\frac{l(t) - 2M +N}{
1 - t\, l(t)\, (1- \frac{t}{(2M-N+2)t-2})} \, 
\quad
\Leftrightarrow
\quad 
l(t) = 
\frac{n(t) + 2M -N}{
1 + t\, n(t)\, (1- \frac{t}{(2M-N+2)t-2})} \, , 
\label{nl rel main}
\end{align}
and as a bonus, we are also able to find expressions for $a,b,d$
in terms of $n$ and $l$, reading
\ba 
a(t,u) \eq  \frac{l(t)-2 M-n(-u)+N}{(2 M-N +1) t u+t-u} \, ,\\
b(t,u) \eq  
\frac{u \, n(-u) (t u-t-u)+t \, n(t) (tu+t+u- u \, n(-u) (t+u))}{(t+u) ((2 M-N +1) t u+t-u)} \,,\\
d(t,u) \eq 
\frac{u \,  l(-u) (t u-t-u)+t \, l(t) ( tu+t+u+ u\, l(-u) (t+u))}{(t+u) ((2 M-N +1) t u+t-u)} \, .
   \ea
 Notice that $a(t,u)$ is well-defined at $u=-t$, and consistent with \eqref{nla rel main},
 while $b$ and $d$ seem singular at that point but the limit $u\to -t$ exists, as it should by definition.
 
 Now that we have established a relation between $n$ and $l$ 
 it is straightforward to pass this on to a relation between $m$ and $k$:
 By formally rewriting $\sum_{n=0}^\infty t^n\, \tr(\bm O^n) = \tr(\tfrac{1}{1-t \bm O})$  (cf. \eqref{tr geo series}) 
 and using the relations in \eqref{NMLK relation}, the following pairwise relations between the generating functions are found:
\be 
   m(t)=\frac{n\big(-\frac{t}{1-M\,t}\big)}{1-M\,t}
    \qquad 
    \Leftrightarrow
    \qquad 
    n(t)=\frac{m\big(-\frac{t}{1-M\,t}\big)}{1-M\,t}\,,
    \label{mn rel}
\ee 
\be 
    k(t)=\frac{l\big(-\frac{t}{1+\frac{N}2\,t}\big)}{1+\frac{N}2\,t}
    \qquad 
    \Leftrightarrow
    \qquad 
    l(t)=\frac{k\big(-\frac{t}{1+\frac{N}2\,t}\big)}{1+\frac{N}2\,t}\,.
    \label{kl rel}
\ee 
From these and \eqref{nl rel main} we finally establish
\begin{align}
    k(t)&=
\frac{(2M-N)\left (1 + \frac{2Mt}{2+N t}\right ) + m\left(\tfrac{2 t}{2 + (2 M +N) t}\right)}{\left (1+\frac{N}{2}t\right )\left (1 + \frac{2Mt}{2+N t}\right ) - \frac{1}{2}t \left (1 + \frac{1+Mt}{1+(1 + M)t}\right ) m\left(\tfrac{2t}{ 2 + (2 M +N) t}\right) } \, ,
\label{km rel}
\\
& \hspace{50mm} \Updownarrow 
\nonumber 
\\
    m(t)&=-
\frac{(2M-N)\left (1 - \frac{Nt}{2-2M t}\right ) - k\left(\tfrac{2t}{2 - (2 M +N) t}\right)}{\left (1-M t\right )\left (1 - \frac{N t}{2-2 M t}\right ) + \frac{1}{2}t \left (1 + \frac{2-N t}{2+(2-N)t}\right ) k\left(\tfrac{2t}{ 2 - (2 M +N) t}\right) } \, .
\label{mk rel}
\end{align}
We notice that the expressions for $k$ and $m$ are symmetric under the exchange $N \leftrightarrow - 2M$ and $m(t) \leftrightarrow - k(t)$, while those for $n$ and $l$ are simply symmetric under $n(t) \leftrightarrow - l(t)$.
This fits into the picture of universality (see, e.g. \cite{Vogel,Mkrtchyan_2012,Avetisyan:2019orp}) and the
$SO_N\underset{N\leftrightarrow -2M}\longleftrightarrow Sp_{2M}$ relation \cite{Mkrtchian:1981bb,Mkrtchyan_2011}.

It is worth
checking the relations at the lowest orders
without using the generating function method.
  The first two within each group read
\begin{align}
    &\tr(\bm N)=MN\,,
    &&\tr(\bm L)=-M N\,,\nn
    &\tr(\bm N^2)=\cC_2[\bm M]+M^2 N\,,
    &&\tr(\bm L^2)=\cC_2[\bm K]+\frac{N^2 M}{2}\,,
 \end{align}
 where $\tr(\bm M)=\tr(\bm K) = 0$ was used.
Next, using
\be
\tr(\bm M^3) =  (\tfrac{N}{2}-1)\,\cC_2[\bm M]\,,
\qquad \tr(\bm K^3)=- (1+M) \,\cC_2[\bm K]
\ee
one finds at the 
third and fourth orders
\begin{align}
    &\tr(\bm N^3)=(1- \tfrac{N}{2}+ 3 M)\,\cC_2[\bm M]+M^3 N\,,\nn
    &\tr(\bm L^3)=(1+M-\tfrac{3 N}2)\,\cC_2[\bm K]-\frac{M N^3}{4}\,,
\end{align}    
 and
 \begin{align}   
     &\tr(\bm N^4)=\cC_4[\bm M]+(6\,M^2+ 4M- 2MN)\,\cC_2[\bm M]+M^4 N\,, \nn
    &\tr(\bm L^4)=\cC_4[\bm K]+(\tfrac{3}2\,N^2-2N-2MN)\,\cC_2[\bm K]+\frac{M N^4}{8}\,.
\end{align} 
With these expressions one can now check the duality relations.
For instance by moving
the $y$ operator in the left end of $\tr(\bm N^n)$ to its right end at $n=2$, we find
\be 
    \tr(\bm N^2)=\tr(\bm L^2)+\tr(\bm L^0)\,\tr(\bm N)+\tr(\bm L)\,\tr(\bm N^0)-\tr(\bm L)\,,
\ee 
while at $n=4$ we find
\begin{align}
   \tr(\bm N^4)
    =\,& \tr(\bm L^4)+\tr(\bm L^0)\,\tr(\bm N^3)+\tr(\bm L)\,\tr(\bm N^2)
    +\tr(\bm L^2)\,\tr(\bm N)+\tr(\bm L^3)\,\tr(\bm N^0)
    \nn
    &
    -\tr(\bm L^3) 
    +\tr(\bm L^0) \tr(\bm  N^2 )
    -\tr(\bm N^0) \tr(\bm N^2) + \tr(\bm N^2)
    \,.
\end{align}
From the above low-order relations it can now be checked that
 \ba
    \cC_2[\bm M] \eq \cC_2[\bm K]+\frac{MN(2M-N+2)}2\,,
    \label{eq:C2_O-Sp_relation} \\
	\cC_4[\bm M] \eq \cC_4[\bm K]
    -\frac{2 M+N}{2}\,
    \cC_2[\bm K] 
\nonumber \\
&&+\frac{M N }{8}(2 M-N+2) \left(
(2M-N)(2M-N+2) +2M N -2N+4
\right),
\ea
which is exactly what one finds directly from \eqref{km rel}.

Using the results of \cite{itoh_2003}, one can derive a relation between generating functions of Casimir operators different from \eqref{km rel}. This is due to relations between odd and even Casimirs of orthogonal and symplectic groups.  
We will comment more on different relations between generating functions in the follow-up paper.

\acknowledgments 

The authors are grateful to Zhirayr Avetisyan and Axel Kleinschmidt for useful remarks and Minoru Itoh for communication on his work \cite{itoh_2003}.
The work of T.B. and E.J. was supported by National Research Foundation (Korea) through the grants 
NRF-2018H1D3A1A02074698 and NRF-2019R1F1A1044065. The work of K.M. was supported in part by Scuola Normale, by INFN (IS GSS-Pi) and by the MIUR-PRIN contract 2017CC72MK\_003. K.M. also thanks Humboldt Foundation and Max-Planck Institute for Gravitational Physics (Albert-Einstein-Institute) for the support in the initial stages of this work.

\appendix

\section{Representations of compact groups}
\label{app:irrep_compact}

Finite-dimensional representations of compact
Lie groups are known to be in one-to-one
correspondence with integral dominant weights.
The latter are weights whose components
in the Dynkin basis are all positive integers.
In other words, finite-dimensional representations
of compact Lie groups $G$ are highest weight
representations, with highest weight of the
form $\bm\alpha = \sum_{k=1}^{{\rm rank}(G)}
\alpha^\omega_k\, \bm\omega_k$ with
$\alpha_k^\omega \in \mathbb N$ the Dynkin
labels and $\bm\omega_k$ the fundamental weights
of $G$. In this appendix, we will review how
these highest weight representations are
realized as spaces of tensors with special
properties. To do so, it will be more convenient
to express the weights of $G$ in the orthonormal
basis $\{\mathsf{e}_k\}_{k=1}^{{\rm rank}(G)}$,
where the vectors $\mathsf{e}_i$ are mutually
orthogonal unit vectors in
$\mathbb R^{{\rm rank}(G)}$. We will write
$\bm\alpha=(\alpha_1, \dots, \alpha_N)$,
for $\alpha_1, \dots, \alpha_N$ the components of
$\bm\alpha$ in this orthonormal basis.

\subsection*{Finite-dimensional representations of $GL_N$}
Let us start with the reductive group $GL_N$.
In the orthonormal basis, the components of
a $GL_N$ integral dominant weight $\bm\alpha$
are all integers (not necessarily positive)
and ordered, i.e.
\begin{equation}
    \alpha_1 \ge \alpha_2 \ge \dots \ge \alpha_N\,.
\end{equation}
Recall that the $GL_N$-irreps appearing in the
decomposition of $V_L = (\mathbb C^N)^{\otimes L}$
are uniquely associated with 
the irreps of the symmetric group $\cS_L$,
defined
by Young diagrams $\bm\ell$,
\begin{equation}
    V_L\rvert_{\cS_L \times GL_N} =
    \bigoplus_{\bm\ell\in{\rm Par}(L|N)}\bm\ell\,
    \otimes\, [\bm\ell]_{GL_N}\,,
\end{equation}
where ${\rm Par}(L|N)$ denotes the set of Young
diagrams with $L$ boxes and at most $N$ rows.
This standard result is known as the Schur--Weyl
duality, and can be seen as another instance of
duality in representation theory, similar to the
dual pair correspondence (see e.g. \cite{Hamermesh1989,Goodman2009,Rowe:2012ym}
and references therein).
The subspace of $V_L$
carrying the $\cS_L$-irrep $\bm\ell$ can be
singled out by the projector
$\mathbb P^{\bm\ell}: V_L \to \bm\ell$, 
 i.e.
the operator that symmetrizes a given tensor
$T^{a_1 \dots a_L}$
so that the resulting tensor
has the symmetry of the Young diagram $\bm\ell$.
In other words, the $GL_N$ representation denoted
$[\bm\ell]_{GL_N}$ and dual to the $\cS_L$-irrep
$\bm\ell$ consists of a tensor with the symmetry of
the Young diagram $\bm\ell$.

For instance, when $L=2$ there are two possible
representation of $\cS_2$ to consider, namely
$\bm\ell=(2)$ and $\bm\ell=(1,1)$, corresponding
to a symmetric and an antisymmetric tensor. The
associated projectors are
\begin{equation}
    \Ylinethick{0.7pt}
    {\mathbb P}_{a_1a_2}^{\tiny\gyoung(;{b_1};{b_2})}
    = \tfrac12\,(\delta^{b_1}_{a_1}
    \delta^{b_2}_{a_2} + \delta^{b_2}_{a_1}
    \delta^{b_1}_{a_2})\,, \qquad
    {\mathbb P}_{a_1a_2}^{\tiny\gyoung(;{b_1},;{b_2})}
    = \tfrac12\,(\delta^{b_1}_{a_1}
    \delta^{b_2}_{a_2} - \delta^{b_2}_{a_1}
    \delta^{b_1}_{a_2})\,.
\end{equation}
For $L=3$, there are three possible Young diagrams
that can appear (for $N>2$), namely $\bm\ell=(3)$,
$\bm\ell=(1,1,1)$ and $\bm\ell=(2,1)$. Only one
standard Young tableau can be attached to the first
two, while two for the last diagram. The associated
projectors read
\begin{equation}
    \Ylinethick{0.7pt}
    {\mathbb P}_{a_1a_2a_3}
    ^{\tiny\gyoung(;{b_1};{b_2};{b_3})}
    = \tfrac1{3!}\, \big(\delta^{b_1}_{a_1}
    \delta^{b_2}_{a_2}\delta^{b_3}_{a_3}
    + \delta^{b_1}_{a_1}\delta^{b_3}_{a_2}
    \delta^{b_2}_{a_3} + \delta^{b_2}_{a_1}
    \delta^{b_1}_{a_2}\delta^{b_3}_{a_3}
    + \delta^{b_2}_{a_1}\delta^{b_3}_{a_2}
    \delta^{b_1}_{a_3} + \delta^{b_3}_{a_1}
    \delta^{b_2}_{a_2}\delta^{b_1}_{a_3}
    + \delta^{b_3}_{a_1}\delta^{b_1}_{a_2}
    \delta^{b_2}_{a_3}\big)\,,
\end{equation}
\begin{equation}
    \Ylinethick{0.7pt}
    {\mathbb P}_{a_1a_2a_3}
    ^{\tiny\gyoung(;{b_1},;{b_2},;{b_3})}
    = \tfrac1{3!}\, \big(\delta^{b_1}_{a_1}
    \delta^{b_2}_{a_2}\delta^{b_3}_{a_3}
    - \delta^{b_1}_{a_1}\delta^{b_3}_{a_2}
    \delta^{b_2}_{a_3} - \delta^{b_2}_{a_1}
    \delta^{b_1}_{a_2}\delta^{b_3}_{a_3}
    + \delta^{b_2}_{a_1}\delta^{b_3}_{a_2}
    \delta^{b_1}_{a_3} - \delta^{b_3}_{a_1}
    \delta^{b_2}_{a_2}\delta^{b_1}_{a_3}
    + \delta^{b_3}_{a_1}\delta^{b_1}_{a_2}
    \delta^{b_2}_{a_3}\big)\,,
\end{equation}
\begin{equation}
    \Ylinethick{0.7pt}
    {\mathbb P}_{a_1a_2a_3}
    ^{\tiny\gyoung(;{b_1};{b_2},;{b_3})}
    = \tfrac13\, \big(\delta^{b_1}_{a_1}
    \delta^{b_2}_{a_2}\delta^{b_3}_{a_3}
    + \delta^{b_2}_{a_1}\delta^{b_1}_{a_2}
    \delta^{b_3}_{a_3} - \delta^{b_2}_{a_1}
    \delta^{b_3}_{a_2}\delta^{b_1}_{a_3}
    - \delta^{b_3}_{a_1}\delta^{b_2}_{a_2}
    \delta^{b_1}_{a_3}\big)\,,
\end{equation}
and
\begin{equation}
    \Ylinethick{0.7pt}
    {\mathbb P}_{a_1a_2a_3}
    ^{\tiny\gyoung(;{b_1};{b_3},;{b_2})}
    = \tfrac13\, \big(\delta^{b_1}_{a_1}
    \delta^{b_2}_{a_2}\delta^{b_3}_{a_3}
    - \delta^{b_2}_{a_1}\delta^{b_1}_{a_2}
    \delta^{b_3}_{a_3} - \delta^{b_3}_{a_1}
    \delta^{b_1}_{a_2}\delta^{b_2}_{a_3}
    + \delta^{b_3}_{a_1}\delta^{b_2}_{a_2}
    \delta^{b_1}_{a_3}\big)\,.
\end{equation}

The action of $GL_N$ on 
$[\bm\ell=(\ell_1,\dots,\ell_p)]_{GL_N}$
with
$p \le N$ 
reads
\begin{equation}
    \rho_{\bm\ell}(g)\, T^{a_1(\ell_1),
    \dots, a_p(\ell_p)} = \prod_{k=1}^p
    \ell_k\, g_b{}^{a_k}\, T^{a_1(\ell_1),
    \dots, b\,a_k(\ell_k-1), \dots,
    a_p(\ell_p)}\,,
\end{equation}
where $g_b{}^c$ denote the components of the
matrix $g \in GL_N$. Correspondingly, the action
of the Lie algebra $\mathfrak{gl}_N$ is given by
\begin{equation}
    X^b{}_c\,T^{a_1(\ell_1), \dots, a_p(\ell_p)}
    = \sum_{k=1}^N\, \ell_k\, \delta^{a_k}_c\,
    T^{a_1(\ell_1), \dots, b\,a_k(\ell_k-1), 
    \dots, a_p(\ell_p)}\,.
\end{equation}
Recall that these generators obey
\begin{equation}
    [X^a{}_b, X^c{}_d] = \delta^c_b\, X^a{}_d
    - \delta^a_d\, X^c{}_b\,,
\end{equation}
from which we can see that $X^a_a$ generate a
Cartan subalgebra, while $X^a{}_b$ for $a<b$
(resp. $a>b$) are raising (resp. lowering)
operators. Let us show that the component
$T^{1(\ell_1), \dots, p(\ell_p)}$ of this tensor
(i.e. the component obtained by setting all the
indices of $a$-th group of symmetric indices to
the value $a$) is the highest weight vector of
this $\mathfrak{gl}_N$ irrep. For this particular
component, the above action simplifies to a
single term,
\begin{equation}
    X^a{}_b\,T^{1(\ell_1), \dots, p(\ell_p)}
    = \ell_b\,T^{1(\ell_1), \dots,
    a\, b(\ell_b-1), \dots, p(\ell_p)}\,,
\end{equation}
when $b \le p$, and vanishes otherwise. First
of all, it is an eigenvector of the
Cartan subalgebra generator $X^a{}_a$ (no
summation implied), with eigenvalue $\ell_a$.
The raising operator are the generators
$X^a{}_b$ with $a < b$, which implies that
their action takes the form
\begin{equation}
    X^a{}_b\,T^{1(\ell_1), \dots, p(\ell_p)}
    = \ell_b\,T^{1(\ell_1), \dots, a(\ell_a), 
    \dots, a\, b(\ell_b-1), \dots, p(\ell_p)}
    =0\,,
\end{equation}
i.e. all the indices of the $a$-th group are
symmetrized with one index of a latter group
(the $b$-th), which identically vanishes due
to the symmetry of the tensor $T$.\footnote{For 
instance, in the case of
$\bm\ell=(2,1)$, the action of a generator
$X^i{}_j$ of $\mathfrak{gl}_N$ reads
\begin{equation}
    X^c{}_d\, T^{a_1a_2,b}
    = 2\,\delta^{(a_1}_d\, T^{a_2)c,b}
    + \delta^b_d\, T^{a_1a_2,c}\,.
\end{equation}
For $a_1=a_2=1$ and $b=2$ and all pairs
$c>d$, the above vanishes: the first term
because of the factor $\delta^1_d$ with $d>1$
and the second term because of the fact
that the total symmetrization of all indices
of $T$ vanishes and hence $T^{11,1}=0$.}
This shows that the representation of $GL_N$
on the space of tensor with the symmetry
of the Young diagram $\bm\ell$ is a highest
weight representation, with highest weight
given by $\bm\ell$ (in the orthonormal basis).

However, this does not exhaust all possible 
finite-dimensional representations. So, what about
representations with highest weights containing some
negative components? It turns out that they can be
realized as a subspace of
\begin{equation}
    V_{L,M} = (\mathbb C^N)^{\otimes L} \otimes 
    (\mathbb C^N{}^*)^{\otimes M}\,,
\end{equation}
i.e. they correspond to spaces of tensors $T$ with
two types of indices: $T^{a_1 \cdots a_L}_{b_1 \cdots b_M}$.

First, let us have a look at $V_{0,1} = \mathbb C^N{}^*$.
As a vector space, this corresponds to the dual space of
the fundamental representation. As such, it also carries
a representation $\rho$ of $GL_N$ defined by
\begin{equation}
    \langle \rho_{\bar f}(g)\,\phi, v \rangle = 
    \langle \phi, \rho_f(g)\,v \rangle\,,
\end{equation}
for $g \in U(N)$, $v \in \mathbb C^N$ and
$\phi \in \mathbb C^N{}^*$, and where $\langle\,,\rangle$
denote the pairing between $\mathbb C^N$ and
$\mathbb C^N{}^*$. More concretely, this means
that
\begin{equation}
    \rho_{\bar f}(g) = \rho_f(g^{-1})^t\,,
\end{equation}
where the exponent ${}^t$ denote the transpose.
The action of $GL_N$ on $V_{0,M}$ is then
simply the tensor product of $\rho_{\bar f}$.
The same arguments as previously show that
$V_{0,M}$ can be decomposed into a direct sum
of $\cS_L \otimes GL_N$ irreps, which are in
bijections. In other words, irreducible
representations of $GL_N$ in $V_{0,M}$
correspond to tensors with the symmetry of
all possible Young diagrams 
$\lowerbcal{m}$ with $M$ boxes
and at most $N$ rows. The corresponding
action of $\mathfrak{gl}_N$ on such tensors
is given by
\begin{equation}
    X^b{}_c\,T_{a_1(\lowercal{m}_1), \dots,
    a_p(\lowercal{m}_p)} = -\sum_{k=1}^N\,
    \lowercal{m}_k\, \delta_{a_k}^b\,
    T_{a_1(\lowercal{m}_1), \dots,
    b\,a_k(\lowercal{m}_k-1), \dots,
    a_p(\lowercal{m}_p)}\,.
\end{equation}
In this case, the component
$T_{N(\lowercal{m}_1), \dots, (N-p)(\lowercal{m}_p)}$
is the highest weight of the representation.
The action of $\mathfrak{gl}_N$ now reads
\begin{equation}
    X^b{}_c\,T_{N(\lowercal{m}_1), \dots,
    1(\lowercal{m}_p)} = -\lowercal{m}_{N+1-b}\,
    T_{N(\lowercal{m}_1), \dots,
    c\,(N+1-b)(\lowercal{m}_b-1), \dots,
    (N+1-p)(\lowercal{m}_p)}\,,
\end{equation}
if $b \le p$ and is trivial otherwise. It is
clear that $T_{N(\lowercal{m}_1), \dots,
(N+1-p)(\lowercal{m}_p)}$ is an eigenvector
of all generators of the Cartan subalgebra,
with weight $(0, \dots, 0, -\lowercal{m}_p,
\dots, -\lowercal{m}_1)$. Moreover, for $b<c$
the previous equation reads
\begin{equation}
    X^b{}_c\,T_{N(\lowercal{m}_1), \dots,
    1(\lowercal{m}_p)} = -\lowercal{m}_{N+1-b}\,
    T_{N(\lowercal{m}_1), \dots,
    c(\lowercal{m}_{N+1-c}), \dots,
    c\,(N+1-b)(\lowercal{m}_b-1), \dots,
    (N+1-p)(\lowercal{m}_p)}=0\,,
\end{equation}
due to the symmetry of $T$. This shows that
tensor on $V_{0,L}$ having the symmetry of
a Young diagram $\lowerbcal{m}$
form a highest weight representation of
$GL_N$ with highest weight $(0, \dots, 0,
-\lowercal{m}_p, \dots, -\lowercal{m}_1)$, i.e.
whose components are all negative.

Finally, consider the case $L=M=1$, i.e.
the space of tensors of the form $T^a_b$,
on which $GL_N$ acts as
\begin{equation}
    \rho_{V_{1,1}}(g)\, T^a_b = g_c{}^a\,
    (g^{-1})_b{}^d\, T^c_d\,,
\end{equation}
for $g \in GL_N$. In particular, one can
notice that the trace of 
$T=\delta_a^b\, T^a_b$ defines a
one-dimensional invariant subspace of
$V_{1,1}$, as $\rho_{V_{1,1}}(g)\, T = T$.
This observation extends to higher rank
tensors, and hence we can conclude that
the traces of any tensor (in the sense
of any contraction of an upper index with
a lower one) define a invariant subspace.
In other words, the irreducible
representations of $GL_N$ in $V_{L,M}$
are composed of traceless tensors.
Added to the previous discussions,
we can conclude that highest weight
representations of $GL_N$, with highest
weight
\begin{equation}
    (\ell_1, \dots, \ell_p, 0, \dots, 0,
    -\lowercal{m}_q, \dots, -\lowercal{m}_1)\,,
\end{equation}
where $p+q \le N$ can be realized as the
space of tensors of the form
\begin{equation}
    T^{a_1(\ell_1), \dots, a_p(\ell_p)}
    _{b_1(\lowercal{m}_1),\dots,b_q(\lowercal{m}_q)}
\end{equation}
i.e. whose upper indices have the symmetry
of the Young diagram 
$\bm\ell=(\ell_1,\dots,\ell_p)$, lower
indices the symmetry of the Young diagram 
$\lowerbcal{m}=(\lowercal{m}_1,\dots,
\lowercal{m}_q)$, and which are traceless.
Notice that the tracelessness  implies
that the tensors with  $p+q \ge N$ 
identically vanish.

It will be useful to record the
expression of the dimension of the
space of tensors with symmetry $\bm\ell$,
\begin{equation}
    {\rm dim} [\bm\ell]_{GL_N} = 
    \prod_{1 \le i < j \le N}
    \frac{\ell_i-\ell_j+j-i}{j-i}\,.
\end{equation}

The finite dimensional representations of $GL_N$ described above provide all unitary irreps of $U(N)$,
which are isomorphic to the unitary irreps of 
$U(1)\times SU(N)$. The $U(1)$ irrep is 
simply given by $|\bm\ell|-|\lowerbcal{m}|$, the number of boxes in $\bm\ell$ minus
the number of boxes in $\lowerbcal{m}$.
An $SU(N)$ irrep is again given by 
a Young diagram now with maximum height $N-1$.
In the following, we demonstrate
how a single Young diagram of $SU(N)$ irrep
can be obtained from
the two Young diagrams $\bm\ell$ and $\lowerbcal{m}$ of a $U(N)$ irrep.
Without the traceless condition, 
each of $\bm\ell$ and $\lowerbcal{m}$
carry an irrep of $SU(N)$, and hence
we get a tensor product representation which
we need to decompose.
To do so, we express the diagram
$\lowerbcal{m}$ in the antisymmetric basis,
as $\lowerbcal{m}=[\lowercal{m}^1, \dots,
\lowercal{m}^r]$ where $\lowercal{m}^k$ are
the height of the $k$-th column (hence
$\lowercal{m}^1=q$ and $r=\lowercal{m}_q$).
The corresponding tensor $\Psi$ now has
the form
\begin{equation}
    \Psi^{i_1(\ell_1), \dots, i_p(\ell_p)}
    _{j^1[\lowercal{m}^1], \dots,
    j^r[\lowercal{m}^r]}\,,
\end{equation}
i.e. $j^k[\lowercal{m}^k]$ denote a group
of $\lowercal{m}^k$ indices which are
antisymmetrized. Then, we define a dual
tensor $\tilde \Psi$ by applying contracting
each group of antisymmetric indices with
$\epsilon^{a[N]}$, the totally antisymmetric
(Levi--Civita) symbol:
\begin{equation}
    \tilde\Psi^{i_1(\ell_1),\dots,i_p(\ell_p);
    j^1[N-\lowercal{m}^1],\dots,
    j^r[N-\lowercal{m}^r]} = \epsilon^{j^1[N]}
    \cdots \epsilon^{j^r[N]}\,
    \Psi^{i_1(\ell_1),\dots,i_p(\ell_p)}
    _{j^1[\lowercal{m}^1],\dots,
    j^r[\lowercal{m}^r]}\,.
\end{equation}
The dual tensor $\tilde \Psi$ has now
two types of upper indices: the first
$p$ groups have the symmetry $\bm\ell$,
while the next $r$ groups (resulting
from dualization) have the symmetry
of the dualized Young diagram
$\tilde{\lowerbcal{m}} = [\tilde{\lowercal{m}}^1,
\dots, \tilde{\lowercal{m}}^r]$ with
$\tilde{\lowercal{m}}^k=N-\lowercal{m}^k$.
In other words, the dual tensor
$\tilde\Psi$ has the symmetry of the
tensor product $\bm\ell \otimes 
\tilde{\lowerbcal{m}}$, which we should
try to decompose. In the symmetric basis,
$\tilde{\lowerbcal{m}} = (\tilde{\lowercal{m}}_1,
\tilde{\lowercal{m}}_2, \dots)$ is given by
(see Figure \ref{fig:dualdiagram}):
\begin{equation}
    \tilde{\lowercal{m}}_k=m_1-m_{N+1-k}\,.
\end{equation}

\begin{figure}[H]
    \centering
    \begin{tikzpicture}[scale=0.7]
    \draw[thick] (-3,3) -- (6,3) -- (6,2.5) -- (-3,2.5) -- (-3,3);
    \node at (0.5,2.75) {\footnotesize$\lowercal{m}_1$};
    \draw[thick] (4,2.5) -- (4,2) -- (-3,2) -- (-3,2.5);
    \node at (0.5,2.25) {\footnotesize$\lowercal{m}_2$};
    \node at (5,2.25) {\footnotesize$\tilde{\lowercal{m}}_{N-1}$};
    \draw[thick] (3,2) -- (3,1.5) -- (-3,1.5) -- (-3,2);
    \node at (0.5,1.75) {\footnotesize$\lowercal{m}_3$};
    \node at (4.5,1.75) {\footnotesize$\tilde{\lowercal{m}}_{N-2}$};
    \draw[thick] (2,1.5) -- (2,1.25);
    \draw[thick, dashed] (2,1.25) -- (0,0.25);
    \draw[thick] (0,0.25) -- (0,0);
    \draw[thick] (0,0) -- (-1,0) -- (-3,0) -- (-3,2);
    \draw[thick] (-3,0) -- (-3,-0.5) -- (-1,-0.5) -- (-1,0);
    \node at (-2,-0.25) {\footnotesize$\lowercal{m}_q$};
    \node at (2,-0.25) {\footnotesize$\tilde{\lowercal{m}}_{N+1-q}$};
    \draw[dashed, gray] (-3,-0.5) -- (-3,-3) -- (6,-3) -- (6,2.5);
    \draw[dashed, gray] (-3,-2.5) -- (6,-2.5);
    \draw[dashed, gray] (-3,-1) -- (6,-1);
    \draw[dashed, gray] (-1,-0.5) -- (6,-0.5);
    \draw[dashed, gray] (0,0) -- (6,0);
    \draw[dashed, gray] (3,1.5) -- (6,1.5);
    \draw[dashed, gray] (4,2) -- (6,2);
    \node at (1,-2) {\footnotesize$\leftarrow\qquad
    \tilde{\lowercal{m}}_1\qquad\rightarrow$};
    \draw[<->] (-4,3) -- (-4,-3);
    \node at (-4.5,0) {\small $N$};
    \end{tikzpicture}
    \caption{A Young diagram
    $\boldsymbol{\lowercal{m}}$
    (black) and its dual $\tilde{\boldsymbol{\lowercal{m}}}$
    (gray).}
    \label{fig:dualdiagram}
\end{figure}

Now recall that $\Psi$ is traceless.
On the dual tensor $\tilde\Psi$, the
traceless condition for $\Psi$
translates to
\begin{equation}
    0 = \tilde\Psi^{\cdots\,[i_k\,\cdots;
    \cdots\,|j^l[N-\lowercal{m}^l]]\,\cdots}\,,
    \label{Psi cond}
\end{equation}
This implies that, when applying
Littlewood--Richardson rule to decompose
$\bm\ell \otimes \tilde{\lowerbcal{m}}$,
no antisymmetrization of an index in
$\bm\ell$ with a column of 
$\tilde{\lowerbcal{m}}$ is allowed.
Such a diagram occurs only when
$p\le N-q$, that is, $p+q\le N$
and is unique. This diagram is
obtained by simply attaching
$\bm\ell$ to the right side of
$\tilde{\lowerbcal{m}}$:
\begin{eqnarray} 
    &&(\tilde{\lowercal{m}}_1+\ell_1,
    \tilde{\lowercal{m}}_1+\ell_2,\ldots)
    \label{lm prod} \\
    &&=\, (\underbrace{\lowercal{m}_1+\ell_1,
    \lowercal{m}_1+\ell_2, \dots,
    \lowercal{m}_{1}+\ell_p}_{p},
    \underbrace{\lowercal{m}_1,\dots,
    \lowercal{m}_1}_{N-p-q},
    \underbrace{\lowercal{m}_1-\lowercal{m}_q,
    \lowercal{m}_1-\lowercal{m}_{q-1},
    \dots,0}_q)\,.
    \nonumber
\end{eqnarray}
To recapitulate,
due to the isomorphism 
$U(N)\cong U(1)\ltimes SU(N)$,
the representations of $U(N)$ correspond to the tensor product
of a $U(1)$ and a $SU(N)$ irrep. More precisely,
\begin{equation}
    [\bm \ell \oslash \lowerbcal{m}]_{U(N)} =
    [|\bm\ell|-|\lowerbcal{m}|]_{U(1)}\otimes 
    [\tilde{\lowerbcal{m}}+\bm\ell]_{SU(N)}\,,
\end{equation}
where by $\tilde{\lowerbcal{m}}+\bm\ell$,
we mean the Young diagram \eqref{lm prod} obtained by attaching 
$\tilde{\lowerbcal{m}}$ and $\bm\ell$.

\subsection*{Finite-dimensional representations of $O(N)$}
Finite-dimensional representations of $O(N)$
can also be obtained by restriction from
the tensor representations of $GL_N$
constructed previously. The orthogonal group
$O(N)$ possesses an invariant tensor, namely
the metric $\delta_{ij}$. As one can raise
and lower indices with it, it is sufficient
to consider tensors with only one kind of
indices. The restriction of $[\bm\ell]_{GL_N}$
to $O(N)$ is however not irreducible: it
contains invariant subspaces generated by
traces taken with respect to $\delta_{ij}$. For
instance, $[(2)]_{GL_N}$ which is the space of
symmetric tensors $T_{ij}$ contains two
irreducible representations of $O(N)$,
corresponding to its trace $T=\delta^{ij}\,
T_{ij}$ and its traceless part $\hat T_{ij}
= T_{ij} - \tfrac1N\, \delta_{ij}\, T$.

 Requiring a tensor to be traceless
imposes some restriction on the possible
symmetry $\bm\ell$ it can have. To see that,
let us have a look at a rank $k+l$ tensor
with the symmetry of the Young diagram
$[k,l]$ in the antisymmetric basis (i.e.
a two-column Young diagram, of respective height $k$ and $l$). It has
\begin{equation}
    \frac{k-l+1}{k+1}\, \binom{N}{k}\,
    \binom{N+1}{l}\,,
\end{equation}
independent components, and its trace
being a tensor with symmetry $[k-1,l-1]$.
As a consequence, the traceless part of
such a tensor has
\begin{equation}
    \tfrac{N!(N+1)!(k-l+1)}{(k+1)!l!(N-k+1)!(N-l+2)!}\,
    \big[(N-k+1)(N-l+2)-(k+1)l\big]\,,
\end{equation}
independent components. This number vanishes
for $k+l=N+1$, and becomes negative when
$k+l$ becomes larger. This means that a
tensor $T$ with shape $[k,l]$ is necessarily
``pure trace'', i.e. of the form
\begin{equation}
    T_{[i_1 \dots i_k], [j_1 \dots j_l]}
    = \delta_{[i_1|[j_1}\,
    U_{i_2 \dots i_k], j_2 \dots j_l]}\,,
\end{equation}
with some tensor $U$ of symmetry $[k-1,l-1]$.
The same conclusion applies to more general
tensor: if the corresponding Young diagram
$\bm\ell=[\ell^1, \ell^2, \dots, \ell^r]$
is such that $\ell^1 + \ell^2 > N$, then its
traceless part identically vanishes.
As a consequence, finite-dimensional
representations of $O(N)$ correspond to
traceless tensor having the symmetry of a
Young diagram $\bm\ell$ such that the sum
of the height of its first two columns is
at most $N$.

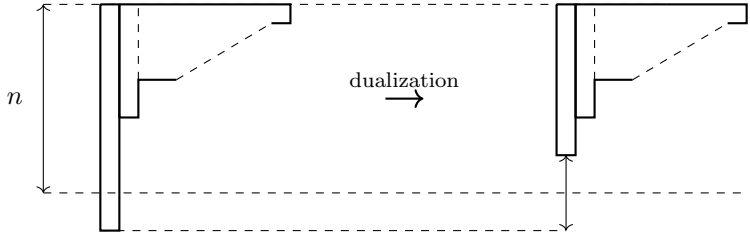
\begin{figure}[H]
    \centering
    \begin{tikzpicture}[scale=0.5]
    \draw[thick] (-7,5) -- (-7,-1) -- (-6.5,-1) -- (-6.5,5) -- (-7,5);
    \draw[thick] (-6.5,5) -- (-6.5,2) -- (-6,2) -- (-6,3) -- (-5,3);
    \draw[dashed] (-5,3) -- (-2.5,4.5);
    \draw[dashed] (-6,3) -- (-6,5);
    \draw[thick] (-2.5, 4.5) -- (-2,4.5) -- (-2,5) -- (-6.5,5);
    \draw[thick] (5,5) -- (5,1) -- (5.5,1) -- (5.5,5) -- (5,5);
    \draw[thick] (5.5,5) -- (5.5,2) -- (6,2) -- (6,3) -- (7,3);
    \draw[dashed] (7,3) -- (9.5,4.5);
    \draw[dashed] (6,3) -- (6,5);
    \draw[thick] (9.5, 4.5) -- (10,4.5) -- (10,5) -- (5.5,5);
    \draw[<->] (-8.5,5) -- (-8.5,0);
    \node at (-9.25,2.5) {\small $n$};
    \draw[<->] (5.25,1) -- (5.25,-1);
    \draw[dashed] (-8.5,5) -- (8,5);
    \draw[dashed] (-8.5,0) -- (10,0);
    \draw[dashed] (-7,-1) -- (5.25,-1);
    \node at (1,3) {\scriptsize dualization};
    \draw[thick,->] (0.5,2.5) -- (1.5,2.5);
    \end{tikzpicture}
    \caption{Example of a Young diagram with a first column
    higher than $n$ (left) and the Young diagram
    obtained after dualization (right).}
    \label{fig:diagram}
\end{figure}

\paragraph{Highest weight of the representation.}
In order to show that the $O(N)$ irrep
consisting of traceless tensor with symmetry
$\bm\ell$ is an highest weight module, it will
be useful to work in the complexified basis
where the metric takes the form
\begin{equation}
    J = 
    \begin{pmatrix}
    0 & \id_n\\
    \id_n & 0 
    \end{pmatrix}\,,
    \quad [\text{if}\,\,N=2n]
    \qquad \text{or} \qquad
    J = 
    \begin{pmatrix}
    0 & \id_n & 0\\
    \id_n & 0 & 0\\
    0 & 0 & 1
    \end{pmatrix}\,,
    \quad [\text{if}\,\,N=2n+1]\,,
\end{equation}
The generators $M_{AB}$ of $\mathfrak{so}(N)$
obey
\begin{equation}
    [M_{AB}, M_{CD}] = J_{BC}\, M_{AD}
    -  J_{AC}\, M_{BD} - J_{BD}\, M_{AC}
    + J_{AD}\, M_{BC}\,,
\end{equation}
where $J_{AB}$ are the components of the
matrix $J$ introduced above. 
The reality condition reads ${M_{AB}}^\dagger=-J_{AC}\,J_{BD}\,M_{CD}$\,.
When $N=2n$,
the index $A$ takes the values $A=\pm a$
with $a=1,\dots,n$ and the only non-vanishing
components of $J$ are $J_{{\sst +}a{\sst -}b}
=\delta_{ab}$. 
When $N=2n+1$, the index $A$
can assume one additional value which we
will denote $*$, and $J_{*A} = \delta_{*A}$.
Splitting the generators $M_{AB}$ accordingly,
the previous commutation relations take the form
\begin{equation}
    [M_{ab}, M_{{\sst-}c{\sst-}d}]
    = \delta_{ad}\, M_{{\sst+}b{\sst-}c}
    -\delta_{bd}\, M_{{\sst+}a{\sst-}c}
    -\delta_{ac}\, M_{{\sst+}b{\sst-}d}
    +\delta_{bc}\, M_{{\sst+}a{\sst-}d}\,,
\end{equation}
\begin{equation}
    [M_{{\st+}a{\st-}b}, M_{{\st+}c{\st+}d}]
    = 2\,\delta_{b[d}\, M_{{\st+}c]{\st+}a}\,,
    \qquad
    [M_{{\st+}a{\st-}b}, M_{{\st-}c{\st-}d}]
    = -2\,\delta_{a[d}\, M_{{\st-}c]{\st-}b}\,,
\end{equation}
\begin{equation}
    [M_{{\st+}a{\st-}b}, M_{{\st+}c{\st-}d}]
    = \delta_{bc}\, M_{{\st+}a{\st-}d}
    - \delta_{ad}\, M_{{\st+}c{\st-}b}\,.
\end{equation}
The last commutator shows that the generators
$M_{{\sst+}a{\sst-}b}$ span a $\mathfrak{u}(n)$
subalgebra. When $N=2n+1$, there are $2n$ 
additional generators, denoted $N_{{\sst\pm}a}
=M_{{\sst\pm}a\,*}$, to take into account:
\begin{equation}
    [M_{{\sst+}a{\sst-}b}, N_{{\sst+}c}]
    = \delta_{bc}\, N_{{\sst+}a}\,, \qquad
    [M_{{\sst+}a{\sst-}b}, N_{{\sst-}c}]
    = -\delta_{ac}\, N_{{\sst-}b}\,,
\end{equation}
\begin{equation}
    [M_{{\sst\pm}a{\sst\pm}b}, N_{{\sst\mp}c}]
    = \delta_{c[b}\, N_{\pm a]}\,, \qquad 
    [N_{\epsilon a}, N_{\varepsilon b}]
    = - M_{\epsilon a\, \varepsilon b}\,,
\end{equation}
where $\epsilon,\varepsilon=\pm$ and all other
commutators vanish. From these commutators, we
can see that $M_{{\sst+}a{\sst-}a}$ generates
a Cartan subalgebra\footnote{Notice that the
Killing form $\kappa$ of $\mathfrak{so}(N)$
is given by $\kappa(M_{AB}, M_{CD}) \propto
J_{A[C}J_{D]B}$ so that the Cartan subalgebra
generators $H_a=M_{{\sst+}a{\sst-}a}$
verify $\kappa(H_a,H_b) = \delta_{ab}$ (upon
choosing the proper normalization). As a 
consequence, their highest weight computed
hereafter will be obtained in the orthonormal
basis.} while $M_{{\sst+}a{\sst+}a}$,
$N_{{\sst+}c}$ and $M_{{\sst+}a{\sst-}b}$ for
$a<b$ (resp. $M_{{\sst-}a{\sst-}a}$ and
$N_{{\sst-}c}$ and $M_{{\sst+}a{\sst-}b}$ for
$a>b$) are raising (resp. lowering) operators.
The action of $\mathfrak{so}(N)$ on a tensor
of symmetry $\bm\ell=(\ell_1, \dots, \ell_p)$,
with $p \le N$, reads
\begin{equation}
    M_{AB}\, T_{C_1(\ell_1), \dots,
    C_p(\ell_p)} = 2\,\sum_{k=1}^p \ell_k\,
    J_{C_k[B} T_{C_1(\ell_1), \dots,
    A]\,C_k(\ell_k{\st-}1), \dots, C_p(\ell_p)}\,.
\end{equation}
We saw previously that if $T$ is traceless,
then $\ell^1+\ell^2 \le N$. This implies that
only the first column can be higher than $n$.
When it is the case, i.e. when $\ell^1 \ge n$,
we can dualize this tensor (as we did in the
previous subsection):
\begin{equation}
    \tilde T_{A_1[\tilde\ell^1], A_2[\ell^2],
    \dots, A_r[\ell_r]}
    = \epsilon^{A_1[N]}\, T_{A_1[\ell^1],
    A_2[\ell^2], \dots, A_r[\ell^r]}\,,
\end{equation}
with $\tilde \ell^1 = N-\ell^1 \le n$ and
$\bm\ell = [\ell^1,\dots,\ell^r]$ in the
antisymmetric basis (i.e. $\ell^1=p$ is the
number of rows and $r=\ell_1$ the number of
columns). The dual tensor $\tilde T$ now has
the symmetry of the dual Young diagram
\begin{equation}
    \tilde{\bm\ell} := [N-\ell^1,\ell^2,\dots,\ell^r]
    = (\ell_1,\dots,\ell_p,
    \underbrace{1,\ldots,1}_{N-2p})\,,
\end{equation}
which, in particular than $n$ rows 
(see Figure \ref{fig:diagram}).
With this in mind, we can now assume (without
loss of generality) that $p \le n$, i.e. consider
tensor corresponding to Young diagrams with less
than $n$ non-empty rows. Now let us show that the
component $T_{{\sst+}1(\ell_1), \dots,
{\sst+}p(\ell_p)}$ is the highest weight vector
of the representation. First of all, we can
immediately see that it is annihilated by all
$M_{{\sst+}a{\sst+}b}$ and $N_{{\sst+}c}$.
Secondly, the action of the generators
$M_{{\sst+}a{\sst-}b}$ on this component
simply takes the form
\begin{equation}
    M_{{\sst+}a{\sst-}b}\, T_{{\sst+}1(\ell_1),
    \dots, {\sst+}p(\ell_p)} = \ell_b\,
    T_{{\sst+}1(\ell_1),\dots,
    {\sst+}a\,{\sst+}b(\ell_b-1),
    \dots, {\sst+}p(\ell_p)}\,.
\end{equation}
When $a=b$, i.e. for a Cartan subalgebra
generator, this reduces to
\begin{equation}
    M_{{\sst+}a{\sst-}a}\, T_{{\sst+}1(\ell_1),
    \dots, {\sst+}p(\ell_p)} = \ell_a\,
    T_{{\sst+}1(\ell_1),\dots,{\sst+}a(\ell_a),
    \dots, {\sst+}p(\ell_p)}\,,
\end{equation}
that is, this component is an eigenvector
of $M_{{\sst+}a{\sst-}a}$ with eigenvalue
$\ell_a$. When $a<b$, i.e. for a raising
operator, this simplifies to
\begin{equation}
    M_{{\sst+}a{\sst-}b}\, T_{{\sst+}1(\ell_1),
    \dots, {\sst+}p(\ell_p)} = \ell_b\,
    T_{{\sst+}1(\ell_1),\dots, {\sst+}a(\ell_a),
    \dots, {\sst+}a\,{\sst+}b(\ell_b-1),
    \dots, {\sst+}p(\ell_p)} = 0\,,
\end{equation}
which vanishes as a consequence of the symmetry
of the tensor $T$. To summarized, we have proved
that the $O(N)$ irrep defined by traceless tensor
with symmetry $\bm\ell$ is an $SO(N)$ highest
weight module with highest weight $\bm\ell$
if the diagram has at most $n$ rows, or highest
weight $\tilde{\bm\ell}$ (dualized diagram) 
otherwise.

\subsection*{Finite-dimensional representations of $Sp(N)$}
Finite-dimensional representations of $Sp(N)$
can also be obtained by restriction from
the tensor representations of $GL_{2N}$
constructed previously. The symplectic group
$Sp(N)$ possesses an invariant tensor, namely
the canonical symplectic matrix $\Omega_{ij}$.
Being non-degenerate, one can raise and lower
indices with it, so that it is sufficient to
consider tensors with one kind of indices.
The restriction of $[\bm\ell]_{GL_{2N}}$
to $Sp(N)$ is not irreducible: it contains
invariant subspaces generated by traces
taken with respect to $\Omega$. For instance,
$[(1,1)]_{GL_{2N}}$ which is the space of
antisymmetric tensors $T_{i,j}$ contains two
irreducible representations of $Sp(N)$,
corresponding to its trace $T=\Omega^{ij}\,
T_{i,j}$ and its traceless part $\hat T_{i,j}
= T_{i,j} + \tfrac1{2N}\, \Omega_{ij}\, T$.

Requiring a tensor to be traceless
imposes some restriction on the possible
symmetry $\bm\ell$ it can have. Consider
for instance a totally antisymmetric tensor
of rank $k$. It has $\binom{2N}{k}$
independent components, and its trace being
a rank $k-2$ antisymmetric tensor has
$\binom{2N}{k-2}$. As a consequence, the
traceless part of a rank $k$ antisymmetric
tensor has
\begin{equation}
    \binom{2N}{k}-\binom{2N}{k-2} = \tfrac{2N!}{k!(2N-k+2)!}\,
    \big[(2N-k+2)(2N-k+1)-k(k-1)\big]\,,
\end{equation}
independent components. This number vanishes
for $k=N+1$, and becomes negative when $k$
becomes larger. This means that an
antisymmetric tensor $T$ of rank $k > N$ is
necessarily ``pure trace'', i.e. of the form
\begin{equation}
    T_{[i_1 \dots i_k]} = \Omega_{[i_1i_2}\,
    U_{i_3 \dots i_k]}\,,
\end{equation}
with some rank $k-2$ antisymmetric tensor $U$.
This arguments extends to tensors with more
general symmetry, leading to the conclusion
that tensors  finite-dimensional
representations of $Sp(N)$ correspond to
traceless tensor having the symmetry of a
Young diagram $\bm\ell$ with at most $N$ rows.

\paragraph{Highest weight of the representation.}
Recall that the algebra $\mathfrak{sp}(N)$ is
spanned by symmetric generators $K_{IJ}=K_{JI}$
with the index $A$ taking $2N$ values and
subject to the commutation relations
\begin{equation}
    [K_{IJ}, K_{KL}] = \Omega_{JK}\, K_{IL}
    + \Omega_{JL}\, K_{IK} + \Omega_{IK}\, K_{JL}
    + \Omega_{IL}\, K_{JK}\,.
\end{equation}
Splitting the index into $A={\st\pm}a$ with
$a=1,\dots,N$, and such that the only non-vanishing
components of the symplectic matrix 
are $\Omega_{{\sst-}a{\sst+}b}=\delta_{ab}$,
the above relations now read
\begin{equation}
    [K_{{\sst+}i{\sst+}j},K_{{\sst-}k{\sst-}l}]
    = \delta_{jk}\, K_{{\sst+}i{\sst-}l}
    + \delta_{jl}\, K_{{\sst+}i{\sst-}k}
    - \delta_{ik}\, K_{{\sst+}j{\sst-}l}
    - \delta_{il}\, K_{{\sst+}j{\sst-}k}\,,
\end{equation}
\begin{equation}
    [K_{{\sst+}i{\sst-}j}, K_{{\sst+}k{\sst+}l}]
    = 2\,\delta_{j(k}\, K_{{\sst+}l){\sst+}i}\,,
    \qquad
    [K_{{\sst+}i{\sst-}j}, K_{{\sst-}k{\sst-}l}]
    = -2\,\delta_{i(k}\, K_{{\sst-}l){\sst-}j}\,,
\end{equation}
\begin{equation}
    [K_{{\sst+}i{\sst-j}},K_{{\sst+}k{\sst-l}}]
    = \delta_{jk}\, K_{{\sst+}i{\sst-}l}
    - \delta_{il}\, K_{{\sst-}j{\sst+}k}\,.
\end{equation}
From this commutators, we can see that 
$K_{{\sst+}i{\sst-}i}$ generate a Cartan
subalgebra,\footnote{Notice that the Killing
form $\kappa$ of $\mathfrak{sp}(N)$ is given
by $\kappa(K_{IJ}, K_{KL}) \propto
\Omega_{I(K}\, \Omega_{L)J}$ so that the
Cartan subalgebra generators
$H_i=K_{{\sst+}i{\sst-}i}$ verify
$\kappa(H_i,H_j)=\delta_{ij}$ (upon choosing
the proper normalization). In other words,
the weights computed later will be
expressed in the orthonormal basis.}
while $K_{{\sst+}j{\sst+}k}$ (resp.
$K_{{\sst+}j{\sst+}k}$) and $K_{{\sst+}l{\sst-}m}$
for $l<m$ (resp. for $l>m$) are raising (resp.
lowering) operators. Notice also that the last
commutation relation shows that
$K_{{\sst+}i{\sst-}j}$ forms a $\mathfrak{u}(N)$
subalgebra in $\mathfrak{sp}(N)$. The action of
$\mathfrak{sp}(N)$ on a tensor of symmetry
$\bm\ell=(\ell_1, \dots, \ell_p)$, with
$p \le N$, is given by
\begin{equation}
    K_{IJ}\, T_{L_1(\ell_1), \dots, L_p(\ell_p)}
    = -2\, \sum_{k=1}^p \ell_k\, \Omega_{L_k(I}\,
    T_{L_1(\ell_1), \dots, J)L_k(\ell_k-1), \dots,
    L_p(\ell_p)}\,.
\end{equation}
Similarly to the previous cases, let us show that
the component $T_{{\sst+}1(\ell_1), \dots,
{\sst+}p(\ell_p)}$ defines a highest weight vector.
First, notice that $K_{{\sst+}i{\sst+}j}$ on this
component is trivial. Second, the action of
$K_{{\sst+}i{\sst-}j}$ reduces to
\begin{equation}
    K_{{\sst+}i{\sst-}j}\, T_{{\sst+}1(\ell_1),
    \dots, {\sst+}p(\ell_p)} = \ell_j\,
    T_{{\sst+}1(\ell_1), \dots, {\sst+}i\,{\sst+}j(\ell_j-1), \dots,
    {\sst+}p(\ell_p)}\,.
\end{equation}
If $i=j$, i.e. we are looking at the action
of the Cartan subalgebra, then $T$ is an
eigenvector of the generator $K_{{\sst+}i{\sst-}i}$
with eigenvalue $\ell_i$. If $i<j$, i.e. we are
acting with a raising operator, then the right
hand side of the previous equation reads
\begin{equation}
    T_{{\sst+}1(\ell_1), \dots, {\sst+}i(\ell_i),
    \dots, {\sst+}i{\sst+}j(\ell_j-1), \dots,
    {\sst+}p(\ell_p)} = 0\,,
\end{equation}
due to the symmetry of $T$. We can therefore
conclude that the $Sp(N)$ irrep carried by
the space of traceless tensors with symmetry
of a Young diagram $\bm\ell$ with at most $N$
rows is a highest weight module whose highest
weight coincides with $\bm\ell$ in the
orthonormal basis.

\section{Real forms of classical Lie groups and Lie algebras}
\label{sec: real form}

A real form $\mathfrak{g}_{\s}$ of a complex Lie algebra $\mathfrak{g}$ can be identified as
a subalgebra of
the realification $\mathfrak{g}^{\mathbb R}$
of $\mathfrak{g}$\,, namely
\begin{equation}
\mathfrak{g}_\s := \{ A \in \mathfrak{g}^{\mathbb R}\,|\,\sigma(A)=-A\}
\end{equation}
where $\sigma:\mathfrak{g}^{\mathbb R}\to\mathfrak{g}^{\mathbb R}$ is an anti-involution, i.e. it satisfies $\sigma([A_1,A_2])=-[\sigma(A_1), \sigma(A_2)]$ and $\sigma^2=1$.

In the following, we 
summarize the list of all real forms
and present the action  $\sigma$ on their generators.
For that, let us define first the following matrices
\be
	\eta=\begin{pmatrix} I_{N_+} & 0 \\ 0 & -\,I_{N_-} \end{pmatrix}\,,
	\qquad
	\O=\begin{pmatrix} 0 & I_N \\ -I_N & 0 \end{pmatrix}\,,
\ee
where $I_N$ is the $N\times N$ identity matrix.

\subsection*{General linear Lie groups and Lie algebras}
The group of invertible $n\times n$ complex
matrices, $GL(N,\mathbb C)$, is a real Lie
group of dimension $2N^2$. The associated Lie
algebra $\mathfrak{gl}(N,\mathbb C)$ is the
commutator Lie algebra of $N\times N$ complex
matrices generated by the matrices
$\mathsf{X}^{A}{}_B$ with the components
$(\mathsf{X}^A{}_B)_C{}^D = \delta^A_C\,
\delta^D_B$. The generators $\mathsf{X}^{A}{}_B$
satisfy the commutation relation,
\be 
    [\mathsf{X}^{A}{}_{B}, \mathsf{X}^{C}{}_D]
    = \delta^C_B\, \mathsf{X}^{A}{}_D
    -\delta^A_D\, \mathsf{X}^C{}_B\,.
\ee
For later use, it will be useful to note
\be 
    Y\, \mathsf{X}^{A}{}_B
    = Y_{C}{}^{A}\, \mathsf{X}^{C}{}_B\,,
    \qquad 
    \mathsf{X}^{A}{}_B\, Y
    = \mathsf{X}^{A}{}_C\, Y_{B}{}^C\,,
\ee 
where $Y_A{}^B$ are the components
of the matrix $Y$.

All real forms of the complex Lie group/algebra
$GL_N$/$\mathfrak{gl}_N$ are subgroups/subalgebras of 
$GL(N,\mathbb C)$/$\mathfrak{gl}(N,\mathbb C)$. 
There are three classes of real forms:
\begin{itemize} 
\item $GL(N,\mathbb R)$\,: the group of $N \times N$ real invertible matrices,
\ba 
    & GL(N,\mathbb R)
    =\{ g\in GL(N, \mathbb C)\,|\, g^*=g\}\,,\nn
    & \mathfrak{gl}(N,\mathbb R)
    =\{A \in \mathfrak{gl}(N,\mathbb C)\,|\,A^*=A\}\,,
\ea 
which is singled out by the anti-involution,
\begin{equation}
        \sigma(A)=-A^*\,.
\end{equation}
Since $\mathsf{X}^{A}{}_B$ are real, the action
of $\sigma$ on the generators $\mathsf{X}^A{}_B$
simply reads
\be 
    \sigma(\mathsf{X}^A{}_B) =
    -\mathsf{X}^A{}_B\,.
\ee

\item $U(N_+,N_-)$\,: the group of $(N_++N_-) \times (N_++N_-)$ complex matrices which preserves the Hermitian form $\eta$ of signature $(N_+,N_-)$,
\ba
    &U(N_+,N_-)=\{g\in GL(N_++N_-,\mathbb C)\,|\,g^\dagger\,\eta\,g=\eta\}\,,
    \nn 
    &\uu(N_+,N_-) =\{ A \in \mathfrak{gl}(N_++N_-,\mathbb C)\,|\, A^\dagger\,\eta+\eta\,A=0\}\,,
\ea
which is singled out by the anti-involution,
\begin{equation}
    \sigma(A)=\eta\,A^\dagger\,\eta\,.
\end{equation}
Since $(X^A{}_B)^\dagger=X^B{}_A$, the action
of $\sigma$ is given by
\be 
    \sigma(\mathsf{X}^A{}_B)
    = \eta\, \mathsf{X}^B{}_A\, \eta
    = \eta_{BD}\,\mathsf{X}^D{}_C\,\eta^{CA}\,,
\ee 
where $\eta_{AB}$ are the components of
the matrix $\eta $\,.

\item $U^*(2N)$\,: the group of $N \times N$
invertible matrices over the quaternions
$\mathbb H$,\footnote{As such, it is also
denoted $GL(N, \mathbb H)$.} which can be
defined as the following subgroup of
$GL(2N,\mathbb C)$:
\ba
   & U^*(2N)=\{g\in GL(2N,\mathbb C)\,|\, g^*=\O \,g\,\O ^{-1}\}\,,\nn 
   & \uu^*(2N) = \{ A \in \mathfrak{gl}(2N,\mathbb C)\,|\, A^*=\O\,A\,\O ^{-1}\}\,,
\ea
and is therefore singled out by the anti-involution,
\begin{equation}
    \sigma(A)=-\O \,A^*\,\O ^{-1}\,.
\end{equation}
The action of the anti-involution $\sigma$ on $X^A{}_B$ reads
\be 
    \sigma(\mathsf{X}^A{}_B)=
    -\Omega_{AC}\,\mathsf{X}^C{}_D\,\Omega^{DB}\,,
\ee 
where $\O_{AB}$ and $\O^{AB}$ are the
components of $\O $ and its inverse:
$\O_{AB}\,\O^{BC}=\delta^C_A$\,.
\end{itemize}

\subsection*{Orthogonal Lie groups and Lie algebras}
The complex orthogonal Lie group and Lie algebras
are defined as
\ba
    && O(N,\mathbb C)=\{g\in GL(N,\mathbb C)\,|\,g^t\,g=I_{N}\}\,,\nn 
    && \so(N,\mathbb C)
      =\{ A \in \mathfrak{gl}(N,\mathbb C)\,|\,
      A^t+A=0\}\,.
\ea
All real forms of $O_N$/$\so_N$ are subgroups/subalgebras
of $O(N,\mathbb C)$/$\so(N,\mathbb C)$.
There are two classes of real forms:

\begin{itemize}
\item $O(N_+,N_-)$\,: the orthogonal group
of indefinite signature, which is defined by
\ba
    &O(N_+,N_-)=\{g\in O(N_++N_-,\mathbb C)\,|\,
    g^*=\eta \,g\,\eta \}\,,\nn 
    &\so(N_+,N_-)
      =\{ A \in \mathfrak{so}(N_++N_-,\mathbb C)\,|\,
      A^*=\eta \,A\,\eta \}\,, 
\ea
and is therefore singled out by the anti-involution,
\begin{equation}
    \sigma(A)=-\eta \,A^*\,\eta \,. 
\end{equation}
 A more familiar description
of $O(N_+,N_-)$ is as the subgroup of $GL(N_++N_-,\mathbb R)$,
which preserves the (flat) metric of signature $(N_+,N_-)$.
To make the link between these two definitions,
we use the isomorphism,
\be 
    \rho_\zeta(A)=\zeta\,A\,\zeta\,,
\ee 
with
\be
	\zeta=\begin{pmatrix} I_{N_+} & 0 \\ 0 & i\,I_{N_-} \end{pmatrix}\,,
\ee
which verifies $\zeta^2=\eta$\,. The Lie algebra 
$\rho_\zeta(\so(N_+,N_-))\cong \so(N_+,N_-)$ 
is equipped with the anti-involution $\tilde \sigma$
induced by $\sigma$\,:
\be 
    \tilde\sigma(A):=\rho_\zeta(\sigma(\rho_\zeta^{-1}(A))=
    -\zeta\,\eta\,(\zeta^{-1}\,A\,\zeta^{-1})^*\,\eta\,\zeta
    =-A^*\,.
\ee     
and
is generated by the matrices
$\mathsf{M}_{AB}=2\,\delta_{C[A}\,\mathsf{X}^{C}{}_{B]}\,,$
whose induced Lie bracket satisfies \eqref{O CR} with $E_{AB}=\eta_{AB}$\,.
The action of the induced anti-involution 
$\tilde\sigma$ on $\mathsf{M}_{AB}$ is simply given by
\be
    \tilde\sigma(\mathsf{M}_{AB})=-\mathsf{M}_{AB}\,.
\ee 

\item $O^*(2N)$\,: the quaternionic orthogonal
group, defined by
\ba
    &O^*(2N)=\{g\in O(2N,\mathbb C)\,|\,
    g^*=\O \,g\,\O ^{-1}\}\,,\nn 
    &\so^*(2n)
      =\{ A \in \mathfrak{so}(2N,\mathbb C)\,|\,
     A^*=\O \,A\,\O ^{-1}\}\,,
\ea
and which is therefore singled out by the anti-involution
\begin{equation}
    \sigma(A) = -\O \,A^*\,\O ^{-1}\,.
\end{equation}
 It is also
convenient to consider another description
for $O^*(2N)$, which can be obtained by
the isomorphism,
\be 
    \rho_U(A)=-i\,U\,A\,U\,,
\ee 
where
\be
	U=\frac1{\sqrt{2}}\begin{pmatrix} I_N & i\,I_N \\ i\,I_N & I_N \end{pmatrix}\,,
	\label{U}
\ee
satisfies
\begin{equation}
    U^2 = i\, J\,, \qquad U\,\Omega\,U=\O\,,
    \label{U identity}
\end{equation}
with
\be 
    J=\begin{pmatrix} 0 & I_N \\ I_N & 0 \end{pmatrix}\,.
    \label{J}
\ee
The components of $J$ will be denoted by $J_{AB}$.
The Lie algebra $\rho_U(\so^*(2N))\cong \so^*(2N)$ is equipped with the anti-involution $\tilde \sigma$
induced by $\sigma$\,:
\be 
    \tilde\sigma(A):=\rho_U(\sigma(\rho_U^{-1}(A)))=
    i\,U\,\O\,(i\,U^{-1}\,A\,U^{-1})^*\,\O^{-1}\,U
    =-\O\,A^*\,\O\,.
\ee 
The induced Lie bracket is given by
$\rho_U([A,B])=
\rho_U(A)\,J\,\rho_U(B)-\rho_U(B)\,J\,\rho_U(A)$
and hence the isomorphic Lie algebra
$\rho_U(\so^*(2N))$ is generated by  the matrices 
$\mathsf{M}_{AB}=2\,\delta_{C[A}\,\mathsf{X}^{C}{}_{B]}$
which obey the commutation relations
\eqref{O CR} with $E_{AB}=J_{AB}$.
The induced anti-involution of $\mathsf{M}_{AB}$ reads
\be 
    \tilde\sigma(\mathsf{M}_{AB})
    =-\O^{AC}\,\mathsf{M}_{CD}\,\O^{DB}\,,
\ee
where $\O^{AB}$ are the components of $\O^{-1}$.
\end{itemize}

\subsection*{Symplectic Lie groups and Lie algebras}
The complex symplectic Lie group and Lie algebras
is defined as the group of $2N\times 2N$ complex matrices preserving the canonical symplectic form $\Omega$,
\ba
    && Sp(2N,\mathbb C)=\{g\in GL(2N,\mathbb C)\,|\,g^t\,\O \,g=\O \}\,,\nn 
    && \syp(2N,\mathbb C)
      =\{ A \in \mathfrak{gl}(2N,\mathbb C)\,|\,
      A^t\,\O +\O \,A=0\}\,.
\ea
Since $\O \,A$ is a symmetric matrix, it is more convenient 
to consider the Lie algebra obtained by the isomorphism,
\be 
    \rho_\O(A)=\O\,A\,.
    \label{O isomorph}
\ee
The isomorphic Lie algebra $\syp(2N,\mathbb C)$ 
is generated by the symmetric matrices $\mathsf{K}_{AB}=2\, \delta_{C(A}\, \mathsf{X}^C{}_{B)}$
with the commutation relations,
\be 
    [\mathsf{K}_{AB},\mathsf{K}_{CD}]=
   2\left( \Omega_{D(A}\, \mathsf{K}_{B)C}
    +\O_{C(A}\, \mathsf{K}_{B)D}\right),
    \label{syp LB}
\ee
where $\O_{AB}$ are the components of $\O$\,.

All real forms of $Sp_{2N}$/$\syp_{2N}$ are subgroups/subalgebras
of $Sp(2N,\mathbb C)$/$\syp(2N,\mathbb C)$.
There are two types of real forms:

\begin{itemize} 
\item $Sp(2N,\mathbb R)$\,: the group of $2N \times 2N$ real matrices 
which preserve the canonical symplectic form $\Omega$,
\ba 
     &Sp(2N,\mathbb R)=\{g\in Sp(2N,\mathbb C)\,|\,g^*=g\}\,,\nn 
    &\syp(2N,\mathbb R)
      =\{ A \in \mathfrak{sp}(2N,\mathbb C)\,|\,
     A^*=A\}\,,
\ea
which is singled out by the anti-involution
\begin{equation}
    \sigma(A)=-A^*\,.
\end{equation}
This anti-involution is compatible with the
isomorphism \eqref{O isomorph}.
In the case of $Sp(2N,\mathbb R)$, it is also
convenient to consider an additional isomorphism
$\rho_U(A)=U\,A\,U$ which induces
the anti-involution $\tilde\sigma$,
\be
    \tilde\sigma(A)=
    -U\,(U^{-1}\,A\,U^{-1})^*\,U
    =J\,A^*\,J\,,
\ee 
where the matrices $U$ and $J$ are defined 
in \eqref{U} and \eqref{J}.
The action of the anti-involution on $\mathsf{K}_{AB}=2\,\delta_{C(A}\,\mathsf{X}^C{}_{B)}$ reads
\be 
    \tilde\sigma(\mathsf{K}_{AB})=
    J\,\mathsf{K}_{AB}\,J=J^{AC}\,\mathsf{K}_{CD}\,J^{DB}\,.
\ee 
Due to the second identity in \eqref{U identity},
the Lie algebra $\rho_U(\syp(2n,\mathbb R))$ has 
the same Lie bracket as \eqref{syp LB}.

\item $Sp(N_+,N_-)$\,, the group of $(N_++N_-) \times (N_++N_-)$ matrices over 
the quaternions that preserve the metric $\eta$ of signature $(N_+,N_-)$, 
which can also be defined as
\ba 
    &Sp(N_+,N_-)=\{
    g\in Sp(2(N_++N_-),\mathbb C)\,|\,
    g^\dagger\,\Upsilon \,g=\Upsilon \}\,,\nn
    &\syp(N_+,N_-)=\{
     A\in \syp(2(N_++N_-),\mathbb C)\,|\,
    A^\dagger\,\Upsilon +\Upsilon \,A=0\}\,,
\ea 
where $\Upsilon$ is defined as
\be
	\Upsilon =\begin{pmatrix} \eta & 0 \\ 0 & 
	\eta \end{pmatrix}\,.
\ee
The corresponding real form is therefore singled out by the anti-involution,
\begin{equation}
    \sigma(A)=\Upsilon \,A^\dagger\,\Upsilon\,.
\end{equation}
The induced anti-involution $\tilde \sigma$
of $\mathsf{K}_{AB}=2\,\delta_{C(A}\,\mathsf{X}^C{}_{B)}$ reads
\be 
     \tilde \sigma(\mathsf{K}_{AB})
    =\O \,\Upsilon \,(\O ^{-1}\,\mathsf{K}_{AB})^t\,
    \Upsilon 
    =\Psi\,\mathsf{K}_{AB}\,\Psi=\Psi^{AC}\,\mathsf{K}_{CD}\,\Psi^{DB}\,,
\ee
where $\Psi$ is defined as
\be 
    \Psi=\O\,\Upsilon=\begin{pmatrix} 0 & \eta \\ -\eta & 
	0 \end{pmatrix}\,,
\ee 
and $\Psi_{AB}$ and $\Psi^{AB}$ are the components of $\Psi$
and $\Psi^{-1}$.
\end{itemize}

\section{Seesaw pairs diagrams}
\label{app:seesaw_diagrams}

In Section
\ref{sec:seesaw_pairs},
we explained how a seasaw diagram 
can be obtained from a dual pair
by systematically looking for
the maximal compact subgroups
of all groups appearing in the process.
Below, we present
such seasaw diagrams for 
all seven irreducible dual pairs listed in Table
\ref{table:list_pairs}.

\begin{itemize}
    \item $\big(GL(M,\mathbb R), GL(N,\mathbb R)\big)$
\begin{mdframed}[linewidth=0.8pt]
\centering
\begin{tikzpicture}[scale=0.9]
\node at (-2.5,1.5) {$Sp(2M,\mathbb R)$};
\draw [<->] (-0.5,1.5) -- (0.5,1.5);
\node at (2.5,1.5) {$O(N)$};
\node [rotate=-45] at (-4.25,0.75) {$\cup$};
\node at (-2.5,0.75) {$\cup$};
\node at (2.5,0.75) {$\cap$};
\node [rotate=-125] at (4.25,0.75) {$\cup$};
\node at (-5.25,0) {$U(M)$};
\node at (-2.5,0) {$GL(M,\mathbb R)$};
\draw [<->] (-0.5,0) -- (0.5,0);
\node at (2.5,0) {$GL(N,\mathbb R)$};
\node at (5.25,0) {$U(N)$};
\node [rotate=45] at (-4.25,-0.75) {$\cup$};
\node at (-2.5,-0.75) {$\cup$};
\node at (2.5,-0.75) {$\cap$};
\node [rotate=125] at (4.25,-0.75) {$\cup$};
\node  at (-2.5,-1.5) {$O(M)$};
\draw [<->] (-0.5,-1.5) -- (0.5,-1.5);
\node at (2.5,-1.5) {$Sp(2N,\mathbb R)$};
\end{tikzpicture}
\end{mdframed}

\item $\big(GL(M,\mathbb C), GL(N,\mathbb C)\big)$
\begin{mdframed}[linewidth=0.8pt]
\centering
\begin{tikzpicture}[scale=0.9]
\node at (-2.5,1.5) {$U(M,M)$};
\draw [<->] (-0.5,1.5) -- (0.5,1.5);
\node at (2.5,1.5) {$U(N)$};
\node [rotate=-45] at (-4.25,0.75) {$\cup$};
\node at (-2.5,0.75) {$\cup$};
\node at (2.5,0.75) {$\cap$};
\node [rotate=-125] at (4.25,0.75) {$\cup$};
\node at (-6,0) {$U(M) \times U(M)$};
\node at (-2.5,0) {$GL(M,\mathbb C)$};
\draw [<->] (-0.5,0) -- (0.5,0);
\node at (2.5,0) {$GL(N,\mathbb C)$};
\node at (6,0) {$U(N) \times U(N)$};
\node [rotate=45] at (-4.25,-0.75) {$\cup$};
\node at (-2.5,-0.75) {$\cup$};
\node at (2.5,-0.75) {$\cap$};
\node [rotate=125] at (4.25,-0.75) {$\cup$};
\node  at (-2.5,-1.5) {$U(M)$};
\draw [<->] (-0.5,-1.5) -- (0.5,-1.5);
\node at (2.5,-1.5) {$U(N,N)$};
\end{tikzpicture}
\end{mdframed}

\item $\big(U^*(2M), U^*(2N)\big)$
\begin{mdframed}[linewidth=0.8pt]
\centering
\begin{tikzpicture}[scale=0.9]
\node at (-2.5,1.5) {$O^*(4M)$};
\draw [<->] (-0.5,1.5) -- (0.5,1.5);
\node at (2.5,1.5) {$Sp(N)$};
\node [rotate=-45] at (-4.25,0.75) {$\cup$};
\node at (-2.5,0.75) {$\cup$};
\node at (2.5,0.75) {$\cap$};
\node [rotate=-125] at (4.25,0.75) {$\cup$};
\node at (-5.5,0) {$U(2M)$};
\node at (-2.5,0) {$U^*(2M)$};
\draw [<->] (-0.5,0) -- (0.5,0);
\node at (2.5,0) {$U^*(2N)$};
\node at (5.5,0) {$U(2N)$};
\node [rotate=45] at (-4.25,-0.75) {$\cup$};
\node at (-2.5,-0.75) {$\cup$};
\node at (2.5,-0.75) {$\cap$};
\node [rotate=125] at (4.25,-0.75) {$\cup$};
\node  at (-2.5,-1.5) {$Sp(M)$};
\draw [<->] (-0.5,-1.5) -- (0.5,-1.5);
\node at (2.5,-1.5) {$O^*(4N)$};
\end{tikzpicture}
\end{mdframed}

\item $\big(U(M_+,M_-), U(N_+,N_-)\big)$
\begin{mdframed}[linewidth=0.8pt]
\begin{tikzpicture}
\node at (-2.5,1.5) {$U(M_+,M_-)^2$};
\draw [<->] (-0.75,1.5) -- (0.5,1.5);
\node at (2.25,1.5) {$U(N_+) \times U(N_-)$};
\node [rotate=-45] at (-4.5,0.75) {$\cup$};
\node at (-2.5,0.75) {$\cup$};
\node at (2.25,0.75) {$\cap$};
\node [rotate=-125] at (4.25,0.75) {$\cup$};
\node at (-5.25,0) {
$U(M_+)^2 \times U(M_-)^2$};
\node at (-2.25,0) {$U(M_+,M_-)$};
\draw [<->] (-0.75,0) -- (0.5,0);
\node at (2,0) {$U(N_+,N_-)$};
\node at (4.75,0) {
$U(N_+)^2 \times U(N_-)^2$};
\node [rotate=45] at (-4.5,-0.75) {$\cup$};
\node at (-2.5,-0.75) {$\cup$};
\node at (2.25,-0.75) {$\cap$};
\node [rotate=125] at (4.25,-0.75) {$\cup$};
\node  at (-2.5,-1.5) {$U(M_+) \times U(M_-)$};
\draw [<->] (-0.75,-1.5) -- (0.5,-1.5);
\node at (2,-1.5) {$U(N_+,N_-)^2$};
\end{tikzpicture}
\end{mdframed}

\item $\big(O(N_+,N_-), Sp(2M,\mathbb R)\big)$
\begin{mdframed}[linewidth=0.8pt]
\centering
\begin{tikzpicture}[scale=0.9]
\node at (-2.5,1.5) {$U(N_+,N_-)$};
\draw [<->] (-0.5,1.5) -- (0.5,1.5);
\node at (2.5,1.5) {$U(M)$};
\node [rotate=-45] at (-4.25,0.75) {$\cup$};
\node at (-2.5,0.75) {$\cup$};
\node at (2.5,0.75) {$\cap$};
\node [rotate=-125] at (4.5,0.75) {$\cup$};
\node at (-6,0) {$U(N_+) \times U(N_-)$};
\node at (-2.5,0) {$O(N_+,N_-)$};
\draw [<->] (-0.5,0) -- (0.5,0);
\node at (2.5,0) {$Sp(2M, \mathbb R)$};
\node at (6,0) {$U(M) \times U(M)$};
\node [rotate=45] at (-4.25,-0.75) {$\cup$};
\node at (-2.5,-0.75) {$\cup$};
\node at (2.5,-0.75) {$\cap$};
\node [rotate=125] at (4.5,-0.75) {$\cup$};
\node  at (-2.5,-1.5) {$O(N_+) \times O(N_-)$};
\draw [<->] (-0.5,-1.5) -- (0.5,-1.5);
\node at (2.75,-1.5) {$Sp(2M,\mathbb R)^2$};
\end{tikzpicture}
\end{mdframed}

\item $\big(O(N,\mathbb C), Sp(2M,\mathbb C)\big)$
\begin{mdframed}[linewidth=0.8pt]
\centering
\begin{tikzpicture}[scale=0.9]
\node at (-2.5,1.5) {$O^*(2N)$};
\draw [<->] (-0.5,1.5) -- (0.5,1.5);
\node at (2.5,1.5) {$Sp(M)$};
\node [rotate=-45] at (-4.25,0.75) {$\cup$};
\node at (-2.5,0.75) {$\cup$};
\node at (2.5,0.75) {$\cap$};
\node [rotate=-125] at (4.25,0.75) {$\cup$};
\node at (-5.5,0) {$U(N)$};
\node at (-2.5,0) {$O(N,\mathbb C)$};
\draw [<->] (-0.5,0) -- (0.5,0);
\node at (2.5,0) {$Sp(2M,\mathbb C)$};
\node at (5.5,0) {$U(2M)$};
\node [rotate=45] at (-4.25,-0.75) {$\cup$};
\node at (-2.5,-0.75) {$\cup$};
\node at (2.5,-0.75) {$\cap$};
\node [rotate=125] at (4.25,-0.75) {$\cup$};
\node  at (-2.5,-1.5) {$O(N)$};
\draw [<->] (-0.5,-1.5) -- (0.5,-1.5);
\node at (2.5,-1.5) {$Sp(4M,\mathbb R)$};
\end{tikzpicture}
\end{mdframed}

\item $\big(Sp(M_+,M_-), O^*(2N)\big)$
\begin{mdframed}[linewidth=0.8pt]
\centering
\begin{tikzpicture}[scale=0.9]
\node at (-2.5,1.5) {$U(2M_+,2M_-)$};
\draw [<->] (-0.5,1.5) -- (0.5,1.5);
\node at (2.5,1.5) {$U(N)$};
\node [rotate=-45] at (-4.25,0.75) {$\cup$};
\node at (-2.5,0.75) {$\cup$};
\node at (2.5,0.75) {$\cap$};
\node [rotate=-125] at (4.25,0.75) {$\cup$};
\node at (-6,0) {$U(2M_+) \times U(2M_-)$};
\node at (-2.5,0) {$Sp(M_+,M_-)$};
\draw [<->] (-0.5,0) -- (0.5,0);
\node at (2.5,0) {$O^*(2N)$};
\node at (5.5,0) {$U(N) \times U(N)$};
\node [rotate=45] at (-4.25,-0.75) {$\cup$};
\node at (-2.5,-0.75) {$\cup$};
\node at (2.5,-0.75) {$\cap$};
\node [rotate=125] at (4.25,-0.75) {$\cup$};
\node  at (-2.5,-1.5) {$Sp(M_+) \times Sp(M_-)$};
\draw [<->] (-0.5,-1.5) -- (0.5,-1.5);
\node at (2.5,-1.5) {$O^*(2N) \times O^*(2N)$};
\end{tikzpicture}
\end{mdframed}
\end{itemize}

\section{Derivation of Casimir relations}
\label{app:CasimiRR}
In the following two subsections
we provide the details of the derivations 
leading to the duality relations, presented
in the main text,
between all Casimir operators of the two 
groups in a dual pair, in terms of their generating functions.

\subsection{Relating different choices of Casimir operators for the same group $GL_N$}
\label{app:Casimir}

Here we derive the relation between the 
generating functions $r(t)$ and $\tilde r(t)$ of the Casimir operators of the same group $GL_N$ spanned by $R_I{}^J$, with $I,J = 1, \ldots, N$,
as introduced in 
\eqref{Casimir 1} and \eqref{Casimir 2}.
In order to do that, we introduce the following notations:
\ba
(\bm R^n)_{I}{}^J\eq R_I{}^{K_1}\,R_{K_1}{}^{K_2}\dots R_{K_{n-1}}{}^{J}\,,\nn 
((\bm R^t)^n)_{I}{}^J\eq R_{K_1}{}^J\,R_{K_2}{}^{K_1}\,R_{K_3}{}^{K_2}\dots R_I{}^{K_{n-1}}\,.
\ea
These two types of operators differ only by ordering and therefore are related as
\be
((\bm R^t)^n)_{I}{}^J=\sum_{m=0}^{n}\, a_{n\,m}\,(\bm R^m)_I{}^J\,,
\ee
for some $a_{n\,m}$ that may also depend on traces of powers of $R$ \eqref{C}. 
Using $((\bm R^t)^{n+1})_J{}^I=((\bm R^t)^n)_K{}^I\,R_J{}^K$
and the commutation relations,
\be
[A_I{}^J,R_K{}^L]=\d_K^J\,A_I{}^L-\d_I^L\,A_K{}^J\,,
\ee
we derive
\be
\sum_{m=0}^{n+1}\,a_{n+1\,m}(\bm R^m)_J{}^I
=\sum_{m=0}^n\,a_{n\,m}\left[(\bm R^{m+1})_J{}^I-N\,(\bm R^m)_J{}^I+\cC_m[\bm R]\,\d_J^I\right]\,.
\ee
The latter equation induces recursion relations for $a_{n\,m}$ in the following form:
\ba
a_{n+1\,m}\eq a_{n,m-1}-N\,a_{n\,m}\quad (m\geq 1)\,,\label{rec1}\nn
a_{n+1\,0}\eq \sum_{m=0}^n\,a_{n\,m}\,\cC_m[\bm R] - N\, a_{n\,0}\,.\label{rec2}
\ea
It is useful to introduce a generating function for $a_{n\,m}$ in the form,
\be
a(x,y)=\sum_{n,m=0}^{\infty}\,a_{n\,m}\,x^n\,y^m\,.
\ee
In terms of the generating function, the recursion relation \eqref{rec1} takes the form:
\be
\frac1{x}(a(x,y)-a(x,0))=y\,a(x,y)-N\,(a(x,y)-a(x,0))\,,
\ee
which is solved as
\be
a(x,y)=\frac{a(x,0)}{1-\tfrac{x}{1+N\,x}\,y}=a(x,0)\,\sum_{k=0}^{\infty}\left(\tfrac{x}{1+N\,x}\,y\right)^k\,.\label{axy}
\ee
The second recursion relation \eqref{rec2} can be written as
\be
\frac{1}{x}(a(x,0)-1)=\tr[a(x,\bm R)]-N\,a(x,0)\,.
\ee
Using \eqref{axy}, we get 
\be
\tr[a(x,\bm R)]=a(x,0)\sum_{k=0}^{\infty}\,\left(\tfrac{x}{1+N\,x}\right)^k\,\cC_k[R]=a(x,0)\,r(\tfrac{x}{1+N\,x})\,.
\ee
Now, \eqref{rec2} can be written as
\be
\frac{1}{x}(a(x,0)-1)=a(x,0)\,r(\tfrac{x}{1+N\,x})-N\,a(x,0)\,,
\ee
which can be solved by
\be
a(x,0)=\frac{1}{1+N\,x-x\,r(\tfrac{x}{1+N\,x})}\,.
\ee
Then,
\be
a(x,y)=\frac{1}{[1+N\,x-x\,r(\tfrac{x}{1+N\,x})][1-\tfrac{x}{1+N\,x}\,y]}=\frac{1}{1+N\,x-x\,r(\tfrac{x}{1+N\,x})}\sum_{k=0}^{\infty}\,(\tfrac{x}{1+N\,x}\,y)^k\,,
\ee
which can be used to get finally
\be
\tilde{r}(x)=\tr[a(x,\bm R)]=\frac{r(\tfrac{x}{1+N\,x})}{1+N\,x-x\,r(\tfrac{x}{1+N\,x})}\,.
\ee
The inverse relation is given by
\be
r(x)=\frac{\tilde{r}(\tfrac{x}{1-N\,x})}{1-N\,x+x\, \tilde{r}(\tfrac{x}{1-N\,x})}\,.
\ee
Note that these relations go to one another, if we change the sign of $N$.

\subsection{
Details on the Casimir relations for $(O_N,Sp_{2M})$}
\label{App:OnSp2M}

Here we derive the relations between Casimir operators of any order
for the dual pair $(O_N,Sp_{2M})\subset Sp(2NM,\mathbb R)$.
We start with repeating relevant definitions of the main text.
Consider $2MN$ operators $y_A^I$, where $A = 1, \ldots, N$ and $I = 1, \ldots , 2M$, satisfying
\begin{equation}
    [y^I_A, y^J_B] = E_{AB}\, \Omega^{IJ}\,,
\end{equation}
where $E_{AB}$ is an arbitrary symmetric flat metric
with eigenvalues $\pm 1$,
and $\O^{IJ}$ is the $2M$ dimensional symplectic metric,
and their inverses $E^{AB}$ and $\O_{IJ}$ are defined by
$$
E^{AB}E_{BC} = \delta^A_C \, , \qquad 
\O^{IJ}\,\O_{JK}=\delta^I_K\,.
$$
The generators of $O_N$ and $Sp_{2M}$ are
realized in terms of these operators by
\be 
    M_{AB}=\O_{IJ}\,y_{[A}^I\,y_{B]}^{J}\,,
    \qquad 
    K^{IJ}=E^{AB}\,y^{(I}_A\,y^{J)}_B\,.
    \label{eq:gen_O-Sp}
\ee 
It is useful to also define
\be
y^A_I := E^{AB}\Omega_{IJ} y_B^J \, ,
\ee
implying the following straightforward to compute, but useful relations
\be
y_A^I = E_{AB}\,\Omega^{IJ}\,y_J^B \, , \quad 
\Omega^{IJ}\,y_J^A = E^{AB}\,y_B^I \, , \quad
E_{AB} \,y^B_I = \Omega_{IJ}\, y_A^J \, ,
\ee
and obeying (note the sign)
\be
[y^A_I, y^B_J] = - E^{AB}\,\O_{IJ} \,  \qquad
[y^A_I, y_B^J] = \delta^A_B\, \delta_I^J \, .
\ee
We also define
\be 
    N^A{}_B:=y^A_I\,y_B^I\,,
    \qquad
    L^I{}_J:=y^I_A\,y^A_J\,,
\ee
and note that
\be
[N^A{}_B, L^I{}_J] 
= 0 \, ,
\label{NLcommute}
\ee
with their relation to 
$M^A{}_B:= E^{AC} M_{CB}$ and $K^I{}_J:= K^{IK}\, \Omega_{KJ}$ being
\be 
   N^A{}_B = -M^{A}{}_B+M\,\delta^A_B\,,
    \qquad 
    L^I{}_J = -\big(K^I{}_J+\frac{N}2\,\delta^I_J\big)\, .
    \label{App:NMLK relation}
\ee

We now introduce the generating functions of the 
$O_N$ and $Sp_{2M}$ Casimirs,
\be 
    m(t)=\sum_{n=0}^\infty t^n\,\cC_n[\bm M]\,,
    \qquad
    k(t)=\sum_{n=0}^\infty t^n\,\cC_n[\bm K]\,,
\ee
\be 
    n(t)=\sum_{n=0}^\infty t^n\,\tr(\bm N^n)\,,
    \qquad
    l(t)=\sum_{n=0}^\infty t^n\,\tr(\bm L^n)\,.
\ee 
The matrix multiplication of $\bm O^n$ for any of the generators $\bm M, \bm N, \bm K, \bm L$ above is taken to be $\bm O^n \equiv O^\alpha{}_{\gamma_1}O^{\gamma_1}{}_{\gamma_2} \ldots O^{\gamma_{n-1}}{}_{\gamma_n}$.
By using the formal rewriting $\sum_{n=0}^\infty t^n \tr(\bm O^n) = \tr(\tfrac{1}{1-t \bm O})$ and the relations in \eqref{App:NMLK relation}, 
the pairwise relations between generating functions of operators of the same algebra,
$m$ and $n$, respectively, $k$ and $l$,
given in the main text, \eqref{mn rel} and \eqref{kl rel},
can be readily found.

To relate the generating functions $n$ and $l$, respectively, $m$ and $k$,
it is useful to introduce the notation,
\be 
    {}^{\alpha}[n]_\beta=y^\alpha_{\gamma_2}\,y_{\gamma_3}^{\gamma_2} \cdots y^{\gamma_{n}}_\beta\,,
    \qquad
    {}_{\alpha}[n]^\beta=y_\alpha^{\gamma_2}\,y^{\gamma_3}_{\gamma_2} \cdots y_{\gamma_{n}}^\beta\,,
\ee 
where the greek indices are either the $O_N$ or $Sp_{2M}$ fundamental
ones,
thus giving rise to four different types of brackets, i.e.
\be 
    {}^I[2n]_J\,,
    \qquad 
     {}^A[2n]_B\,,
    \qquad 
     {}^I[2n+1]_A\,,
     \qquad 
     {}^A[2n+1]_I\,,
     \label{fourbracket}
\ee 
with the opposite contractions of the $y$ operators  related by
\ba
    &{}_I[2n]^J=(-1)^{n}\,\O_{II'}\,{}^{I'}[2n]_{J'}\,\O^{J'J}\,,
    \qquad 
    &{}_A[2n]^B=(-1)^{n}\,
    E_{AA'} 
    {}^{A'}[2n]_{B'}\,
    E^{B'B}
    \,,\nn 
    &{}_I[2n+1]^A=(-1)^n\,
    \O_{IJ}\,{}^J[2n+1]_B\,
    E^{BA}
    \,,
     \qquad 
     &{}_A[2n+1]^I=(-1)^{n+1}\,
    E_{AB}
    \,{}^B[2n+1]_J\,\Omega^{JI}\,.\nn\label{[] rel}
\ea 
Note, that ${}_A[2n]^B$ and ${}_I[2k]^K$, belonging to groups of the dual pair,
commute, implying the following relation:
\be
^I[2n+1]_A\, {}_I[2l+1]^A={}_A[2l+1]^I\,{}^A[2n+1]_I\,.\label{AI}
\ee
Thus, the four brackets in \eqref{fourbracket}
give rise to three independent generating functions
\ba
    && a(t,u)=\sum_{k,l=0}^\infty
    t^k\,u^l\,{}^I[2k+1]_A\,{}_I[2l+1]^A,
    \nn  
     && b(t,u)=\sum_{k,l=0}^\infty
    t^k\,u^l\,{}^B[2k]_A\,{}_B[2l]^A\,,
   \qquad 
     d(t,u)=\sum_{k,l=0}^\infty
    t^k\,u^l\,{}^I[2k]_J\,{}_I[2l]^J\,,
\ea
with
$$
{}^\alpha [0]_\beta := \delta^\alpha_\beta \, , \quad 
{}_\alpha[0]^\beta := \delta_\alpha^\beta \, ,
$$
which enter in the definition of $b$ and $d$. 
We note that
\ba
a(t,0)=\tfrac1{t}(l(t)-l(0)) \, , \qquad
b(t,0) &=& n(t) \, , \qquad 
d(t,0) = l(t) \, ,
\nn
a(0,u)=\tfrac{1}{u}(n(-u)-n(0))\, , \qquad
b(0,u)&=&n(-u)\,, \qquad 
d(0,u)=l(-u)\,.
\label{tu0}
\ea
Likewise, their derivatives evaluated at $t=0$ or $u=0$ are related to derivatives of $l$ and $n$.
But as we will see now, it is at the point $u=-t$ that we need to understand these generating functions.

To relate the Casimirs of $O_N$ to those of $Sp_{2M}$,
we first move the $y$ operator 
in the left end of $\tr(\bm N^n)$
to its right end to find
\be
\tr(\bm N^n)=
    \tr(\bm L^n)+\sum_{i=0}^{n-1}
    \tr(\bm L^i)\,\tr(\bm N^{n-1-i})
    -\sum_{i=0}^{n-2} 
    (-1)^{n-2-i}\,{}^I[2i+1]_A\,{}_I[2(n-2-i)+1]^A
    \label{os rel 1}
\ee 
valid for $n\geq 1$, and with the last term understood as zero for $n=1$.
The identity is equivalent to
\be
    n(t)-N=l(t)-2M+
    t\,l(t)\,n(t)
    -t^2\,a(t,-t)\, ,
    \label{nla rel}
\ee 
making it evident that 
another relation for $a(t,-t)$ in terms of $n$ and $l$ is needed.
We therefore now derive relations between $a$, $b$, and $d$.

By moving the $y$ operator in 
the left end of ${}^I[2k+1]_A$
to its right end, we find
\begin{align}
    {}^I[2k+1]_A=& 
    {}^B[2k]_A\,y^I_B 
    -\sum_{i=1}^k
    \tr(\bm N^{i-1})\,{}^I[2(k-i)+1]_A\,
    -\sum_{i=1}^{k}(-1)^{i}\,{}_B[2i-1]^I\,{}^B[2(k-i)]_A\, .
    \label{os rel 2-MM}
\end{align}
Considering the last term, 
we can pass $2i-2$ out of the $2i-1$ $y$-operators in ${}_B[2i-1]^I$ through ${}^B[2(k-i)]_A$, since they commute, 
thus getting 
\ba
    {}^I[2k+1]_A\,{}_I[2l+1]^A\eq 
    {}^B[2k]_A\,{}_B[2l+2]^A
    -\sum_{i=1}^k
    \tr(\bm N^{i-1})\,{}^I[2(k-i)+1]_A\,{}_I[2l+1]^A\nn
    &&
    {-\sum_{i=1}^k}(-1)^{i}\,{}^I[2(k-i)+1]_A\,{}_I[2(l+i-1)+1]^A\,.
    \label{os rel 2}
\ea
Multiplying by $t^k u^l$ and summing over all $k \in \mathbb{N}$ and $l \in \mathbb{N}_0$, 
as well as making use of the following relations that will be used repeatedly,
\begin{align}
    \sum_{k=1}^\infty \sum_{i=1}^{k} a_{k,i} = 
    \sum_{m,n=0}^\infty a_{m+n+1,n+1} \,  ,
    \label{sum rel 1}
\end{align}
and
\begin{align}
    \sum_{m,n=0}^\infty 
    x^m (-y)^n a_{m+n} 
    =
    \sum_{j=0}^\infty x^j \frac{1-\left (-\frac{y}{x}\right )^{j+1}}{1+\frac{y}{x}} a_j
    =
    \frac{x}{x+y} \sum_{j=1}^\infty \left (x^j + \frac{y}{x}(-y)^j\right)a_j , 
    \label{sum rel 2}
\end{align}
valid for $|y/x| \neq 1$ (since use of the geometric series is needed),
we find that \eqref{os rel 2} can be equivalently expressed as:
\begin{align}
    a(t,u)-a(0,u)=
    &\frac1{u} \left (
    b(t,u)-b(0,u)-b(t,0)+b(0,0)
    \right )
    -t\,n(t)\,a(t,u)
    \nonumber \\
    &
    +
   \frac{t}{t+u} \left (
u a(t,u)+ t a(t,-t)
\right) \, ,  \qquad
\text{for $|u/t| \neq 1$.}
\label{ab rel 1}
\end{align}
Notice that $u = -t$ is not included in the region of validity of this expression, meaning that it cannot be simply evaluated at $u=-t$. It turns out, however, that its limit at $u=-t$ is well-defined and continuous, but leads to a differential expression for $a(t,-t)$,
\be
t \left (1 - t + t\, n(t) + t^2 \frac{\partial}{\partial u} \right )
a(t,-t) = n(t)  - b(t,-t)\,,
\ee
which we will not make use of. 
Instead we establish more relations away from $u=-t$
to find an expression for $n$ and $l$ at $u=0$.

By moving the $y$ operator in 
the left end of ${}_B[2k+2]^A$
to the left of ${}^B[2l]_A$, we find
\ba
    {}^B[2k+2]_A\,{}_B[2l]^A\eq 
    {}^I[2k+1]_A\,{}_I[2l+1]^A
    +\sum_{i=0}^k
     \tr(\bm L^{i})\,
   {}^B[2(k-i)]_A\,{}_B[2l]^A\,\nn
    &&+\sum_{i=0}^{k-1} (-1)^i {}_I [2i +1]^B {}^I [2(k-i)-1]_A \,{}_B[2l]^A\,.
    \label{os rel 3-MM}
\ea
By splitting the middle bracket of the last term into $2(k-i-1)$ $y$-operators, ${}^I [2(k-i-1)]_J$, and one $y$-operator and moving them through the left and right brackets respectively, one gets
\begin{align}
    {}^B[2k+2]_A\,{}_B[2l]^A
    =&\,  {}^I[2k+1]_A\,{}_I[2l+1]^A+\sum_{i=0}^{k}\tr(\bm L^{i})\,{}^B[2(k-i)]_A\,{}_B[2l]^A
    \nonumber \\
    &
    -\sum_{i=1}^k\,(-1)^{l+k+i}\,{}^I[2l+2i-1]_A\,{}_I[2(k-i)+1]^A
    \nonumber \\
    &
    -\sum_{i=1}^k\sum_{j=0}^{l-1}(-1)^{k+i-j}\,{}^I[2i+2j-1]_A\,{}_I[2(k+l-i-j)-1]^A
    \nonumber \\
    &
    -\sum_{i=1}^k\sum_{j=0}^{l-1}(-1)^{l+i+k}\,{}^I[2i+2j-1]_A\,{}_I[2(k-i)+1]^A\,\tr(\bm N^{l-j-1})\,.\label{ba rel}
\end{align}
This equation can be equivalent expressed as
\begin{align}
\frac1{t}\left[b(t,u)-b(0,u)\right]=&\, 
a(t,u)+l(t)\,b(t,u)-\frac{t}{t+u}\left[u\,a(-u,-t)+t\,a(t,-t)\right]\nn
&
-\frac{t\,u}{(t+u)^2}\left[u^2\, a(-u,u)+t\,u\,(a(t,u)+a(-u,-t))+t^2\,a(t,-t)\right]\nn
&
+\frac{t\,u}{t+u}\left[u\, a(-u,-t)+t\,a(t,-t)\right]\,n(-u)\,.\label{ab rel 2}
\end{align}

We continue to give relations with less details. 
First,
\begin{align}
   {}^I [2k+1]_A \, {}_I [2l+1]^A = 
   &\, {}^I [2(k+1)]_J \, {}_I [2l]^J
   + \sum_{i=1}^l (-1)^i    {}^I [2k+1]_A  \, {}_I [2(l-i)+1]^A  \tr(\bm L^{i-1})
   \nonumber \\
   &
   + \sum_{i=1}^l (-1)^i   {}^I [2(k+i-1)+1]_A  \, {}_I [2(l-i)+1]^A  \, ,
\end{align}
is equivalent to
\begin{align}
a(t,u)-a(t,0) 
= &\, \frac{1}{t} (d(t,u)-l(t)-l(-u) + 2M )
\nonumber \\
&
- u a(t,u)l(-u) - \frac{u}{t+u} ( t a(t,u) + u a(-u, u)) \, .
\label{ad rel 1}
\end{align}
Next,
%
\begin{align}
{}^I[2k]_J \, {}_I [2l+2]^J 
=
&\, {}^I [2k+1]_A \,{}_I [2l+1]^A 
+ \sum_{i=0}^l {}^I[2k]_J \,{}_I [2(l-i)]^J (-1)^i\, \tr(\bm N^i) 
\nonumber \\
&
+ \sum_{i=1}^l (-1)^{i +k+l}
{}^I [2(l-i)+1]_A \,{}_I [2(k+i-1)+1]^A
\nonumber \\
&
+ \sum_{i=1}^l \sum_{j=1}^{k} (-1)^{i +j+l}\,
\tr(\bm L^{k-j})\,
{}^I [2(l-i)+1]_A \, {}_I[2(j+i-2)+1]^A
\nonumber \\
&
+ \sum_{i=1}^l \sum_{j=1}^{k} (-1)^{l+i+j} 
{}^I [2(k-j+l-i)+1]_A \,{}_I [2(j+i-2)+1]^A \, ,
\end{align}
is equivalent to 
\begin{align}
\frac{1}{u} (d(t,u)-d(t,0)) = &\, a(t,u) + d(t,u)n(-u)+\frac{u}{t+u} ( u a(-u,u) + t a(-u,-t))
\nonumber \\
&-\frac{tu}{(t+u)^2} ( t^2 a(t,-t) + tu a(-u,-t) + tu a(t,u) + u^2 a(-u,u) ) 
\nonumber \\
&-\frac{t\,u\, l(t)}{t+u} ( u a(-u,u) + t a(-u,-t)) \, .
\label{ad rel 2}
\end{align}
Finally, we are in a position to derive the duality relations
given in \eqref{nl rel main} and \eqref{km rel}-\eqref{mk rel}:
Eq.~\eqref{nla rel} provides an expression for $a(t,-t)$ in terms of $n$ and $l$ only. Using this in 
\eqref{ab rel 1} and \eqref{ab rel 2} to derive an expression for $a(t,u)$, and inserting in \eqref{ad rel 1} to derive a relation for $d(t,u)$, we finally get from
\eqref{ad rel 2} an equation only in terms of $n$ and $l$.
This equation turns out to be regular and nontrivial at $u=0$, yielding \eqref{nl rel main}.
Then from \eqref{mn rel} and \eqref{kl rel} one readily gets the relations \eqref{km rel} and \eqref{mk rel} relating finally $m(t)$ with $k(t)$.

\bibliographystyle{JHEP}
\bibliography{biblio}
	
\end{document}